# Baryons as Chiral Solitons in the Nambu–Jona-Lasinio Model†

R. Alkofer, H. Reinhardt and H. Weigel‡

Institute for Theoretical Physics

Tübingen University

Auf der Morgenstelle 14

D-72076 Tübingen, Germany

†Supported in part by the Deutsche Forschungsgemeinschaft (DFG) under contract number Re–856/2-2.

‡ Supported by a Habilitanden–scholarship of the Deutsche Forschungsgemeinschaft (DFG).

# Abstract

The description of baryons as chiral solitons of the Nambu–Jona–Lasinio (NJL) model is reviewed. A motivation for the soliton description of baryons is provided from large $N_C$ QCD. Rigorous results on the spontaneous breaking of chiral symmetry in QCD are discussed. It is then argued that the NJL model provides a fair description of low–energy hadron physics. The NJL model is therefore employed to mimic the low–energy chiral flavor dynamics of QCD. The model is bosonized by functional integral techniques and the physical content of the emerging effective meson theory is discussed. In particular, its relation to the Skyrme model is established.

The static soliton solutions of the bosonized NJL model are found, their properties discussed, and the influence of various meson fields studied. These considerations provide strong support of Witten's conjecture that baryons can be understood as soliton solutions of effective meson theories. The chiral soliton of the NJL model is then quantized in a semiclassical fashion and various static properties of the nucleon are studied. The dominating $1/N_C$ corrections to the semiclassically quantized soliton are investigated. Time–dependent meson fluctuations off the chiral soliton are explored and employed to estimate the quantum corrections to the soliton mass. Finally, hyperons are described as chiral solitons of the NJL model. This is done in both, the collective rotational approach of Yabu and Ando as well as in the bound state approach of Callan and Klebanov.

# Contents

# 1 Introduction

It is generally accepted that Quantum Chromo Dynamics (QCD) is the theory of strong interactions (for an introduction see *e.g.* refs. [1, 2]). Besides coming in three colors the fermion fields of QCD, the quarks, also carry flavor. The interactions of QCD are flavor blind but sensitive to color. QCD is an asymptotically free theory which means that the forces between quarks become weak for small quark separations, or equivalently, large momentum transfers. This allows one to quantitatively calculate observables of strong interaction physics, which are sensitive to the short distance behavior of QCD, by perturbative techniques. As a matter of fact, the predicted scaling violations have been verified to a high accuracy at existing accelerators [1]. According to present knowledge QCD is the only renormalizable theory that can account for these scaling violations.

The same self–interactions of gluons which give rise to asymptotic freedom lead to a strong quark–quark interaction for medium and small energies. Using additionally the empirical fact that neither quarks nor gluons have been detected as "free" particles has lead to the confinement hypothesis: Only singlets of the gauge group appear as physical particles. Especially, the quarks belonging to the fundamental representation and the gluons being in the adjoint representation can never be observed directly. The perturbative result[a] that the coupling constant increases as the momentum transfer becomes small, or the distance large, is in accordance with the confinement hypothesis. Unfortunately, this behavior of the coupling constant excludes perturbative calculations for low energies. Thus it is still unproven that QCD is really a confining theory. Even worse, until today properties of hadrons have not been calculated from QCD without making use of severe assumptions or simplifications.

There is probably one exception to this statement: With the help of Monte–Carlo techniques attempts have been made to calculate hadron properties directly from QCD using a discrete lattice (for an introduction to lattice gauge theories see *e.g.* refs. [3] or [4]). Despite the fact that these non–perturbative calculations should enable the determination of every physical quantity the results often deviate very strongly from the experimental values. There are several reasons for this. Even by using "high–performance" computers the possible lattice sizes are still modest, especially, if one wants to include dynamical fermions. Furthermore, a calculation using massless quarks is impossible. And there are still open conceptual questions concerning the continuum limit of lattice theory.

Given this state of affairs it is natural to resort to effective models of strong interactions. These are intended to mimic the low–energy behavior of QCD as closely as possible. For this purpose the approximate chiral symmetry of QCD [5] provides a very useful guideline. Hadron phenomenology has left no doubt that this symmetry is broken dynamically by strong interactions. Requiring the known pattern of explicit, spontaneous and anomalous chiral symmetry breaking puts significant constraints on possible models for the strong interactions of quarks. Additional guidance can be obtained if one generalizes QCD to a gauge theory with an arbitrary number of colors $N_C$. This is because for large $N_C$ QCD reduces to an effective theory of infinitely many weakly interacting mesons and glueballs [6]. Unfortunately, this effective meson theory cannot be constructed explicitly. Nevertheless, Witten was able to give arguments that within this effective theory baryons emerge as soliton solutions [7].

Although Witten's conjecture has never been proven rigorously the soliton picture of baryons has turned out quite successful in recent years. The starting points have been phenomenological effective meson theories, which possess soliton solutions. The most popular ones perhaps are the Skyrme model [8, 9] and the gauged $\sigma$-model [10, 11]. Investigations within these models have satisfactorily explained the wealth of spectroscopic baryon data, see *e.g.* ref. [12] for a recent compilation of references on soliton models for baryons. It is worthwhile to mention that many of the early difficulties encountered for the Skyrme model, like *e.g.* form factors being too soft [13, 10], the missing intermediate range attraction in the nucleon–nucleon force [14] or the linearly rising phase shifts in pion nucleon scattering [11], have found satisfactory solutions. Partly, this was achieved by using effective meson Lagrangians [15, 16] describing the meson physics better than the original simple Skyrme Lagrangian. On the other hand, this increased complexity also generated more ambiguity in the effective action.

[a] See *e.g.* Chapter 3.4.2 in ref.[2].

At that point naturally the question arises whether a more microscopic picture of the soliton can provide some guide for choosing the effective meson theory. Even more, one may aim at a justification of the soliton picture of baryons in general. Therefore not only a microscopic realization of the soliton picture in terms of quark degrees of freedom is wanted but the dynamics of the model should be able, at least in principal, to determine its favored picture of the baryon. In this sense the Nambu–Jona-Lasinio (NJL) model [17] is unique. First of all, it is simple enough that such complicated field configuration like solitons can be determined self–consistently, see Chapter 6. Second, it contains the correct chiral symmetry breaking pattern and reproduces a lot of meson properties like masses, decay constants, scattering lengths *etc.*, using only a few input parameters, see *e.g.* [18, 19] and references therein. Third, and more important, it allows for two complementary pictures of baryons, namely either as ordinary three–valence–quark bound state or as chiral soliton. The latter will be the subject of this review. In the other approach the baryon wave function is obtained as a solution of a Faddeev equation using diquarks as intermediate building blocks [20, 21]. Using functional integral techniques one can derive a generating functional which allows one to treat both pictures in one quark theory without any ambiguities or double counting [20]. Such a theory contains the relativistic quark model and the Skyrmion as asymptotic limits. Preliminary results indicate that such a hybrid model can display unexpected features, *e.g.* the diquark mass is drastically reduced in a soliton background [22].

Obviously, the simplicity of the NJL model which makes it suitable for such complicated investigations is also its most severe drawback. The NJL model is non–renormalizable and only uniquely defined if the necessary regularization prescription is specified. This introduces an ultraviolet cut–off which indicates the range of applicability of the model. The results for several observables differ in various regularization schemes [18]. Even worse, sometimes the qualitative behavior changes when altering the regularization prescription. Fortunately, the situation is not as dramatic for most of the observables, or can be understood from the deficiencies of the regularization procedure like *e.g.* missing gauge invariance. The other serious disadvantage of the NJL model is the absence of confinement, or more precisely, the appearance of two–quark (or quark–antiquark) thresholds. These cause unphysical imaginary parts in correlation functions for large (time–like) momenta. Therefore the results of the NJL model are restricted to low energies not only by the ultraviolet cut–off introduced via the regularization but also by the two–quark thresholds.

This review is devoted to the soliton description of baryons within the NJL model. It is not the primary goal of such investigations to reproduce all the phenomenological successes of Skyrme type models. The basic motivation has rather been, and is still, to increase our understanding how the soliton emerges, and to better understand its relation to the underlying quark dynamics. Especially, the conceptional easy and direct access to the quark degrees of freedom allows one to study questions which escape our considerations when starting from a purely mesonic model. On the other hand, in order to arrive at physical baryons one has to use techniques which are well known from the Skyrmion: semiclassical quantization [9] (cranking [23]), calculation of quantum corrections [24, 25], generalization to three flavors by either collective quantization [26, 27, 28] or bound state approach [29, 30], and so on. These subjects will be discussed in detail in this review, see Chapter 7. They should be considered as a basis for model calculations where one wants to describe baryons as chiral solitons and additionally wants to have access to the quark degrees of freedom in one consistent frame.

The organization of this review is as follows: In Chapter 2 we will present some rigorous results from QCD which are relevant for the soliton description of baryons. After a short summary of the large $N_C$ arguments for baryons being solitons we will outline some low–energy theorems based on chiral symmetry and the chiral anomaly. In Chapter 3 we will describe the NJL model and its bosonization [31]. In Section 3.3 a very important property of the NJL model, the dynamical breaking of chiral symmetry, is described. We will also use a local chiral rotation to display the hidden gauge symmetry [16] of the bosonized model [32]. Chapter 4 is devoted to the effective meson theory of the NJL model. Hereby, the gradient expansion does not only yield approximate meson masses and coupling constants but also reveals the relation to the Skyrme model. The determination of meson masses with the help of Bethe–Salpeter equations is explained in Section 4.3. Section 4.4 contains some comments on the chiral anomaly in the effective meson theory. In Chapter 5 the generic aspects of chiral solitons are discussed: the topological properties, the emergence of the soliton and its semiclassical quantization. Chapter 6 is the central piece of this review. After giving the energy functional of the static NJL soliton we discuss the self–consistent solutions for different meson fields included (or neglected). Chapter 7 embodies the description of baryons as NJL solitons. This also includes a comprehensive explanation of the cranking method for two and three flavors as well as the description of time–dependent meson fluctuations off the static soliton. In Chapter 8 we give a short summary. Some lengthy formulas which are nevertheless necessary to make this review reasonably self–contained are given in five appendices.

# 2 Rigorous results from QCD

In this chapter we will present some rigorous QCD results which are relevant to hadron physics and in particular to the soliton description of baryons. As discussed in the introduction QCD cannot be treated perturbatively at low energies. As a consequence there are very few such results. They are either based on considerations assuming the number of colors $N_C$ to be large or on the chiral symmetry of massless QCD and its spontaneous breaking. As we will see both provide substantial arguments for the chiral soliton picture of baryons.

## 2.1 Large $N_C$ QCD

In the low–energy regime there is no obvious expansion parameter to treat QCD perturbatively. However, in the seventies 't Hooft [6] and Witten [7] demonstrated that generalizing QCD from the gauge group SU(3) to SU($N_C$) with $N_C$ being large, $1/N_C$ might be considered as an implicit expansion parameter. Naturally, one might wonder whether this provides a sound basis for a perturbative analysis since $N_C = 3$ in the real world. Nevertheless, we will see that this idea is very fruitful, especially it is the essential motivation for identifying baryons with solitons of meson fields.

Assuming confinement in the large $N_C$ world it can be shown that QCD with $N_C \to \infty$ and $g^2 N_C$ fixed has a limit which can be described as keeping only planar Feynman diagrams [6]. The reason is that non–planar diagrams or diagrams with gluon handles are suppressed by $N_C^{-2}$, and the ones with quark loops by $N_C^{-1}$. Simple power counting arguments using 't Hooft's diagrammatic analysis reveal that the correlation function of a color singlet operator is dominated by planar diagrams with exclusive gluon insertions and a quark loop on the exterior. Furthermore, in leading order in the $1/N_C$ expansion this correlation function cannot factorize, *i.e.* it is saturated only by color singlet intermediate states. This then leads to 't Hooft's most important result on large $N_C$ QCD: In this limit QCD reduces to a theory of (infinitely many) weakly interacting mesons and glueballs. Their masses are of order $N_C^0$, their mutial interaction is suppressed by powers of $1/N_C$.

The mesons are stable objects in the limit $N_C \to \infty$. Their decay rates are of order $1/N_C$, their cross sections of order $N_C^{-2}$. This is very gratifying because it provides a natural explanation for the existence of narrow meson resonances in nature. Indeed, it is not obvious at all that *e.g.* the decay $\rho \to 2\pi$ is narrow enough to detect the $\rho$ meson experimentally. Even more, the ratio of this decay width to the $\rho$ meson mass very roughly follows the $1/N_C$ counting. To be more specific we consider an $n$–point correlation function of a color singlet operator. The summation of planar diagrams leads, among others, to an effective meson self–coupling which at tree level takes the form [7]

$$\mathcal{L}_n = N_C^{-\frac{n}{2}+1} \lambda_{i_1 \ldots i_n} \Phi_{i_1} \ldots \Phi_{i_n} . \tag{2.1}$$

Rescaling the meson fields $\Phi_i$ according to $\Phi_i = \sqrt{N_C}\varphi_i$ leads to

$$\mathcal{L}_n = N_C \lambda_{i_1 \ldots i_n} \varphi_{i_1} \ldots \varphi_{i_n} . \tag{2.2}$$

From eq. (2.2) one can then conclude that the generating functional of QCD becomes in leading order of the $1/N_C$ expansion

$$Z_{QCD} = \int \mathcal{D}[A, \bar{q}, q] e^{i\mathcal{A}_{QCD}} \quad \stackrel{N_C \to \infty}{\longrightarrow} \quad \int \mathcal{D}\varphi e^{iN_C \mathcal{A}[\varphi]} \tag{2.3}$$

where $\mathcal{A}[\varphi]$ is an effective meson action that involves infinitely many mesons and is of the generic form (summation over repeated indices is assumed)

$$\mathcal{A}[\varphi] = \int d^4x \left( \frac{1}{2}(\partial_\mu \varphi_i)^2 - \frac{1}{2!} m_{ij}\varphi_i\varphi_j + \frac{1}{3!}\lambda_{ijk}\varphi_i\varphi_j\varphi_k + \ldots \right). \tag{2.4}$$

$1/N_C$ corrections are then given in terms of meson loop terms.

Using the large $N_C$ diagrammatics Witten conjectured [7] that baryons behave as soliton solutions of the effective theory (2.4). Assuming again that quarks are confined and that $g^2 N_C$ is a constant the mass of a baryon composed of $N_C$ quarks in their antisymmetric ground state is given by

$$M_B = N_C(m_q + T) + \frac{1}{2}N_C(N_C - 1)g^2 V \approx N_C \mathcal{F}(g^2 N_C, m_q) \tag{2.5}$$

to leading order in the $1/N_C$ expansion. Besides the masses $m_q$ and the kinetic energy $T$ of the individual quarks also the one–gluon–exchange is included. By combinatoric arguments it can be shown that the $n$–gluon–exchange behaves like $N_C(N_C g^2)^n$ for $N_C \gg n$. Therefore the function $\mathcal{F}(g^2 N_C, m)$ whose leading order is $N_C^0$ parametrizes the $1/N_C$ corrections in a smooth fashion. Since the strength of the meson coupling is proportional to $1/N_C$ the baryon mass $M_B$ is proportional to the inverse of this meson coupling. As the baryon radius is proportional to the inverse of the quark kinetic energy up to suppressed binding effects it is independent of $N_C$ in leading order. This is the typical behavior of a soliton field! These results (as well as the fact that the meson–baryon and baryon–baryon interactions are of order $N_C^0$ and $N_C^1$, respectively) are the basis for Witten's conjecture that baryons emerge as solitons of an effective meson theory.

Obviously, such a behavior can never be obtained in perturbation theory. Unfortunately, the effective meson theory (2.4) cannot be constructed explicitly.[a] Nevertheless, this kind of reasoning provides some insight in the wealth of experimental results. The large $N_C$ arguments account for suppression of exotics, confirms Zweig's rule and Regge phenomenology. Furthermore, it is consistent with phenomenological models like *e.g.* the meson exchange models.

On the other hand, to use the $1/N_C$ expansion also quantitatively one has to resort to effective models. The most prominent example for soliton models of baryons is the Skyrme model [8, 9], for a recent review see ref. [12]. In the following chapters we will present the description of baryons as solitons within the NJL model. Besides others it has the virtue that Witten's conjecture may be tested by the dynamics of the model. Nevertheless, one might also raise the question whether it is not more natural to describe baryons as three–quark bound states if one starts from an effective quark theory. Also here the $1/N_C$ expansion might give some hints. Starting from an effective quark interaction described by a current–current coupling of color octet quark currents (see eq. (3.2)) fierzing into attractive channels and subsequent functional integral hadronization leads to an effective theory which contains besides mesons also baryon fields composed of a diquark and a quark [20]. The interesting fact now is that in this effective theory the quark–quark, which is responsible for diquark formation, interaction is suppressed by $1/N_C$ as compared to the antiquark–quark interaction, *i.e.* in the limit $N_C \to \infty$ these explicit baryon fields are removed from the theory [35] and one has to find the baryons in the left–over effective meson theory which, by the way, is formally identical to the effective meson theory obtained by bosonizing the NJL model. This effective theory will be discussed in some detail in chapter 4, and its soliton solutions are exactly the NJL solitons.

[a] In contrast to the situation in four space–time dimensions $QCD_2$ can be bosonized exactly [33, 34].

## 2.2 Chiral symmetry and low–energy theorems

Besides the exact color symmetry, the QCD Lagrangian possesses an additional symmetry in the limit of vanishing quark current masses, the chiral symmetry. In order to display this symmetry it is convenient to split the quark fields in right– and left–handed components. These are defined by

$$q_R = P_R q, \quad q_L = P_L q \tag{2.6}$$

where

$$P_{R,L} = \frac{1}{2}(1 \pm \gamma_5) \tag{2.7}$$

are the corresponding projectors. Then the QCD Lagrangian taken at zero current masses decouples into a sum of two Lagrangians containing only right–(left–) handed fields, respectively. These two Lagrangians are invariant under global unitary flavor transformations of the corresponding right–(left–) handed fields, *i.e.* the QCD Lagrangian is invariant under $U_L(N_f) \times U_R(N_f)$ where $N_f$ is the number of flavors under consideration. The decomposition into semi–simple subgroups

$$U_L(N_f) \times U_R(N_f) \cong U_{L+R}(1) \times U_{L-R}(1) \times SU_L(N_f) \times SU_R(N_f)$$

allows one to discuss the associated conserved charges. The invariance under $U_{L+R}(1)$ is responsible for the conservation of baryon number whereas $U_{L-R}(1)$ is subject to an anomaly [36]. The implications of this anomaly will be discussed in the subsequent section in more detail.

The invariance under unitary transformations belonging to the subgroup $SU_L(N_f) \times SU_R(N_f)$ implies the conservation of the $N_f^2 - 1$ currents

$$\begin{aligned} J^a_{L\mu}(x) &= \bar{q}_L(x)\frac{\lambda^a}{2}\gamma_\mu q_L(x) \\ J^a_{R\mu}(x) &= \bar{q}_R(x)\frac{\lambda^a}{2}\gamma_\mu q_R(x) \end{aligned} \tag{2.8}$$

where the matrices $\lambda^a$ are the generators of $SU(N_f)$. The corresponding conserved charges

$$\begin{aligned} Q^a_L &= \int d^3x J^a_{L0}(x) \\ Q^a_R &= \int d^3x J^a_{R0}(x) \end{aligned} \tag{2.9}$$

fulfill commutation relations which are identical to the one of the generators of the group $SU_L(N_f) \times SU_R(N_f)$. However, they mix under parity, $\mathcal{P} Q_L \mathcal{P}^{-1} = Q_R$. Parity eigenstates are given by the currents which are related to the diagonal subgroup $SU_{L+R}(N_f)$ and the coset $(SU_L(N_f) \times SU_R(N_f))/SU_{L+R}(N_f)$:

$$\begin{aligned} V^a_\mu(x) &= J^a_{R\mu}(x) + J^a_{L\mu}(x) = \bar{q}(x)\frac{\lambda^a}{2}\gamma_\mu q(x) \\ A^a_\mu(x) &= J^a_{R\mu}(x) - J^a_{L\mu}(x) = \bar{q}(x)\frac{\lambda^a}{2}\gamma_\mu \gamma_5 q(x). \end{aligned} \tag{2.10}$$

Under parity these currents transform like vectors or axial–vectors, respectively. Furthermore, their time components fulfill the equal–time commutation relations

$$\begin{aligned}
[V_0^a(\boldsymbol{x},t), V_0^b(\boldsymbol{y},t)] &= if^{abc}V_0^c(\boldsymbol{x},t)\delta^{(3)}(\boldsymbol{x}-\boldsymbol{y}) \\
[V_0^a(\boldsymbol{x},t), A_0^b(\boldsymbol{y},t)] &= if^{abc}A_0^c(\boldsymbol{x},t)\delta^{(3)}(\boldsymbol{x}-\boldsymbol{y}) \\
[A_0^a(\boldsymbol{x},t), A_0^b(\boldsymbol{y},t)] &= if^{abc}V_0^c(\boldsymbol{x},t)\delta^{(3)}(\boldsymbol{x}-\boldsymbol{y})
\end{aligned} \tag{2.11}$$

where $f^{abc}$ are the structure constants of the Lie algebra $SU(N_f)$. These commutation relations are known as current algebra [37]. As the left hand side is quadratic in the currents whereas the right hand side is linear these relations also fix the normalization of the currents. This is the basis of the phenomenologically successful current algebra sum rules.

Chiral symmetry is explicitly broken by the current masses. Nevertheless it is a good approximation for the two light flavors (up and down) since their current masses are of the order of a few MeV. For the strange quark whose current mass is of the same order of magnitude as the fundamental QCD scale, $\Lambda_{QCD}$, chiral symmetry is certainly not as good a symmetry. In spite of this, assuming approximate chiral symmetry for strange quarks is useful for a wide range of applications. For the heavy quarks the opposite type of expansion, namely in powers of $1/m$, has proven to be successful [38]. Heavy quarks can be neglected in low energy processes as a consequence of the Appelquist–Carrazzone theorem [39] which states that particles decouple whose mass is much larger than the typical energy scale for the process under consideration. We will therefore treat only the cases of two or three flavors in the following.

Chiral symmetry is spontaneously broken implying the existence of pseudoscalar (would–be) Goldstone bosons. These are identified with the isotriplett of pions or for three flavors with the octet of pseudoscalar mesons $\pi$, K and $\eta$. A further consequence is the dynamical generation of a non–perturbative quark mass. It is expected that this quark mass, which is often labeled as constituent quark mass, is of the order of $\Lambda_{QCD}$. Furthermore, a finite non–perturbative quark mass implies a non–zero quark condensate which is defined in terms of the full quark propagator

$$\langle \bar{q}q \rangle = -i \lim_{y\to x^+} \mathrm{tr} S_F(x,y). \tag{2.12}$$

Note that the quark condensate $\langle \bar{q}q \rangle$ is a gauge invariant quantity. Therefore it can be evaluated in any gauge. In a covariant gauge the quark propagator is of the form

$$S_F(q) = \frac{i}{\not{q}A(q^2) - B(q^2) + i\epsilon}. \tag{2.13}$$

In perturbation theory in the chiral limit one has $B(q^2) = 0$. Therefore the quark condensate would vanish due to the vanishing Dirac trace of $\gamma_\mu$. On the other hand, a dynamically generated $B(q^2) \neq 0$ implies a non–vanishing condensate.

Using $\langle \bar{q}q \rangle \neq 0$ and the Goldstone theorem one can prove that the axial currents $A_\mu^a$ (2.10) couple the Goldstone bosons to the vacuum. Denoting the one particle states of Goldstone bosons with momentum $p$ by $|\pi^a(p)\rangle$ one obtains

$$\langle 0|A_\mu^a(x)|\pi^b(p)\rangle = if^{ab}p_\mu e^{-ipx}. \tag{2.14}$$

The $f^{ab}$ are non–vanishing constants. In the isospin symmetric case they are proportional to the unit matrix, $f^{ab} = \delta^{ab} f_a$. One usually determines their numerical values from weak pion (kaon) decay, *e.g.* $\pi \to \mu\bar{\nu}$. $f_\pi$ ($f_K$) is therefore called pion (kaon) decay constant. For a compilation of recent experimental data see page 1443 of ref. [40]. For the purpose of this review it suffices to know that $f_\pi = 93\text{MeV}$ and $f_K/f_\pi = 1.22$.

Acting with the derivative operator on eq. (2.14) and using the Klein–Gordon equation for the pion state ($p^2 = m_\pi^2$) yields

$$\langle 0|\partial^\mu A_\mu^a(x)|\pi^b(p)\rangle = \delta^{ab} f_\pi m_\pi^2 e^{-ipx}. \tag{2.15}$$

The conservation of the axial current implies either $f_\pi = 0$ or $m_\pi = 0$. These two possibilities correspond to the Wigner–Weyl or Nambu–Goldstone realization of chiral symmetry, respectively. As both quantities are finite chiral symmetry has to be explicitly broken, *i.e.* one has to use finite current quark masses in order to describe nature. To further proceed eq. (2.15) is elevated to an operator identity. Let us first introduce the pion field operator $\phi_\pi^a(x)$ and choose its normalization with respect to the one pion state to be

$$\langle 0|\phi_\pi^a(x)|\pi^b(p)\rangle = \delta^{ab} e^{-ipx}. \tag{2.16}$$

Then the operator identity

$$\partial^\mu A_\mu^a(x) = f_\pi m_\pi^2 \phi_\pi^a(x) \tag{2.17}$$

implies the relation (2.15).[b] Eq. (2.17) is known as the PCAC (Partially Conserved Axialvector Current) hypothesis. Assuming additionally that meson and baryon form factors can be extrapolated smoothly from the corresponding mass shells allows one to relate different hadronic observables depending on both, weak and strong interaction parameters. One famous example is the Goldberger–Treiman relation

$$f_\pi g_{\pi NN} = m_N g_A \tag{2.18}$$

which connects the axial coupling of the nucleon $g_A$ with the pion nucleon coupling constant $g_{\pi NN}$ ($m_N$ is the nucleon mass). This relation is experimentally fulfilled within ten percent. This (in)accuracy sheds some light on the usefulness of the PCAC hypothesis.

Using the PCAC hypothesis one furthermore deduces [5]

$$\delta^{ab} m_\pi^2 f_\pi^2 = i \int d^4x \langle 0|\delta(x_0)[A_0^a(x), \partial^\mu A_\mu^b(0)]|0\rangle \tag{2.19}$$

in the limit of vanishing meson momentum. This may be rewritten in terms of the (partially conserved) axial charges $Q_5^a$ and the Hamiltonian $\mathcal{H}(x)$ as

$$\delta^{ab} m_\pi^2 f_\pi^2 = \langle 0[Q_5^a, [Q_5^b, \mathcal{H}(0)]]|0\rangle. \tag{2.20}$$

As these charges generate infinitesimal axial transformations it is obvious that only chiral symmetry breaking terms in the Hamiltonian contribute to the double commutator. In the case of three flavors the symmetry breaking mass term is given by ($\lambda^0 = \sqrt{2/3}\ \mathbb{1}$)

$$m_u^0 \bar{u}u + m_d^0 \bar{d}d + m_s^0 \bar{s}s = c_0 \bar{q}\frac{\lambda^0}{2}q + c_3 \bar{q}\frac{\lambda^3}{2}q + c_8 \bar{q}\frac{\lambda^8}{2}q \tag{2.21}$$

[b] Note that eq. (2.17) is not derived from (2.15). One has also to assume that the matrix elements $\langle 0|\partial^\mu A_\mu^a(x)|X\rangle$ for all states $X$ except the one pion state vanish.

where the symmetry breaking parameters $c_i$ are functions of the current quark masses

$$\begin{aligned} c_0 &= \frac{1}{\sqrt{6}}(m_u^0 + m_d^0 + m_s^0) \\ c_3 &= \frac{1}{2}(m_u^0 - m_d^0) \\ c_8 &= \frac{1}{2\sqrt{3}}(m_u^0 + m_d^0 - 2m_s^0). \end{aligned} \tag{2.22}$$

The commutators (2.20) may be computed with the help of the explicit expressions (2.9) and (2.10) of the axial charges. Using the algebra of the Gell–Mann matrices and especially the relation

$$[\bar{q}\frac{\lambda^a}{2}\gamma_0\gamma_5 q, \bar{q}\frac{\lambda^b}{2}q] = -id^{abc}\bar{q}\frac{\lambda^c}{2}\gamma_5 q \tag{2.23}$$

(the $d^{abc}$ are the symmetric structure constants of SU(3)) one obtains the famous Gell-Mann–Renner–Oakes relations [41]

$$\begin{aligned} f_\pi^2 m_\pi^2 &= \frac{1}{2}(m_u^0 + m_d^0)\langle 0|\bar{u}u + \bar{d}d|0\rangle \\ f_K^2 m_K^2 &= \frac{1}{2}(m_u^0 + m_s^0)\langle 0|\bar{u}u + \bar{s}s|0\rangle \\ f_\eta^2 m_\eta^2 &= \frac{1}{6}(m_u^0 + m_d^0)\langle 0|\bar{u}u + \bar{d}d|0\rangle + \frac{4}{3}m_s^0\langle 0|\bar{s}s|0\rangle. \end{aligned} \tag{2.24}$$

For the case of equal quark condensates

$$\langle 0|\bar{u}u|0\rangle = \langle 0|\bar{d}d|0\rangle = \langle 0|\bar{s}s|0\rangle$$

the definition of the decay constants (2.14) implies that all decay constants are equal. For this special case eqs. (2.24) also yield the Gell-Mann–Okubo mass relation [42] $4m_K^2 = 3m_\eta^2 + m_\pi^2$ which has been observed empirically. In addition one can extract the ratio of current quark masses

$$\frac{m_u^0 + m_d^0}{2m_s^0} = \frac{m_\pi^2}{2m_K^2 - m_\pi^2} \approx \frac{1}{25}. \tag{2.25}$$

Taking into account isospin breaking $m_u^0 \neq m_d^0$ and electromagnetic corrections the relations (2.24) can be refined. Nowadays the values of all current masses can be estimated quite reliably. However, only the ratios of current masses are uniquely defined because these are scale invariant. Citing an absolute value of a current mass hence necessitates reference to a scale. At 1 GeV commonly accepted values for the current masses are [43]

$$\begin{aligned} m_u^0(1\text{GeV}) &\approx 5\text{MeV} \\ m_d^0(1\text{GeV}) &\approx 9\text{MeV} \\ m_s^0(1\text{GeV}) &\approx 160\text{MeV}. \end{aligned} \tag{2.26}$$

One should, however, note that these data are to some extend model dependent because they are extracted form low–energy meson phenomenology [44].

## 2.3 Chiral anomaly and the QCD vacuum

The properties of the QCD vacuum are still poorly understood. From hadron phenomenology one deduces that non–vanishing condensates have to exist. Examples are the gluon and the quark condensates, $\langle G^a_{\mu\nu}G^{a\mu\nu}\rangle$ and $\langle\bar{q}q\rangle$. The relation of the latter to the spontaneous breaking of chiral symmetry has been discussed in the preceding section. In this section we will look at it from another point of view: The quark condensate is related to the mean density of eigenvalues of the quark Dirac operator. To see this we consider the quark propagator for a fixed gauge field $A_\mu^{\rm fixed}$

$$(S_F(x,y))_{A_\mu^{\rm fixed}} = \langle Tq(x)\bar{q}(y)\rangle_{A_\mu^{\rm fixed}} = \sum_n \frac{u_n(x)u_n^\dagger(y)}{m^0 - i\lambda_n} \tag{2.27}$$

where $u_n(x)$ and $\lambda_n$ are eigenfunctions and eigenvalues of the Euclidean Dirac operator

$$\not{D}u_n(x) = \lambda_n u_n(x). \tag{2.28}$$

Except for the zero modes the eigenfunctions occur in pairs of opposite chirality with corresponding eigenvalues $\pm\lambda_n$. Therefore one obtains from eq. (2.27)

$$\frac{1}{V}\int_V d^4x \langle\bar{q}(x)q(x)\rangle_{A_\mu^{\rm fixed}} = -\frac{2m^0}{V}\sum_{\lambda_n>0}\frac{1}{(m^0)^2+\lambda_n^2} \tag{2.29}$$

where the zero–mode contribution has been neglected. To arrive at the quark condensate one has to average over all gauge field configurations and then take the limit $V\to\infty$. In this limit the spectrum becomes dense and gives a non–vanishing condensate if the mean number of eigenvalues in the interval $d\lambda$ is proportional to the volume:

$$\langle\bar{q}q\rangle = -2m^0\int_0^\infty d\lambda \frac{\rho(\lambda)}{(m^0)^2+\lambda^2} \tag{2.30}$$

where $\rho(\lambda)$ is the mean spectral density. Taking now the limit $m^0\to 0$ one arrives at [45]

$$\langle\bar{q}q\rangle = -\pi\rho(0). \tag{2.31}$$

*i.e.* the quark condensate is related to the level density at zero virtuality, $\lambda = 0$.

In deriving this result the infinite volume limit is very important. In a finite volume there is no spontaneous breaking of chiral symmetry, and therefore the quark condensate would vanish. The spectrum is discrete for finite $V$ and the sum in eq. (2.29) does not develop an infrared singularity. If the chiral limit had been taken at finite $V$ the chiral symmetry would have been restored. Obviously, the two limits $m^0\to 0$ and $V\to\infty$ are not interchangeable.

In order to analyze the situation at finite volume one has to consider the dimensionless quantity $x := -Vm^0\langle\bar{q}q\rangle$. (Eq. (2.31) implies that the level spacing for small $\lambda$ is $\Delta\lambda \approx \pi/V\langle\bar{q}q\rangle$.) Hence the denominator $(m^0)^2+\lambda^2$ varies only slowly for neighboring levels if $x \gg 1$. In this case it is a good approximation to replace the sum in eq. (2.29) by an integral, *i.e.* eq. (2.30) stays true as long as the quark current mass is larger than $1/V\langle\bar{q}q\rangle$. This suggests that spontaneous chiral symmetry breaking is related to the appearance of small eigenvalues of $\not{D}$ such that $\lambda_n \propto 1/V$.

Recently the spectrum of this operator has been investigated [46] by studying the role played by the winding number $\nu$ of the vacuum gauge field configuration. A topologically

non–trivial gluon field configuration necessarily gives rise to quark zero modes. For small (or vanishing) quark masses these zero modes tend to suppress the fermion determinant like $(m^0)^{|\nu|N_f}$ ($N_f$ is the number of light flavors). However, as emphasized in ref. [46] this suppression of the winding number $\nu$ configuration is always accompanied by an enhancement proportional to $V^{|\nu|N_f}$. Hence, in the physical situation $x \gg 1$ there is no suppression at all.

A further very interesting result obtained in ref. [46] is a set of sum rules for the spectrum of the Dirac operator $\not{D}$. For a gluon field configuration with winding number $\nu$ there are $|\nu|$ zero modes of $\not{D}$. The few lowest non–zero eigenvalues are proportional to $1/V\langle\bar{q}q\rangle$ as anticipated from the discussion above. Their distribution is sensitive to the winding number, and the levels are pushed up if $|\nu|$ increases. The sum rules of ref. [46] relate the inverse moments of the eigenvalue distribution to the quark condensate. *E.g.* the lowest one is

$$\sum_n{}' \frac{1}{\lambda_n^2} = \frac{1}{4}\left(V\langle\bar{q}q\rangle\right)^2 \frac{1}{|\nu| + N_f}. \tag{2.32}$$

These sum rules reflect the fact that for a finite volume the eigenvalues with $-1/V\langle\bar{q}q\rangle \lesssim \lambda_n \ll \Lambda_{\rm QCD}$ are the one related to spontaneous chiral symmetry breaking and the occurrence of the quark condensate.

On the basis of these results one can conclude that the winding number is irrelevant as long as $x \gg 1$ which is most likely the case in the real world. However, the winding number density fluctuations, *i.e.* the topological susceptibility

$$\chi = \frac{\langle\nu^2\rangle}{V} = \frac{1}{(32\pi^2)^2}\int d^4x \langle\tilde{G}G(x)\tilde{G}G(0)\rangle, \tag{2.33}$$

has measurable consequences for hadron physics: it generates the $\eta'$ mass. Phrased otherwise, despite the fact that the winding number is irrelevant the axial U(1) symmetry is still broken in an anomalous fashion, and this is reflected by the QCD vacuum. The strength of this breaking is determined by the topological susceptibility $\chi$. Also here Leutwyler and Smilga [46] found an astonishing result,

$$\chi = \frac{-\tau m^0\langle\bar{q}q\rangle}{N_f\tau - m^0\langle\bar{q}q\rangle} \tag{2.34}$$

where $\tau$ is the would–be topological susceptibility in the absence of quark fields. If one considers large $N_C$ and a fixed quark mass, $\tau$ has a finite limit whereas $\langle\bar{q}q\rangle \propto N_C$. Then one obtains $\chi = \tau$, *i.e.* in the large $N_C$ world the topological susceptibility would be given only by gluons. However, for $N_C = 3$ and a small current quark mass ($m^0\langle\bar{q}q\rangle \ll \tau$) the topological susceptibility is dominated by the quark condensate: $\chi = -m^0\langle\bar{q}q\rangle/N_f$ (for unequal quark masses $m^0/N_f$ is replaced by the reduced mass). The mean square winding number is large, $\langle\nu^2\rangle = -Vm^0\langle\bar{q}q\rangle/N_f$, and for an infinite volume all winding numbers are equally likely.

The physical implications of these considerations become more transparent when one considers the anomalous Ward identity[c]

$$\partial_\mu j_5^\mu = 2im^0 j_5 - \frac{N_f}{16\pi^2}\tilde{G}^a_{\mu\nu}G^{a\mu\nu} \tag{2.35}$$

where

$$j_5^\mu = \langle\bar{q}\gamma^\mu\gamma_5 q\rangle \quad \text{and} \quad j_5 = \langle\bar{q}\gamma_5 q\rangle \tag{2.36}$$

[c] For the simpler case of an abelian anomaly the derivation of the anomalous Ward identity is presented in Appendix A.

denote the axial singlet and pseudoscalar currents of the quarks, respectively. Considering the correlator of the axial singlet current this anomalous Ward identity may be used to derive an expression for the $\eta'$ mass [47]

$$m_{\eta'}^2 = \frac{2}{f_\pi^2}\left(N_f\tau - m^0\langle\bar{q}q\rangle\right) \approx \frac{2}{f_\pi^2}N_f\tau. \tag{2.37}$$

Since for $N_C = 3$ the topological susceptibility $\chi$ is governed by the quark condensate one might have anticipated this to be the case for the $\eta'$ mass, too, thereby contradicting the result of ref. [47]. Note, however, that the $\eta'$ mass is still dominated by the winding number density fluctuations of the purely gluonic theory.

Summarizing this chapter we would like to emphasize the following points: First, considering the limit of a large number of colors $N_C$ indicates that baryons may be described as solitons (Witten's conjecture). Unfortunately, the explicit expression of the effective meson theory (2.4) is unknown. Second, from the approximate chiral symmetry of QCD and its spontaneous breaking we know that low energy hadron physics is dominated by the would–be Goldstone bosons of chiral symmetry, the pions. Thus, at least, these meson fields have to appear in an *ansatz* for the effective meson theory. Third, due to the chiral anomaly in the flavor singlet channel we know that the chiral symmetry breaking is an inherent property of the small eigenvalues of the quark Dirac operator in the QCD vacuum.

These ideas will at least partially be applied in the proceeding chapters. We will present a model which is meant as a simplified version of the low–energy quark dynamics of QCD, the NJL model [17]. It displays dynamical breaking of chiral symmetry and yields quite a successful description of meson physics. Having noticed this common feature with the underlying theory we will again take advantage of the large $N_C$ arguments and attempt a description of baryons as solitons in the NJL model.

# 3 The Nambu–Jona–Lasinio model

In this chapter we introduce the NJL model as an effective chirally invariant theory of quark flavor dynamics. Originally, it was proposed to describe the pion as a massless bound state of the nucleon and the anti–nucleon [17]. Nambu and Jona-Lasinio studied a local four–fermion interaction in the scalar–isoscalar and pseudoscalar–isovector channel. By construction the original NJL model possesses a global U(1) × SU(2)$_L$× SU(2)$_R$ symmetry. Nowadays, it is common to call all chirally invariant models with local four–fermion (or even six–fermion) interactions an NJL model.

## 3.1 Description of the model

To be specific we will consider the model described by the Lagrangian

$$\begin{aligned}\mathcal{L}_{\rm NJL} = \bar{q}(i\partial\!\!\!/ - \hat{m}^0)q \quad &+ \quad 2G_1 \sum_{i=0}^{N_f^2-1} \left( (\bar{q}\frac{\lambda^i}{2}q)^2 + (\bar{q}\frac{\lambda^i}{2}i\gamma_5 q)^2 \right) \\ &- \quad 2G_2 \sum_{i=0}^{N_f^2-1} \left( (\bar{q}\frac{\lambda^i}{2}\gamma_\mu q)^2 + (\bar{q}\frac{\lambda^i}{2}i\gamma_5\gamma_\mu q)^2 \right) . \end{aligned} \tag{3.1}$$

Here $q$ denotes the quark spinors and $\hat{m}^0$ the current quark mass matrix. The matrices $\lambda^i/2$ are the generators of the flavor group ($\lambda^0 = \sqrt{2/N_f 1\!\!1}$), $N_f$ being the number of flavors under consideration. Note that the coupling constants $G_1$ and $G_2$ have dimension [energy]$^{-2}$. These coupling constants may take different values, $G_1 \neq G_2$, without spoiling the global chiral U$_L(N_f)$ × U$_R(N_f)$ symmetry, *i.e.* the two sums in eq (3.1) are independently chirally invariant.

The special form of the Lagrangian (3.1) can be motivated from the one–gluon–exchange. In the local limit of the one–gluon–exchange the quark interaction is given by the current–current interaction

$$j_{a\mu} j_a^\mu \tag{3.2}$$

where $j_a^\mu$ is the color octet flavor singlet current of the quarks,

$$j_a^\mu = \bar{q}\frac{\lambda_c^a}{2}\gamma^\mu q, \quad a = 1, \ldots, N_c^2 - 1 = 8, \tag{3.3}$$

the $\lambda_c^a/2$ being the generators of the color group SU(3) in the fundamental representation. Fierzing the interaction (3.2) into the color singlet channel leads to the interaction of eq. (3.1) with the additional relation $G_2 = G_1/2$ [20]. Since we are considering the NJL model as an effective theory we will relax this condition. As mentioned in section 2.1 the interaction (3.2) also contains quark–quark interactions leading to diquarks. These are, however, suppressed for large $N_C$ [35].

The invariance of the Lagrangian (3.1) under chiral rotations

$$\begin{aligned} q \to U_V q, \quad \bar{q} \to \bar{q}U_V^\dagger, \quad U_V = \exp{(i\theta_V)}\,, \quad \theta_V = \theta_V^i \frac{\lambda^i}{2} \\ q \to U_A q, \quad \bar{q} \to \bar{q}U_A, \quad U_A = \exp{(i\theta_A)}\,, \quad \theta_A = \theta_A^i \frac{\lambda^i}{2}\gamma_5 \end{aligned} \tag{3.4}$$

in the limit $m^0 \to 0$ is easily verified. Expressing the quark fields in terms of right– and left–handed quark fields defined by (*cf.* section 2.2)

$$q_R = P_R q, \quad q_L = P_L q \tag{3.5}$$

the Lagrangian (3.1) decouples in a sum of two Lagrangians containing only right–(left–) handed fields, respectively. Only the mass term

$$\mathcal{L}_{\text{mass}} = -\bar{q}\hat{m}^0 q = -(\bar{q}_R \hat{m}^0 q_L + \bar{q}_L \hat{m}^0 q_R) \tag{3.6}$$

couples right– to left–handed fields and thus breaks the chiral symmetry.

The local four–fermion interaction of (3.1) with a dimensionful coupling constant is obviously not renormalizable. Therefore it is only completely defined when supplemented with a regularization scheme in order to cut off momentum integrals and thus avoiding ultraviolet divergencies. In QCD these divergencies would have been absorbed in the renormalization procedure. On an operational level one may interpret the occurrence of the cut–off as a very crude way of mimicing the asymptotic freedom of QCD. Various schemes have been discussed in the literature[a]: O(3) and O(4) invariant sharp cut–offs, Pauli–Villars regularization and so on. Here we will exclusively use the proper time regularization scheme proposed by Schwinger [48]. This procedure has the advantage of being gauge invariant if (external) gauge fields are coupled to the model. Especially, for the calculation of electromagnetic and weak form factors this scheme is superior to sharp cut–offs because it allows one to appropriately manipulate the momentum space integrals. A further benefit of this regularization scheme is the fact that it may be defined at the level of the action rather than being defined via its application to Feynman integrals. This automatically guarantees that different quantities are regularized in a consistent manner. We will apply this regularization to the bosonized action, see below.

## 3.2 Bosonization

The aim is to rewrite the quark (fermion) theory (3.1) into an effective meson (boson) theory. It is convenient to use a compact notation

$$\Lambda_\alpha = \frac{\lambda^i}{2} \otimes \Gamma_a, \quad i = 0, \ldots, N_f^2 - 1, \quad \Gamma_a \in \{\mathbf{1}, i\gamma_5, i\gamma_\mu, i\gamma_\mu\gamma_5\} \tag{3.7}$$

and

$$Q^{\alpha\beta} = \begin{cases} 4G_1\delta^{\alpha\beta} & \text{for } \Gamma_a \in \{\mathbf{1}, i\gamma_5\} \\ 4G_2\delta^{\alpha\beta} & \text{for } \Gamma_a \in \{\gamma_\mu, \gamma_\mu\gamma_5\} \end{cases}. \tag{3.8}$$

The auxiliary field $\Phi = \Phi_\alpha \Lambda_\alpha$, which is introduced via the identity,

$$\exp\left(-\frac{i}{2}\int \bar{q}\Lambda_\alpha q Q^{\alpha\beta}\bar{q}\Lambda_\beta q\right) = \int \mathcal{D}\Phi \exp\left(-\frac{i}{2}\int \Phi_\alpha (Q^{-1})^{\alpha\beta}\Phi_\beta - i\int \Phi_\alpha \bar{q}\Lambda_\alpha q\right), \tag{3.9}$$

contains therefore (in the case of three flavors) nonets of scalar, pseudoscalar, vector and axial–vector meson fields. Using eq. (3.9) the generating functional

$$Z_{NJL} = \int \mathcal{D}q\mathcal{D}\bar{q} \exp(-i\int d^4x \mathcal{L}_{NJL})$$

[a] For a recent review see ref.[18]

may be written as [31]

$$\begin{aligned} Z_{NJL} &= \int \mathcal{D}\Phi \exp\left(-\frac{i}{2}\int \Phi Q^{-1}\Phi\right) Z_F[\Phi], \\ Z_F[\Phi] &= \int \mathcal{D}q\mathcal{D}\bar{q} \exp\left(-i\int \bar{q}(i\partial\!\!\!/ - \hat{m}^0 - \Phi\Lambda)q\right) \; . \end{aligned} \tag{3.10}$$

For the interpretation of $Z_F[\Phi]$ we note that it is equivalent to

$$Z_F[\Phi] = \lim_{T\to\infty} \langle 0|e^{-\hat{h}T}|0\rangle \tag{3.11}$$

where $T$ denotes a large Euclidean time interval and $\hat{h} = \int d^3x q^\dagger h q$ is the second quantized form of an one–particle Dirac Hamiltonian. In Minkowski space this Hamiltonian is given by

$$h = \boldsymbol{\alpha}\cdot\boldsymbol{p} + \beta(\hat{m}^0 + \Phi) \; , \quad \boldsymbol{p} = \frac{1}{i}\boldsymbol{\partial} \; , \tag{3.12}$$

as can be verified from the relation $i\partial\!\!\!/ - \hat{m}^0 - \Phi = \beta(i\partial_t - h)$.

For subsequent considerations it is convenient to remove the current quark mass from the Dirac operator by shifting $\Phi \to \Phi - \hat{m}^0$. This yields the functional

$$Z_{NJL} \;=\; \int \mathcal{D}\Phi \; e^{-\frac{i}{2}\int d^4x\left(\Phi-\hat{m}^0\right)Q^{-1}\left(\Phi-\hat{m}^0\right)} \int \mathcal{D}q\mathcal{D}\bar{q} \; e^{-i\int d^4x \bar{q}(i\partial\!\!\!/-\Phi)q}. \tag{3.13}$$

Furthermore we decompose the generic meson field $\Phi$ into irreducible Lorentz tensors

$$\Phi = S + i\gamma_5 P - iV\!\!\!/ - iA\!\!\!/\gamma_5. \tag{3.14}$$

$S$ is a scalar, $P$ a pseudoscalar, $V$ a vector and $A$ an axial–vector field. All these fields are flavor matrices, *i.e.* $S = S^i(\lambda^i/2)$ etc.

Using eq. (3.14) the chirally invariant interaction term of the NJL model is written as

$$\frac{1}{2}(\Phi - \hat{m}^0)Q^{-1}(\Phi - \hat{m}^0) = \frac{1}{2G_1}\mathrm{tr}((S - \hat{m}^0)^2 + P^2) + \frac{1}{2G_2}\mathrm{tr}(V_\mu V^\mu + A_\mu A^\mu). \tag{3.15}$$

For subsequent considerations it is also convenient to introduce the angular decomposition of the scalar and pseudoscalar meson fields by defining a complex field $M$

$$M = S + iP = \xi_L^\dagger \; \Sigma \; \xi_R, \tag{3.16}$$

which in turn defines the Hermitian field $\Sigma$ and unitary fields $\xi_L$ and $\xi_R$. The decomposition (3.16) is not unique; $\xi_L$ and $\xi_R$ are rather related by a "gauge" condition.

In eq. (3.13) the quark field appears bilinearly in the exponent and can therefore be integrated out. Using that Det = exp Tr log one obtains

$$\begin{aligned} Z_{NJL} &= \int \mathcal{D}\Phi e^{i\mathcal{A}[\Phi]}, \\ \mathcal{A}[\Phi] &= -\frac{1}{2}\int d^4x(\Phi - \hat{m}^0)Q^{-1}(\Phi - \hat{m}^0) + \mathrm{Tr}\log(i\partial\!\!\!/ - \Phi). \end{aligned} \tag{3.17}$$

Note that the action $\mathcal{A}[\Phi]$ is a non–linear, even non–polynomial function of the meson field $\Phi$. Even more, the term $\mathrm{Tr}\log(i\partial\!\!\!/ - \Phi)$ is non–local. The quantum theory defined by eq. (3.17)

is, however, equivalent to the underlying NJL model defined by the Lagrangian (3.1). On the other hand, the generating functional (3.17) has the advantage that it may be treated in a semiclassical approximation because the vacuum expectation value of the bosonic field $\Phi$ can be different from zero, and in general will be, whereas the vacuum expectation value of the fermionic quark field is necessarily vanishing in the absence of external fermion sources.

Symmetry currents are constructed by adding external gauge fields $a_\mu^{(v)} = a_\mu^{(v)i}\lambda^i/2$ and $a_\mu^{(a)} = a_\mu^{(a)i}\lambda^i/2$ to the action for the vector ($v$) and axial–vector ($a$) symmetries, respectively

$$q\,(i\partial\!\!\!/ - \Phi)\,q \rightarrow q\left(i\partial\!\!\!/ - i a\!\!\!/^{(v)} - i a\!\!\!/^{(a)} - \Phi\right)q. \tag{3.18}$$

The symmetry currents are then identified as the terms coupling linearly to the external gauge fields

$$j_\mu^{(v,a)} = \left.\frac{\delta Z}{\delta a^{(v,a)\mu}}\right|_{a^{(v,a)\mu}=0}. \tag{3.19}$$

The gauge fields may be eliminated from the fermion part of the action by transforming the (axial) vector fields accordingly

$$V_\mu \rightarrow V_\mu + a_\mu^{(v)} \quad \text{and} \quad A_\mu \rightarrow A_\mu + a_\mu^{(a)}. \tag{3.20}$$

This allows one to straightforwardly compute the derivatives in eq (3.19) yielding the current field identities [37]

$$j_\mu^{(v)} = \frac{1}{2G_2}V_\mu \quad \text{and} \quad j_\mu^{(a)} = \frac{1}{2G_2}A_\mu. \tag{3.21}$$

## 3.3 Dynamical breaking of chiral symmetry

The vacuum expectation value (VEV) of the auxiliary field $\Phi$ is found from the stationary point of the action (3.17):

$$\langle\Phi_\alpha\rangle = -Q^{\alpha\beta}\langle\bar{q}\Lambda_\beta q\rangle. \tag{3.22}$$

Especially, the quark condensate $\langle\bar{q}q\rangle$ is related to the VEV of the scalar field via the coupling constant $G_1$. Alternatively the effective action $\mathcal{A}[\Phi]$ (3.17) yields the Dyson–Schwinger equation

$$\Phi_\alpha(x) = -Q^{\alpha\beta}\mathrm{tr}(G_\Phi(x,x)\Lambda_\beta). \tag{3.23}$$

The solution of this equation determines the VEV of the meson field $\Phi$. Hereby $G_\Phi$ is the quark propagator in the background of the $\Phi$-field. It is defined by

$$G_\Phi^{-1}(x,y) = (i\partial\!\!\!/ - \Phi)\delta(x-y). \tag{3.24}$$

Before being able to derive explicit expressions for (3.23) in the NJL model a regularization scheme has to be imposed. Using the decomposition (3.14) the effective action (3.17) may be cast into the form

$$\begin{aligned}
\mathcal{A} &= \mathcal{A}_F + \mathcal{A}_m, \\
\mathcal{A}_F &= \mathrm{Tr}\log(iD\!\!\!\!/\,) = \mathrm{Tr}\log\left(i\partial\!\!\!/ + iV\!\!\!\!/ + iA\!\!\!/\gamma_5 - (P_R M + P_L M^\dagger)\right), \\
\mathcal{A}_m &= \int d^4x\left(-\frac{1}{4G_1}\mathrm{tr}(M^\dagger M - m^0(M+M^\dagger) + (m^0)^2) + \frac{1}{4G_2}\mathrm{tr}(V_\mu V^\mu + A_\mu A^\mu)\right).
\end{aligned} \tag{3.25}$$

As this action is equivalent to the non–renormalizable NJL model it is only completely defined if a regularization scheme is provided. As already stated in subsection 3.1 we will use Schwinger's proper time regularization[48] which introduces an $O(4)$-invariant cut-off $\Lambda$ after continuation to Euclidean space. For this regularization procedure it is necessary to consider the real and imaginary part of $\mathcal{A}_F$ separately

$$\begin{aligned}\mathcal{A}_F &= \mathcal{A}_R + \mathcal{A}_I, \\ \mathcal{A}_R &= \frac{1}{2}\mathrm{Tr}\log(\not{D}_E^\dagger \not{D}_E), \\ \mathcal{A}_I &= \frac{1}{2}\mathrm{Tr}\log((\not{D}_E^\dagger)^{-1}\not{D}_E).\end{aligned} \tag{3.26}$$

The real part $\mathcal{A}_R$ diverges for large momenta $p$ whereas the imaginary part $\mathcal{A}_I$ does not contain ultraviolet divergencies, *i.e.* it is finite without regularization.[b] Therefore one has the option of keeping $\mathcal{A}_I$ unregularized, or to regularize it in a way consistent with the regularization of $\mathcal{A}_R$. Note that this defines two different models.

For the real part of the action the proper time regularization consists in replacing the logarithm by a parameter integral

$$\mathcal{A}_R \rightarrow -\frac{1}{2}\int_{1/\Lambda^2}^{\infty}\frac{ds}{s}\mathrm{Tr}\exp\left(-s\not{D}_E^\dagger \not{D}_E\right), \tag{3.27}$$

which for $\Lambda \rightarrow \infty$ reproduces the logarithm up to an irrelevant constant.[c] Since the operator $\not{D}_E^\dagger \not{D}_E$ is Hermitian and positive definite this integral is well defined.

For the issues discussed in this section it is sufficient to only inspect $\mathcal{A}_R$. Varying the regularized effective action with respect to the scalar and pseudoscalar fields yields the Dyson–Schwinger or gap equations

$$\begin{aligned}\langle \Sigma_{ij}\rangle &= \delta_{ij} m_i, \\ m_i &= m_i^0 - 2G_1\langle \bar{q}q\rangle_i, \\ \langle \bar{q}q\rangle_i &= -m_i^3\frac{N_c}{4\pi^2}\Gamma(-1, m_i^2/\Lambda^2)\end{aligned} \tag{3.28}$$

which is the regularized version of eq. (3.23) specialized to the case of the scalar meson field. The dynamically generated constituent quark masses $m_i$, $i = u, d$, (and also the quark condensates $\langle \bar{q}q\rangle_i$ ) are equal, $m_u = m_d$ and $\langle \bar{u}u\rangle = \langle \bar{d}d\rangle$, respectively, if and only if the quark current masses are equal, $m_u^0 = m_d^0$. Throughout this review we will restrict ourselves to this isospin symmetric limit, and in the following we will use the notation $m := m_u = m_d$ and $m^0 := m_u^0 = m_d^0$. Of course, for the strange quark we will consider larger current masses according to (2.25) leading to larger constituent masses. For these quantities we will keep the flavor index and denote them $m_s^0$ and $m_s$, respectively.

[b]For a more detailed discussion of a correct definition of $\mathcal{A}_I$ and its properties see sect. 6.1.

[c]The functions

$$\Gamma(u, x) = \int_x^\infty d\tau \tau^{u-1} e^{-\tau}$$

are known as incomplete $\Gamma$-functions. Especially,

$$\Gamma(0, x) = -\log x + \gamma + \mathcal{O}(x) \quad \text{for} \quad x \rightarrow 0^+$$

where $\gamma = 0.57721$ is Euler's constant.

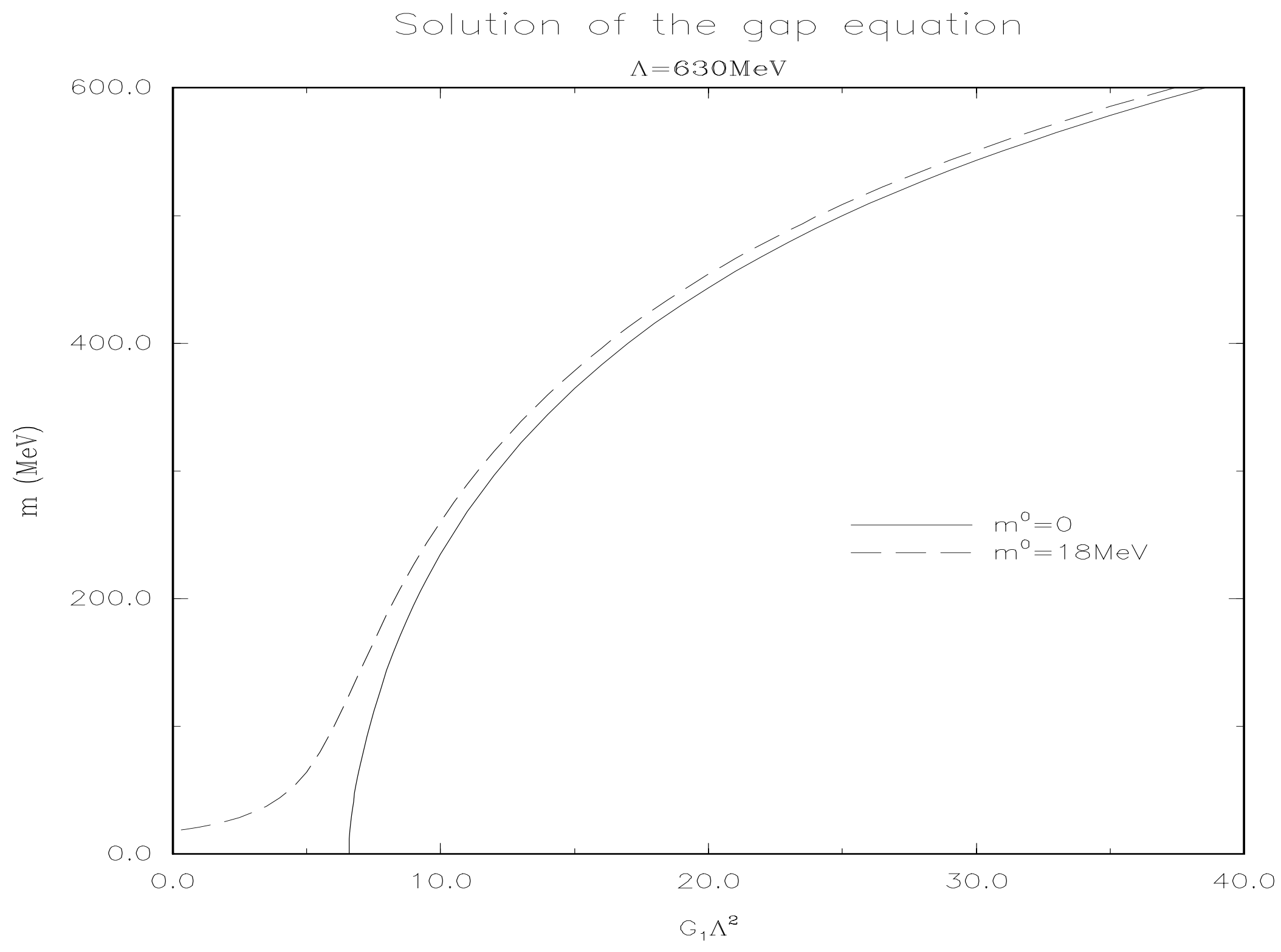

Figure 3.1: The solution of the gap equation (3.28) for vanishing current mass $m^0 = 0$ (solid line) and $m^0 = 18$MeV (dashed line). For the chosen value of the cut-off, $\Lambda = 630$MeV, a constituent quark mass $m = 400$MeV reproduces the phenomenological value of the pion decay constant, $f_\pi = 93$MeV.

As can be seen from figure 3.1 in the chiral limit the quark condensate and therefore also the quark constituent mass is zero when the coupling constant $G_1$ stays below a critical value. Above this critical value the trivial solution coexists with a non–trivial one. Examining the effective potential for a constant scalar field in the chiral limit[d]

$$\mathcal{V}(\Sigma) = \frac{1}{2G_1}\Sigma^2 + \frac{N_c}{16\pi^2}\left(\Sigma^4\Gamma(0,\frac{\Sigma^2}{\Lambda^2}) - (\Sigma^2 - \Lambda^2)\Lambda^2 e^{-\Sigma^2/\Lambda^2}\right) \tag{3.29}$$

one concludes that the non–trivial solution is energetically favored.

### 3.4 Chiral rotation and hidden local symmetry

The freedom in the choice of $\xi_{L,R}$ in the parametrization (3.16) of the scalar and pseudoscalar field reflects the local hidden symmetry $\mathrm{SU}(3)_h$. Under $\mathrm{SU}(3)_L \times \mathrm{SU}(3)_R \times \mathrm{SU}(3)_h$ the fields $\xi_{L,R}$ and $\Sigma$ transform as

$$\xi_L(x) \;\;\rightarrow\;\; h(x)\xi_L(x)L^\dagger(x),$$

[d]The first part follows trivially from $\mathcal{A}_m$ whereas the second part can be obtained from $\mathcal{A}_R$ by evaluating the functional trace.

$$\begin{aligned} \xi_R(x) &\rightarrow h(x)\xi_R(x)R^\dagger(x), \\ \Sigma(x) &\rightarrow h(x)\Sigma(x)h^\dagger(x) \end{aligned} \tag{3.30}$$

where

$$L \in \mathrm{SU}(3)_L, R \in \mathrm{SU}(3)_R \quad \text{and} \quad h \in \mathrm{SU}(3)_h. \tag{3.31}$$

The additional degrees of freedom contained in $\xi_{L,R}$ can be gauged away as in the usual Higgs mechanism [16]. Later on we will adopt the unitary gauge

$$\xi_R = \xi_L^\dagger = \xi \tag{3.32}$$

where $\xi$ is an element of the coset-space $\mathrm{SU}(3)_L \times \mathrm{SU}(3)_R/\mathrm{SU}(3)_V$. Gauge conditions like (3.32) furthermore fix $h(x)$ in terms of $L(x)$, $R(x)$ and $\xi(x)$.

For subsequent considerations it is convenient to perform a chiral rotation from the current quarks $q_{L,R}$ to the constituent quarks (*cf.* eq (2.6))

$$\chi_{L,R} = \xi_{L,R} q_{L,R}. \tag{3.33}$$

Under a chiral rotation of the original quark fields

$$q_L \rightarrow L q_L, \quad q_R \rightarrow R q_R \tag{3.34}$$

the constituent quark fields transform according to the hidden gauge symmetry

$$\chi_{L,R} \rightarrow h(x)\chi_{L,R}. \tag{3.35}$$

After the chiral rotation

$$\bar{q} i \not{D} q = \bar{\chi} i \tilde{\not{D}} \chi \tag{3.36}$$

the chirally rotated Dirac-operator

$$i\tilde{\not{D}} = \mathcal{T} i \not{D} \mathcal{T}^\dagger, \quad \text{with} \quad \mathcal{T} = P_R \xi_R + P_L \xi_L \tag{3.37}$$

becomes

$$i\tilde{\not{D}} = i(\not{\partial} + \tilde{\not{V}} + \tilde{\not{A}}\gamma_5) - \Sigma. \tag{3.38}$$

It contains the chirally rotated vector and axial–vector fields

$$\tilde{V}_\mu \mp \tilde{A}_\mu = \xi_{L,R}(V_\mu \mp A_\mu + \partial_\mu)\xi_{L,R}^\dagger. \tag{3.39}$$

These fields transform under a left-right-transformation according to the hidden symmetry group

$$\tilde{V}_\mu \mp \tilde{A}_\mu \rightarrow h(x)(\tilde{V}_\mu \mp \tilde{A}_\mu + \partial_\mu)h^\dagger(x). \tag{3.40}$$

From these relations it follows that the vector field transforms as a gauge field of the hidden local symmetry $\mathrm{SU}(3)_h$ defining a covariant derivative

$$D^\mu = \partial^\mu + \tilde{V}^\mu. \tag{3.41}$$

On the other hand the chirally rotated axial–vector field transforms homogeneously under the hidden local symmetry

$$\tilde{A}^\mu(x) \to h(x)\tilde{A}^\mu(x)h^\dagger(x). \tag{3.42}$$

In terms of the chirally rotated fields the Lagrangian becomes [31, 32]

$$\begin{aligned} \mathcal{L} = \bar{\chi}\left(i(\partial\!\!\!/ + \tilde{V\!\!\!/} + \tilde{A\!\!\!/}\gamma_5) - \Sigma\right)\chi \quad &- \quad \frac{1}{4G_1}\mathrm{tr}_F\left[\Sigma^2 - \hat{m}_0\left(\xi_L^\dagger\Sigma\xi_R + \xi_R^\dagger\Sigma\xi_L\right)\right] \\ &- \quad \frac{1}{4G_2}\mathrm{tr}_F\left[(\tilde{V}_\mu - v_\mu)^2 + (\tilde{A}_\mu - a_\mu)^2\right]. \end{aligned} \tag{3.43}$$

Here we have defined the vector and axial–vector fields

$$v_\mu = \frac{1}{2}(\xi_R\partial_\mu\xi_R^\dagger + \xi_L\partial_\mu\xi_L^\dagger) \tag{3.44}$$

$$a_\mu = \frac{1}{2}(\xi_R\partial_\mu\xi_R^\dagger - \xi_L\partial_\mu\xi_L^\dagger) \tag{3.45}$$

which are induced by the chiral rotation. The Lagrangian (3.43) is the basis for the investigation described in section 6.2.5.

The chiral rotation of the quark fields does not leave the fermionic integration measure invariant but introduces a non-local Jacobian [49]

$$\int DqD\bar{q} = J\int D\chi D\bar{\chi}, \tag{3.46}$$

which contains the anomaly. Its physical implications will be discussed in section 4.4.

# 4 Effective meson theory

In the previous chapter the NJL model has been converted into an effective meson theory. In the present chapter we will examine its implications for meson physics.

## 4.1 Gradient expansion

The quanta of the small amplitude fluctuations of the composite meson fields $\Phi$ (see section 3.2) around their vacuum expectation values represent the mesons. For the extraction of these (free) mesons the effective action has to be expanded up to second order in the fields $\Phi$. This will be done in section 4.3 and leads to the Bethe–Salpeter equation. At the moment we will concentrate on a simplified description of the mesons at low energies. In this section we will follow ref. [31].

The effective meson action obtained in the previous section by bosonizing the NJL model is a highly non-local object due to the presence of the quark determinant, $\mathrm{Tr}\log(i\partial\!\!\!/-\Phi)$. Since $[i\partial\!\!\!/,\Phi]\neq 0$ this term contains infinitely high derivatives of the meson field. Obviously, for the determination of the low-energy (*e.g.* static) properties of the mesons we do not need to keep all the higher order derivatives but a gradient expansion of the quark determinant seems to be appropriate. In leading order gradient expansion of the real part of the effective meson action one finds

$$\begin{aligned}\mathcal{A}_R &= \int d^4x\Big\{-C_0(m/\Lambda)+C_2(m/\Lambda)\Big[\frac{1}{4}\mathrm{tr}(\nabla^\nu M^\dagger\nabla_\nu M)-\frac{1}{2}\mathrm{tr}(F^V_{\mu\nu}F^{V\mu\nu})-\frac{1}{2}\mathrm{tr}(F^A_{\mu\nu}F^{A\mu\nu})\Big]\\ &\qquad+\frac{1}{4G_2}\mathrm{tr}\left(V_\mu V^\mu+A_\mu A^\mu\right)\Big\}+\dots \end{aligned}\tag{4.1}$$

where

$$\nabla^\nu M=\partial^\nu M+[V^\nu,M]-\gamma_5\{A^\nu,M\}\tag{4.2}$$

denotes the covariant derivative of the scalar – pseudoscalar field $M$. Furthermore,

$$\begin{aligned}F^V_{\mu\nu} &= \partial_\mu V_\nu-\partial_\nu V_\mu+[V_\mu,V_\nu]+[A_\mu,A_\nu]\\ F^A_{\mu\nu} &= \partial_\mu A_\nu-\partial_\nu A_\mu+[A_\mu,V_\nu]+[V_\mu,A_\nu]\end{aligned}\tag{4.3}$$

are the vector and axial–vector parts of the field strength tensor. The quantity $C_2(m/\Lambda)$ depends only on the ratio $m/\Lambda$ and diverges logarithmically as $\Lambda\to\infty$. Eq. (4.1) defines the gauged linear $\sigma$-model. To cast this model into its standard form one has to perform field renormalizations and redefinitions in order to remove the $\pi-A_1$ and $\sigma-V$ mixings from the Lagrangian. One then finds for the vector and axial–vector masses

$$m_V^2=\frac{g_V^2}{4G_2}\quad\text{and}\quad m_A^2=m_V^2+6m^2\tag{4.4}$$

where

$$g_V=\left[\frac{N_c}{24\pi^2}\Gamma\left(0,\frac{m^2}{\Lambda^2}\right)\right]^{-1/2}\tag{4.5}$$

is the universal vector meson coupling constant. Furthermore the scalar meson mass is given by $m_\sigma^2 = 4m^2 + m_\pi^2$ (see also eq. (4.46)). The pseudoscalar meson mass is given by

$$m_\pi^2 = \frac{m_0 m}{G_1 f_\pi^2}. \tag{4.6}$$

The resulting effective meson Lagrangian satisfies the well–known and phenomenologically very successful current algebra results[a]. This comes with no surprise since our starting quark theory has strict chiral symmetry for $\hat{m}^0 = 0$. From eqs (4.5) and (4.6) one furthermore finds the relation

$$m_V^2 = a g_V^2 f_\pi^2 \ , \quad a = \left(1 - \frac{m_V^2}{m_A^2}\right)^{-1} \tag{4.7}$$

which for $a = 2$ represents the KSRF–relation [50]. Furthermore, for this "magic" value of $a$ Weinberg's relation [51] $m_A = \sqrt{2} m_V$ also holds true. All these low energy theorems are quite reasonably reproduced by the experimental data.

So far we have discarded the imaginary part of the effective meson action ($\mathcal{A}_I$ in eq (3.26)). This part allows for a similar gradient expansion where again subsequent terms are suppressed by powers of the ratio $m/\Lambda$. As already mentioned in chapter 3 the imaginary part of the effective meson action stays finite when the cut-off is removed. For $\Lambda \to \infty$ only the leading order term of the gradient expansion survives and this term is then given by the (gauged) Wess–Zumino–Witten action [52] (see section 4.4)

$$\mathcal{A}_I = \mathcal{A}_{WZW} + \cdots \tag{4.8}$$

Therefore in leading order gradient expansion we find for the effective meson action on the chiral circle ($\Sigma = \langle \Sigma \rangle = \hat{m}$)

$$\mathcal{A} = \mathcal{A}_{nl\sigma} + \mathcal{A}_{WZW} \tag{4.9}$$

where $\mathcal{A}_{nl\sigma}$ denotes the gauged nonlinear $\sigma$–model displayed in eq. (4.1). This model Lagrangian has stable chiral solitons provided that the (axial–) vector mesons are retained[b].

## 4.2 Relation to the Skyrme model

For the soliton description of baryons the chiral field is crucial. Let us therefore concentrate now on the chiral field while setting all the other meson fields to their vacuum value. Aside from the pseudoscalar mass term the effective meson action of the chiral field is then given by the quark loop $\mathrm{Tr}\log(i\partial\!\!\!/ - m(U)^{\gamma_5})$ where $U = \xi_L^\dagger \xi_R$ is the chiral field, see eq. (3.16). In leading order its derivative expansion is given by the nonlinear $\sigma$-model Lagrangian

$$\mathcal{L}_{nl\sigma} = -\frac{f_\pi^2}{4}\mathrm{tr} L_\mu L^\mu, \quad L_\mu = U^\dagger \partial_\mu U \tag{4.10}$$

in agreement with eq. (4.1). As is well–known this model does not allow for stable solitons as can be seen by applying Derrick's theorem [53]. To obtain stable solitons the non-linear $\sigma$-model has to be supplemented by stabilizing higher order derivative terms like the Skyrme term [8], which has four derivatives. One would expect that these stabilizing terms emerge in higher order of the gradient expansion of the effective meson action. In next to leading order of the derivative expansion one finds in the limit $\Lambda \to \infty$ [31]

$$\mathcal{L}^{(4)} = \frac{N_C}{32\pi^2}\mathrm{tr}\left\{\frac{1}{12}([L_\mu, L_\nu])^2 - \frac{1}{3}(\partial_\mu L^\mu)^2 + \frac{1}{6}(L_\mu L^\mu)^2\right\} \tag{4.11}$$

[a]For a compilation of current algebra results see ref.[37] and references therein.

[b]For a review on mesonic solitons with vector mesons see ref. [10].

The first term is in fact the desired Skyrme term which shows up with a coefficient $e = 2\pi$. This value favorably compares with the phenomenological value $e = 5.45$ obtained by Adkins *et al.* [9] by fitting the Skyrmion mass[c] (after semiclassical quantization, *cf.* section 7.1) to the nucleon mass. Unfortunately the last two terms destabilize the soliton. The last term leads to a negative definite energy for any chiral field configuration and forces the soliton to collapse. The second term which provides a tachyonic pole in the pion propagator is even more dangerous because it leads to a vacuum instability [54]. Even if one goes to the next (sixth) order gradient expansion one does not find stable solitons [55]. This certainly signals the break-down of the naïve gradient expansion in the soliton sector of the effective action of the chiral field. These rather frustrating results had for some time given rise to the fear that the chiral soliton of the effective action might not at all posses soliton solutions.

Some hints for the existence of stable chiral solitons are provided by the heat kernel expansion of the effective chiral action (3.25). The heat kernel expansion represents a semi-classical type of asymptotic expansion which corresponds to a resummation of the gradient expansion [56]. In next to leading order this expansion gives, besides the forth order terms given by equation (4.11), also a term with six derivatives [31]

$$\mathcal{L}^{(6)} = \Gamma(2, \frac{m^2}{\Lambda^2})\frac{1}{32\pi^2}\frac{1}{30m^2}\mathrm{tr}\left((\partial^\mu \Box U^\dagger)(\partial_\mu \Box U)\right) \tag{4.12}$$

which in fact overcomes the forth order terms and stabilizes the chiral soliton [57].

Let us also mention that the next to leading order terms of the derivative expansion considerably improve the prediction for the $\pi\pi$-scattering length towards their experimental values [58]. In fact the coefficients of these terms which are predicted by the gradient expansion of the quark loop are in rather good agreement [31] with the phenomenological values determined in chiral perturbation theory [59]. Thus the destabilizing terms are obviously relevant for the low energy meson physics.

Nevertheless, we will show below that the Skyrme model does arise as low energy approximation to the non–local effective meson theory provided one includes the vector and axialvector mesons in an appropriate manner. The importance of the vector mesons for the stabilization of the chiral soliton against the collapse should come with no surprise since the scalar and axialvector mesons are much heavier than the pseudoscalar mesons and therefore should control the short distance behavior of the soliton at low energies.

For the study of the soliton sector of the effective meson theory it is obviously advantageous to regard the chirally rotated vector and axialvector fields introduced in section 3.4 as the physical fields. The corresponding treatment in the soliton sector will be discussed at length in subsection 6.2.5. Note that after the chiral rotation described in section 3.4 all higher than leading order gradient terms of the chiral field $U$ are the fully absorbed into the rotated vector and axialvector fields. The elimination of these gradients is obviously crucial in The soliton sector since the gradient expansion seems not to converge for the chiral field, at least not in next-to-leading order and next-to-next-to-leading order. On the other hand for the fields of the heavier vector and axialvector mesons a gradient expansion might still be appropriate at low energies, which, however, has to be justified *a posteriori*.

[c] Note that Atkins et al. considered $f_\pi$ as a free parameter. Their fit to baryon masses provided $f_\pi = 64$MeV.

In terms of the rotated fields the leading order of the derivative expansion the vacuum part of the action is given by the following Lagrangian [32]

$$\mathcal{L} = \tilde{\mathcal{L}}_{nl\sigma} + \mathcal{L}_{WZW} \tag{4.13}$$

where

$$\begin{aligned} \tilde{\mathcal{L}}_{nl\sigma} &= \frac{1}{2g_V^2}\mathrm{tr}\left[\left(\tilde{F}_V\right)^2 + \left(\tilde{F}_A\right)^2\right] - \frac{a}{a-1}f_\pi^2\mathrm{tr}\tilde{A}^2 \\ &\quad -\frac{m_V^2}{g_V^2}\mathrm{tr}\left[\left(\tilde{V}-v\right)^2 + \left(\tilde{A}-a\right)^2\right] + \frac{1}{4}f_\pi^2 m_\pi^2 \mathrm{tr}\left(U + U^\dagger - 2\right). \end{aligned} \tag{4.14}$$

Note that we have assumed the chiral circle condition $\Sigma = \hat{m}$. $\mathcal{L}_{WZW}$ is the Wess–Zumino–Witten [52, 15] term expressed now in terms of the rotated fields $\tilde{V}_\mu, \tilde{A}_\mu$. The vector fields $v_\mu$ and $a_\mu$ have been defined in eqs (3.44,3.45). Furthermore the field strength tensors $(\tilde{F}_V)_{\mu\nu}$, $(\tilde{F}_A)_{\mu\nu}$ are defined in equation (4.3) with the unrotated fields replaced by the rotated fields $\tilde{V}_\mu, \tilde{A}_\mu$.

Since the (axial-) vector meson masses $m_V, m_A$ are large compared to the pion mass it is tempting to integrate out the (axial-)vector fields $\tilde{V}_\mu, \tilde{A}_\mu$ in the above Lagrangian in the static limit. Then one might consider the kinetic energy $\left((\tilde{F}_V)^2_{\mu\nu}, (\tilde{F}_A)^2_{\mu\nu}\right)$ and higher order terms as well as the Wess-Zumino term as perturbations. The resulting equation of motion yield [32]

$$\tilde{V}_\mu = v_\mu , \tag{4.15}$$

$$\tilde{A}_\mu = \frac{a-1}{a}a_\mu . \tag{4.16}$$

For these field configurations the field strength tensors reduce to

$$\tilde{F}_V^{\nu\mu} = \left((\frac{a-1}{a})^2 - 1\right)[a^\mu, a^\nu] \tag{4.17}$$

$$\tilde{F}_A^{\mu\nu} = \frac{a-1}{a}\left([\partial^\mu + v^\mu, a^\nu] + [a^\mu, \partial^\nu + v^\nu]\right) = 0 \tag{4.18}$$

and the Lagrangian (4.14) precisely becomes the Skyrme Lagrangian

$$\mathcal{L} = -\frac{1}{4}f_\pi^2\mathrm{tr}L_\mu L^\mu + \frac{1}{32e^2}\mathrm{tr}[L_\mu, L_\nu] + \frac{1}{4}m_\pi^2 f_\pi^2 \mathrm{tr}(U + U^\dagger - 2) \tag{4.19}$$

with

$$e = g_V \frac{a^2}{|2a-1|}. \tag{4.20}$$

Thus in the static limit of the vector and axial–vector mesons the total effective meson Lagrangian in fact becomes the celebrated Skyrme model [8].

Since the lowest lying axial–vector mesons ($m_A \approx 1260$ MeV) is considerably heavier than the lowest lying vector mesons ($m_\rho \approx m_\omega \approx 770$MeV) one might discard the axial–vector mesons at low energies. Then one obtains

$$e = g_V \tag{4.21}$$

Using the experimental value of $g_V \simeq 6.0$ [40] for the universal vector coupling constant, the Skyrme parameter is not far from its phenomenological value $g_V = 5.45$ obtained by Adkins *et al.* [9] by fitting the nucleon and $\Delta$-masses.

In the derivation of the Skyrme model we have used the static limit for the vector and axial–vector mesons considering both the kinetic energy as well as the Wess–Zumino–Witten term as perturbations. When the static equations of motion (4.15,4.16) are used the Wess–Zumino–Witten term vanishes. Alternatively, one could, however, also include the Wess–Zumino–Witten term into the definition of the static equation of motion and leaving only the kinetic energy as a perturbation. This adds a current-current interaction to the Skyrme Lagrangian

$$\mathcal{L}_6 = -\frac{1}{2}\epsilon_6^2 B_\mu B^\mu \tag{4.22}$$

from the coupling of the $\omega$-meson to the topological current $B_\mu$ in the Wess–Zumino–Witten term. In the context of the NJL model the coupling strength is obtained to be [60]

$$\epsilon_6^2 = \frac{6\pi^2 N_C}{m_\rho^2 \Gamma(0, m^2/\Lambda^2)}. \tag{4.23}$$

The current–current interaction (4.22) modifies the short distance behavior of the soliton resulting in a reasonable description of the short range behavior of the central potential in the nucleon–nucleon interaction [61].

Alternatively, if the (axial) vector mesons are kept as dynamical fields there will be no reason to neglect the Wess–Zumino–Witten action (contained in the full Lagrangian) in comparison to the normal parity terms. In fact, the Wess–Zumino–Witten term is then absolutely necessary to ensure the existence of stable solitons. On the other hand, if one integrates out the (axial-) vector mesons in the static approximation in the way it was done above, the Wess–Zumino–Witten term vanishes for the corresponding static field configurations of two flavors. In this case stability of the soliton is provided by the induced Skyrme term.

## 4.3 Bethe–Salpeter equations

In this section we will discuss the determination of the meson masses as functions of the NJL model parameters. However, this time we will go beyond the gradient expansion, *i.e.* we will take into account the gradients of the fields in an exact manner. This can be done by extracting the Bethe–Salpeter equations for the mesons starting with the regularized action, see eqs. (3.25) and (3.27). In order not to disguise the used method by technical difficulties we will display it for the case of pions in the isospin limit only. For the other cases we will only cite the results. For details we refer the reader to Appendix A of ref.[62][d].

The physical meson excitations are given by small amplitude fluctuations around the translational invariant vacuum field configuration obtained by solving (3.28). Expanding the effective action in Euclidean space

$$\begin{aligned}
\mathcal{A} &= \mathcal{A}_m + \mathcal{A}_F \\
\mathcal{A}_m &= \int d^4x \left(-\frac{1}{4G_1}\mathrm{tr}(M^\dagger M - m_0(M+M^\dagger) + m_0^2)\right) \\
\mathcal{A}_F &= -\frac{1}{2}\int_{1/\Lambda^2}^\infty \frac{ds}{s}\mathrm{Tr}\exp\left(-s\not{D}_E^\dagger \not{D}_E\right) \\
\not{D}_E^\dagger \not{D}_E &= \partial^2 + [i\not{\partial}, P_R M + P_L M^\dagger] + P_R M^\dagger M + P_L M M^\dagger
\end{aligned} \tag{4.24}$$

[d]Furthermore, we will restrict ourselves to proper time regularization.

up to second order in the pion fluctuations of the pseudoscalar field $\Theta(x) = 2\pi(x)$ allows us to extract the inverse propagator for the pseudoscalar mesons.[e] Since we are not interested in the fluctuations of the scalar field we substitute it by the vacuum expectation value (VEV) $\langle\Sigma\rangle = m\mathbb{I}$. The complex field $M$ (3.16) is then given by

$$M = mU, \quad U = me^{i\Theta(x)} = me^{i2\pi(x)}. \tag{4.25}$$

Obviously, the VEV of the pseudoscalar field $\Theta(x)$ is zero. The chiral field $U$ is expanded for small–amplitude fluctuations of the pseudoscalar meson field $\pi = \pi^i(\tau^i/2)$ as

$$U = e^{2i\pi} = 1 + 2i\pi - 2\pi^2 + \dots . \tag{4.26}$$

The factor 2 is hereby introduced for later convenience.

As the stationary point of the action occurs for $\pi = 0$ there is no linear term. First, we will consider the mass term of the mesonic action $\mathcal{A}_m$

$$\mathcal{A}_m = -\frac{1}{2G_1}\int d^4x \left((m - m^0)^2 + 4m^0 m \, \pi^2\right) + \mathcal{O}(\pi^3). \tag{4.27}$$

Introducing the Fourier transform of the fluctuating field, $\pi(q)$, the bilinear term of (4.27) is obtained to be:

$$\int \frac{d^4q}{(2\pi)^4} \sum_{i=1}^{3} \frac{1}{2}\pi^i(-q)\pi^i(q)\frac{m^0 m}{G_1}. \tag{4.28}$$

In order to expand the fermion determinant $\mathcal{A}_F$ we rewrite the operator

$$\not{D}_E^\dagger \not{D}_E = A_0 + A_1 + A_2 + \dots \tag{4.29}$$

where $A_k$ is of $k$-th order in the field $\pi$. Using eqs. (4.24), (4.25) and (4.26) we obtain

$$\begin{aligned} A_0 &= \partial^2 + m^2 \\ A_1 &= 2\gamma_5 m(\not{\partial}\pi) \\ A_2 &= 0. \end{aligned} \tag{4.30}$$

The last relation is valid in the isospin limit only. Obviously, only derivatives of $\pi$ can occur as a consequence of the chiral invariance of $\mathrm{Det}\not{D}_E^\dagger \not{D}_E$. Using now

$$\begin{aligned} \mathcal{A}_F &= -\frac{1}{2}\int_{1/\Lambda^2}^{\infty} \frac{ds}{s} \mathrm{Tr}\exp\left(-s\not{D}_E^\dagger \not{D}_E\right) \\ &= -\frac{1}{2}\int_{1/\Lambda^2}^{\infty} \frac{ds}{s}\mathrm{Tr}e^{-sA_0} + \frac{1}{2}\int_{1/\Lambda^2}^{\infty} ds \int_0^1 d\zeta \mathrm{Tr}e^{-s\zeta A_0} A_2 e^{-s(1-\zeta)A_0} \\ &\quad -\frac{1}{2}\int_{1/\Lambda^2}^{\infty} ds s \int_0^1 d\zeta \int_0^{1-\zeta} d\eta \mathrm{Tr}e^{-s\eta A_0} A_1 e^{-s(1-\zeta-\eta)A_0} A_1 e^{-s\zeta A_0} + \mathcal{O}(\pi^3) \end{aligned} \tag{4.31}$$

we can systematically expand this part of the action. The flavor (or isospin) trace only gives an overall factor 2, the color trace a factor $N_c$ and the Dirac trace a factor 4. The functional trace will be done using momentum eigenstates $\pi(q)$. As $A_2 = 0$ in the isospin limit we have to calculate the following parameter integral

$$\begin{aligned} &\int_0^1 d\zeta \int_0^{1-\zeta} d\eta \Big( e^{-s(\zeta+\eta)(k^2+m^2)} e^{-s(1-\zeta-\eta)((k+q)^2+m^2)} \\ &\qquad\qquad + e^{-s(\zeta+\eta)(k^2+m^2)} e^{-s(1-\zeta-\eta)((k+q)^2+m^2)} \Big) \\ = &\int_0^1 d\alpha e^{-s\alpha(k^2+m^2)} e^{-s(1-\alpha)((k+q)^2+m^2)}. \end{aligned} \tag{4.32}$$

[e] At this order the imaginary part the action does not contribute.

Since the momentum $k$ is only appearing in this exponential we may shift $k \to k-(1-\alpha)q$ in the second term without changing the value of the integral yielding

$$\int \frac{d^4k}{(2\pi)^4} \int_0^1 d\alpha \; e^{-s((k+(1-\alpha)q)^2+\alpha(1-\alpha)q^2+m^2)} = \int \frac{d^4k}{(2\pi)^4} \int_0^1 d\alpha \; e^{-s(k^2+\alpha(1-\alpha)q^2+m^2)}. \tag{4.33}$$

Using

$$4N_c \int \frac{d^4k}{(2\pi)^4} e^{-sk^2} = \frac{N_c}{4\pi^2} \tag{4.34}$$

as well as the definition of the incomplete $\Gamma$–function, the term bilinear in pseudoscalar fields arising from $\mathcal{A}_F$ reads

$$\int \frac{d^4q}{(2\pi)^4} \sum_{i=1}^{3} \frac{1}{2} \pi^i(-q)\pi^i(q) \; (-q^2) m^2 \frac{N_c}{4\pi^2} \int_0^1 d\alpha \Gamma(0, [m^2+\alpha(1-\alpha)q^2]/\Lambda^2). \tag{4.35}$$

This allows us now to extract the inverse pion propagator,

$$D_\pi^{-1}(q^2) = -\frac{m^0 m}{G_1} - \Pi(q^2), \tag{4.36}$$

where the polarization operator $\Pi(q^2)$ is given by

$$\begin{aligned} \Pi(q^2) &= q^2 f^2(q^2) \\ f^2(q^2) &= m^2 \frac{N_c}{4\pi^2} \int_0^1 dx \; \Gamma\Big(0, [m^2 + x(1-x)q^2]/\Lambda^2\Big). \end{aligned} \tag{4.37}$$

The Bethe–Salpeter equation which determines the physical meson masses $m_\pi$ is equivalent to the condition that the meson propagator has a pole:

$$D_\pi^{-1}(q^2 = -m_\pi^2) = 0. \tag{4.38}$$

Note that $f^2(q^2 = -m_\pi^2)$ then is the corresponding meson decay constant. We want to emphasize here that the Bethe–Salpeter equation (4.36) – (4.38) is the one in ladder approximation and that no other approximations as *e.g.* gradient (subsection 4.1) or heat kernel expansions have been made. This also implies that the decay constants are evaluated on the corresponding meson mass shell. The pion decay constant is then given by

$$f_\pi^2 = m^2 \frac{N_c}{4\pi^2} \int_0^1 dx \Gamma\Big(0, [m^2 - x(1-x)m_\pi^2]/\Lambda^2\Big). \tag{4.39}$$

Table 4.1: Mass parameters fixed in the meson sector of the NJL model. The kaon decay constant $f_K$ is predicted.

| $m$ (MeV) | $m_s$(MeV) | $m_s^0/m^0$ | $f_K$(MeV) |
|---|---|---|---|
| 350 | 577 | 23.5 | 104.4 |
| 400 | 613 | 22.8 | 100.3 |
| 450 | 650 | 22.5 | 97.4 |
| 500 | 687 | 22.3 | 95.5 |

One clearly sees that in the chiral limit ($m_\pi = 0$) the expression

$$f_\pi^2 = m^2 \frac{N_c}{4\pi^2} \Gamma(0, (m/\Lambda)^2) \tag{4.40}$$

calculated by means of a derivative expansion becomes exact. Using eqs. (4.36) and (4.37) it is trivial to show that in the chiral limit $m^0 = 0$ the Bethe–Salpeter equation (4.38) is solved by setting $q^2 = m_\pi^2 = 0$, *i.e.* the pions are Goldstone bosons. The mass shell condition (4.38) can be expressed as

$$m_\pi^2 f_\pi^2 = \frac{m^0 m}{G_1} = 2m^0 \langle \bar{u}u \rangle \frac{m}{m - m^0} \approx 2m^0 \langle \bar{u}u \rangle \tag{4.41}$$

where we used the gap equation (3.28) to express the coupling constant $G_1$ in terms of the quark condensates. This approximate relation is the Gell-Mann–Renner–Oakes relation [41], see eq. (2.24).

The generalization of the above calculation to the case of unequal quark masses, and especially the case of three flavors, can be found in Appendix A of ref.[62]. For the inverse propagator for the pseudoscalar mesons one obtains

$$D_{ij,kl}^{-1}(q^2) = \Big( - \frac{(m_i^0 + m_j^0)(m_i + m_j)}{4G_1} - \Pi_{ij}(q^2) \Big) \delta_{il}\delta_{kj}. \tag{4.42}$$

The polarization operator $\Pi_{ij}(q^2)$ is given by

$$\Pi_{ij}(q^2) = q^2 f_{ij}^2(q^2) + (m_i - m_j)^2 f_{ij}^2(q^2) - \frac{1}{4}(m_i^2 - m_j^2)\Big( \frac{\langle \bar{q}q \rangle_i}{m_i} - \frac{\langle \bar{q}q \rangle_j}{m_j} \Big) \tag{4.43}$$

wherein

$$f_{ij}^2(q^2) = \frac{1}{4}(m_i + m_j)^2 \frac{N_c}{4\pi^2} \int_0^1 dx \; \Gamma\Big(0, [(1-x)m_i^2 + x m_j^2 + x(1-x)q^2]/\Lambda^2\Big) \tag{4.44}$$

while the quark condensates are defined in eqs. (3.28). Obviously, these expressions coincide with the one given above if the flavor symmetric limit is chosen. For the kaon one simply uses $i = s$ and $j = u(d)$, *e.g.* the kaon decay constant can be obtained from eq. (4.44) by this choice of indices and evaluation at $q^2 = -m_K^2$ which in turn is obtained as the root of (4.42).

Numerical results are displayed in table 4.1. As input serves $m_\pi = 135$MeV, $m_K = 495$MeV and $f_\pi = 93$MeV. The up constituent quark mass is chosen as independent variable. One observes that the NJL model in the proper time regularization scheme underestimates the experimental value $f_K = 113$MeV.

For the other mesons (scalars, vectors and axialvectors) we will only give the propagators for the case of two flavors in the isospin symmetric limit. For the inverse propagator of the scalar mesons $\sigma$ and $a_0$ one obtains

$$D_\sigma^{-1}(q^2) = -\frac{m^0 m}{G_1} - (q^2 + 4m^2) f(q^2). \tag{4.45}$$

Note the similarity to the pion propagator (4.36). As a consequence the scalar meson mass can be written as

$$m_\sigma^2 = 4m^2 + m_\pi^2 \frac{f^2(-m_\pi^2)}{f^2(-m_\sigma^2)} \approx 4m^2 + m_\pi^2 \tag{4.46}$$

where for the last relation we have assumed $f^2(q^2)$ (4.37) a slowly varying function. However, there is a problem. From (4.46) one sees that the scalar meson mass lies above the quark–antiquark threshold, *i.e.* $m_\sigma^2 > 4m^2$. In this case $f^2(-m_\sigma^2)$, or equivalently $D_\sigma^{-1}$, acquires an imaginary part[f]

$$\mathrm{Im}(D_\sigma^{-1}) = (q^2 + 4m^2)\frac{N_c}{4\pi} m^2 \sqrt{1 + 4m^2/q^2} \quad \text{for} \quad 4m^2 \leq -q^2 \leq 4(m^2 + \Lambda^2). \tag{4.47}$$

Therefore the extraction of meson masses via the Bethe–Salpeter equation are obscured by the unphysical quark–antiquark threshold which is present in the (non–confining) NJL model.

For the calculation of the vector meson propagator it is convenient to split the expressions in the expansion of the fermion determinant in longitudinal and transverse ones. A naïve extraction of the longitudinal part of the inverse propagator yields

$$m^2 \Gamma(-1, m^2/\Lambda^2) - \int_0^1 dx \left( (m^2 + x(1-x)q^2)\Gamma(-1, Y) + 2(x - \frac{1}{2})^2 q^2 \Gamma(0, Y) \right) \tag{4.48}$$

where

$$Y = (m^2 + x(1-x)q^2)/\Lambda^2.$$

This expression identically vanishes as can be shown by a formal expansion in powers of $q^2$ and evaluating the Feynman parameter integrals for each power of $q^2$ separately. Note that it is not sufficient to argue that the proper time regularization is gauge invariant (and therefore the longitudinal part has to vanish). The small amplitude expansion for vector mesons which keeps only terms up to second order in the vector (gauge boson) field but all derivatives does explicitly violate (local) gauge invariance.

As the expression (4.48) vanishes only the mesonic action $\mathcal{A}_m$ contributes to the longitudinal part of the inverse vector meson propagator. The latter is given by

$$(D^{-1})^{\mu\nu}(q) = \left(\delta^{\mu\nu} - \frac{q^\mu q^\nu}{q^2}\right)\left(\frac{q^2}{g^2(q^2)} - \frac{1}{4G_2}\right) - \frac{q^\mu q^\nu}{q^2}\frac{1}{4G_2} \tag{4.49}$$

where

$$\frac{1}{g^2(q^2)} = \frac{N_c}{4\pi^2}\int_0^1 dx x(1-x)\Gamma(0, (m^2 + x(1-x)q^2)/\Lambda^2). \tag{4.50}$$

[f] For $-q^2 \geq 4(m^2 + \Lambda^2)$ additional (even more unphysical) imaginary parts appear.

The vector meson mass is determined by the on–shell condition

$$m_V^2 = \frac{1}{4G_2} g^2(q^2 = -m_V^2) \tag{4.51}$$

and $g_V^2 = g^2(q^2 = -m_V^2)$ is the universal vector meson coupling constant taken at the on–shell mass.

Including the axialvector meson changes the formulas for the pion propagator. This is due $\pi - a_1$ mixing, *i.e.* the occurrence of a term

$$f^2(q^2) i q_\mu \pi(q) A_\mu(-q)$$

in the small amplitude expansion. This modifies the inverse pion propagator:

$$D_\pi^{-1}(q^2) = q^2 \frac{f^2(q^2)}{1 + 4G_2 f^2(q^2)} - \frac{m^0 m}{G_1}. \tag{4.52}$$

Especially, the pion decay constant is then given by

$$f_\pi^2 = \frac{f^2(-m_\pi^2)}{1 + 4G_2 f^2(-m_\pi^2)} \tag{4.53}$$

where the pion mass is determined by the root of eq. (4.52). As a further consequence of $\pi - a_1$ mixing the axialvector propagator is different from the vector one. The axial–vector mass,

$$m_A^2 = g^2(q^2 = -m_A^2) \left( \frac{1}{4G_2} + f^2(q^2 = -m_A^2) \right), \tag{4.54}$$

is significantly larger than the vector mass and generally lies above the quark–antiquark threshold.

This subsection may be summarized as follows: The propagators for the scalar and pseudoscalar fields may generically be written as

$$D(q^2) = \frac{Z(q^2)}{q^2 + m^2(q^2)} \tag{4.55}$$

while those for the vector and axialvector fields read

$$D^{\mu\nu}(q^2) = \frac{Z(q^2)}{q^2 + m^2(q^2)} \left( \delta^{\mu\nu} - \frac{q^\mu q^\nu}{m^2(q^2)} \right) \tag{4.56}$$

with the functions $Z(q^2)$ and $m^2(q^2)$ given in table 4.2. For the scalar and axialvector mesons their masses lie above the quark–antiquark threshold making their determination via the Bethe–Salpeter equation doubtful. It should be remarked that a procedure, which extrapolates the polarization tensors from below to above the quark–antiquark threshold, has been proposed in ref.[63].

| Meson | $Z^{-1}(q^2)$ | $m^2(q^2)$ |
|---|---|---|
| $\sigma$,$a_0$ | $f^2(q^2)$ | $4m^2 + m^0 m/(G_1 f^2(q^2))$ |
| $\pi$ | $f^2(q^2)/(1 + 4G_2 f^2(q^2))$ | $m^0 m/(G_1 f^2(q^2))$ |
| $\omega$,$\rho$ | $1/g^2(q^2)$ | $g^2(q^2)/4G_2$ |
| $a_1$ | $1/g^2(q^2)$ | $g^2(q^2)(1/4G_2 + f^2(q^2))$ |

Table 4.2: The functions $Z(q^2)$ and $m^2(q^2)$ which determine the meson propagators, see eqs. (4.55) and (4.56). The quantities $f^2(q^2)$ and $g^2(q^2)$ are given in eqs. (4.37) and (4.50), respectively.

## 4.4 Chiral anomaly

In this section we will discuss how the chiral anomaly (see section 2.3) is represented in the effective meson theory (3.25). Defining the quantum theory via path integrals the anomalous symmetry breaking is realized by the non–invariance of the functional integral measure [49]; despite the fact that the classical action appearing as 'weight' factor under the path integral is invariant under the considered transformation. The integral measure

$$\mathcal{D}q\mathcal{D}\bar{q} \tag{4.57}$$

is not invariant under chiral rotations

$$\begin{aligned} q(x) &\rightarrow \chi(x) = (U(x))^{\gamma_5}\, q(x) = e^{i\Theta(x)\gamma_5} q(x)\,, \quad \Theta = \Theta^a \left(\frac{\lambda^a}{2}\right)_F \\ \bar{q}(x) &\rightarrow \bar{\chi}(x) = \bar{q}(x)\,(U(x))^{\gamma_5} \end{aligned} \tag{4.58}$$

but rather acquires a phase $J(\Theta) \neq 1$

$$\mathcal{D}\chi\mathcal{D}\bar{\chi} = J(\Theta)\mathcal{D}q\mathcal{D}\bar{q}\,. \tag{4.59}$$

This implies that for the effective meson theory (3.25) the anomaly is contained in the fermion determinant $\mathcal{A}_F$. The proof of anomalous chiral symmetry breaking, reflected by the relation $J(\Theta) \neq 1$, is given in Appendix A for the simplified case of a Dirac operator containing besides the usual kinetic and mass term only a vector field

$$i\not{D} = i(\not{\partial} + \not{V}) - m^0 =: i\not{d} - m^0. \tag{4.60}$$

In this case one obtains

$$J(\Theta) = \exp\left(-i\frac{N_c}{8\pi^2}\int d^4x\Theta(x)\mathrm{tr}_F(\tilde{F}F)\right) \tag{4.61}$$

where $F_{\mu\nu}$ is the field strength tensor corresponding to the vector field $V_\mu$ and $\tilde{F}_{\mu\nu}$ is the dual tensor. Note that the Jacobian $J(\Theta)$ is a pure phase factor, $|J(\Theta)| = 1$, and therefore contributes to the imaginary part of the effective action. In the presence of a vector field this factor will in general differ from one.

The expression (4.61) is the Jacobian for infinitesimal Abelian chiral rotations. Additionally, we want to know the Jacobian $J(\Theta)$ for finite non–abelian chiral transformations. It can be calculated by functional integration of the differential anomaly

$$J(\Theta) = \int \mathcal{D}\Theta \frac{\delta J(\Theta)}{\delta\Theta} = \exp(i\mathcal{A}_{\mathrm{WZW}}). \tag{4.62}$$

The term $\mathcal{A}_{\mathrm{WZW}}$ represents the logarithm of the integrated anomaly and is the Wess–Zumino–Witten (WZW) term [52]. In general it is a very complicated functional. We will therefore discuss in more detail only the special case of a flavor singlet vector meson ($\omega$–meson) which couples to the baryon number and a SU(2) chiral field,

$$\begin{aligned} V_\mu(x) &= \omega_\mu(x)\mathbb{1}_F \\ U(x) &= e^{i\Theta(x)}, \quad \Theta(x) = \Theta^a(x)\frac{\tau^a}{2}. \end{aligned} \tag{4.63}$$

The WZW term is then given by

$$\mathcal{A}_{\mathrm{WZW}} = N_c \int d^4x\ \omega_\mu(x) B^\mu(x) \tag{4.64}$$

where

$$B^\mu(x) = \frac{1}{24\pi^2}\epsilon^{\mu\nu\kappa\lambda}\mathrm{tr}(L_\nu L_\kappa L_\lambda), \quad L_\nu = U^\dagger \partial_\nu U \tag{4.65}$$

is the winding number current related to the chiral field $U(x)$. Due to its topological nature the current $B^\mu$ is conserved independent of the explicit form of the chiral field $\Theta(x)$.

In order to reveal the physical nature of this current we consider the quark generating functional treating the vector field $V_\mu$ as an external source

$$\begin{aligned} \exp(\mathcal{A}_F) &= \int \mathcal{D}q\mathcal{D}\bar{q} \exp\left(\int d^4x \bar{q}(i(\partial\!\!\!/ + V\!\!\!/) - \hat{m}^0)q\right) \\ &= J(\Theta)\exp(\tilde{\mathcal{A}}[V]) \\ &= \exp\left(iN_c \int d^4x V_\mu(x) B^\mu(x) + \tilde{\mathcal{A}}[V]\right) \end{aligned} \tag{4.66}$$

where $\tilde{\mathcal{A}}[V]$ contains the non–anomalous terms of the effective action. From the first equation we obtain

$$\left(\frac{\delta \mathcal{A}_F}{i\delta V_\mu(x)}\right)_{V=0} = \langle \bar{q}(x)\gamma^\mu q(x)\rangle =: j_B^\mu(x) \tag{4.67}$$

*i.e.* the baryon current of the quarks. The last equation allows us to relate the baryon current to the topological current

$$j_B^\mu(x) = B^\mu(x) + \ldots \tag{4.68}$$

where the dots indicate the contributions from the "normal" terms $\tilde{\mathcal{A}}[V]$. This leads to the following interpretation: The baryon current of the original fermion (quark) theory is carried by the topological current of the chiral field in the bosonized effective meson theory, at least in leading order gradient expansion. As mentioned above this topological current is conserved independent of the specific form of the chiral field whereas the baryon current of the underlying quark theory is the (conserved) Noether current of flavor singlet phase transformations. So, the dynamical property of the quark theory has turned into a purely topological property in the effective theory.

*It should be remarked that continuing to Euclidean space is equivalent to using Feynman boundary conditions in Minkowski space. This implies that the generating functional corresponds to the "vacuum to vacuum" transition amplitude. For infinitely large Euclidean times* $j_B^\mu$ *thus represents the vacuum part of the quark baryon current only.*

This immediately leads to the question how the pseudoscalar meson field $\Theta(x)$ which is bosonic and spinless can describe properties of baryons which are fermions with spin $s = 1/2$. We will see in section 5.2 that a sufficiently strong topological non–trivial chiral field with winding number $n$ polarizes the vacuum or Dirac sea of the quarks so strongly that $n$ of the valence quark levels are bound tightly enough that they join the negative Dirac sea: their energy is negative. As the physical vacuum is defined as the state with lowest energy all negative energy levels are occupied in the vacuum. The physical vacuum carries a non–vanishing baryon number if quark states are bound in the Dirac sea. So it is not really the chiral field itself which carries the baryon number but rather the polarized vacuum. As we cannot observe the vacuum but only the polarizing meson field we relate the baryon number to the chiral field. In that sense meson fields carry baryonic charge. This fact is the underlying feature for the description of the baryons as chiral solitons as will be discussed in more detail in the following chapter.

Let us close this section with a comment. Using the invariance of the classical action and the Jacobian (4.61) the anomalous Ward identity

$$\partial_\mu j_5^\mu = 2im^0 j_5 - i\frac{N_c}{8\pi^2}\int d^4x \mathrm{tr}_F\langle(\tilde{F}F)\rangle \tag{4.69}$$

can straightforwardly be derived (see Appendix A). Note that even for $m^0 = 0$ the axial singlet current $j_5^\mu$ is not conserved if vector fields are present despite the fact that the classical Lagrangian is invariant under chiral rotations for $m^0 = 0$. The axial anomaly as formulated by the anomalous Ward identity (4.69) is known as Adler–Bell–Jackiw anomaly [36]. Usually it is required that in fundamental (gauge) theories anomalies should be not present or should be canceled by other effects. As we are considering effective low–energy models anomalies can be present. They even have measurable consequences as *e.g.* the decay $\pi^0 \to 2\gamma$.

Historically, the anomaly was discovered in perturbation theory. The one–loop diagram which is responsible for the decay $\pi^0 \to 2\gamma$ is shown in figure 4.1. The path integral derivation of the anomaly is a non–perturbative one and demonstrates that higher order terms do **not** contribute to the anomaly. If one calculates the Adler–Bell–Jackiw anomaly in a renormalizable theory (as QCD) in which the cut–off $\Lambda$ is taken to infinity at the end of the calculation only the triangle diagram figure 4.1 contributes. However, if one works within an effective non–renormalizable theory where the ratio $m/\Lambda$ has to be kept finite, not only the triangle but all higher order diagrams contribute. Remember that the limit $\Lambda \to \infty$, *i.e.* $\tau \to 0$, (see Appendix A) was necessary in order to make the contributions from $h_n$, $n \geq 3$ vanish. In the literature [64] there are claims that the NJL model is not appropriate for the treatment of the anomaly and the $\pi^0$ decay because the triangle diagram gives only about two thirds of the experimentally determined decay amplitude using a finite cutoff. However, together with the finite cut–off also higher order terms should be included and presumably yield the missing third of the decay amplitude [65]. Note also that not regularizing the imaginary part of the action leads to a suppression of higher order terms for the $\pi^0$ decay, *i.e.* in this case the triangle diagram gives already the correct decay amplitude.

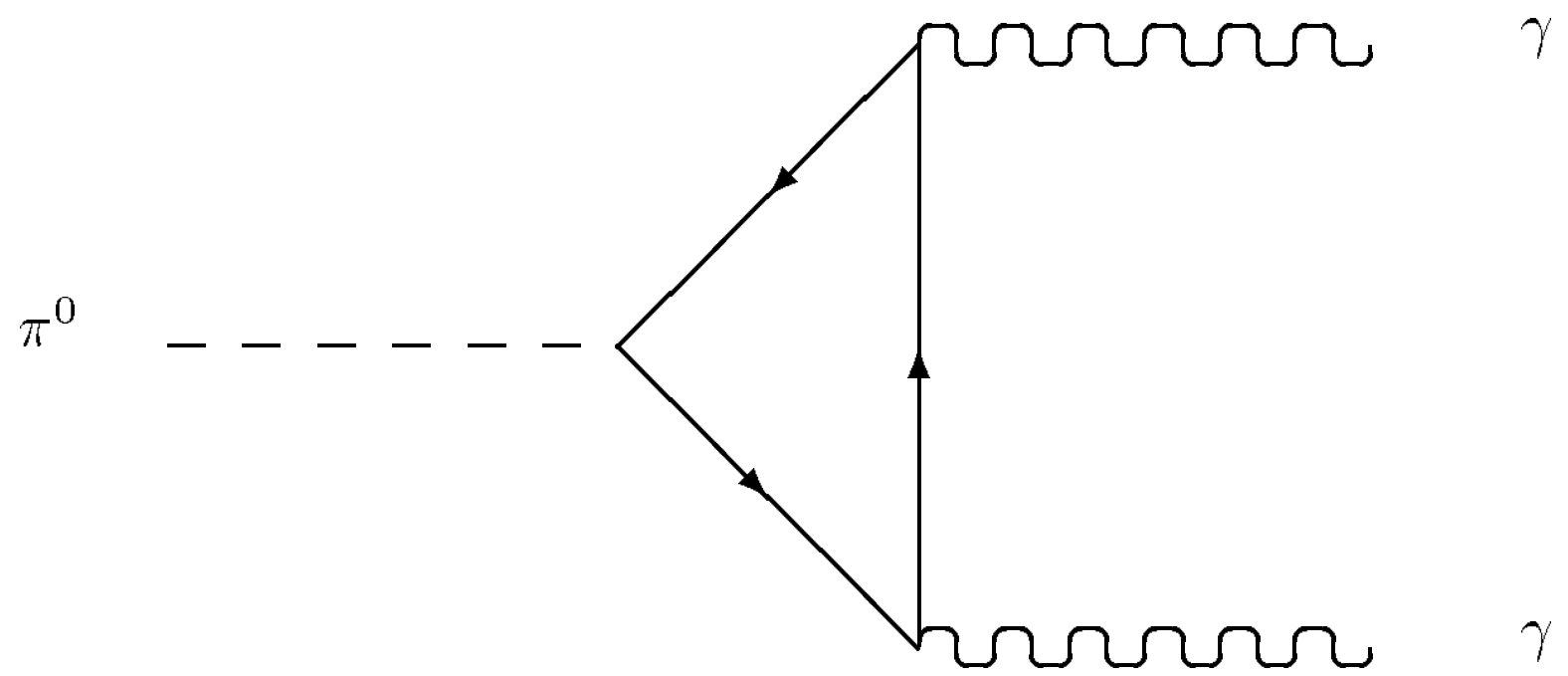

Figure 4.1: The triangle diagram which is responsible for the anomaly.

# 5 Chiral Solitons

This chapter is devoted to a brief discussion on the interpretation of solitons as baryons. Also an outlook is provided on calculations which are performed at length in chapters 6 and 7. For these introductory remarks on soliton physics we assume that some effective chirally invariant meson theory is given which allows for stable soliton solutions. When studying low energy properties it is suggestive to take into account only those mesons with the largest Compton wave–length, *i.e.* with the smallest mass. As explained in the preceding chapters these are the pseudoscalar would–be Goldstone bosons $\pi(x)$ of spontaneous chiral symmetry breaking. Again the non–linear realization will be adopted

$$U(x) = \exp\left(i\Theta(x)\right) \tag{5.1}$$

which defines the chiral field $U(x)$.

## 5.1 Topological properties

For two flavors a static chiral field maps the coordinate space into the group of isospin

$$U : \mathbb{R}^3 \rightarrow SU(2). \tag{5.2}$$

Solitons are finite energy field solutions to the Euler–Lagrange equations. Finiteness of the energy requires the energy density to vanish for $r = |\boldsymbol{r}| \rightarrow \infty$. This implies that the soliton configuration $U(x)$ asymptotically approaches a constant value (independent of the orientation $\hat{\boldsymbol{r}}$) which, as a consequence of chiral symmetry, can be chosen to be unity:

$$U(x) \rightarrow 1\!\!1 \quad \text{for} \quad r \rightarrow \infty. \tag{5.3}$$

Hence all points at spatial infinity are identified thereby compactifying $\mathbb{R}^3$ to a three dimensional sphere $S^3$. Since the group manifold of $SU(2)$ is also $S^3$ a static chiral field configuration with the boundary condition (5.3) represents a mapping from $S^3$ to $S^3$

$$U : S^3 \rightarrow S^3. \tag{5.4}$$

These mappings are distinguished by the winding number $\nu \in \mathbb{Z}$ which is a topologically invariant functional of the chiral field

$$\nu[U] = \int d^3x B^0(x). \tag{5.5}$$

Here $B^\mu(x)$ is the topological current (4.65).

A non-trivial mapping $U_0(\boldsymbol{r})$ is given by the so-called hedgehog *ansatz* (*cf.* subsection 6.2.1)

$$U_0(\boldsymbol{r}) = \exp\left(i\boldsymbol{\tau}\cdot\hat{\boldsymbol{r}}\Theta(r)\right). \tag{5.6}$$

For this field configuration the isospin vector points into the radial direction. Uniqueness of the chiral field at $r \rightarrow 0$ requires the chiral angle to obey the boundary condition

$$\Theta(r=0) = -n\pi, \qquad n \in \mathbb{Z}. \tag{5.7}$$

For the hedgehog configuration (5.6) the spatial components of the topological current vanish while

$$B^0(\boldsymbol{r}) \;=\; \frac{1}{2\pi^2}\Theta'(r)\frac{\sin^2\Theta(r)}{r^2}. \tag{5.8}$$

Without loss of generality one may choose $\Theta(r \to \infty) = 0$ to satisfy (5.3) yielding

$$\nu[U] = n. \tag{5.9}$$

Let us emphasize at this point that the winding number $\nu = n$ is a purely topological property of the chiral field, which *a priori* is not related to any physically relevant quantity.

## 5.2 Emergence of the soliton

In general the effective meson theory results from integrating out the quark and gluon degrees of freedom contained in QCD. Then the baryons, which are originally built from $N_C$ quarks, emerge as solitons of meson degrees of freedom. In order to explain how the baryonic character is encoded in the topological properties of the soliton we examine the spectrum of constituent quarks in the background of the chiral field. The corresponding Dirac Hamiltonian reads

$$h = \boldsymbol{\alpha}\cdot\boldsymbol{p} + \beta m (U)^{\gamma_5}, \tag{5.10}$$

where $m$ denotes the mass of the constituent quarks. The hedgehog field configuration (5.6) violates the spin and isospin symmetries[a], *i.e.*

$$[h, \boldsymbol{j}] = -[h, \boldsymbol{t}] \neq 0, \tag{5.11}$$

but preserves the grand spin

$$\boldsymbol{G} = \boldsymbol{j} + \boldsymbol{t} \tag{5.12}$$

symmetry, *i.e.* $[h, \boldsymbol{G}] = 0$. Hence the eigenfunctions of $h$ carry good grand spin and parity. In order to examine the spectrum of (5.10) we parametrize the chiral angle by [66]

$$\Theta(r) = -n\pi \begin{cases} 1 - \frac{2r}{3a} & \text{for } r \le a \\ \frac{a^2}{3r^2} & \text{for } r \ge a \end{cases} . \tag{5.13}$$

This special form is motivated by the fact that for small $r$ the chiral angle is linear whereas in the chiral limit it is proportional to $1/r^2$ for large $r$. The chiral angle and its derivative are continuous at the matching point $r = a$, which parametrizes the spatial extension of the chiral angle. Figure 5.1 shows the eigenvalues $\epsilon_\nu$ of the Dirac-Hamiltonian (5.10) for a chiral field with winding number $n = 1$ as a function of the strength of the chiral field measured by $a \cdot m$. As the strength of the chiral field increases, the lowest valence quark state in the $G^\pi = 0^+$ channel becomes strongly bound and eventually joins the Dirac sea. Figure 5.2 shows the spectrum of constituent quarks in the background field of a hedgehog with winding number $n = 2$. In this case a second valence quark state $G^\pi = 0^-$ becomes bound in the Dirac sea for sufficiently large $a \cdot m$. In general, strong chiral fields with winding number $n$ bind exactly $n$ valence quark orbits in the negative Dirac sea. Obviously the baryon charge of the valence quarks gets encoded in the topological structure of the soliton when describing baryons in the framework of effective meson theories.

[a] The quantities $\boldsymbol{j}$ and $\boldsymbol{t}$ denote the single quark total angular momentum and isospin operators, respectively.

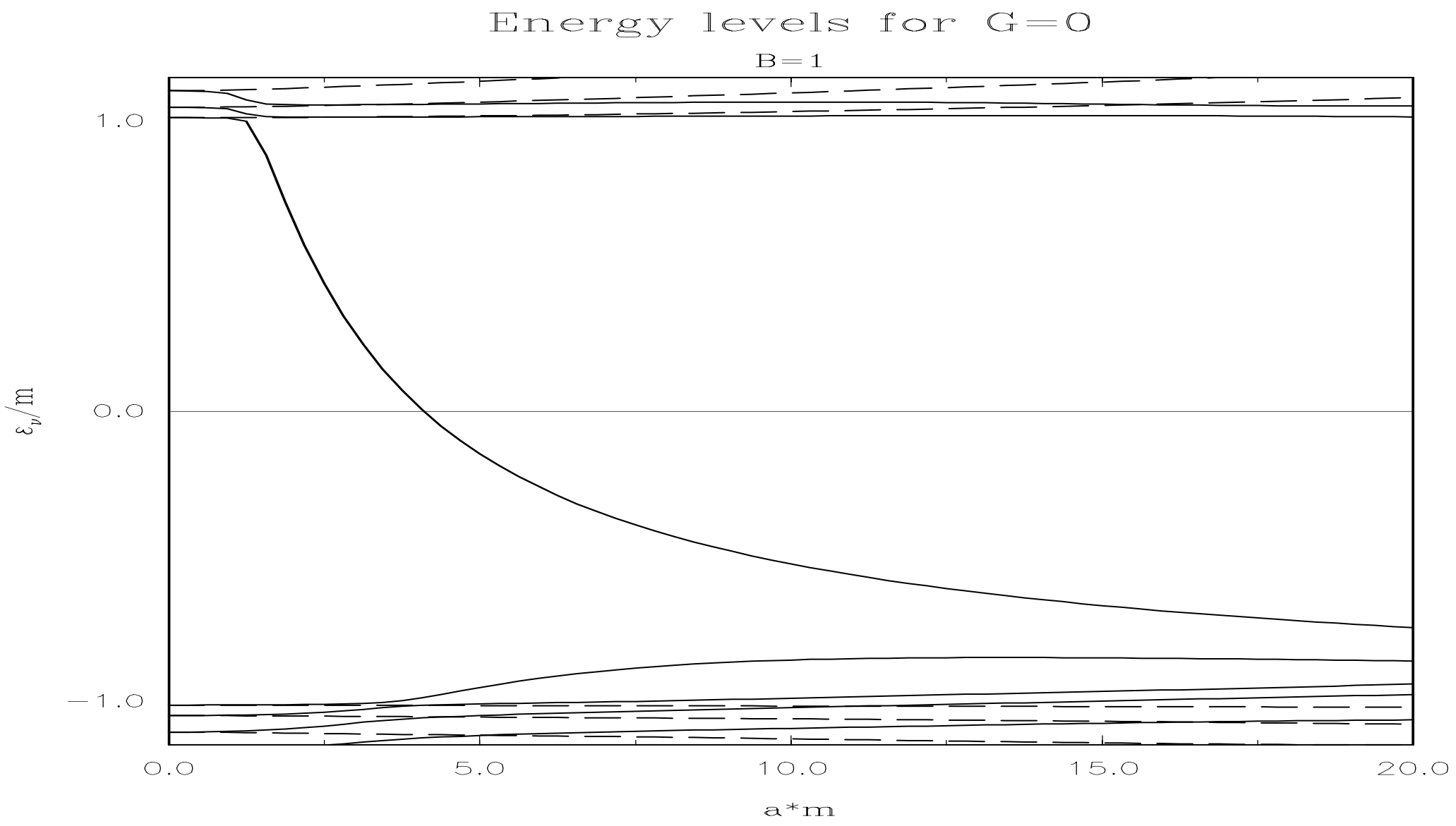

Figure 5.1: The eigenvalues of the Dirac Hamiltonian (5.10) in the background of a chiral field (5.13). Displayed are the lowest eigenvalues in the $0^+$ (full lines) and $0^-$ (dashed lines) channels.

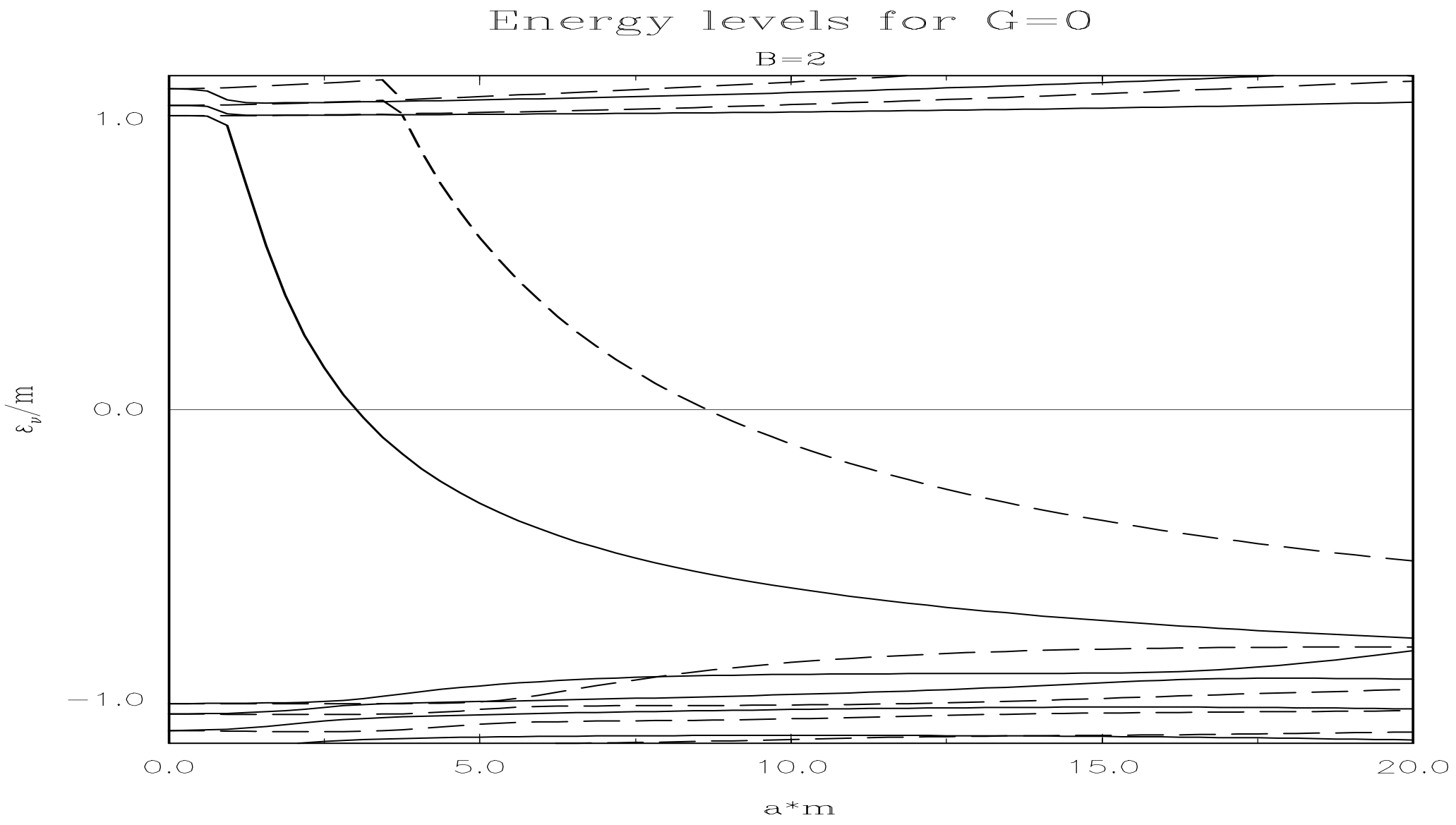

Figure 5.2: Same as figure 5.1 for $n = 2$.

By definition, all positive energy states are empty in the vacuum configuration while all negative energy states are occupied. Hence for sufficiently strong chiral fields the vacuum possesses a baryonic charge which is identical to the winding number. As the polarizing meson field is observable rather than the vacuum, the baryonic properties of the charged vacuum are attributed to this meson cloud. This justifies the statement that baryons emerge as solitons in the effective meson theory. Strictly speaking, however, the baryonic charge is carried by the polarized Dirac vacuum. To make this scenario more explicit it is helpful to consider the baryon number carried by the vacuum, which is defined as the asymmetry of the Dirac spectrum (see eq (6.12))

$$B^{\rm vac} = -\frac{1}{2}\sum_\nu {\rm sgn}(\epsilon_\nu). \tag{5.14}$$

For a vanishing chiral field the Dirac spectrum is symmetric and therefore $B^{\rm vac} = 0$. However, as the strength of chiral field increases the energy eigenvalue of the valence quark orbits eventually reverse their signs leading to $B^{\rm vac} = n$.

These studies provide the following reasoning for the solitonic picture of baryons: The valence quarks of the baryons with baryon number $B$ generate a strongly localized chiral field of winding number $n = B$. In the background of this field the valence quarks are bound in such a way that the baryonic charge is carried by the polarized vacuum, *i.e.* $B^{\rm vac} = n$. Once the valence quarks have joined the Dirac sea, the polarizing chiral meson cloud is interpreted as the baryon.

## 5.3 Semiclassical quantization

As already mentioned after eq (5.10) neither spin nor isospin represent good quantum numbers of the soliton. Here we will briefly describe a treatment which generates states carrying quantum numbers of physical baryons. For details we refer to section 7.1.

Time–independent rotations in coordinate– and/or isospace do not alter the energy of the soliton. In order to perform the semiclassical quantization one requires time dependent solutions to the equations of motion. These are approximated by allowing these rotations to adiabatically vary in time. Due to the grand spin symmetry the rotations in coordinate– and iso–space are equivalent. It therefore suffices to write

$$U(\boldsymbol{r},t) = R(t)U_0(\boldsymbol{r})R^\dagger(t), \tag{5.15}$$

where $R(t)$ denotes an SU(2) isospin matrix. Defining the angular velocity $\boldsymbol{\Omega}$ by

$$R^\dagger(t)\frac{\partial}{\partial t}R(t) = \frac{i}{2}\boldsymbol{\Omega}\cdot\boldsymbol{\tau}, \tag{5.16}$$

the energy functional $E[U]$ of the rotating soliton configuration can be expanded in powers of $\boldsymbol{\Omega}$ [b]

$$E[U = RU_0R^\dagger] = E[U_0] + \frac{1}{2}\Omega_i\Theta^{ij}[U_0]\Omega_j + \cdots. \tag{5.17}$$

[b]Since isospin breaking is neglected the rotation matrix $R$ does not appear explicitly.

The term linear in $\boldsymbol{\Omega}$ is missing due to time reflection symmetry which holds for the isospin symmetric two flavor model. The moment of inertia tensor

$$\Theta^{ij}[U_0] = \left.\frac{\partial^2 E[U]}{\partial\Omega_i \partial\Omega_j}\right|_{\boldsymbol{\Omega}=0} = \delta^{ij}\alpha^2 \tag{5.18}$$

is diagonal because no direction in isospace is distinguished. The canonical quantization of the coordinates $R$ corresponds to the replacement

$$\alpha^2 \boldsymbol{\Omega} \rightarrow \boldsymbol{J}, \tag{5.19}$$

with $\boldsymbol{J}$ being the collective spin operator. The quantized energies of the baryons are finally obtained to be

$$E[U] = E[U_0] + \frac{J(J+1)}{2\alpha^2} \tag{5.20}$$

where $J(J+1)$ is the eigenvalue of $\boldsymbol{J}^2$.

Due to the grand spin symmetry of the hedgehog the spin of the quantized soliton coincides with the collective isospin $\boldsymbol{I}$ up to a rotation. The quantized hedgehog soliton therefore yields a tower (rotational band) of states with $I = J$. Generalizing this treatment to three flavor models shows that for an odd (even) number of colors the spin is constrained to half–integer (integer) values [52, 26, 27]. For $N_C = 3$ the nucleon ($I = J = \frac{1}{2}$) and the $\Delta(I = J = \frac{3}{2})$ resonance thus represent ground and first excited states, respectively.

Both, the classical soliton mass and the moment of inertia are of order $N_C$. Therefore the classical soliton mass $E[U_0]$ and the rotational energy $I(I+1)/2\alpha^2$ are of order $N_C$ and $1/N_C$, respectively. However, a systematic expansion of the baryon energy in powers of $1/N_C$

$$E[U] = E^{(-1)} + E^{(0)} + E^{(1)} + \cdots \tag{5.21}$$

contains in addition to (5.20) a term of order $(1/N_C)^0$, $E^{(0)}$. This term, which presumably dominates over the rotational energy, is generated from small amplitude mesonic fluctuations off the soliton and is associated with the quantum corrections to $E[U_0]$. Naïvely one would expect

$$E^{(0)} = \frac{1}{2}\left(\sum_i \omega_i - \sum_i \omega_i^{(0)}\right), \tag{5.22}$$

where $\omega_i$ and $\omega_i^{(0)}$ refer to the eigen–frequencies of the meson fluctuations in presence and absence of the soliton, respectively. However, the expression (5.22) is UV divergent and thus subject to regularization. Nevertheless, (5.22) is illuminating since it indicates that the dominant contribution to $E^{(0)}$ stems from the zero–modes ($\omega_i = 0$) because there are no counterparts in the absence of the soliton. Moreover, the contributions of the zero–modes is negative and hence causes a substantial reduction of the classical mass. This is a desired feature since the classical mass commonly overestimates the baryon masses once the parameters are fitted to mesonic data (see chapter 4). Recently, it has been shown for the Skyrme model that a *renormalized* version of (5.22) indeed predicts baryon masses which well agree with experimental data [24, 25]. The analogous calculation [67] for the NJL model is presented in section 7.4.

# 6 Static solitons of the Nambu–Jona-Lasinio model

In the present section we will discuss the emergence as well as properties of static solitonic meson configurations within the NJL model. The most important ingredient of the effective meson action of the bosonized NJL model is the fermion determinant (3.25). In practice we have to evaluate this determinant in the presence of non–perturbative meson field configurations. This determinant is given in terms of the eigenvalues of the Dirac operator $\not{D}$. However, since $\det(i\beta) = 1$ one may equally well consider

$$i\beta\not{D} = i\partial_t - h \tag{6.1}$$

which introduces the one–particle Hamilton operator $h$. For the time being we consider $h$ to be static and Hermitian. The discussion of more general cases will be postponed to later subsections. For static fields the eigenvalues of (6.1) separate into the eigenvalues of $i\partial_t$ and $h$. The fermion fields assume anti–periodic boundary conditions on the time interval $T$. Therefore the eigenvalues of $i\partial_t$ are given by the Matsubara frequencies $\Omega_n = (2n+1)\pi/T$ with $n = 0, \pm 1, \pm 2, \ldots$. Denoting furthermore the eigenvalues of $h$ by $\epsilon_\nu$ the fermion determinant is obtained as the product[a] [68, 70]

$$\mathrm{Det}\,(\not{D}) = \prod_\nu \prod_n \left(\frac{2n+1}{T}\pi - \epsilon_\nu\right) = \mathcal{C} \prod_\nu \prod_{n\ge 0} \left(1 - \left[\frac{\epsilon_\nu T}{(2n+1)\pi}\right]^2\right), \tag{6.2}$$

since for static configurations $i\partial_t$ and $h$ may be diagonalized simultaneously. The constant

$$\mathcal{C} = \prod_\nu \prod_{n\ge 0} \left(-\left[\frac{2n+1}{T}\pi\right]^2\right) \tag{6.3}$$

does not depend on the dynamical properties of the system and may hence be absorbed into the integration measure. The product over $n$ is readily carried out

$$\begin{aligned} \mathrm{Det}\,(\not{D}) &= \mathcal{C} \prod_\nu \cos\left(\frac{\epsilon_\nu T}{2}\right) \\ &= \tilde{\mathcal{C}} \exp\left[\frac{i}{2}\sum_\nu |\epsilon_\nu| T\right] \prod_\nu \left(1 + \exp\left[-iT|\epsilon_\nu|\right]\right). \end{aligned} \tag{6.4}$$

With the introduction of occupation numbers $\eta_\nu = 0, 1$ the product over $\nu$ may finally be expressed as

$$\prod_\nu \left(1 + \exp\left[-iT|\epsilon_\nu|\right]\right) = \sum_{\{\eta_\nu\}} \exp\left[-iT\sum_\nu \eta_\nu |\epsilon_\nu|\right], \tag{6.5}$$

where the sum goes over all possible combinations of $\eta_\nu = 0, 1$. Then the fermion determinant acquires the form

$$\mathrm{Det}\,(\not{D}) = \tilde{\mathcal{C}} \exp\left[i\mathcal{A}_0\right] \sum_{\{\eta_\nu\}} \exp\left[i\mathcal{A}_V^{\{\eta_\nu\}}\right] \tag{6.6}$$

[a] See also chapter 9 of ref. [69].

which provides a natural decomposition into vacuum

$$\mathcal{A}_0 = \frac{1}{2} T \sum_\nu |\epsilon_\nu| \tag{6.7}$$

and valence (anti–) quark

$$\mathcal{A}_V^{\{\eta_\nu\}} = -T \sum_\nu \eta_\nu |\epsilon_\nu| = -T E_V^{\{\eta_\nu\}} \tag{6.8}$$

contributions to the fermion determinant.

In order to equip the occupation numbers $\eta_\nu$ with a physical meaning let us consider the baryon number current which is defined as the average

$$j^\mu(x) = \langle \bar{q}(x)\gamma^\mu q(x)\rangle = \frac{i\delta}{\delta V_\mu(x)} \log \, \mathrm{Det}\,(\not{D} - i\not{V}) \Big|_{V^\mu(x)=0}. \tag{6.9}$$

Treating $V_\mu$ as a perturbation in the eigenvalue problem $(h + \beta\not{V})\psi_\nu = \epsilon_\nu \psi_\nu$ reveals that

$$\frac{\delta \epsilon_\nu}{\delta V_\mu(x)}\Big|_{V^\mu(x)=0} = \psi_\nu^\dagger(x)\beta\gamma^\mu\psi_\nu(x) \tag{6.10}$$

with $\psi_\nu(x)$ and $\epsilon_\nu$ being the eigenstates and –values of $h$. It is then easy to see that according to (6.6) the baryon number current is additive in vacuum and valence parts [70]

$$\begin{aligned} j^\mu(x) &= j_0^\mu(x) + j_V^\mu(x) \\ j_0^\mu(x) &= -\frac{1}{2}\sum_\nu \mathrm{sgn}(\epsilon_\nu)\bar{\psi}_\nu(x)\gamma^\mu\psi_\nu(x) \\ j_V^\mu(x) &= \sum_\nu \eta_\nu \mathrm{sgn}(\epsilon_\nu)\bar{\psi}_\nu(x)\gamma^\mu\psi_\nu(x). \end{aligned} \tag{6.11}$$

Taking into account that the eigenfunctions $\psi_\nu$ are properly normalized one obtains the baryon number

$$B = \sum_\nu \left(\eta_\nu - \frac{1}{2}\right) \mathrm{sgn}(\epsilon_\nu). \tag{6.12}$$

Thus, considering a special set $\{\eta_\nu\}$ of occupation numbers confines the system to a sector with definite baryon number.

Up to now we have ignored the fact that the vacuum part of the fermion determinant (6.7) is divergent and hence needs regularization. We will address this problem in the next subsection when extracting the energy functional from $\mathcal{A}_0$. Furthermore, a more rigorous treatment of symmetry currents for the regularized theory will be presented in chapter 7.

## 6.1 The energy functional

We have already observed that the contribution to the energy functional due to the explicit occupation of the valence quark orbits is given by (*cf.* eq (6.8))

$$E_V = \sum_\nu \eta_\nu |\epsilon_\nu| \tag{6.13}$$

with $\epsilon_\nu$ being the eigenvalues of a one–particle Hermitian Dirac Hamiltonian $h$. This quantity may be extracted from the Dirac operator via eq (6.1).

The vacuum (or ground state) contribution $E_0$ to the energy is extracted from the functional integral by continuing to Euclidean times $\tau = ix_0$ and observing that for large Euclidean time intervals, $T \to \infty$, the ground state provides the dominant contribution, *i.e.*

$$\lim_{T\to\infty} \mathrm{Det}(i\beta \not{D}_E) \propto \exp\left(-E_0 T\right). \tag{6.14}$$

Here $\not{D}_E$ denotes the Euclidean Dirac operator which is obtained from $\not{D}$ (6.1) by analytic continuation. It is important to note that in Euclidean space $\tau$ has to be considered a real quantity.

An Hermitian one–particle Dirac Hamiltonian has been the starting–point of the preceding considerations. We will waive this assumption from now on and allow $h$ to contain anti–Hermitian parts as well. Nevertheless, the Euclidean Dirac operator for static meson fields is decomposed into a temporal part and a static Euclidean Hamiltonian

$$i\beta \not{D}_E = -\partial_\tau - h. \tag{6.15}$$

Commonly the non–Hermiticity of $h$ stems from continuing the time components of (axial) vector fields to Euclidean space: $V_0 \to -iV_4$ and $A_0 \to -iA_4$. The Euclidean Dirac Hamiltonian then reads

$$h = \boldsymbol{\alpha}\cdot\boldsymbol{p} + V_4 + \gamma_5 A_4 + i\boldsymbol{\alpha}\cdot\boldsymbol{V} + i\gamma_5\boldsymbol{\alpha}\cdot\boldsymbol{A} + \beta(P_R M + P_L M^\dagger). \tag{6.16}$$

Furthermore our notation for the (axial) vector meson implies to take $V_\mu$ and $A_\mu$ anti–Hermitian. Especially the anti-Hermiticity of $V_4$ and $A_4$ causes the eigenvalues of (6.16) to be complex $\epsilon_\nu = \epsilon_\nu^R + i\epsilon_\nu^I$. Hence the eigenvalues $\lambda_{n,\nu}$ of the operator $\partial_\tau + h$ are given by

$$\lambda_{n,\nu} = -i\Omega_n + \epsilon_\nu = -i\Omega_n + \epsilon_\nu^R + i\epsilon_\nu^I. \tag{6.17}$$

Let us (for the moment) ignore regularization and perform manipulations analogous to those described above for the special case of $h$ being Hermitian. The fermion determinant in Euclidean space can again be shown to separate into vacuum and valence contributions

$$\mathrm{Det}\,(\not{D}_E) = \tilde{\mathcal{C}}\,\exp\left(-TE_0\right)\sum_{\{\eta_\nu\}}\exp\left(-T\sum_\nu \eta_\nu\bar{\epsilon}_\nu\right) \tag{6.18}$$

with the vacuum energy $E_0 = -(N_C/2)\sum_\nu \bar{\epsilon}_\nu$. The quantities $\bar{\epsilon}_\nu$ are defined in terms of the real and imaginary parts of the eigenvalues of $h$

$$\bar{\epsilon}_\nu = |\epsilon_\nu^R| + i\;\mathrm{sgn}(\epsilon_\nu^R)\epsilon_\nu^I. \tag{6.19}$$

The contribution of the valence orbits to the (Euclidean) energy functional can be read off (6.18) as

$$E_V = E_V^R + iE_V^I = N_C\sum_\nu \eta_\nu|\epsilon_\nu^R| + iN_C\sum_\nu \eta_\nu \mathrm{sgn}(\epsilon_\nu^R)\epsilon_\nu^I \tag{6.20}$$

in the case that $h$ contains anti–Hermitian parts. We have also made explicit the dependence on $N_C$.

At this point a few remarks on the analytic properties of the eigenvalues $\epsilon_\nu$ are in order. Considering the generalization of the Euclidean Dirac Hamiltonian (6.16) defined by substituting $V_0 \to zV_4$ and $A_0 \to zA_4$ the eigenvalues $\epsilon_\nu$ actually become functions of the complex variable $z$ [71, 72]. It has been observed that for physically motivated field configurations these eigenvalues are analytic in $z$. This has been achieved by numerically verifying that the Laurent series for $\epsilon_\nu(z)$ indeed reduce to Taylor series[71]. The relevant Cauchy integrals are computed by parametrizing the path $\zeta(\varphi) = \delta e^{i\varphi} - z_0$. Here $z_0$ refers to the center of the Laurent expansion while a variation of $\delta$ provides a means to extract the radius of convergence. Then the eigenvalues of the generalized Dirac Hamiltonian are computed along the path $(0 < \varphi < 2\pi)$ and substituted into the Cauchy integrals. The eigenvalues are then found to exhibit an analytic structure for a radius of convergence of the order unity or even larger. This analyticity is not *a priori* clear[b] because the eigenvalues represent roots of the characteristic polynomial, which carries a large degree (infinitely large in the continuum case). For the relevant field configuration it has furthermore been shown that $\epsilon_\nu(z)^* = \epsilon_\nu(z^*)$. Obviously all quantities which depend on $\epsilon_\nu^R(z,z^*) = (\epsilon_\nu(z) + \epsilon_\nu(z^*))/2$ and $\epsilon_\nu^I(z,z^*) = -i(\epsilon_\nu(z) - \epsilon_\nu(z^*))/2$ separately are no longer analytic functions of $z$. Such quantities are *e.g.* $\bar{\epsilon}_\nu$. Next we consider the special parametrization $z = e^{i\varphi}$, which for $0 \le \varphi \le \pi/2$ describes the path connecting Euclidean and Minkowski spaces. In case the sign of $\epsilon_\nu^R(z,z^*)$ is reversed along this path the analytic structure of $E_0$ is obviously destroyed. In ref.[71] it has, however, been demonstrated that the total energy for a subsystem with unit baryon number, $B = 1$, can be written as

$$\frac{N_C}{2}\left(\epsilon_{\rm val} - \sum_{\nu \neq {\rm val}} \bar{\epsilon}_\nu\right). \tag{6.21}$$

The valence quark level (val) refers to the state with the smallest module $|\epsilon_\nu^R|$. Thus a change of sgn $\left(\epsilon_{\rm val}^R\right)$ does actually not destroy the analytic properties. As the other states ($\nu \neq$ val) vary only mildly along the path $z = e^{i\varphi}$, analyticity is formally maintained for the unregularized energy functional when constraining oneself to configurations with unit baryon number. Unfortunately, a different type of non–analyticity exists. Level crossings along the path connecting Euclidean and Minkowski spaces will appear if the time components $V_4$ and/or $A_4$ are strong enough. This makes a definition of a Minkowski energy functional by analytic continuation impossible. Thus, in order to attach physical significance to a given field configuration one has to check that no such level crossing occurs. When these level crossings are avoided the unregularized energy functional is analytic and the continuation forth and back from Euclidean to Minkowski spaces can be performed. In this context it is then evident that models like the chiral quark model with the $\omega$–meson included [74] indeed exhibit analytical structures.

Next we have to face the problem of regularization which has been ignored for the static energy functional up to here. Eq (6.18) demonstrates that for $T \to \infty$ only the vacuum (ground) state contributes to the functional integral since only the term with all $\eta_\nu = 0$ survives in the sum (6.18). However, the expression for $E_0$ is divergent and thus needs regularization. As in the study of the meson sector (*cf.* chapter 4) we will employ the proper–time regularization scheme [48] which substitutes the logarithm by a parameter integral (3.27). Unfortunately, this procedure is only applicable when the argument of the logarithm is positive. We therefore decompose the contribution of the fermion determinant to the mesonic action into real ($\mathcal{A}_R$)

[b] See *e.g.* appendix C of ref.[73].

and imaginary ($\mathcal{A}_I$) parts

$$\mathcal{A}_F = \mathcal{A}_R + \mathcal{A}_I = \frac{1}{2}\mathrm{Tr}\log\left(\not{D}_E^\dagger \not{D}_E\right) + \frac{1}{2}\mathrm{Tr}\log\left((\not{D}_E^\dagger)^{-1}\not{D}_E\right). \qquad (6.22)$$

In terms of the eigenvalues $\lambda_{n,\nu}$ (6.17) this decomposition is expressed as

$$\mathcal{A}_R = \frac{1}{2}\sum_{\nu,n}\log\left(\lambda_{n,\nu}\lambda^*_{n,\nu}\right) \qquad \text{and} \qquad \mathcal{A}_I = \frac{1}{2}\sum_{\nu,n}\log\left(\frac{\lambda_{n,\nu}}{\lambda^*_{n,\nu}}\right). \qquad (6.23)$$

Note that at this point the rules for manipulating the logarithm have been used in the sense that Tr $\log(\not{D}_E)$ and Tr $\log(\not{D}_E^\dagger)$ have been computed independently. This obviously represents an approximation because the traces of $h$ and $h^\dagger$ are associated with different Hilbert spaces. In ref.[75] it has been demonstrated that such a treatment may induce an error which is of quadratic or higher order in the time component of the (axial) vector fields.

The proper–time prescription can obviously be applied to this form of $\mathcal{A}_R$ yielding

$$\mathcal{A}_R = \frac{1}{2}\sum_{\nu,n}\log\left((\Omega_n - \epsilon^I_\nu)^2 + (\epsilon^R_\nu)^2\right) \qquad (6.24)$$

$$\to -\frac{1}{2}\sum_{\nu,n}\int_{1/\Lambda^2}^\infty \frac{ds}{s}\exp\left\{-s\left((\Omega_n - \epsilon^I_\nu)^2 + (\epsilon^R_\nu)^2\right)\right\}. \qquad (6.25)$$

According to the above discussion the expression (6.24) can only be considered as an approximation to the action in the presence of $V_4$ and/or $A_4$ rather than being exact.

For large Euclidean time intervals ($T \to \infty$) the sum over $n$ in (6.25) may now be replaced by an integral

$$\mathcal{A}_R = -\frac{T}{2}\sum_\nu \int_{-\infty}^\infty \frac{dz}{2\pi}\int_{1/\Lambda^2}^\infty \frac{ds}{s}\exp\left\{-s\left(z^2 + (\epsilon^R_\nu)^2\right)\right\} \qquad (6.26)$$

where we have furthermore shifted the integration variable[c] $z - \epsilon^I_\nu \to z$. As mentioned above, the limit $T \to \infty$ allows one to extract the vacuum contribution to the real part of the energy functional from $\mathcal{A}_R \to -TE^R_0$ which coincides with the corresponding expression derived from a Hermitian Hamiltonian [70]:

$$\begin{aligned} E^R_0 &= \frac{N_C}{4\sqrt{\pi}}\sum_\nu |\epsilon^R_\nu|\Gamma\left(-\frac{1}{2}, \left(\frac{\epsilon^R_\nu}{\Lambda}\right)^2\right) \\ &= \frac{N_C}{2}\sum_\nu\left\{\frac{\Lambda}{\sqrt{\pi}}\exp\left(-\left(\frac{\epsilon^R_\nu}{\Lambda}\right)^2\right) - |\epsilon^R_\nu|\mathcal{N}_\nu\right\} \end{aligned} \qquad (6.27)$$

where

$$\mathcal{N}_\nu = \frac{1}{\sqrt{\pi}}\Gamma\left(\frac{1}{2}, \left(\frac{\epsilon^R_\nu}{\Lambda}\right)^2\right) = \mathrm{erfc}\left(\left|\frac{\epsilon^R_\nu}{\Lambda}\right|\right) \qquad (6.28)$$

are the "vacuum occupation numbers" in the proper time regularization scheme, which for $\Lambda \to \infty$ reduce to $\mathcal{N}_\nu = 1$.

[c] Since the integral (6.26) converges absolutely, the sum over $\nu$ and the integral over $z$ may be exchanged.

For the imaginary part (6.23) we obtain (again to be considered with some caution as eq (6.24))

$$\mathcal{A}_I = \frac{1}{2}\left(\sum_\nu \sum_{n=-\infty}^{\infty} \log(\lambda_{\nu,n}) - \sum_\nu \sum_{n=-\infty}^{\infty} \log(\lambda^*_{\nu,n})\right) = \frac{1}{2}\sum_\nu \sum_{n=-\infty}^{\infty} \log\frac{i\Omega_n - \epsilon_\nu}{i\Omega_n - \epsilon^*_\nu} \qquad (6.29)$$

where we have reversed the sign in the first sum over the integer variable $n$. Next we express $\mathcal{A}_I$ in terms of a parameter integral

$$\mathcal{A}_I = \frac{1}{2}\sum_\nu \sum_{n=-\infty}^{\infty} \int_{-1}^{1} d\lambda \frac{-i\epsilon^I_\nu}{i\Omega_n - \epsilon^R_\nu - i\lambda\epsilon^I_\nu}. \qquad (6.30)$$

In analogy to (6.27) we may carry out the temporal trace in the limit $T \to \infty$:

$$\mathcal{A}_I = \frac{-i}{2}\sum_\nu \int_{-1}^{1} d\lambda\, T\ \mathcal{P}\int_{-\infty}^{\infty} \frac{dz}{2\pi} \frac{\epsilon^I_\nu}{i(z-\lambda\epsilon^I_\nu) - \epsilon^R_\nu}. \qquad (6.31)$$

Care has to be taken when performing the $z$–integration because only its principle value ($\mathcal{P}$) is properly defined. Next the shift in the integration variable $z - \lambda\epsilon^I_\nu \to z$ is performed. This, of course, also effects the boundaries

$$\mathcal{A}_I = \frac{-i}{2}\sum_\nu \int_{-1}^{1} d\lambda\, T \lim_{\mathcal{M}\to\infty} \int_{-\mathcal{M}-\lambda\epsilon^I_\nu}^{\mathcal{M}-\lambda\epsilon^I_\nu} \frac{dz}{2\pi} \frac{\epsilon^I_\nu}{iz - \epsilon^R_\nu}. \qquad (6.32)$$

Taking into account that $\lambda$ covers a symmetric range of integration, which allows one to reverse the sign in the $\lambda$ integral without changing the associated boundaries, the shift appearing in the $z$–integral boundaries can be shown not to contribute as $\mathcal{M} \to \infty$. This yields

$$\mathcal{A}_I = \frac{-i}{2}\sum_\nu \int_{-1}^{1} d\lambda\, T \lim_{\mathcal{M}\to\infty} \int_{-\mathcal{M}}^{\mathcal{M}} \frac{dz}{2\pi} \frac{\epsilon^I_\nu}{iz - \epsilon^R_\nu}. \qquad (6.33)$$

Now the integral over the parameter $\lambda$ may be done. Due to the principle value prescription for the $z$ integration the terms of odd powers in $z$ cancel. This results in a convergent expression

$$\mathcal{A}_I = \frac{-i}{2} T \sum_\nu \epsilon^I_\nu \int_{-\infty}^{\infty} \frac{dz}{2\pi} \frac{-2\epsilon^R_\nu}{z^2 + (\epsilon^R_\nu)^2}. \qquad (6.34)$$

It should be mentioned that the divergence, which disappears in the principle value prescription, has the drastic consequence that the energy functional is not invariant when shifting the $\omega$ field by a constant $\omega_c$. In that case the imaginary part ot the Euclidean action is augmented by $TN_C\omega_c B$ where $B$ denotes the baryon number of the original configuration. This result is very gratifying, as will be seen below.

Although $\mathcal{A}_I$ is finite in the principle value formulation, the proper time regularization may be imposed by expressing the integrand as a parameter integral [76]

$$\frac{1}{z^2 + (\epsilon^R_\nu)^2} \to \int_{1/\Lambda^2}^{\infty} ds\, \exp\left\{-s\left(z^2 + (\epsilon^R_\nu)^2\right)\right\} \qquad (6.35)$$

which does obviously not diverge as $\Lambda \to \infty$. Completing the evaluation of $\mathcal{A}_I$ in analogy to eqs. (6.25-6.27) we find for the contribution of the Dirac sea to the imaginary part of the Euclidean energy $E^I_0$ from $\mathcal{A}_I \overset{T\to\infty}{\longrightarrow} -iTE^I_0$

$$E^I_0 = \frac{-N_C}{2}\sum_\nu \epsilon^I_\nu \mathrm{sgn}(\epsilon^R_\nu) \begin{cases} 1, & \mathcal{A}_I \ \text{not regularized} \\ \mathcal{N}_\nu, & \mathcal{A}_I \ \text{regularized} \end{cases}. \qquad (6.36)$$

The upper case in eq (6.36) corresponds to the limit $\Lambda \to \infty$. Obviously only the real part of the one-particle energy eigenvalue is relevant for the regularization of $\mathcal{A}_I$. Eq (6.36) represents a regularization for $\mathcal{A}_I$ that only involves quantities which are strictly positive definite.

The Euclidean "energy" functional

$$E^R + iE^I = E_0^R + E_V^R + i\left(E_0^I + E_V^I\right) \qquad (6.37)$$

exhibits the interesting feature that, as a quark state becomes part of the vacuum, both real ($E^R$) and imaginary ($E^I$) are continuous under the condition that the baryon number (6.12) remains unchanged. For a state $|\nu\rangle$ to become part of the vacuum means that $\epsilon_\nu^R$ changes sign. Assume that for the configuration under consideration one special orbit, say $|\mathrm{val}\rangle$, has $\eta_{\mathrm{val}} = 1$ as long as $\epsilon_{\mathrm{val}}^R > 0$. In order to stay in the same baryon number sector $\eta_{\mathrm{val}}$ has to vanish as $\epsilon_{\mathrm{val}}^R$ reverses its sign (*cf.* eq (6.12)). Thus we demand $\eta_{\mathrm{val}} = \left(1 + \mathrm{sgn}(\epsilon_{\mathrm{val}}^R)\right)/2$. Noting that $\mathrm{erfc}(0) = 1$ one easily verifies that for $\epsilon_{\mathrm{val}} \approx 0$ the terms which depend on the sign of $\epsilon_{\mathrm{val}}^R$ cancel in the sum (6.37). From this discussion it is also obvious that the valence quark orbit occupation numbers $\eta_\nu$ are dynamical quantities which functionally depend on the soliton configuration.

The determination of the Euclidean energy functional is now completed and we have to obtain a Minkowski energy functional from this. In the discussion proceeding eq (6.20) we have already remarked that the one–particle energy eigenvalues $\epsilon_\nu(z)$ can be considered analytic in the complex variable $z$. Unfortunately, such a statement was shown [71] not to hold for the energy functional

$$E(z, z^*) = E^R(z, z^*) + iE^I(z, z^*). \qquad (6.38)$$

Here the dependence on $z$ and its complex conjugate $z^*$ is due to the implicit dependencies $\epsilon_\nu^{R,I}(z, z^*)$. The appearance of $z^*$ is caused by the fact that the regularization treats real and imaginary parts of the action in the complex plane separately. Remember that (formally) the unregularized action has been observed to be analytic in $z$. The functional (6.38) has been investigated with respect to its analytical properties [71]. It has been found that the Laurent series centered at the Euclidean point ($z = i$) has vanishing radius of convergence and the coefficients of the singular terms are non–vanishing. Stated more drastically: The analytic continuation of (6.37) does not exist. Hence the Minkowski energy functional for static, non–perturbative field configurations involving time components of (axial) vector fields cannot be obtained by means of analytical continuation.

It has then been argued [71] that the generalized energy functional (6.38) can well be approximated by $E^R(i, -i) + zE^I(i, -i)$ at the Euclidean point ($z = i$). This approximation gains further support by the fact that the regularization as discussed above is not without ambiguities at the quadratic order in the time components of the (axial) vector fields (see section 6.2.4). Adopting this approximation the analytic continuation to the Minkowski point ($z = 1$) is trivial, yielding

$$E = E_0^R + E_0^I + E_V^R + E_V^I + E_m. \qquad (6.39)$$

The mesonic contribution $E_m$ is straightforwardly obtained by substituting the soliton configuration into the purely mesonic part of the action $\mathcal{A}_m$. One might wonder whether there is a sign ambiguity for $E_{0,V}^I$ in the definition (6.39). This is not the case for the energy functional of a self–consistent soliton configuration because the eigenvalues of $h$ are analytical and hence such a sign ambiguity may be absorbed in the definition of the time components $V_4$ and $A_4$.

Let us next mention two consistency conditions which have to be satisfied by the Minkowski energy functional. As already noted after eq (6.34) shifting the $\omega$ field by a constant amount ($\omega_c$) leads to a change of the (unregularized and thus analytical) energy functional of $iN_C B\omega_c$. We demand this to hold for the regularized energy functional in Minkowski space as well. As a matter of fact this condition is a consequence of global gauge invariance (*i.e.* a constant $V_0$ behaves like a chemical potential). Furthermore, the current field identity (3.21) for the baryon current imposes a normalization on the profile function of the isoscalar $\omega$ meson. This normalization is obtained by integrating the stationary condition for this field over the whole coordinate space. As this stationary condition is obtained from extremizing the Minkowski space energy functional we have available a second consistency condition. It is important to note that the energy functional (6.39) satisfies these two consistency conditions; at least when the imaginary part remains unregularized. Thus (6.39) provides a well suited object for the investigation of soliton solutions [76, 73, 75].

At this point it should be mentioned that two other approaches to include time components of vector mesons are discussed in the literature. In ref. [77] these components are not continued according to the discussion preceding (6.16). In that case the energy functional in the whole $z$–plane depends on the eigenvalue $\epsilon_\nu(z)$ only and is thus analytical. However, due to the regularization the global gauge invariance as well as the current field identity for the baryon number current are lost in Minkowski space. This approach has therefore to be abandoned because of physical reasons. In the second treatment [78, 79] the time components have indeed been continued. These authors have expressed the Euclidean energy functional (6.37) in terms of $\epsilon_\nu(z)$ and its complex conjugate $\epsilon_\nu^*(z) = \epsilon_\nu(z^*)$. Then the energy functional depends on $z$ and $z^*$: $E = E(z, z^*)$. It should be evident from the preceding discussions that the dependence on $\epsilon_\nu^*(z)$ and thus $z^*$ purely originates from regularization and treating real and imaginary parts of the action differently. As in Euclidean space $z = i = -z^*$ the authors of ref.[78, 79] have then set $\tilde{E} = E(z, -z)$ and claimed that the continuation of $\tilde{E}$ to $z = 1$ yielded the Minkowski energy functional. It should be noted that this approach preserves the global gauge invariance. However, as already indicated above, the continuation of $E(z, z^*)$ does not exist and hence the treatment of ref.[78, 79] has to be considered as a physically motivated definition but not as a mathematically correct derivation of the Minkowski energy functional.

Very recently it has been demonstrated that counting powers of the time components of vector fields in the eigenvalues of the Dirac Hamiltonian does not correctly reproduce the corresponding expansion for $\mathcal{A}_R$ [71]. The reason is that the manipulations leading to the sum (6.24) are in fact ill–defined as already indicated because the traces of $h$ and $h^\dagger$ are computed in different Hilbert spaces. This topic will further be illuminated in subsection 6.2.4.

## 6.2 Self–consistent solutions

We denote meson field configurations which extremize the Minkowski energy functional (6.39) self–consistent. Let $\varphi$ represent the whole set of meson fields involved in the soliton configuration. Then the stationary condition may be expressed as

$$\frac{\delta E}{\delta \varphi} = 0. \tag{6.40}$$

The mesonic part of the energy functional, $E_m$, is at least quadratic in $\varphi$ while the Dirac operator provides a Yukawa type coupling of $\varphi$ to the quark fields. Thus (6.40) relates the meson fields to the quark fields. The latter diagonalize the static Hamiltonian $h$ and the associated eigenvalues again enter the energy functional. It is this interplay between eigenstates and –values of $h$ which leads to the notion of self–consistency. Commonly the self–consistent soliton solution is numerically obtained with the use of an iterative procedure [83, 80]. A test profile is employed to diagonalize $h$. The resulting eigenvalues and –vectors are then substituted into the stationary condition (6.40) yielding an updated profile function. This updated profile serves again as input for $h$. This procedure is repeated until convergence is gained.

### 6.2.1 The pseudoscalar hedgehog

The simplest meson field configuration which allows for soliton solutions involves the chiral field only. This configuration has already been employed in chapter 5 to discuss general properties of the soliton. Then all vector meson fields are set to zero and for the pseudoscalar field the celebrated hedgehog *ansatz* is assumed

$$M = m \exp\left(i\boldsymbol{\tau}\cdot\hat{\boldsymbol{r}}\Theta(r)\right). \tag{6.41}$$

Obviously the scalar fields are constrained to the chiral circle $\Sigma = \langle\Sigma\rangle = m$. The radial function $\Theta(r)$ is commonly referred to as the chiral angle. Substituting this *ansatz* into the Dirac Hamiltonian (6.16) yields

$$h \;=\; \boldsymbol{\alpha}\cdot\boldsymbol{p} + \beta m\left(\cos\Theta(r) + i\gamma_5\boldsymbol{\tau}\cdot\hat{\boldsymbol{r}}\ \sin\Theta(r)\right). \tag{6.42}$$

As $h$ is Hermitian the eigenvalues $\epsilon_\nu$ obtained from

$$h\Psi_\nu = \epsilon_\nu\Psi_\nu \tag{6.43}$$

are real. Technically, the discretization of the eigenvalues $\epsilon_\nu$ is achieved by restricting the coordinate space $I\!R^3$ to a spherical cavity of radius $D$ and demanding certain boundary conditions at $r = D$. Eventually the continuum limit $D \to \infty$ has to be considered. We relegate the discussion of the special form of the boundary conditions to appendix B where the coordinate representations of the states $|\nu\rangle$ are given. It is nevertheless worthwhile to discuss the structure of these states. Due to the special form of the hedgehog (6.41) the Dirac Hamiltonian commutes with the grand spin operator

$$\boldsymbol{G} = \boldsymbol{j} + \frac{\boldsymbol{\tau}}{2} = \boldsymbol{l} + \frac{\boldsymbol{\sigma}}{2} + \frac{\boldsymbol{\tau}}{2} \tag{6.44}$$

where $\boldsymbol{j}$ labels the total spin and $\boldsymbol{l}$ the orbital angular momentum. $\boldsymbol{\tau}/2$ and $\boldsymbol{\sigma}/2$ denote the isospin and spin operators, respectively. The eigenstates of $h$ are then as well eigenstates of $\boldsymbol{G}$. The latter are constructed by first coupling spin and orbital angular momentum to the total spin which is subsequently coupled with the isospin to the grand spin [81]. The resulting states are denoted by $|ljGM\rangle$ with $M$ being the projection of $\boldsymbol{G}$. These states obey the selection rules

$$j = \begin{cases} G + 1/2, & l = \begin{cases} G+1 \\ G \end{cases} \\ & \\ G - 1/2, & l = \begin{cases} G \\ G-1 \end{cases} \end{cases} \;. \tag{6.45}$$

As the one–particle eigenenergies $\epsilon_\nu$ are real, the Euclidean energy does not possess an imaginary part. The Minkowski energy functional arising from eq (6.27) is then given by

$$\begin{aligned} E[\Theta] &= N_C \sum_\nu \eta_\nu |\epsilon_\nu| + \frac{N_C}{4\sqrt{\pi}} \sum_\nu |\epsilon_\nu| \Gamma\left(-\frac{1}{2}, \left(\frac{\epsilon_\nu}{\Lambda}\right)^2\right) + E_m \\ &\quad - \frac{N_C}{4\sqrt{\pi}} \sum_\nu |\epsilon_\nu^0| \Gamma\left(-\frac{1}{2}, \left(\frac{\epsilon_\nu^0}{\Lambda}\right)^2\right) \end{aligned} \tag{6.46}$$

with the mesonic part

$$E_m = 4\pi m_\pi^2 f_\pi^2 \int drr^2 \left(1 - \cos\Theta(r)\right). \tag{6.47}$$

Also the energy functional associated with the trivial meson configuration $\Theta \equiv 0$ is subtracted in (6.46) and (6.47). The stationary condition $\delta E[\Theta]/\delta\Theta(r)$ is obtained with the help of the chain rule. Whence we require the functional derivative of $\epsilon_\nu$

$$\frac{\delta \epsilon_\nu}{\delta\Theta(r)} = m \int d\Omega\ \Psi_\nu^\dagger(\boldsymbol{r})\beta\left(-\sin\Theta(r) + i\gamma_5 \boldsymbol{\tau}\cdot\hat{\boldsymbol{r}}\ \cos\Theta(r)\right)\Psi_\nu(\boldsymbol{r}). \tag{6.48}$$

Then the stationary condition becomes the equation of motion for $\Theta$ [83]

$$\cos\Theta(r)\ \mathrm{tr}\int d\Omega\ \rho_S(\boldsymbol{r},\boldsymbol{r}) i\gamma_5 \boldsymbol{\tau}\cdot\hat{\boldsymbol{r}} = \sin\Theta(r)\ \left\{\mathrm{tr}\int d\Omega\ \rho_S(\boldsymbol{r},\boldsymbol{r}) - \frac{4\pi}{N_C}\frac{m_\pi^2 f_\pi^2}{m}\right\} \tag{6.49}$$

where the traces are over flavor and Dirac indices only. According to the sum (6.46) the scalar quark density matrix $\rho_S(\boldsymbol{x},\boldsymbol{y}) = \langle q(\boldsymbol{x})\bar{q}(\boldsymbol{y})\rangle$ is decomposed into valence quark and Dirac sea parts:

$$\begin{aligned} \rho_S(\boldsymbol{x},\boldsymbol{y}) &= \rho_S^V(\boldsymbol{x},\boldsymbol{y}) + \rho_S^0(\boldsymbol{x},\boldsymbol{y}) \\ \rho_S^V(\boldsymbol{x},\boldsymbol{y}) &= \sum_\nu \Psi_\nu(\boldsymbol{x})\eta_\nu\bar{\Psi}_\nu(\boldsymbol{y})\mathrm{sgn}(\epsilon_\nu) \\ \rho_S^0(\boldsymbol{x},\boldsymbol{y}) &= \frac{-1}{2}\sum_\nu \Psi_\nu(\boldsymbol{x})\mathrm{erfc}\left(\left|\frac{\epsilon_\nu}{\Lambda}\right|\right)\bar{\Psi}_\nu(\boldsymbol{y})\mathrm{sgn}(\epsilon_\nu). \end{aligned} \tag{6.50}$$

As indicated above the stationary condition relates the soliton profile function $\Theta(r)$ to the eigenfunctions $\Psi_\nu$ of the Dirac Hamiltonian $h$.

We are now at the point to discuss the numerical solutions in the unit baryon number sector ($B = 1$). For the eigenvectors (B.4-B.6) one can easily verify that $\mathrm{tr}\int d\Omega\ \rho_S(\boldsymbol{r},\boldsymbol{r})i\gamma_5\boldsymbol{\tau}\cdot\hat{\boldsymbol{r}}\Big|_{r=0} = 0$. This transfers to the boundary condition for the chiral angle: $\Theta(0) = l\pi$. On the other hand, for $r \to \infty$ (6.49) has a solution with $\Theta = 0$. In the ($B = 1$) sector it turns out that the soliton profiles with the boundary conditions $\Theta(0) = -\pi$ and $\Theta(\infty) = 0$ (modulu $2\pi$) extremize the static energy functional (6.46). It should be mentioned that for configurations which do not obey these boundary conditions the energy remains finite although it is not a minimum. This is in contrast to topological soliton models [82] where the energy diverges for infinitesimal deviations from these boundary conditions. The NJL soliton is not a topological one!

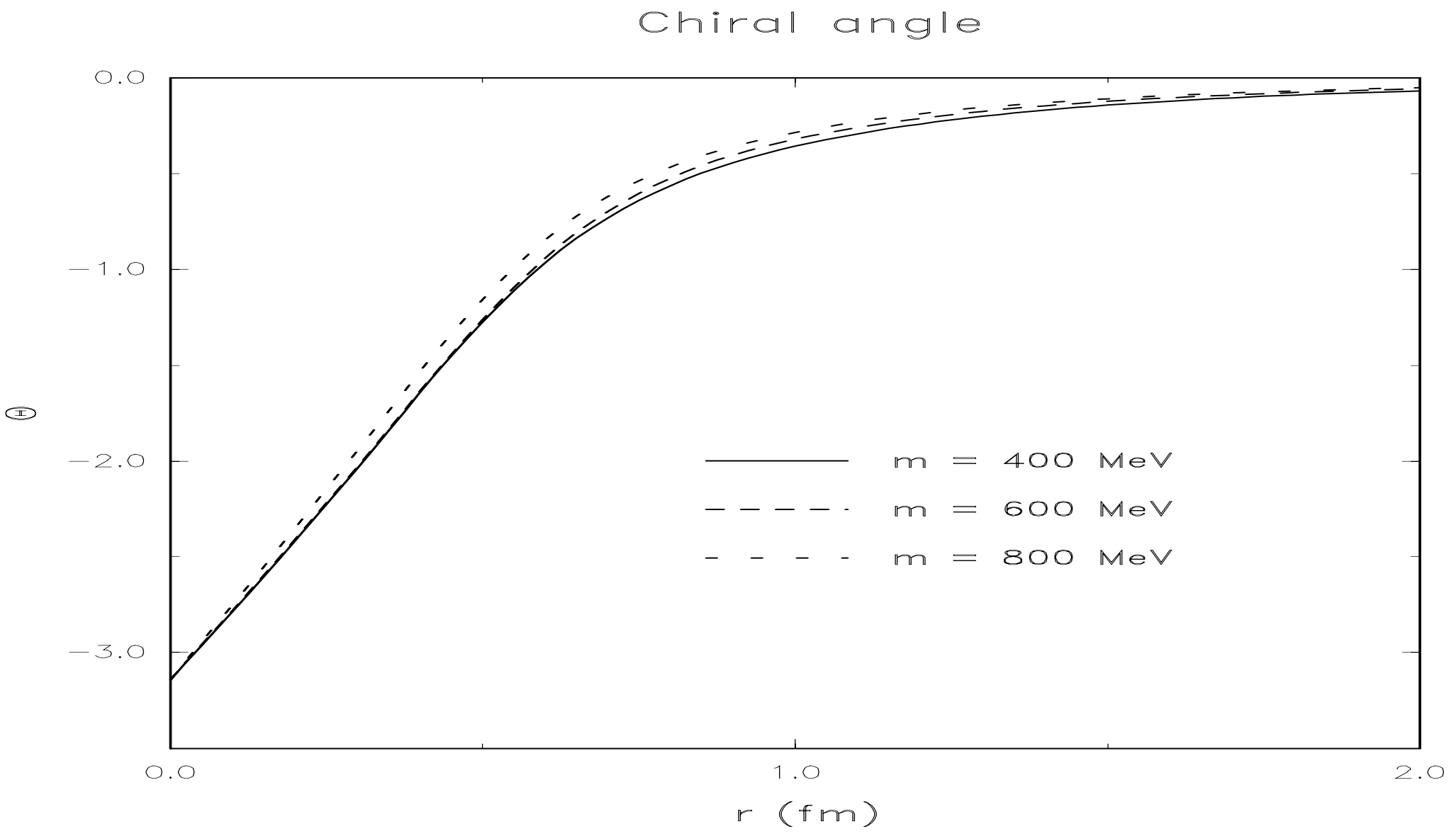

Figure 6.1: The radial dependence of the self–consistent profile function $\Theta(r)$ for various constituent quark masses $m$. The pion mass is $m_\pi = 135$ MeV.

Table 6.1: The soliton energy $E_{\rm tot}$ as well as its various contributions according to the sum (6.46) as functions of the constituent quark mass $m$. The pion mass is taken to be $m_\pi =$ 135MeV. All numbers are in MeV.

| $m$ | 350 | 400 | 500 | 600 | 700 | 800 |
|---|---|---|---|---|---|---|
| $E_{\rm tot}$ | 1236 | 1239 | 1221 | 1193 | 1161 | 1130 |
| $E_V$ | 745 | 633 | 460 | 293 | 121 | -55 |
| $E_0$ | 459 | 571 | 728 | 869 | 1012 | 1103 |
| $E_m$ | 31 | 34 | 33 | 31 | 28 | 26 |

Self–consistent solutions were numerically obtained for $m \ge 325$MeV [83]. In Figure 6.1 the radial dependence of the self–consistent profile is displayed for various values of the constituent quark mass $m$. The profile function obviously exhibits only a very mild dependence on $m$. The soliton energy $E_{\rm tot}$, *i.e.* the energy functional corresponding to the self–consistent chiral angle, shows the same characteristics as can be seen from table 6.1. Obviously the soliton mass is larger than the three quark threshold as long as $m \lesssim 420$MeV. In table 6.1 also the various contributions to the soliton energy are displayed. For small constituent quark masses the explicit occupation of the valence quark orbit ($\eta_{\rm val} = 1$) provides the dominant contribution to the energy. As $m$ increases the situation is reversed. For $m > 750$MeV $E_V$ even becomes negative. According to our previous discussions this implies $\eta_{\rm val} = 0$. Stated otherwise: The valence quark becomes part of the polarized vacuum which then carries the baryon number. This situation actually corresponds to the Skyrmion picture of the baryon. Furthermore, the contribution $E_m$ of the purely mesonic part of the action to the energy is shown in table (6.1). This quantity can actually be interpreted as the pion–nucleon $\Sigma$ term at zero momentum transfer $\Sigma_{\pi N}(q^2 = 0)$. The results for $E_m$ seem to be somewhat lower than the data extracted from a critical examination of the existing data on $\pi N$ scattering [84]: $\Sigma_{\pi N}(q^2 = 0) \approx 45\text{MeV}$. It should, however, be noted that $\Sigma_{\pi N}(q^2)$ is claimed [84] to have a strong momentum dependence which makes the extraction of the $\Sigma_{\pi N}(q^2 = 0)$ from $\pi N$ scattering data rather involved.

### 6.2.2 Beyond the chiral circle

So far the scalar mesons have been constrained to their vacuum expectation value, the constituent quark mass $m$. The reason for choosing this *ansatz* is not only that of simplification but rather the appearance of a special instability of the soliton configuration when the scalar field is allowed to be space dependent; at least in the proper time regularization scheme [77, 85]. In this case one can prove the existence of a meson field configuration, which has vanishing energy. When varying the spatial extension of this configuration below a critical size the system passes from a unit baryon number to a zero baryon number configuration. This variation can be parametrized as ($M = \sigma + \boldsymbol{\pi} \cdot \boldsymbol{\tau}$)

$$\sigma(r) = \phi_0[1 + Wf(\frac{r}{R})\cos\Theta(\frac{r}{R})], \qquad \boldsymbol{\pi}(\boldsymbol{r}) = \hat{\boldsymbol{r}}\phi_0 Wf(\frac{r}{R})\sin\Theta(\frac{r}{R}) \tag{6.51}$$

with $W \cdot (R\phi_0)^\alpha = \text{const}$. In ref.[85] a Wood-Saxon shape was chosen for $f(r/R)$ and a linear profile for $\Theta(r/R)$. The constant $\phi_0 \sim m$ characterizes the vacuum expectation value of the scalar field. For very narrow configurations, *i.e.* $R \to 0$, the mesonic part of the NJL soliton energy, which is maximally quadratic in the scalar field $\phi = \sqrt{\sigma^2 + \boldsymbol{\pi}^2}$ (to which we will refer to as the chiral radius), can be shown to vanish for $\alpha < 3/2$. Thus only the fermion determinant contributes to the energy in this case. For such a value of $\alpha$ the spectrum of the Dirac Hamiltonian has the interesting feature that the valence quark level gets transferred from the lower boundary of the positive Dirac spectrum to the upper boundary of negative Dirac spectrum. All other levels in the $G^\pi = 0^+$ channel follow as a consequence of "avoided crossings" [85]. Obviously the baryon number is carried by the Dirac sea and hence the soliton energy is solely given by the vacuum part $E_0$. This is plotted as a function of $R \cdot \phi_0$ in figure 6.2 for $\alpha = 4/3$. Evidently the total energy is identical to zero in the limit $R \to 0$. However, since these localized meson field configurations carry no baryon number for small $R$, this instability is unphysical indicating merely that for $R \to 0$ the system changes from the $B = 1$ to the $B = 0$ sector. This is also indicated by the fact that these configurations assume zero energy, which is the energy of the ground state in the $B = 0$ sector.

One way to circumwent these problems is *e.g.* indicated the possibility to mock up the trace anomaly of QCD in effective meson theories. In such an approach the scalar dilaton field $\chi \sim \langle G_{\mu\nu}G^{\mu\nu}\rangle$ is incorporated as an order parameter to absorb the mass dimension of the parameters [87]. Especially one may introduce a coupling to the chiral radius $\phi$ such that a fourth order term appears in the meson part of the energy [88, 86]

$$E_m = 4\pi \int drr^2 a^2 \left(\frac{1}{2}\phi^4(r) - \eta\chi_0^2\phi^2(r) - \text{const}\right). \tag{6.52}$$

Here $\chi_0$ denotes the vacuum expectation value of the dilaton field. The two new parameters $a$ and $\eta$ are related via the (extended) gap equation. Hence one may equally well consider $\eta\chi_0^2$ as the only free parameter (besides the constituent quark mass $m$). The meson energy with the dilation included (6.52) only vanishes for $\alpha < 3/4$ when the parametrization (6.51) is adopted [86]. Glancing again at figure 6.2 one observes that in this case the energy does not vanish for $R \to 0$ but rather becomes very large. Indeed, the solution to the self–consistent problem exists and is displayed in figure 6.3. From this figure one also recognizes that the collapse appears when the fourth order interaction is switched off. As a consequence of the gap equation the limit $a \to 0$ corresponds to $\eta\chi_0^2/\phi_0^2 \to -\infty$. Figure 6.3 also demonstrates that for increasing coupling to the dilaton field, the chiral angle, $\Theta$ gets more concentrated at $r = 0$. This reflects the bag formation caused by the dilaton field [89].

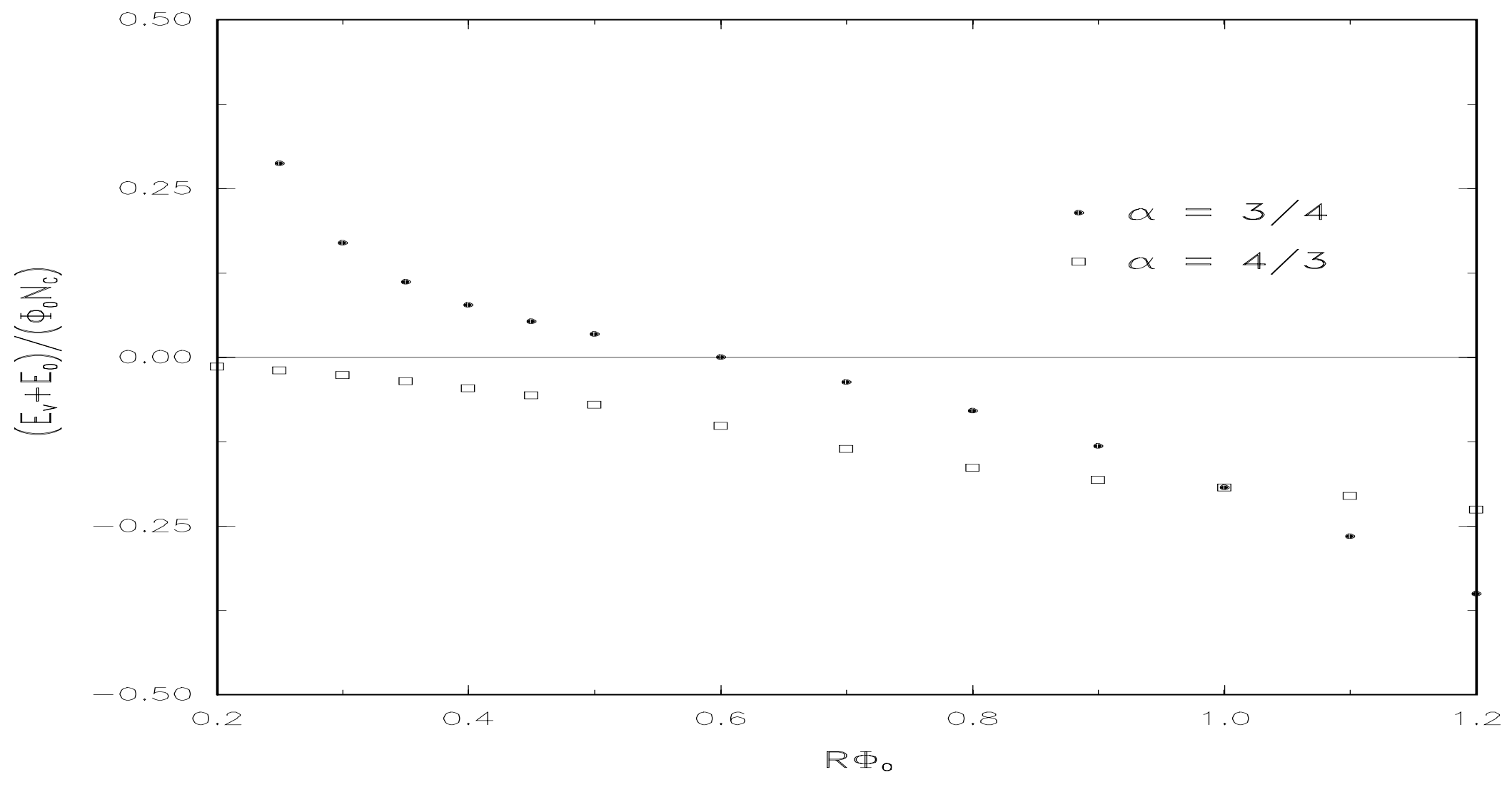

Figure 6.2: The energy of the fermion determinant, $E_V + E_0$, for the variational meson field configuration (6.51). The parameter $\alpha$ is defined after eq (6.51). (Figure taken from ref.[86].)

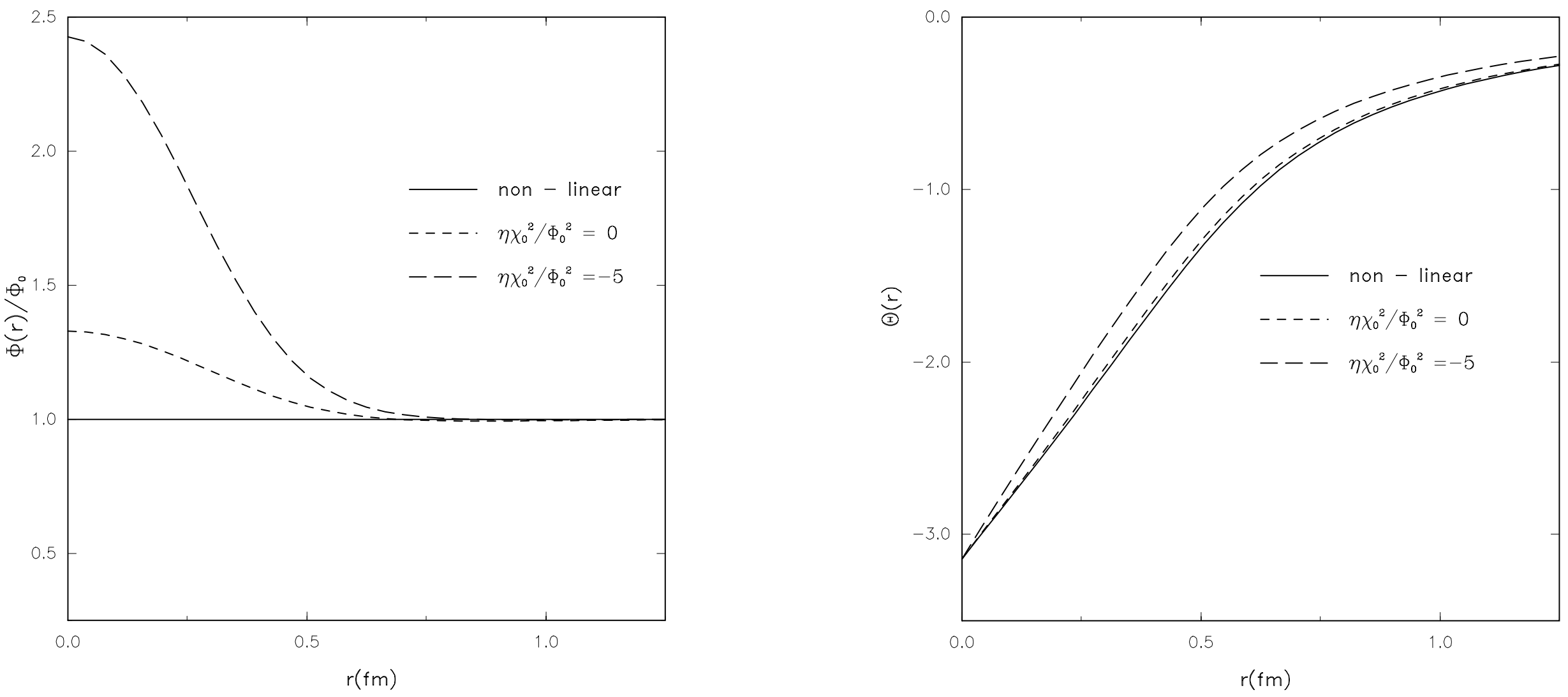

Figure 6.3: The self–consistent soliton configuration with the scalar field included. "Non–linear" refers to the case with the scalar fields fixed at their vacuum expectation values. (Figure taken from ref.[86].)

The above considerations made use of an extended definition of the energy functional in order to avoid the collapse of the meson profiles. As the net result of the limit $R \to 0$ in the configuration (6.51) for $\alpha = 3/4$ is to shift all eigenvalues $\epsilon_\nu$ in the $G^\pi = 0^+$ channel by one level, the vanishing energy is just a matter of regularization which ignores the asymmetries of the spectrum of the Dirac Hamiltonian at large energies. However, the baryon number (6.12) is sensitive to these large energies because no regularization is involved. Taking the point of view that the imaginary part of the Euclidean action should also be regularized, automatically leads to a regularized baryon number $B_\Lambda$

$$B_\Lambda[\phi,\Theta] = \sum_\nu \mathrm{sgn}(\epsilon_\nu)\left(\eta_\nu - \frac{1}{2}\mathrm{erfc}\left(\left|\frac{\epsilon_\nu}{\Lambda}\right|\right)\right). \qquad (6.53)$$

In the same way as the energy vanishes, $B_\Lambda$ goes to zero for the above considered limit $R \to 0$. In ref.[90] therefore the idea has been pursued to constrain $B_\Lambda$ to unity in order to avoid the collapse. Then the energy functional (6.46) is supplemented by a Lagrange multiplier for $B_\Lambda$

$$E[\phi,\Theta] \to E[\phi,\Theta] + \lambda\left(B_\Lambda[\phi,\Theta] - 1\right)^2 + \frac{\zeta}{2}\lambda^2. \qquad (6.54)$$

In this case no quartic term in $\phi$ is needed in the mesonic part of the action. The last term in (6.54) is introduced for convenience; eventually the limit $\zeta \to 0$ should be assumed. At first place it should be noted that for the chiral soliton with the scalar fields constrained to the vacuum expectation values (see subsection 6.2.1) $B_\Lambda$ is not exactly unity; but rather only 0.97. Even by including the constraint (6.54) this number cannot be increased significantly as long as $\phi$ is not allowed to be space dependent. Taking $\phi = \phi(r)$ to be a radial function indeed increases $B_\Lambda$. Numerically, however, a solution with $B_\Lambda = 1$ has not been found for finite[d] $D$. For finite $D$ always a non–vanishing lower limit for $\zeta$ exists below which no solution has been found. This limit can be translated into a value for $B_\Lambda$ which slightly deviates from unity but approaches it as $D$ increases. The associated self–consistent soliton profiles are displayed in figure 6.4. The most important result is, of course, that stable solutions do exist when the regularized baryon number is fixed. Furthermore, one observes that the large distance ($r \ge 1$fm) behavior of the profiles is almost uneffected by the constraint (6.54). However, in the vicinity of the origin drastic changes occur. For $B_\Lambda \to 1$ the chiral radius, $\phi$ vanishes at the origin, *i.e.* the chiral symmetry is restored. Simultaneously, the slope of the chiral angle, $\Theta$ increases and eventually goes to infinity. In ref.[90] it has also been shown that constraining the regularized baryon number leads to a valence quark dominated picture of the soliton. Especially, the corresponding energy eigenvalue approaches the one obtained in the baryon number zero sector. Nevertheless, the valence quark wave–function was found to be strongly localized.

A further approach to prevent the meson profiles from collapsing is to include vector mesons [79, 75]; especially because the $\omega$–meson provides a sizable repulsion which is supposed to stabilize the system. This will also be discussed in the proceeding subsection.

[d] $D$ denotes the radius of the spherical box for the numerical calculations, *cf.* the discussion after eq (6.43).

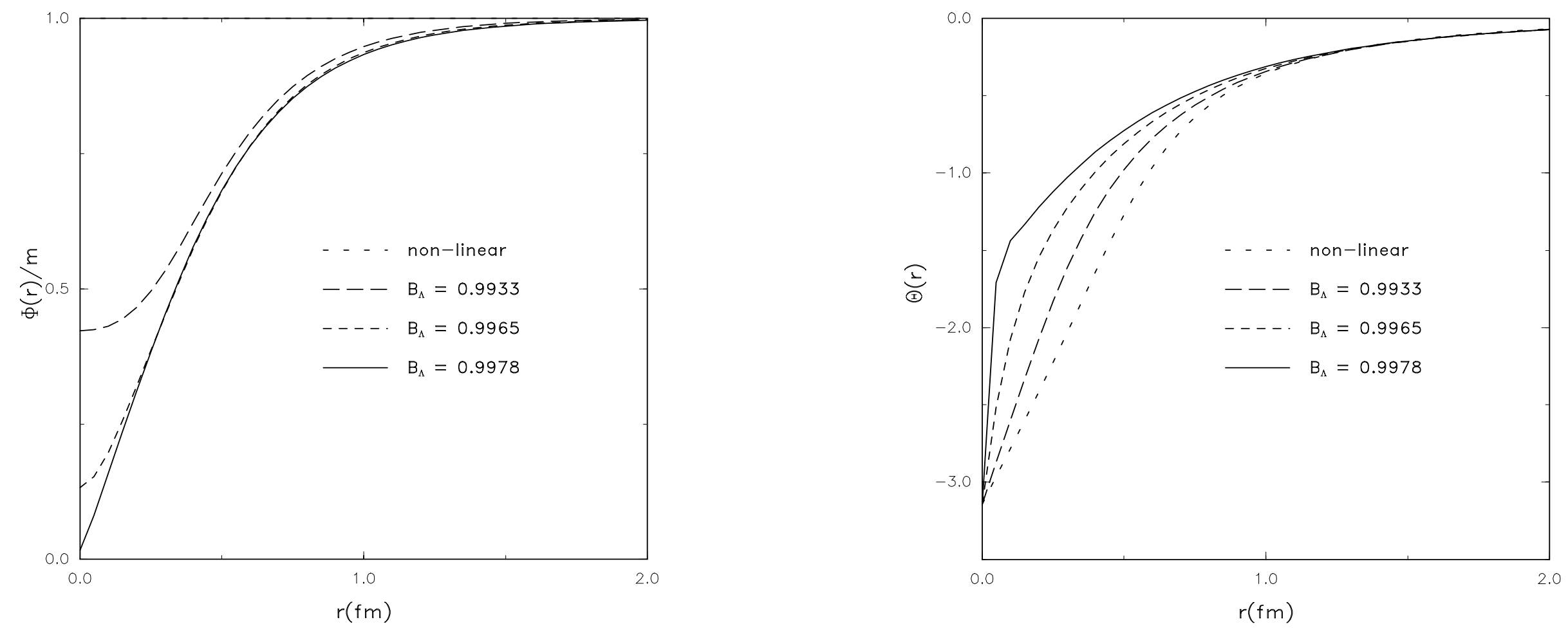

Figure 6.4: The self–consistent soliton configuration with the scalar field included and the regularized baryon number $B_\Lambda$ (6.53) being constrained. Different values for $B_\Lambda$ correspond to various radii of the spherical box used for the numerical calculations. "Non–linear" denotes the case without constraint and the scalar field at its vacuum expectation values. (Figure taken from ref.[90].)

### *6.2.3 Inclusion of (axial) vector mesons*

In recent years the NJL soliton has experienced several types of extensions. In the present section we will describe the effects of various vector and axial–vector meson fields on the soliton solution and put some emphasis on the historical development. Proceeding in this manner we are also enabled to illuminate the effects of different (axial) vector meson fields separately. A convenient parametrization of the anti–Hermitian fields $V_\mu$ and $A_\mu$ is given by separating isoscalar and isovector parts

$$V_\mu = i\omega_\mu + i\boldsymbol{\rho}_\mu \cdot \frac{\boldsymbol{\tau}}{2}\,, \qquad A_\mu = if_{1\mu} + i\boldsymbol{a}_{1\mu} \cdot \frac{\boldsymbol{\tau}}{2}. \tag{6.55}$$

The Euclidean Dirac Hamiltonian for the most general grand spin zero field configuration consistent with the parity properties of the meson fields is given in eq (6.63). Additionally the complete set equations of motion resulting from extremizing the Minkowski energy functional (6.39) is displayed in appendix C. The Hamiltonian corresponding to the subsystems which will subsequently be discussed below can easily be obtained from (6.63) by setting those fields, which are not involved, to their vacuum expectation values. The detailed numerical results will be presented at the end of this section. This allows us to directly compare the influence of the various fields.

In ref.[91] the inclusion of the $\rho$–meson via the Wu–Yang *ansatz*

$$\rho_{ia} = \epsilon_{ika}\hat{r}_k G(r) \tag{6.56}$$

has been studied. Except for the pseudoscalar field, for which the hedgehog shape (6.41) was adopted, all other fields were set to their vacuum expectation values. Although this configuration does not exhibit global chiral invariance it allows one to investigate the attractive character of the $\rho$–meson. Indeed, the size of the soliton, as measured by the baryonic root mean squared radius, was found to decrease. Furthermore the value for $m$, above which soliton solutions do exist, decreased to 270MeV. It should be stressed that the incorporation of the $\rho$–meson neither alters any of the parameters involved nor introduces additional ones. This, however, is not the case when also the $a_1$–meson is included. As the $\pi - a_1$ mixing changes the relation between the pion decay constant, $f_\pi$, and the cut–off, $\Lambda$, the value of $\Lambda$ for a given constituent quark mass, $m$, is significantly increased [31], see also eq (4.53). For details the reader is referred to chapter 4. Needless to remark that the inclusion of the $a_1$ meson is mandatory to preserve chiral symmetry. The *ansatz* for the $a_1$ meson which is consistent with the grand spin zero assumption for the static fields and exhibits the pseudovector character involves two radial functions

$$a_{1ia} = \delta_{ia} H(r) + \hat{r}_i \hat{r}_a F(r). \tag{6.57}$$

Self–consistent solutions were found for $m \geq 300$MeV [92, 93]. From the equation of motion for $H(r)$ (*cf.* eq (C.9)) one easily verifies that $H(0)$ is non–vanishing. Thus the axial–vector meson has a direct influence on the valence quark wave–function which is localized at the origin. As a result, the energy of the valence quark state is significantly lowered and actually becomes negative. From the discussion in section 6.1 it is then obvious that the total baryon charge is carried by the polarized Dirac sea and no orbit is explicitly occupied, *i.e.* all $\eta_\nu \equiv 0$, in the unit baryon number sector. The baryon charge is solely due to $j_0^\mu$ in eq (6.11). This result strongly supports the Skyrmion picture of baryons as can be seen from the following considerations. The gradient expansion of $j_0^\mu$ yields in leading order the topological current (4.65) [94] which, in the Skyrme model, is identified with the baryon current[e]. *A priori* the point of view that a distorted Dirac sea is responsible for a non–vanishing baryon charge is integrated in the Skyrme model. This point of opinion is commonly referred to as Witten's conjecture. Obviously, it is supported by the NJL model [92]. Furthermore, the Skyrmion phenomenology has established that short range effects are described by either explicit quark degrees of freedom or (axial) vector mesons [95, 96]; the "or" being exclusive. In the NJL model, however, this matter of opinion represents a direct result!

Up to now the $\omega$–meson has been left out, mostly for technical reasons. As discussed in section 6.1 it introduces a non–Hermitian Dirac Hamiltonian (*cf.* eq (6.63)) and, more troublesome, in a non–perturbative approach a unique extraction of a Minkowski energy functional is not available. However, the incorporation of this meson is extremely desirable from the point of vector meson dominance (the $\omega$–meson has significant influence on the isoscalar radius) and the current field identities (3.21) (the isoscalar part of the electromagnetic current is proportional to the $\omega$ field). Additionally the repulsive character of the $\omega$–meson is expected to provide stability even when the scalar field is allowed to vary in space. We will therefore report on investigations which are based on the physically motivated definition (6.39) for the Minkowski energy functional.

First attempts in this direction have concerned the $\pi - \omega$ system with all other fields put to their vacuum expectation values [78, 76]. Only a radial function for the time–component of the isoscalar vector meson is allowed by the grand spin symmetry

$$\omega_\mu = \omega(r)\delta_{\mu 4}. \tag{6.58}$$

[e]The topological current (4.65) can also be obtained as the Noether current associated with the singlet vector symmetry once the Skyrme model is extended to flavor SU(3) and the Wess–Zumino term is included.

As the Hamiltonian is non–Hermitian we have to distinguish between left and right eigenstates

$$h|\Psi_\nu\rangle = \epsilon_\nu|\Psi_\nu\rangle \quad \langle\tilde{\Psi}_\nu|h = \epsilon_\nu\langle\tilde{\Psi}_\nu| \qquad i.e.\ h^\dagger|\tilde{\Psi}_\nu\rangle = \epsilon_\nu^*|\tilde{\Psi}_\nu\rangle \tag{6.59}$$

with the normalization condition $\langle\tilde{\Psi}_\mu|\Psi_\nu\rangle = \delta_{\mu\nu}$. In general it is possible to separate the non–Hermitian part by writing

$$h = h_\Theta + i\omega \tag{6.60}$$

with $h_\Theta$ being Hermitian. Assuming the basis (B.4,B.5) it is easy to see that the matrix elements of $\omega$ are real and symmetric. This further implies $|\tilde{\Psi}_\nu\rangle = |\Psi_\nu^*\rangle$. It is then useful to decompose the eigenstates into real and imaginary parts $|\Psi_\nu\rangle = |\Psi_\nu^R\rangle + i|\Psi_\nu^I\rangle$. Here the superscript refers to real (R) and imaginary (I) parts of the expansion coefficients $V_{\nu k}$ in eq (B.6). Taking account of the fact that $|\tilde{\Psi}_\nu\rangle = |\Psi_\nu^R\rangle - i|\Psi_\nu^I\rangle$ allows one to extract real and imaginary parts of the one–particle energy eigenvalues

$$\begin{aligned}
\epsilon_\nu^R &= \langle\Psi_\nu^R|h_\Theta|\Psi_\nu^R\rangle - \langle\Psi_\nu^I|h_\Theta|\Psi_\nu^I\rangle - \langle\Psi_\nu^I|\omega|\Psi_\nu^R\rangle - \langle\Psi_\nu^R|\omega|\Psi_\nu^I\rangle, \\
\epsilon_\nu^I &= \langle\Psi_\nu^R|\omega|\Psi_\nu^R\rangle - \langle\Psi_\nu^I|\omega|\Psi_\nu^I\rangle + \langle\Psi_\nu^I|h_\Theta|\Psi_\nu^R\rangle + \langle\Psi_\nu^R|h_\Theta|\Psi_\nu^I\rangle.
\end{aligned} \tag{6.61}$$

These relations are suitable to evaluate the functional derivatives of $\epsilon_\nu^{(R,I)}$ with respect to the meson fields. *E.g.* one finds

$$\frac{\delta\epsilon_\nu^I}{\delta\omega(r)} = r^2 \int \frac{d\Omega}{4\pi} \left(\langle\boldsymbol{r}|\Psi_\nu^R\rangle\langle\Psi_\nu^R|\boldsymbol{r}\rangle - \langle\boldsymbol{r}|\Psi_\nu^I\rangle\langle\Psi_\nu^I|\boldsymbol{r}\rangle\right). \tag{6.62}$$

Expressions like this enter the equations of motion for the soliton profile functions (*cf.* appendix C). As can be inferred from the equation of motion for the $\omega$–meson (C.6) this profile function is directly related to the baryon density. As a matter of fact this causes a non–vanishing $\omega$ field. In particular, the integral $\int d^3r\ \omega = g_V^2/4m_V^2$ is fixed when $\mathcal{A}_I$ is not regularized and satisfies the normalization condition on the $\omega$ field imposed by the current field identities (3.21).

Numerically, solutions have been found for $m \geq 350$MeV, although the determination of a lower bound has not been the central issue in ref.[76]. More importantly it has to be remarked that these solutions were obtained for the physical $\omega$ meson mass, $m_\omega = 770$MeV. This is in contrast to the treatment discussed in ref.[78][f] where stable solutions appeared only when $m_\omega$ was chosen about four times as large. In subsection 6.2.4 we give a possible explanation for the non–existence of solutions in that treatment.

Next it should be noted that the value of the constituent quark mass $m$ at which the valence quark energy changes its sign is at 545MeV and thus considerably lower than in the purely pionic system. This result is easy to understand since the repulsive character of the $\omega$ meson yields a larger extension of the soliton profile which causes the valence quark to be more strongly bound. This repulsive effect is also observed when evaluating the baryonic radius although its extraction is somewhat hampered by finite size effects. These effects do not show up in the determination of the energy, however, higher moments of the $\omega$ profile function may be obscured. It has also been ascertained that regularization of the imaginary part of the action does not lead to qualitative changes of the above presented results.

In the next step all vector mesons were included [73]. This embraces the *ansätze* (6.56,6.57) and (6.58), only the isoscalar–scalar field being kept at its vacuum expectation value. The above discussed structure of the Euclidean Dirac Hamiltonian is not altered, however, it is more complex because a larger number of fields is involved. In this case the interesting question arises whether or not Witten's conjectures remains valid. The previous explorations seem to indicate that effects on the valence quark energy eigenvalue of the isovector and isoscalar (axial) vector mesons add up coherently. Numerically self–consistent solutions to this extended problem have been found in the interval 300MeV$\le m \le$400MeV although the authors of ref.[73] do not exclude the existence of solutions in a larger range. Indeed a further decrease of the valence quark energy eigenvalue, $\epsilon^R_{\rm val}$ has been observed, *e.g.* for $m = 350$MeV $\epsilon^R_{\rm val}$ changes from -133MeV to -154MeV when supplementing the $\pi - \rho - a_1$ system by the $\omega$ meson. This, however, is only the case when the imaginary part of the action is subject to regularization; when this regularization is discared $\epsilon^R_{\rm val}$ increases even slightly as an effect of including $\omega$. On the whole, however, the influence of regularizing the imaginary part has been found to be small.

[f] See also section 6.1 for the discussion of this approach.

Very recently the energy functional (6.39) has been studied for the case that also the isoscalar–scalar field is space dependent [75]. This introduces the additional radial function $\phi(r)$, *cf.* subsection 6.2.2. As this represents the most general case we use this opportunity to display the associated Euclidean Dirac Hamiltonian

$$\begin{aligned} h &= \boldsymbol{\alpha}\cdot\boldsymbol{p} + i\omega(r) + m\phi(r)\beta\left(\cos\Theta(r) + i\gamma_5\boldsymbol{\tau}\cdot\hat{\boldsymbol{r}}\sin\Theta(r)\right) \\ &\quad +\frac{1}{2}(\boldsymbol{\alpha}\times\hat{\boldsymbol{r}})\cdot\boldsymbol{\tau}G(r) + \frac{1}{2}(\boldsymbol{\sigma}\cdot\hat{\boldsymbol{r}})(\boldsymbol{\tau}\cdot\hat{\boldsymbol{r}})F(r) + \frac{1}{2}(\boldsymbol{\sigma}\cdot\boldsymbol{\tau})H(r). \end{aligned} \qquad (6.63)$$

As long as the scalar field $\phi$ has not been included as a dynamical degree of the freedom the question of whether or not regularizing the imaginary part only played a subleading role. Taking, however, $\phi$ to be space dependent leads to a completely different situation. As already seen in subsection 6.2.2 a field configuration which is sharply peaked at $r = 0$ yields a vanishing soliton energy. This remains true even when vector mesons are included as long as no mechanism prevents these vector mesons from being zero. There is no such mechanism for the static $\rho$ and $a_1$ fields but the static $\omega$ field is directly correlated to the baryon number density. Unfortunately, for the sharply peaked field configuration (6.51) also the regularized baryon number vanishes and thus also $\omega \equiv 0$ is allowed in this case.

When the imaginary part is not regularized stable solutions have been obtained for $m \gtrsim 350$ MeV [75]. Then the $\omega$ meson provides sufficient repulsion to prevent the scalar field from collapsing. The lower bound for $m$ is somewhat larger (410MeV) when $\rho$ and $a_1$ fields are ignored. Actually the deviation of the scalar field from its vacuum expectation values is almost negligible. The alternative method of stabilization (6.54) has not been studied for the case that (axial) vector mesons are included.

In table 6.2 we summarize the numerical results for the various cases discussed above. There the specific contributions to the energy are compared for the constituent quark mass $m = 350$MeV.

Figure 6.5 is well suited to discuss the effects of various mesons fields on the chiral angle. One observes that the $\rho$ meson is repulsive at small distances and attractive for $r \ge 0.3$fm. A much stronger attraction is provided by the axial–vector meson. Also the repulsive character of the $\omega$–meson is reproduced by the NJL model. In the inner region of the soliton $\lesssim$1.5fm the scalar field provides a small repulsion while at larger distances (not shown in figure 6.5) the scalar field causes the chiral angle to swell slightly. Although this attraction appears to be quite weak it indicates that the NJL soliton with scalar mesons contains at least some intermediate range attraction called for by the central potential of the nucleon–nucleon interaction. Unfortunately, no quantitative statement on this subject can be made presently in the NJL

Table 6.2: The soliton energy for various treatments of the NJL soliton. The meson fields listed in the first line represent those meson profiles which are allowed to deviate from their vacuum expectation values. All numbers are evaluated for a constituent quark mass $m = 350$MeV and $m_\pi = 135$MeV.

| | $\pi$ | $\pi, \rho$ | $\pi, \omega$ | $\pi, \rho, a_1$ | $\pi, \omega, \rho, a_1$ | $\phi, \pi, \omega, \rho, a_1$ |
|---|---|---|---|---|---|---|
| $E$ (MeV) | 1236 | 966 | 1404 | 1011 | 1144 | 1125 |
| $E_0^R$ (MeV) | 459 | 584 | 546 | 616 | 558 | 2271 |
| $E_0^I$ (MeV) | 0 | 0 | -14 | 0 | 186 | 177 |
| $E_{\rm m}$ (MeV) | 31 | 173 | 0 | 395 | 400 | -1323 |
| $\epsilon_{\rm val}^R/m$ | 0.71 | 0.20 | 0.56 | -0.37 | -0.27 | -0.28 |
| $\epsilon_{\rm val}^I/m$ | 0 | 0 | 0.27 | 0 | 0.16 | 0.15 |

model.

### *6.2.4 Quadratic expansion for time components of vector fields*

The preceding discussions concerning the analytic continuation of the action in the presence of the $\omega$ meson were based on the analysis of the analytic properties of the eigenvalues of the Euclidean Dirac Hamiltonian $h$ (6.63). It has been assumed that the expansion of these eigenvalues in powers of the $\omega$–field can straightforwardly be transferred to the action functional. However, it has recently been demonstrated that this is actually not the case [71]. In order to prove the identity one had to assume that $h$ and $h^\dagger$ can be diagonalized simultaneously. Starting off at

$$\not{D}_E \not{D}_E^\dagger = -\partial_\tau^2 + h_\Theta^2 + 2i\omega_4\partial_\tau + i\left[h_\Theta, \omega_4\right] + \omega_4^2 \tag{6.64}$$

and imposing the proper–time prescription at the operator level

$$\mathcal{A}_R \rightarrow -\frac{1}{2}{\rm Tr} \int_{1/\Lambda^2}^\infty \frac{ds}{s} \exp\left\{-\partial_\tau^2 + h_\Theta^2 + 2i\omega_4\partial_\tau + i\left[h_\Theta, \omega_4\right] + \omega_4^2\right\} \tag{6.65}$$

the energy functional (6.25) can only be obtained with the above mentioned assumption of simultaneous diagonalizability because the term linear in $\partial_\tau$ originates from $\left(h - h^\dagger\right)\partial_\tau$. One may obtain eq (6.25) from (6.65) by substituting the coefficient of $\partial_\tau$ with $2i\epsilon_\nu^I$. The before–mentioned ambiguities, which arise from the inadequate application of the rules for manipulating the logarithm to derive (6.24), are avoided when starting from (6.65) without further assumptions. It is also obvious that because of these assumptions different expansion schemes will lead to different results. It is the purpose of the present section to point out the differences between the approach which relies on expanding the eigenvalues $\epsilon_\nu$ of $h$ on the one hand and the expansion of the operator (6.65) on the other; both expansions are understood in terms of $\omega_4$.

Employing techniques which have been worked out in the context of the semi–classical quantization of the soliton (section 7.1) [70] and been extended to the treatment of small amplitude fluctuations off the soliton (see section 7.3) [97] one obtains a Minkowski energy

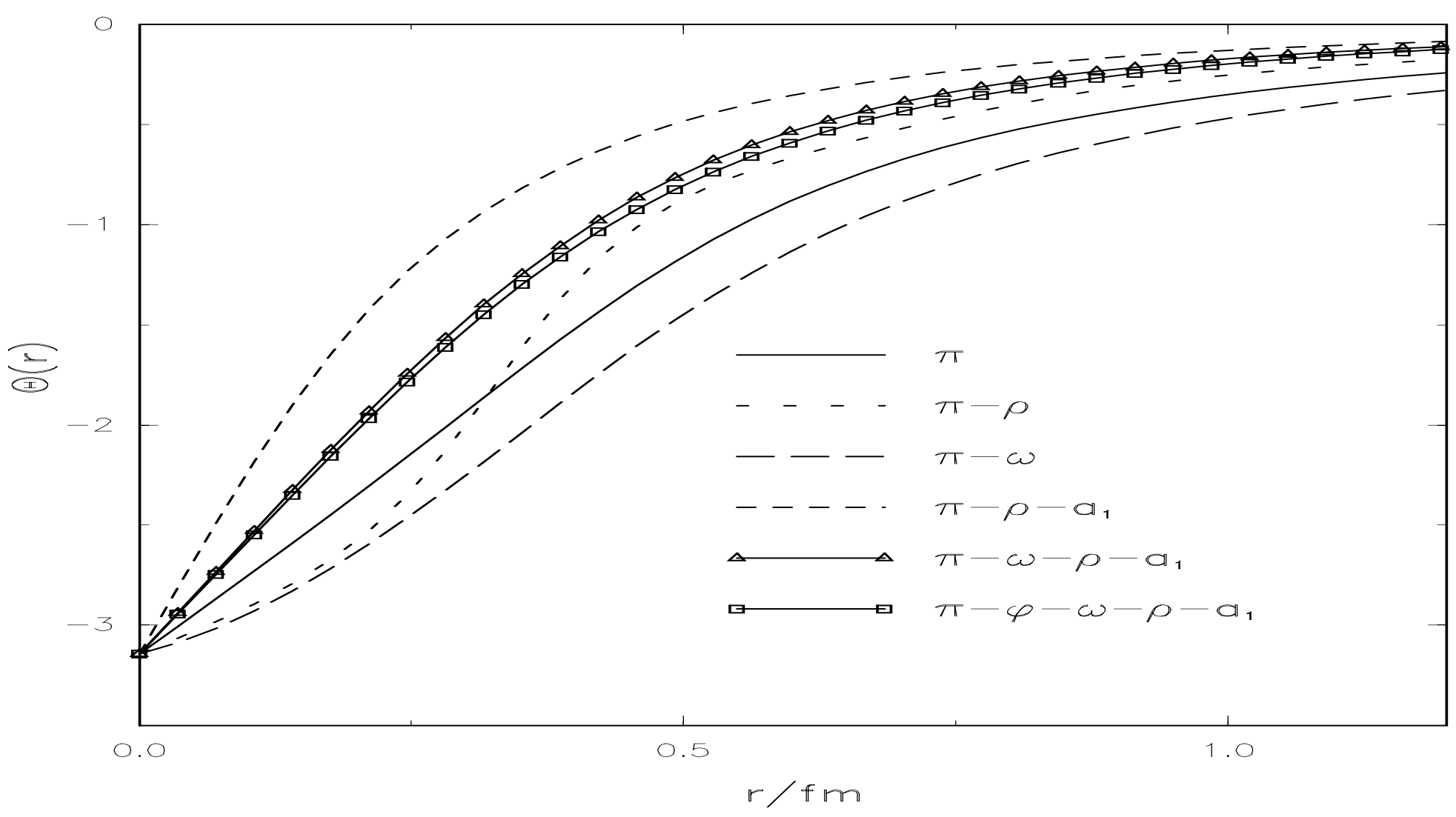

Figure 6.5: The profile function $\Theta(r)$ for the self–consistent chiral angle in various approaches to the NJL model. Those fields which are allowed to be space dependent are indicated. The parameters used are $m = 350$MeV and $m_\pi = 135$MeV.

functional up to second order in the $\omega$ field [71]

$$\begin{aligned} E_2^{\rm Mink}[\Theta,\omega_0] &= E_0[\Theta] + N_C\eta_{\rm val}\langle{\rm val}|\omega_0|{\rm val}\rangle - \frac{N_C}{2}\sum_\nu {\rm sgn}(\epsilon_\nu^0){\rm erfc}\left(\left|\frac{\epsilon_\nu^0}{\Lambda}\right|\right)\langle\nu|\omega_0|\nu\rangle \qquad (6.66)\\ &-N_C\eta_{\rm val}\sum_{\nu\ne{\rm val}}\frac{|\langle\nu|\omega_0|{\rm val}\rangle|^2}{\epsilon_\nu^0-\epsilon_{\rm val}^0} - \frac{N_C}{2}\sum_{\nu\mu} f\left(\epsilon_\nu^0,\epsilon_\mu^0;\Lambda\right)|\langle\nu|\omega_0|\mu\rangle|^2 - \frac{4\pi}{3}\frac{m_\omega^2 f_\pi^2}{m^2}\int dr r^2\omega_0^2. \end{aligned}$$

Here $E_0[\Theta]$ refers to the energy functional associated with the Hermitian part of the Hamiltonian $h_\Theta$. Furthermore $\epsilon_\nu^0$ and $|\nu\rangle$ denote the eigenvalues and –vectors of $h_\Theta$. Also the mass terms associated with the $\rho$ and $a_1$ mesons are included in $E_0[\Theta]$. The imaginary part of the action has been assumed in regularized form. If one wishes to abandon this regularization the complementary error function in (6.66) has to be replaced by unity. Finally the regularization function

$$f\left(\epsilon_\mu,\epsilon_\nu;\Lambda\right) = \frac{1}{2}\frac{{\rm sgn}(\epsilon_\mu){\rm erfc}\left(\left|\frac{\epsilon_\mu}{\Lambda}\right|\right) - {\rm sgn}(\epsilon_\nu){\rm erfc}\left(\left|\frac{\epsilon_\nu}{\Lambda}\right|\right)}{\epsilon_\mu-\epsilon_\nu} - \frac{\Lambda}{\sqrt{\pi}}\frac{e^{-(\epsilon_\mu/\Lambda)^2}-e^{-(\epsilon_\nu/\Lambda)^2}}{\epsilon_\mu^2-\epsilon_\nu^2} \qquad (6.67)$$

has a smooth limit [70]

$$\lim_{\epsilon_\mu\to\epsilon_\nu} f\left(\epsilon_\mu,\epsilon_\nu;\Lambda\right) = 0. \qquad (6.68)$$

It is actually because of this limit that the consistency conditions for energy functional, which are discussed in section 6.1, are satisfied by the functional (6.66). As was to be expected this regulator function is identical to the one for the moment of inertia [70]. In both cases the real part of the action is expanded in terms of static and anti–Hermitian operators (in the case of the moment of inertia these are the isospin generators).

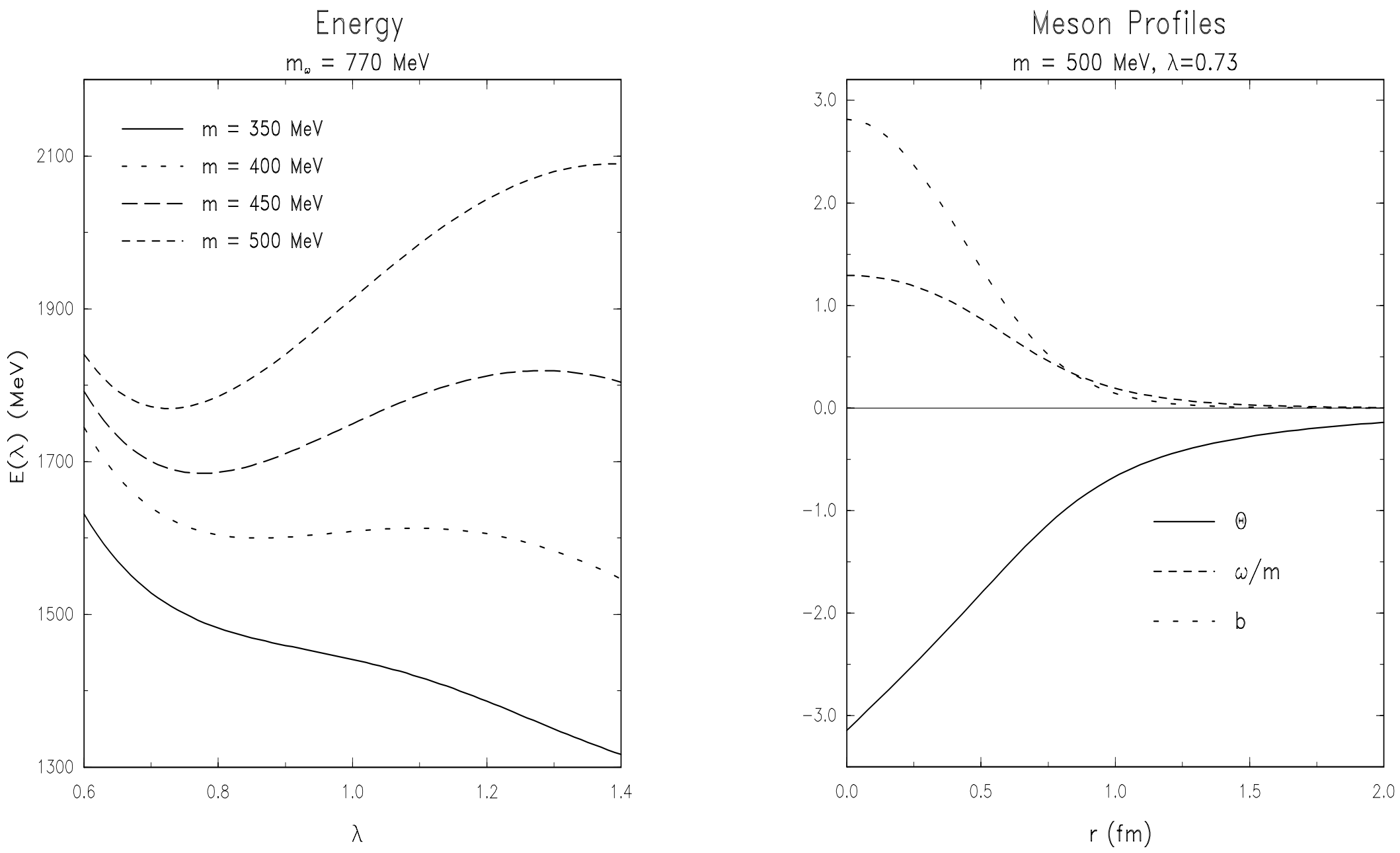

Figure 6.6: Left: The total energy as a function of the scaling parameter $\lambda$ defined in eq (6.69) for various values of the constituent quark mass $m$. Right: The profile functions which minimize the energy functional (6.66) under variation of the scaling parameter $\lambda$. The quark baryonic density $b \sim q^\dagger q$ is artificially scaled such that the spatial integrals over $\omega/m$ and $b$ coincide. Here the constituent quark mass is assumed to be 500MeV. (Figure taken from ref.[71].)

In ref.[71] several parametrical descriptions have been adopted for the profiles other than $\omega(r)$ and subsequently the latter has been determined from the stationary condition corresponding to (6.66). As an example we quote the scaling *ansatz*

$$\Theta_\lambda(r) = \Theta_{\rm s.c.}(\lambda r) \tag{6.69}$$

for the model consisting only of the pion and the $\omega$ meson. In eq (6.69) $\Theta_{\rm s.c.}(r)$ refers to the self–consistent soliton solution to the problem without the $\omega$ meson. Subsequently the stationary condition associated with the functional (6.66) was solved in order to obtain the profile function $\omega_\lambda(r)$. This procedure of first parametrizing the chiral angle and subsequently solving for the $\omega$ profile exactly, is also justified by the fact that commonly the stationary condition for $\omega(r)$ represents a constraint rather than only an Euler–Lagrange equation of motion. Then the profile functions as well as the energy functional depend on the parameter $\lambda$. This dependence is displayed in figure 6.6. This figure indicates that the functional (6.66) indeed possesses a unique local minimum for constituent quark masses $m \gtrsim 400$MeV. This figure also contains the profile function which minimize (6.66) when $m = 500$MeV is adopted. Obviously the $\omega$ profile gets support at larger distances than does the quark baryon current $b \sim q^\dagger q$. This confirms the repulsive character of the $\omega$ meson. The minimal energy values are given as functions of $m$ in table 6.3. Also shown are the contributions due to the linear and quadratic terms in $\omega$. Note the virial factor, -2, between these two pieces. The presence of the $\omega$ meson then increases the soliton energy by about $300-400{\rm MeV}$. This contribution grows with the constituent quark mass. It should be noted that the quadratic term splits into two pieces, one from the purely mesonic part of the action ($-296{\rm MeV}$ for $m=500{\rm MeV}$) and one stemming from the expansion of $\mathcal{A}_R$ ($-129{\rm MeV}$ for $m=500{\rm MeV}$). The smallness of the last quantity compared to the total energy gives evidence for the fact that this expansion of $\mathcal{A}_R$ in terms of $\omega_4$ converges quickly. Furthermore, it has been shown that including the $\omega$ meson via the expansion (6.66) indeed provides a sizable repulsion, a feature commonly attributed to this meson field. This is expressed by the fact that the value, which $\lambda$ assumes at the minimum of the energy, is significantly smaller than one.

Table 6.3: Soliton energy $E_2^{\rm Mink}$ (6.66) when $\Theta$ and $\omega$ are the only space dependent fields for various constituent quark masses. Also shown are the contributions from the terms linear (lin.) and quadratic (quad.) in $\omega$. All numbers are in MeV. (Results are taken from ref.[71].)

| $m$ | $E_2^{\rm Mink}$ | lin. | quad. |
|---|---|---|---|
| 400 | 1585 | 650 | -325 |
| 450 | 1685 | 734 | -367 |
| 500 | 1769 | 856 | -428 |

The approach having just been described has to be compared with an expansion of the eigenvalues of the whole Dirac Hamiltonian $h_\Theta + i\omega_4$ (6.60) in Euclidean space. Applying standard perturbation techniques results in

$$\epsilon_\nu^R = \epsilon_\nu^0 - \sum_{\mu\ne\nu} \frac{\langle\nu|\omega_4|\mu\rangle\langle\mu|\omega_4|\nu\rangle}{\epsilon_\nu^0-\epsilon_\mu^0}\ldots \qquad \text{and} \qquad \epsilon_\nu^I = i\langle\nu|\omega_4|\nu\rangle\ldots \quad . \tag{6.70}$$

Substitution of this expansion into the expression for real and imaginary parts of the Euclidean energy (6.27 and 6.36) and subsequent continuation to Minkowski space yields at second order an energy functional similar to (6.66), however, with $f\left(\epsilon_\mu,\epsilon_\nu;\Lambda\right)$ replaced by

$$\tilde{f}\left(\epsilon_\nu,\epsilon_\mu;\Lambda\right) = \begin{cases} \frac{1}{2}\frac{{\rm sgn}(\epsilon_\nu){\rm erfc}\left(\left|\frac{\epsilon_\nu}{\Lambda}\right|\right)-{\rm sgn}(\epsilon_\mu){\rm erfc}\left(\left|\frac{\epsilon_\mu}{\Lambda}\right|\right)}{\epsilon_\nu-\epsilon_\mu}, & \epsilon_\nu\ne\epsilon_\mu \\ 0, & \epsilon_\nu=\epsilon_\mu \end{cases} . \tag{6.71}$$

We will refer to the corresponding energy functional by $\tilde{E}_2$. In ref.[71] it has been demonstrated that for a well motivated field configuration $\tilde{E}_2$ deviates only by about 3% from (6.37) which results from diagonalizing (6.60).

However, the regulator functions $f$ and $\tilde{f}$ differ significantly. First of all, they only agree in the limit $\Lambda\to\infty$. As the cut–off $\Lambda$ is quite small this difference is sizable numerically. Next it has to be noted that $\tilde{f}$ is discontinuous as $\epsilon_\mu\to\epsilon_\nu$. This has the unpleasant consequence that the $\omega$ contribution to the energy is not positive definite for configurations which satisfy the stationary condition. Furthermore, the second order contribution associated with $\tilde{f}$ is $-553{\rm MeV}$; to be compared with the above mentioned $-129{\rm MeV}$ for $f$. Most importantly, however, $\tilde{E}_2$ does not possess a local minimum. This explains why for empirical parameters the authors of ref.[78] did not find a self–consistent solution in the model containing only the pion and the $\omega$ degrees of freedom.

These investigations demonstrate that counting powers of $\omega_4$ in the Euclidean energy functional (6.37) may not be the correct approach to derive a Minkowski energy functional in the presence of isoscalar field $\omega$. Rather the functional (6.66) appears to be the appropriate starting point. This statement is also supported by previous calculations in the context of the semi–classical quantization of the chiral soliton as well as other nucleon observables (see chapter 7). These computations all impose the proper–time regularization at the action level in Euclidean space (6.65).

### 6.2.5 Local chiral rotation

In sections 3.4 and 4.2 the effectiveness of the chiral transformation $\tilde{\Psi} = \mathcal{T}\Psi$ with $\mathcal{T} = P_L\xi_L + P_R\xi_R$ has become evident when exploring the meson sector of the NJL model. In particular, it provides the link between the massive Yang–Mills and hidden gauge approaches to vector mesons. It is therefore interesting to illuminate the role of this transformation in the soliton sector as well. For this study the unitary gauge $\xi_R = \xi_L^\dagger = \xi$ will be adopted. This implies $\mathcal{T}(\Theta) = \cos(\Theta/2) + i\gamma_5\boldsymbol{\tau}\cdot\hat{\boldsymbol{r}}\,\sin(\Theta/2)$. The Dirac Hamiltonian for the chirally rotated quark fields $\tilde{\Psi}$ then becomes [32, 98]

$$\begin{aligned} h_R = \mathcal{T}(\Theta)h\mathcal{T}^\dagger(\Theta) &= \boldsymbol{\alpha}\cdot\boldsymbol{p} + \beta m - \frac{1}{2}(\boldsymbol{\sigma}\cdot\hat{\boldsymbol{r}})(\boldsymbol{\tau}\cdot\hat{\boldsymbol{r}})\left(\Theta'(r) - \frac{1}{r}\sin\Theta(r)\right) \\ &\quad -\frac{1}{2r}(\boldsymbol{\sigma}\cdot\boldsymbol{\tau})\sin\Theta(r) - \frac{1}{r}\boldsymbol{\alpha}\cdot(\hat{\boldsymbol{r}}\times\boldsymbol{\tau})\sin^2\left(\frac{\Theta(r)}{2}\right). \end{aligned} \tag{6.72}$$

Again one observes that the chiral field has been eliminated from the mass term at the expense of induced vector mesons. Before discussing the physical implications of this transformation in the soliton sector a few remarks on its technical feasibility are in order. First of all, one observes that the coordinate singularities do not disappear at $r = 0$ as long as $\Theta(r = 0) = l\pi$. These singularities appear since $\mathcal{T}(\Theta)$ is topologically distinct from the unit transformation. For $\mathcal{T}(2\Theta)$ these coordinate singularities are absent, however, the spectrum obtained from diagonalizing $\mathcal{T}(2\Theta)h\mathcal{T}^\dagger(2\Theta)$ in a <u>finite</u> basis is not identical to that of $h$. In the $G^\pi = 0^+$ channel the lowest state is missing while an additional one shows up at the upper boundary in momentum space. This shift of eigenstates is repeated for each $n$ in $\mathcal{T}(2n\Theta)$, with $n \geq 2$. Hence, one observes another reflection of the topological character of the transformation $\mathcal{T}$. Secondly, one should remember that for the diagonalization (6.43) certain boundary conditions on the eigenstates have been imposed (*cf.* the discussion after eq (B.6)). It should be obvious that these boundary conditions are affected by the chiral rotation $\mathcal{T}$ which is not uniquely defined at $r = 0$. With the corresponding redefinition of the basis spinors the coordinate singularities are canceled. The numerical diagonalization

$$h_R\tilde{\Psi}_\nu = \tilde{\epsilon}_\nu\tilde{\Psi}_\nu \tag{6.73}$$

then indeed yields $\tilde{\epsilon}_\nu = \epsilon_\nu$ and $\tilde{\Psi}_\nu = \mathcal{T}^\dagger(\Theta)\Psi_\nu$ with $\epsilon_\nu$ and $\Psi_\nu$ being the solutions to the eigenvalue problem (6.43).

This formalism has in turn been employed to examine the situation when the chiral rotation is performed on the (axial) vector fields [31]

$$\begin{aligned} \tilde{V}_\mu + \tilde{A}_\mu &= \xi_R(\partial_\mu + V_\mu + A_\mu)\xi_R^\dagger, \\ \tilde{V}_\mu - \tilde{A}_\mu &= \xi_L(\partial_\mu + V_\mu - A_\mu)\xi_L^\dagger. \end{aligned} \tag{6.74}$$

Table 6.4: Contributions to the energy for self–consistent solution in various treatments of the axial–vector meson in the NJL model. Those meson fields which are allowed to be space dependent are indicated. The constituent quark mass $m$ =400MeV is common. All numbers are in MeV. (Results are taken from ref.[98].)

| | $\pi-\rho$ | $\pi-\rho-a_1$ | $\pi-\tilde{V}$ |
|---|---|---|---|
| $\epsilon_{\rm val}$ | 313 | -222 | -351 |
| $E^{\rm det}$ | 711 | 543 | 240 |
| $E_m$ | 149 | 393 | 175 |
| $E_{\rm tot}$ | 861 | 937 | 415 |

and the transformed axial–vector field is neglected, *i.e.* $\tilde{A}_\mu = 0$. This *ansatz* does not violate the local chiral symmetry in contrast to the approach $A_\mu = 0$, *cf.* eq (3.42). This configuration is of special interest since this approximation has frequently been applied to extended Skyrme models[g]. The important question arises which (or whether at all any) features of the unrotated axial–vector field ($A_\mu$) are maintained. For these studies the effects of the $\omega$ meson have been ignored. After carrying over the chiral rotation onto the quark spinors only vector meson degrees of freedom are contained in the Dirac Hamiltonian. For these the Wu-Yang *ansatz* has been assumed yielding [98]

$$h_R = \boldsymbol{\alpha}\cdot\boldsymbol{p} + \beta m + \frac{G(r)}{2r}\boldsymbol{\alpha}\cdot(\hat{\boldsymbol{r}}\times\boldsymbol{\tau}). \tag{6.75}$$

The coupling of the vector and pseudoscalar fields is then completely contained in the mesonic part of the energy functional [98]

$$E_{\rm m} = \frac{\pi}{G_2}\int dr \left\{(G(r)+1-\cos\Theta(r))^2 + \frac{1}{2}r^2\left(\Theta'(r)\right)^2 + \sin^2\Theta(r)\right\}. \tag{6.76}$$

Obviously $\Theta(r=0) = -\pi$ implies that $G(r=0) = -2$. One thus has to deal with a Dirac Hamiltonian which contains a topologically non–trivial vector meson field. Thus (6.75) is singular and cannot be treated using the standard basis [81] but rather by employing techniques which are analogous to those developed to diagonalize the rotated Hamiltonian (6.72).

It has been demonstrated that self–consistent solutions which minimize the total energy exist [98]. Here the total energy is defined as the sum $E_{\rm tot} = E_m + E^{\rm det}$ with $E^{\rm det}$ being the contribution of the fermion determinant (6.27) in terms of the eigenvalues of (6.75). Also the valence quark contribution has to be added if necessary to accommodate unit baryon number. In table 6.4 the results of these calculations are compared with two different treatments of the axial–vector degrees of freedom in the unrotated formulation. Regarding the strong binding of the valence quark state it is clear that this important effect, which is commonly asserted to $A_\mu \ne 0$, is retained in the model with $\tilde{A}_\mu = 0$. Furthermore, the valence quark wave–functions exhibits features which are associated with a localized antiquark as *e.g.* a dominating lower component. Unfortunately the total energy comes out quite low making an application of this model for the description of baryons doubtful. In ref.[98] is has also been demonstrated that the smallness of the total energy is linked to the missing repulsion in this model. This is also reflected by the meson profile functions which possess only a very small spatial extension. Again this is an effect originating from the presence of an axial–vector field in the unrotated frame, $A_\mu \neq 0$. Furthermore the effects asserted to the $\omega$–meson were modeled by the inclusion of a $6^{\rm th}$ order term (4.22) usually used in the Skyrme model to simulate the $\omega$–meson. Then a sizable repulsion was obtained and the energy of the valence quark orbit decreased even more to approximately the negative constituent mass [98]. Thus this treatment provides a strong support of Skyrme type models which rely on the assumption that the valence quark has joined the negative Dirac sea.

[g] For a compilation of relevant articles see ref.[12].

# 7 Baryons

In the preceding sections we have examined the soliton solutions in the NJL model. Except for the baryon number these solutions do not carry the quantum numbers of physical baryons. In particular, these solutions are neither eigenstates of the angular momentum $\boldsymbol{J}$ nor of isospin $\boldsymbol{I}$. As these solitons have vanishing grand spin ($G = 0$) they represent linear combinations of states with $|\boldsymbol{J}| = |\boldsymbol{I}|$. In order to describe physical baryons the soliton configuration has to be projected onto states with good angular momentum and isospin quantum numbers. As already mentioned in section 5.3 this is commonly achieved by introducing collective coordinates which describe the orientation of the soliton in coordinate– and isospace. Subsequently these coordinates are canonically quantized yielding the angular momentum and isospin operators. As the grand spin symmetry of the hedgehog (6.41) causes the equivalence of rotations in coordinate– and isospace only one set of collective coordinates is needed. For convenience, one chooses the isospin orientation. This procedure has first been applied by Adkins, Nappi and Witten to the SU(2) Skyrmion [9]. In the proceeding section we will explain the analogous treatment for the NJL chiral soliton of pseudoscalar fields [70]. The generalization to three flavors is more involved not only due to the fact that the moment of inertia tensor (5.18) is no longer proportional to the unit matrix but also because of symmetry breaking being present. Two complementary approaches ([62] and [99]) will be presented in subsection 7.5.

## 7.1 Quantization of the chiral soliton

The time–dependent collective coordinates describing the isospin orientation of the soliton are parametrized with the help of an $2 \times 2$ unitary matrix $R(t)$

$$M(\boldsymbol{r}, t) = R(t) M_0(\boldsymbol{r}) R^\dagger(t) \tag{7.1}$$

where $M_0(\boldsymbol{r})$ denotes the static hedgehog configuration (6.41). The scalar–pseudoscalar part of $\mathcal{A}_m$ does not contain any time derivatives. Thus it is independent of collective coordinates as long as symmetry breaking is ignored. Then the dependence of the action on $R(t)$ and its time derivatives completely originates from the fermion determinant

$$\mathrm{Tr}\,\log\left(i\beta\partial\!\!\!/ - \beta\left[P_L M^\dagger(\boldsymbol{r}, t) + P_R M(\boldsymbol{r}, t)\right]\right). \tag{7.2}$$

The functional trace is most conveniently evaluated by transforming to the flavor rotating system $q \rightarrow q' = Rq$ [70]

$$\begin{aligned}
&\mathrm{Tr}\,\log\left(i\beta\partial\!\!\!/ - \beta\left[P_L M^\dagger(\boldsymbol{r}, t) + P_R M(\boldsymbol{r}, t)\right]\right) \\
&\qquad = \mathrm{Tr}\,\log\left(i\beta\partial\!\!\!/ - \frac{1}{2}\boldsymbol{\tau}\cdot\boldsymbol{\Omega} - \beta\left[P_L M_0^\dagger(\boldsymbol{r}) + P_R M(\boldsymbol{r})\right]\right) \\
&\qquad = \mathrm{Tr}\,\log\left(i\partial_t - \frac{1}{2}\boldsymbol{\tau}\cdot\boldsymbol{\Omega} - h\right)
\end{aligned} \tag{7.3}$$

which introduces the angular velocities

$$R^\dagger(t)\frac{\partial}{\partial t}R(t) = \frac{i}{2}\boldsymbol{\tau}\cdot\boldsymbol{\Omega}. \tag{7.4}$$

In eq (7.3) $h$ denotes the static Dirac Hamiltonian defined in eq (6.42). In the adiabatic approximation, *i.e.* when the time derivatives of $\boldsymbol{\Omega}$ are neglected, one might define an intrinsic Hamiltonian $h' = h + \frac{1}{2}\boldsymbol{\tau}\cdot\boldsymbol{\Omega}$ which could formally be treated analogously to the preceding calculations as long as regularization is ignored. It is then obvious that the fermion determinant can be decomposed into vacuum and valence quark parts. Unfortunately, the eigenvalues of $h'$ are not known and thus a perturbation expansion in $\boldsymbol{\Omega}$ has to be carried out. For the valence quark part this is a straightforward application of standard perturbation theory resulting in

$$\mathcal{A}_V = -TE_V + \frac{T}{2}\sum_{i,j=1}^{3}\Theta_{ij}^{\rm val}\Omega_i\Omega_j + \mathcal{O}\left(\boldsymbol{\Omega}^4\right). \tag{7.5}$$

In a symmetric two flavor model only even powers of $\boldsymbol{\Omega}$ are allowed. $\Theta_{ab}^{\rm val}$ denotes the valence quark contribution to the moment of inertia [70]

$$\Theta_{ij}^{\rm val} = \frac{N_C}{2}\sum_{\mu\nu}\eta_\mu(1-\eta_\nu)\frac{\langle\mu|\tau_i|\nu\rangle\langle\nu|\tau_j|\mu\rangle}{\epsilon_\nu-\epsilon_\mu} = \frac{N_C}{2}\eta_{\rm val}\sum_{\mu\neq{\rm val}}\frac{\langle{\rm val}|\tau_i|\mu\rangle\langle\mu|\tau_j|{\rm val}\rangle}{\epsilon_\mu-\epsilon_{\rm val}}. \tag{7.6}$$

The latter equality arises because only the valence quark orbit is occupied. The fermion determinant gives also rise to a vacuum contribution to the moment of inertia $\Theta_{ij}^{\rm vac}$ [70]. In order to extract this part from the fermion determinant one again has to continue to Euclidean space and consider the limit $T\to\infty$. In this context it is important to note that $\boldsymbol{\Omega}$ represents the time component of an induced vector field. Thus $\boldsymbol{\Omega}$ has to be continued as

$$\boldsymbol{\Omega} \to i\boldsymbol{\Omega}_E \tag{7.7}$$

with $\boldsymbol{\Omega}_E$ being Hermitian. Due to isospin symmetry only even powers of $\boldsymbol{\Omega}_E$ appear in an expansion. Thus we only need to consider the real part of the fermion determinant. It is useful to define

$$K(s,z) = \exp\left(-s(-\partial_\tau - h')(-\partial_\tau - h')^\dagger\right)\Big|_{\partial_\tau\to iz} = \exp\left(-sA(z)\right) \tag{7.8}$$

where we have indicated that the temporal part of the trace is substituted by a spectral integral over $z$ for $T\to\infty$. This allows us to express the real part of the fermion determinant

$$\mathcal{A}_R = -\frac{1}{2}\int_{1/\Lambda^2}^{\infty}\frac{ds}{s}\int_{-\infty}^{\infty}\frac{dz}{2\pi}\,{\rm tr}\,K(s,z) \tag{7.9}$$

in the proper–time regularization. In (7.9) the trace includes spatial and internal degrees of freedom. Furthermore,

$$A(z) = (-\partial_\tau - h')(-\partial_\tau - h')^\dagger\Big|_{\partial_\tau\to iz} = \left(z+\frac{i}{2}\boldsymbol{\tau}\cdot\boldsymbol{\Omega}_E\right)^2 + \frac{1}{2}\left[\boldsymbol{\tau}\cdot\boldsymbol{\Omega}_E, h\right] + h^2. \tag{7.10}$$

In order to extract the moment of inertia we first expand $K(s,z)$ up to second order in $\boldsymbol{\Omega}_E$ and define the coefficient of the quadratic term as

$$K_{ab}^2(s,z) = \frac{\partial^2 K(s,z)}{\partial\Omega_E^a\partial\Omega_E^b}\bigg|_{\boldsymbol{\Omega}_E=0}. \tag{7.11}$$

This derivative may be expressed with the help of a Feynman parameter integral

$$\begin{aligned}\mathrm{tr}\left(K^2_{ab}(s,z)\right) &= s^2\int_0^1 dx\ \mathrm{tr}\left[\left(iz\tau_a+\left[\frac{\tau_a}{2},h\right]\right)K^0(s(1-x),z)\left(iz\tau_b+\left[\frac{\tau_b}{2},h\right]\right)K^0(sx,z)\right]\\ &\qquad +\mathrm{tr}\left(\left\{\frac{\tau_a}{2},\frac{\tau_b}{2}\right\}K^0(s,z)\right)\end{aligned} \tag{7.12}$$

where $K^0(s,z)=\exp\left(-s(z^2+h^2)\right)$ denotes the zeroth–order heat kernel. The spectral integral in (7.9) is of Gaussian type and may be carried out straightforwardly. The remaining trace is performed using the eigenstates of $h$ (6.43). This then allows one to also carry out the Feynman parameter integral. The vacuum part of the moment of inertia may finally be extracted from $\mathcal{A}_R=-TE_0-\frac{T}{2}\Theta^{\mathrm{vac}}_{ab}\Omega^a_E\Omega^b_E+\mathcal{O}\left(\boldsymbol{\Omega}^4_E\right)$ in the limit $T\to\infty$ [70]

$$\Theta^{\mathrm{vac}}_{ab}=\frac{N_C}{4}\sum_{\mu\nu}f_\Theta(\epsilon_\mu,\epsilon_\nu;\Lambda)\langle\mu|\tau_a|\nu\rangle\langle\nu|\tau_b|\mu\rangle \tag{7.13}$$

with the cut-off function (which actually is the same as in (6.67)) given by

$$f_\Theta(\epsilon_\mu,\epsilon_\nu;\Lambda)=\frac{\Lambda}{\sqrt{\pi}}\frac{e^{-(\epsilon_\mu/\Lambda)^2}-e^{-(\epsilon_\nu/\Lambda)^2}}{\epsilon_\nu^2-\epsilon_\mu^2}-\frac{\mathrm{sgn}(\epsilon_\nu)\mathrm{erfc}\left(\left|\frac{\epsilon_\nu}{\Lambda}\right|\right)-\mathrm{sgn}(\epsilon_\mu)\mathrm{erfc}\left(\left|\frac{\epsilon_\mu}{\Lambda}\right|\right)}{2(\epsilon_\mu-\epsilon_\nu)}. \tag{7.14}$$

Due to isospin invariance no direction in isospace is distinct and the total moment of inertia is isotropic

$$\alpha^2[\Theta]\delta_{ab}=\left(\Theta^{\mathrm{val}}_{ab}+\Theta^{\mathrm{vac}}_{ab}\right). \tag{7.15}$$

$\alpha^2$ is a functional of the chiral angle $\Theta$ since the eigenstates $|\nu\rangle$ and –values $\epsilon_\nu$ functionally depend on $\Theta$. Numerical results for $\alpha^2$ may *e.g.* be found in refs.[100, 101][a]. Up to quadratic order in $\boldsymbol{\Omega}$ the collective Lagrangian $L(\boldsymbol{\Omega})$ in Minkowski space is finally given by

$$L(\boldsymbol{\Omega})=-E[\Theta]+\frac{1}{2}\alpha^2[\Theta]\boldsymbol{\Omega}^2. \tag{7.16}$$

One easily verifies that the infinitesimal change of the meson fields $M(\boldsymbol{r},t)$ under spatial rotations may be written as

$$[M(\boldsymbol{r},t),\boldsymbol{r}\times\boldsymbol{\partial}]=\frac{\partial\dot{M}(\boldsymbol{r},t)}{\partial\boldsymbol{\Omega}}. \tag{7.17}$$

Thus the total spin is given by the Noether charges

$$\boldsymbol{J}=\int d^3r\ \mathrm{tr}\ \left\{\frac{\partial\mathcal{L}(M,\partial_\mu M)}{\partial\dot{M}}\frac{\partial\dot{M}}{\partial\boldsymbol{\Omega}}+\mathrm{h.c.}\right\}=\frac{\partial L}{\partial\boldsymbol{\Omega}}=\alpha^2\boldsymbol{\Omega}. \tag{7.18}$$

The hedgehog structure of the soliton relates isospin $\boldsymbol{I}$ and spin $\boldsymbol{J}$ via the adjoint representation $D_{ij}=\frac{1}{2}\mathrm{tr}\left(\tau_iR\tau_jR^\dagger\right)$ of $R$

$$I_i=-\sum_{j=1}^3 D_{ij}J_j. \tag{7.19}$$

[a] Our numerical results agree with ref.[100] but disagree with ref.[101].

In this respect the spin of the soliton may be regarded as the isospin in the isorotating frame and vice versa. As in the Skyrme model the resulting collective Hamiltonian is quantized as a rigid top

$$H = \boldsymbol{J} \cdot \frac{\partial L}{\partial \boldsymbol{\Omega}} - L = E + \frac{1}{2\alpha^2}\boldsymbol{J}^2 = E + \frac{1}{2\alpha^2}\boldsymbol{I}^2 \tag{7.20}$$

which yields a tower of baryons with identical spin and isospin like *e.g.* the nucleon or the $\Delta$ resonance. We will postpone the discussion of numerical results for the baryon masses $M_B = E + \frac{J(J+1)}{2\alpha^2}$ to section (7.5.1) where the application of this approach to three flavors is discussed. However, the quantization (7.18) is equally important for the study of static nucleon properties which will be discussed in the next section. Then frequent use will be made of one more relation between collective coordinates and operators [9]

$$I_i J_j = -\frac{3}{4} D_{ij} \tag{7.21}$$

which, however, is valid only when sandwiched between nucleon states because then $\boldsymbol{J}^2 = \boldsymbol{I}^2 = 3/4$.

## 7.2 Static nucleon properties

In this section we will discuss the results for nucleon observables such as magnetic moments and charge radii (subsection 7.2.1) and the axial charge of the nucleon, $g_A$ (subsection 7.2.2). As these quantities correspond to certain moments of symmetry currents the calculations are performed in two steps. Firstly, these currents are constructed from the NJL model action (3.25). Secondly, matrix elements of these currents with respect to nucleon states are evaluated using the apparatus developed in section (7.1).

The results of such calculations have already been reported in another review article [102]. Therefore we only briefly report and comment on these results. The reader may consult that reference for more details.

The symmetry currents $j_\Gamma^\nu(x)$ are extracted by introducing external gauge fields $a_\nu^\Gamma(x)$ and identifying their linear coupling to the meson fields, see also eqs (3.18) and (3.19). $\Gamma$ denotes the symmetries under consideration. These are: (1) the vector symmetry in electromagnetic direction $\mathcal{Q} = \frac{1}{2}\tau_3 + \frac{1}{6}$ and (2) the axial symmetry. Formally we write

$$j_\Gamma^\nu(x) = \left.\frac{\delta \mathcal{A}[\varphi, a_\rho^\Gamma]}{\delta a_\nu^\Gamma(x)}\right|_{a_\mu^\Gamma=0} \tag{7.22}$$

where $\varphi$ denotes the set of meson fields involved. In the present case this only refers to the chiral field (7.1). The gauge fields $a_\nu^\Gamma(x)$ only appear in the fermion determinant

$$\begin{aligned} j_\Gamma^\nu(x) &= \frac{\delta}{\delta a_\nu^\Gamma(x)} \mathrm{Tr}\ \log\ \left.\left(\not{D} - iM(\boldsymbol{r},t) + a_\rho^\Gamma \gamma^\rho\right)\right|_{a_\mu^\Gamma=0} \\ &= \frac{\delta}{\delta a_\nu^\Gamma(x)} \mathrm{Tr}\ \log\ \left.\left(i\partial_t - \frac{1}{2}\boldsymbol{\tau}\cdot\boldsymbol{\Omega} - h - iR^\dagger \beta a_\rho^\Gamma \gamma^\rho R\right)\right|_{a_\mu^\Gamma=0} \end{aligned} \tag{7.23}$$

where the transformation to the flavor rotating frame has been performed (*cf.* section (7.1)). As demonstrated for the baryon number current (6.11) the currents are additive in valence and vacuum parts. The valence quark part $j^{\nu}_{\rm val,\Gamma}(x)$ is obtained by a perturbative expansion of the single particle eigenvalues of $h + \frac{1}{2}\boldsymbol{\tau}\cdot\boldsymbol{\Omega} + R^{\dagger}\beta a^{\Gamma}_{\rho}\gamma_{\rho}R$ up to linear order in both $\boldsymbol{\Omega}$ and $a^{\Gamma}_{\rho}$. This ends up to

$$
\begin{aligned}
j^{\nu}_{\rm val,\Gamma}(x) \;=\; \eta_{\rm val} N_C \frac{\delta}{\delta a^{\Gamma}_{\nu}(x)} \Bigg[ & \langle {\rm val}|R^{\dagger}\beta a^{\Gamma}_{\rho}\gamma^{\rho}R|{\rm val}\rangle \\
& + \sum_{\mu\ne{\rm val}} \left\{\frac{1}{2}\Omega_a, \frac{\langle {\rm val}|\tau_a|\mu\rangle\langle\mu|R^{\dagger}\beta a^{\Gamma}_{\rho}\gamma^{\rho}R|{\rm val}\rangle}{\epsilon_{\mu}-\epsilon_{\rm val}}\right\}\Bigg]\Bigg|_{a^{\Gamma}_{\mu}=0}
\end{aligned}
\tag{7.24}
$$

where "val" labels the valence quark orbit. The evaluation of the vacuum part of the currents $j^{\nu}_{\rm vac,\Gamma}(x)$ proceeds in a fashion similar to the computation of the moment of inertia [70] (see section 7.1), *i.e.* we again have to continue to Euclidean space and consider the limit $T\to\infty$. However, in the present case also the imaginary part of the Euclidean action contributes. Besides the angular velocities $\boldsymbol{\Omega}$ also the time component of the external gauge field has to be continued as $a^{\Gamma}_0 \to i a^{\Gamma}_4$, with $a^4_{\Gamma}$ Hermitian. Furthermore, in order to apply the proper time regularization scheme, one has to distinguish between real and imaginary parts of the fermion determinant. For the real part the regularization is defined in eq (6.25) while for the imaginary part a procedure analogous to (6.35) is employed. The part of the fermion determinant which does not contain the angular velocity receives contributions proportional to $a^k_{\Gamma}$ ($k=1,..,3$) from the real part and proportional to $a^4_{\Gamma}$ from the imaginary part. Taking into account that $\boldsymbol{\tau}\cdot\boldsymbol{\Omega}_E$ (7.7) also behaves like a time component of a vector field one finds that the parts which are linear in the Euclidean angular velocity have $\Omega^a_E\; a^4_{\Gamma}$ stemming from the real part and $\Omega^a_E\; a^k_{\Gamma}$ from the imaginary part. Finally, the total expression for the vacuum part of the current is continued back to Minkowski space

$$
\begin{aligned}
j^{\nu}_{\rm vac,\Gamma}(x) \;=\; -\frac{N_C}{2}\frac{\delta}{\delta a^{\Gamma}_{\nu}(x)} \Bigg[ & \sum_{\mu} {\rm sgn}(\epsilon_{\mu}){\rm erfc}\left(\left|\frac{\epsilon_{\mu}}{\Lambda}\right|\right)\langle\mu|R^{\dagger}\beta a^{\Gamma}_{\rho}\gamma^{\rho}R|\mu\rangle \\
& -\sum_{\mu\nu} f_{\Theta}(\epsilon_{\mu},\epsilon_{\nu};\Lambda)\left\{\frac{1}{2}\Omega_a, \langle\mu|\tau_a|\nu\rangle\langle\nu|R^{\dagger}\beta a^{\Gamma}_{\rho}\gamma^{\rho}R|\mu\rangle\right\}\Bigg]\Bigg|_{a^{\Gamma}_{\mu}=0} .
\end{aligned}
\tag{7.25}
$$

The regulator function $f_{\Theta}$ is defined in eq (7.14). Its appearance in (7.25) is not accidental but rather guarantees the proper normalization of the nucleon charge when the quantization rule (7.18) is employed. The expressions (7.24,7.25) were obtained in ref. [101] for the evaluation of magnetic moments and the axial charge of the nucleon.

### *7.2.1 Electromagnetic properties of the nucleon*

In order to extract the electromagnetic current we put $a^{\Gamma}_{\nu}(x) = a^{\rm e.m.}_{\nu} = \mathcal{Q}a_{\mu}(x)$ where $\mathcal{Q}$ is defined above eq (7.22). Obviously $\mathcal{Q}$ contains isovector $\sim \tau_3/2$ and isoscalar $\sim 1/6$ parts. Accordingly, the electromagnetic $j^{\mu}_{\rm e.m.}(x)$ current may be decomposed into isovector $j^{\mu}_V(x)$ and isoscalar $j^{\mu}_S(x)$ contributions.

Let us first discuss the magnetic moment operator. As in all static soliton models these are obtained as the spatial integral of the space components of $j^{\mu}_{\rm e.m.}(x)$ via $\boldsymbol{\mu} = (1/2)\int d^3r\,\boldsymbol{r}\times\boldsymbol{j}_{\rm e.m.}(x)$. It is then obvious that $\boldsymbol{\mu}$ involves the matrix elements of $\boldsymbol{r}\times\beta\boldsymbol{\gamma} = \boldsymbol{r}\times\boldsymbol{\alpha}$ between the eigenstates $\Psi_\mu$ of $h$ (6.43). It is most convenient to consider $\mu_z$. For the isovector part of the magnetic moment one has $R^\dagger \tau_3 R = D_{3i}\tau_i$. For the isoscalar part the explicit dependence on $R$ drops out. The quantization rules (7.21) and (7.18) (for isovector and –scalar parts, respectively) are used to replace the explicit dependence on the collective coordinates by collective operators which act on nucleon states with spin–projection $J_3 = 1/2$. It is then obvious that the magnetic moments operator can be written as $\mu = \frac{1}{2}\mu_S + I_3\mu_V$. Both $\mu_S$ and $\mu_V$ may be decomposed into valence and vacuum parts according to (7.24,7.25) $\mu_{S,V} = \mu_{S,V}^{\rm val} + \mu_{S,V}^{\rm vac}$. Explicit evaluation shows that only the first terms on the $RHS$ of eqs (7.24,7.25) contribute to $\mu_V$

$$\begin{aligned} \mu_V^{\rm val} &= -N_C\eta_{\rm val}\frac{M_N}{3}\langle{\rm val}|\tau_3\left(\boldsymbol{r}\times\boldsymbol{\alpha}\right)_3|{\rm val}\rangle \\ \mu_V^{\rm vac} &= N_C\frac{M_N}{6}\sum_\mu\langle\mu|\tau_3\left(\boldsymbol{r}\times\boldsymbol{\alpha}\right)_3|\mu\rangle{\rm sgn}(\epsilon_\mu){\rm erfc}\left(\left|\frac{\epsilon_\mu}{\Lambda}\right|\right) \end{aligned} \tag{7.26}$$

where $M_N = 939$MeV denotes the experimental nucleon mass[b]. On the other hand the second terms on the $RHS$ of eqs (7.24,7.25) contribute to $\mu_S$

$$\begin{aligned} \mu_S^{\rm val} &= -N_C\eta_{\rm val}\frac{M_N}{2\alpha^2}\sum_{\mu\neq{\rm val}}\frac{\langle{\rm val}|\tau_3|\mu\rangle\langle\mu|\left(\boldsymbol{r}\times\boldsymbol{\alpha}\right)_3|{\rm val}\rangle}{\epsilon_\mu-\epsilon_{\rm val}} \\ \mu_S^{\rm vac} &= -N_C\frac{M_N}{4\alpha^2}\sum_{\mu\nu}f_\Theta(\epsilon_\mu,\epsilon_\nu;\Lambda)\langle\mu|\tau_3|\nu\rangle\langle\nu|\left(\boldsymbol{r}\times\boldsymbol{\alpha}\right)_3|\mu\rangle. \end{aligned} \tag{7.27}$$

Table 7.1: The magnetic moments in nucleon magnetons as functions of the constituent quark mass $m$ compared to the experimental data. (Taken from ref.[101]).

| $m$(MeV) | 350 | 400 | 500 | 600 | 700 | 800 | expt. |
|---|---|---|---|---|---|---|---|
| $\mu_S$ | 0.68 | 0.61 | 0.54 | 0.51 | 0.48 | 0.46 | 0.88 |
| $\mu_V$ | 2.87 | 2.64 | 2.29 | 1.99 | 1.75 | 1.56 | 4.70 |
| $\mu_p$ | 1.78 | 1.63 | 1.42 | 1.25 | 1.12 | 1.01 | 2.79 |
| $\mu_n$ | -1.10 | -1.02 | -1.38 | -0.74 | -0.68 | -0.55 | -1.91 |

In all cases use has been made of the fact that the choice of the $z$–component for the magnetic moment operator projects out the $z$–component in matrix elements like $\langle\mu|\tau_i|\nu\rangle$ after summing over the grand spin projection.

In table 7.1 the numerical results as obtained in ref. [101] for the isovector and –scalar moments are displayed for various constituent quark masses $m$, the only free parameter in the baryon sector of the model. Both $\mu_S$ and $\mu_V$ are seen to decrease with increasing $m$. Furthermore the corresponding values for the proton and nucleon magnetic moments $\mu_{p,n} = (1/2)(\mu_S \pm \mu_V)$ are compared to the experimental data. Although the isoscalar part of the magnetic moment is reasonably well reproduced, the isovector part comes out too small. It should be added that in the calculations of ref. [103] a somewhat larger $\mu_V$ is obtained since these authors do not regularize terms which stem from the imaginary part of the action. We will comment on a possible solution to the problem of too small a $\mu_V$ in section 7.2.3.

In order to obtain the charge radii one needs the second moment of the electromagnetic charge density $\langle r^2\rangle = \int d^3r r^2 j^0_{\rm e.m.}(x)$. Again a decomposition into isoscalar and –vector parts

[b] The magnetic moments are measured in nucleon magnetons

Table 7.2: The mean squared charge radii (in $\mathrm{fm}^2$) as functions of the constituent quark mass $m$ compared to the experimental data. (Taken from ref.[103]). In parenthesis the results for a regularized imaginary part are given.

| $m$(MeV) | 370 | 420 | 450 | expt. |
|---|---|---|---|---|
| $\langle r^2\rangle_S$ | 0.63 (0.64) | 0.52 (0.50) | 0.48 (0.45) | 0.62 |
| $\langle r^2\rangle_V$ | 1.07 | 0.89 | 0.84 | 0.86 |
| $\langle r^2\rangle_p$ | 0.85 (0.86) | 0.71 (0.70) | 0.66 (0.65) | 0.74 |
| $\langle r^2\rangle_n$ | -0.22(-0.22) | -0.18(-0.19) | -0.18(-0.19) | -0.12 |

can be carried out

$$\langle r^2\rangle = \frac{1}{2}\langle r^2\rangle_S + I_3\langle r^2\rangle_V \tag{7.28}$$

with

$$\langle r^2\rangle_S = \frac{N_C}{3}\left\{\eta_{\rm val}\langle {\rm val}|r^2|{\rm val}\rangle - \frac{1}{2}\sum_\mu \langle\mu|r^2|\mu\rangle {\rm sgn}(\epsilon_\mu){\rm erfc}\left(\left|\frac{\epsilon_\mu}{\Lambda}\right|\right)\right\} \tag{7.29}$$

$$\langle r^2\rangle_V = \frac{N_C}{6\alpha^2}\left\{\eta_{\rm val}\sum_{\mu\ne{\rm val}}\frac{\langle{\rm val}|r^2\boldsymbol{\tau}|\mu\rangle\cdot\langle\mu|\boldsymbol{\tau}|{\rm val}\rangle}{\epsilon_\mu-\epsilon_{\rm val}}\right.$$
$$\left. +\frac{1}{2}\sum_{\mu\nu} f_\Theta(\epsilon_\mu,\epsilon_\nu;\Lambda)\langle\mu|r^2\boldsymbol{\tau}|\nu\rangle\cdot\langle\nu|\boldsymbol{\tau}|\mu\rangle\right\}. \tag{7.30}$$

These expressions correspond to the case that the imaginary part of the Euclidean action is regularized. Noting that $\langle r^2\rangle_V$ originates from those terms in (7.24,7.25) which are linear in $\boldsymbol{\Omega}$ it should be obvious that only $\langle r^2\rangle_S$ receives contributions from the imaginary part. Choosing not to regularize this part then corresponds to replacing the complementary error function in (7.29) by unity. In that case numerical results are also available [103] which are displayed in table 7.2. Obviously the question of regularizing the imaginary part only plays a minor role. For the range of $m$ displayed in table 7.2 the valence contributions strongly dominate. Especially the vacuum contribution to $\langle r^2\rangle_S$ is only about 1-2%, however it increases with $m$ for $\langle r^2\rangle_S$ and $\langle r^2\rangle_V$ [103].

Nucleon Compton scattering provides access to electromagnetic polarizabilities which thus may be taken as a measure for the square of dipole transitions between baryon states. Technically, the polarizabilities may be extracted from the response of the soliton to external electromagnetic fields. For this purpose the action for the rotating soliton (7.1) is expanded up to quadratic order in the source $a_\mu^\Gamma = A_\mu\mathcal{Q}$, *i.e.* one higher order than for the current (7.23). It is sufficient to only consider homogeneous electromagnetic fields in this context: $A_\mu = (Ez, \frac{1}{2}\epsilon_{ijk}x_jB_k)_\mu$. When expanding the action the electric (magnetic) polarizabilities $\alpha$ ($\beta$) are then read off from the coefficients of $E^2(B^2)$. In the NJL model so far only the isoscalar electric polarizability of the nucleon, $\alpha_{I=0} = \frac{1}{2}(\alpha_p+\alpha_n)$ has been investigated [104]. This quantity is sensitive to the large distance behavior of the chiral angle $\Theta(r)$. Furthermore, the leading term in a chiral expansion is known [105] to be proportional to $g_A^2$ ($g_A$ denotes the axial charge of the nucleon.). The authors of ref.[104] also took into account the subleading (in $1/N_C$) contributions to $\alpha_{I=0}$, which unfortunately are subject to ordering ambiguities in the quantization procedure and might be artificial. In any event, these contributions have the desired effect of rendering the experimental value for $g_A$ (see section 7.2.3). For the constituent quark mass $m = 420$MeV the isoscalar electric polarizability was found to be $\alpha_{I=0} = 19 \times 10^{-4}\text{fm}^3$ to which the subleading terms contribute about $3 \times 10^{-4}\text{fm}^3$. Although this result overestimates the experimental value $(9.6 \pm 1.8 \pm 2.2)\text{fm}^3$ by about a factor of 2 it is still significantly smaller than in other soliton models which predict $g_A$ correctly as *e.g.* the $\sigma$–model [105]. It should be added that the calculation of ref.[104] contains the further simplification that the $1/N_C$ subleading terms have been approximated by the leading order expression of a gradient expansion. The numerical result $\alpha_{I=0} = 19 \times 10^{-4}\text{fm}^3$ may further be lowered when one takes account of the quantum character of the pion fields. In that case the seagull contribution to $\alpha_{I=0}$ has been claimed to be absent as a consequence of current conservation [106]. In the calculation of ref.[104] the seagull terms contributes about 60%, though.

Table 7.3: The axial charge of the nucleon $g_A$ as a function of the constituent quark mass $m$.

| $m$(MeV) | 350 | 400 | 500 | 600 | 700 | 800 | expt. |
|---|---|---|---|---|---|---|---|
| $g_A$ | 0.80 | 0.76 | 0.70 | 0.66 | 0.62 | 0.59 | 1.26 |

### *7.2.2 Axial charge of the nucleon*

In order to extract the axial properties of the nucleon we put $a_\nu^\Gamma(x) = a_\nu(x)^a \gamma_5 \frac{\tau_a}{2}$. Eqns (7.24,7.25) then provide the axial current $j_5^{\mu,a}$. Due to isospin invariance the axial charge $g_A$ of the nucleon may be obtained as the matrix element of $2j_5^{3,3}$ between proton states with spin projections $+\frac{1}{2}$. Considering the angular velocities $\boldsymbol{\Omega}$ as commuting c–numbers one observes that there are no contributions to $g_A$ from the terms linear in $\boldsymbol{\Omega}$ (see, however, subsection 7.2.3). Then one only needs to evaluate the matrix element $\langle p \uparrow | D_{33} | p \uparrow \rangle = -1/3$ as a consequence of (7.21). Further use of $\beta\gamma_3\gamma_5 = \sigma_3$ yields [108]

$$g_A = -\frac{N_C}{3}\left\{\eta_{\rm val}\langle{\rm val}|\sigma_3\tau_3|{\rm val}\rangle - \frac{1}{2}\sum_\mu \langle\mu|\sigma_3\tau_3|\mu\rangle {\rm sgn}(\epsilon_\mu){\rm erfc}\left(\left|\frac{\epsilon_\mu}{\Lambda}\right|\right)\right\}. \tag{7.31}$$

As $g_A$ is obtained to be the matrix element of spatial components of a current and does not contain any $\boldsymbol{\Omega}$ dependence it is obvious that $g_A$ completely stems from the real part of the fermion determinant and thus necessarily undergoes regularization. It should be noted that in the chiral limit ($m_\pi = 0$) the $RHS$ of eq (7.31) acquires an additional factor (3/2) corresponding to the symmetric zero momentum transfer limit [9].

One may as well employ PCAC (2.17) in order to relate $g_A$ to the profile function $\Theta(r)$ [9, 109]

$$g_A = \frac{4\pi}{3} f_\pi^2 \lim_{R\to\infty} R^3 \frac{\partial\Theta}{\partial r}\Big|_{r=R} - \frac{8\pi}{9} f_\pi^2 m_\pi^2 \int dr r^3 {\rm sin}\Theta. \tag{7.32}$$

As a matter of fact these two formulas for evaluating $g_A$ provide an excellent check on the accuracy of the numerical results which are displayed in table 7.3. The results extracted from eqs. (7.31) and (7.32) differ by 1–2% only. Obviously the experimental value is underestimated by about 40% for all values of the constituent quark mass.

The calculation of nucleon axial charges has been extended to those corresponding including strange quark currents. Then there are two more axial charges of the nucleon which recently received special attention in the context of the "proton spin puzzle" [107]. The first one, $g_A^8$ is obtained by assuming $a_\nu^\Gamma(x) = a_\nu^8(x)\gamma_5\lambda_8/2$ in the definition of the currents (7.22) with $\lambda_a$ being the Gell–Mann matrices. The second, $g_A^0$ is related to the axial singlet current defined via $a_\nu^\Gamma(x) = a_\nu^0(x)\gamma_5/3$.

In order to compute $g_A^8$ and $g_A^0$ the model has to be extended to flavor SU(3). This is achieved most straightforwardly by choosing the collective coordinates $R(t)$ in eq (7.72) to be $3 \times 3$ unitary matrices. This is often referred to as the collective approach to include strangeness in chiral soliton models. It will be discussed extensively in section 7.5.1. Here we will not go into detail but rather remark two essential differences to the two flavor case. First, one has to take into account that the SU(3) nucleon wave–function differs from the SU(2) case especially in the flavor symmetric case. One consequence is that $\langle p \uparrow | D_{33} | p \uparrow \rangle = -7/30$ [110] *i.e.* a significant reduction from (7.21). Thus the axial charge for nucleon $\beta$ decay is altered when considering an SU(3) collective quantization even without changing the chiral soliton. We will refer to this axial charge by $g_A^3$. Secondly one has to include the flavor symmetry breaking arising from the different quark masses. This gives rise to further changes of the nucleon wave–function as will be described below[c]. Furthermore the extension to three flavors leads to additional terms for the currents not contained in (a three flavor generalization of) (7.24,7.25) [112]. These effects have been taken into account in a perturbative scheme with the strange current quark mass as the perturbation parameter while non–strange and strange constituent quark masses were identified [113]. This scheme differs somewhat from the more elaborate one discussed below, however, as the baryon spectrum comes out similar (when physical values for the parameters are assumed) there is enough reason to take the results on the axial charges seriously. From $g_A^3$, $g_A^8$ and $g_A^0$ one may finally extract the contributions of the individual flavors to the axial charges[d] $\triangle u$, $\triangle d$ and $\triangle s$. Assuming a constituent quark mass $m = 420\text{MeV}$ the authors of ref.[113] obtain

$$\triangle u = 0.64 \quad \triangle d = -0.24 \quad \triangle s = -0.02. \tag{7.33}$$

These should be compared with the "experimental" data given by Ellis and Karliner [114]

$$\triangle u = 0.81 \quad \triangle d = -0.44 \quad \triangle s = -0.12. \tag{7.34}$$

The absolute values for $\triangle u$ and $\triangle d$ turn out to be somewhat too small as a result of the too small $g_A^3$. When comparing (7.33) with (7.34) one should be somewhat careful since the extraction of the latter from semi–leptonic hyperon decays involves the assumption of flavor symmetric baryon wave–functions [115]. Within such a framework usually the absolute value of $\triangle s$ comes out too large. This assumption has been dropped in the collective approach. From studies in Skyrme type models it is well known that nevertheless the branching ratios for the semileptonic hyperon decays are well described in the collective approach together with a small $|\triangle s|$ [116]. As most chiral soliton models [96, 44] the NJL model reasonably explains the smallness of the quark contribution to the proton spin $\triangle u + \triangle d + \triangle s \approx 0.38$.

### *7.2.3 Remarks on $1/N_C$ corrections*

In all preceding considerations we have considered the angular velocities $\boldsymbol{\Omega}$ as commuting c-numbers. For the computation of nucleon observables $\boldsymbol{\Omega}$ is replaced by the spin operator

[c] For infinitely large strange quark masses one has $g_A^3 = g_A$, *i.e.* the two flavor limit [111].
[d] *E.g.* $g_A^3 = \triangle u - \triangle d$.

$\boldsymbol{J}$ according to (7.18). However, in non–leading contributions ordering ambiguities exist and different results may be obtained because of non–vanishing commutators

$$[\Omega_i, D_{ab}] \rightarrow \frac{1}{\alpha^2}[J_i, D_{ab}] = \frac{i}{\alpha^2}\sum_{m=1}^{3}\epsilon_{ibm}D_{am}. \tag{7.35}$$

This substitution leads to non–vanishing matrix elements between nucleon states

$$\epsilon_{lmn}\langle N|D_{il}\Omega_m D_{jn}|N\rangle \rightarrow \frac{i}{\alpha^2}\delta_{ij}. \tag{7.36}$$

The prefactor on the right hand side is ambiguous, in particular, it is vanishing in the ordering assumed in section 7.1. Eqn (7.36) then yields additional contributions to $\mu_V$ and $g_A$ which we denote by $\mu_V^{(1)}$ and $g_A^{(1)}$, respectively. These corrections are subleading in an $1/N_C$ expansion because the moment of inertia $\alpha^2$ is of order $N_C$. Since the starting point for the computation of the baryon observables is completely classical the ordering between the angular velocities and the rotations matrices in the prescription (7.36) is ambiguous. Adopting a "natural" ordering [117], which is suggested by the perturbation expansion of the quark wave function in the flavor rotating frame

$$R(t)\left[\Psi_\mu + \sum_{\nu\neq\mu}\Psi_\nu\frac{\langle\mu|\boldsymbol{\tau}\cdot\boldsymbol{\Omega}|\nu\rangle}{\epsilon_\nu-\epsilon_\mu}\right] \longrightarrow R(t)\left[\Psi_\mu + \sum_{\nu\neq\mu}\Psi_\nu\frac{\langle\mu|\boldsymbol{\tau}\cdot\boldsymbol{J}/\alpha^2|\nu\rangle}{\epsilon_\nu-\epsilon_\mu}\right] \tag{7.37}$$

the numerical results for $\mu_V^{(1)}$ and $g_A^{(1)}$ are about 30% of the leading order [118]. Thus it is suggestive but not obvious that the series has already converged at this order. Similar calculations have been performed [113] for the three flavor axial couplings discussed in the preceding subsection. For these quantities the quantization prescription (7.35) leads to $1/N_C$ corrections of the same order.

At the time when these corrections were first observed they were highly welcome since they helped to solve (among others) the problem of the too small $g_A$ (see table 7.3). Nevertheless the applicability of (7.35) from the very beginning remains doubtful, especially since a special ordering of the collective coordinates and operators has to be adopted. In order to obtain some restrictions on possible orderings the compatibility of the associated $1/N_C$ corrections to $g_A$ with symmetries of the underlying model have been examined.

As the $1/N_C$ corrections appear to be a special feature of the semiclassical quantization procedure the self–consistent soliton profile remains uneffected. Thus $g_A$ evaluated with eqn (7.32) assumes the leading order value. The correction $g_A^{(1)}$ may then be interpreted as to violate PCAC by 30% which, of course, represents an undesired breaking of one of the fundamental symmetries of the model. It has been shown that this problem can be solved by simply adopting PCAC as the equation of motion instead of the stationary condition (6.40) [109]. This then leads to an extended soliton profile function causing the moment of inertia and the radii to increase. Unfortunately, until now this equation of motion has not been derived from an action principle.

Furthermore, requiring the proper behavior of the symmetry currents under particle conjugation may provide additional constraints on orderings of the collective coordinates. This has recently been done in the framework of the chiral quark model [119], which at least for the valence quark part of the action is similar to the NJL soliton model. It has been shown [120] that the axial current transforms under this symmetry properly only when an ordering is chosen such that the correction $g_A^{(1)}$ actually is zero. For example the Hermitian ordering prescription

$$D_{ab}\Omega_j \longrightarrow \frac{1}{2\alpha^2}\left(D_{ab}J_j + J_j D_{ab}\right). \tag{7.38}$$

leads to symmetry currents which transform properly under particle conjugation while there are no corrections to $g_A$ and $\mu_V$ at the subleading order in the $1/N_C$ expansion. In ref.[112] an expression for $g_A^{(1)}$ has been presented, which involves the absolute value $1/(|\epsilon_\mu - \epsilon_\nu|)$ rather than $1/(\epsilon_\mu - \epsilon_\nu)$ as in the chiral quark model. Hence this result for $g_A^{(1)}$ complies with particle conjugation. It should, however, be remarked that the alternative quantization prescription (7.38) leads to vanishing $1/N_C$ corrections in the NJL model as well.

### 7.3 Meson fluctuations off the chiral soliton

In order to explore baryon properties like *e.g.* $\pi - N$ scattering[e], electro–magnetic polarizabilities (beyond the isoscalar electric channel) [105] or for the investigation of quantum fluctuations it is mandatory to go beyond the zero–mode quantization described in section (7.1). For applications like these time–dependent fluctuations off the soliton have to be included and quantized canonically. This formalism is also of relevance for the discussion of quantum corrections to the soliton mass [25] (section 7.4) as well as the description of hyperons in the bound state approach [29] (compare subsection 7.5.2).

In order to formally introduce the fluctuations we consider the ansatz [97]

$$M = \xi_0 \xi_f \langle\Sigma\rangle \xi_f \xi_0. \tag{7.39}$$

Here $\xi_0 = \exp\left(\frac{i}{2}\boldsymbol{\tau}\cdot\hat{\boldsymbol{r}}\ \Theta(r)\right)$ denotes the hedgehog soliton configuration while the space–time dependent fluctuations $\eta_a(x)$ are given by

$$\xi_f(x) = \exp\left(i\sum_{a=1}^{8}\eta_a(x)\lambda_a/2\right). \tag{7.40}$$

The main task now is to expand the action up to quadratic order in the fluctuations $\eta_a$. In order to avoid problems with stability the meson fields have been constrained to the chiral circle. Although the parametrization (7.39) deviates from the commonly adopted unitary gauge $M = \xi_L^\dagger\langle\Sigma\rangle\xi_R$, $\xi_L^\dagger = \xi_R$ it has proven to be convenient when matrix elements between eigenstates of the static Hamiltonian (6.42) are computed. This can be observed easily by considering the Euclidean Dirac operator $\not{D}_E$

$$i\beta\not{D}_E = -\partial_\tau - \boldsymbol{\alpha}\cdot\boldsymbol{p} - \mathcal{T}\beta\left(\xi_f\langle\Sigma\rangle\xi_f P_R + \xi_f^\dagger\langle\Sigma\rangle\xi_f^\dagger P_L\right)\mathcal{T}^\dagger \tag{7.41}$$

wherein $\tau = ix_0$ is the Euclidean time. The unitary matrix

$$\mathcal{T} = \xi_0 P_L + \xi_0^\dagger P_R \tag{7.42}$$

[e]For a review on $\pi - N$ scattering in soliton models see ref.[11].

contains all the information on the chiral soliton. Thus, whenever a matrix element involving the fluctuations has to be evaluated it can be simplified by chirally transforming the states between which the operator is sandwiched.

As the first step towards expanding the action we write the Euclidean Dirac operator as

$$i\beta \not{D}_E = -\partial_\tau - h = -\partial_\tau - \left(h_{(0)} + h_{(1)} + h_{(2)} + \cdots\right) \tag{7.43}$$

wherein the subscript labels the power of the meson fluctuations. The dots indicate powers of the fluctuations larger than two. Upon noting that $h_{(0)}$ as given by (6.42) is time independent one obtains for the argument of the real part of the fermion determinant $\mathcal{A}_R$ (6.22)

$$\not{D}_E^\dagger \not{D}_E = -\partial_\tau^2 + h_{(0)}^2 - [\partial_\tau, h_{(1)}] + \{h_{(1)}, h_{(0)}\} - [\partial_\tau, h_{(2)}] + \{h_{(2)}, h_{(0)}\} + h_{(1)}^2 + \cdots . \tag{7.44}$$

When the valence quark contribution as well as the terms originating from the mesonic part of the action are included it is obvious that the zeroth–order (in the fluctuations) just renders the static energy functional while the expression linear in $\eta_a(x)$ vanishes subject to the equation of motion for the static soliton. Therefore only the second order expression is of interest for the current discussion. The corresponding contribution from the real part of the fermion determinant reads

$$\begin{aligned}\mathcal{A}_R^{(2)} \;=\; & \frac{1}{2}\mathrm{Tr}\int_{1/\Lambda^2}^{\infty} ds \hat{K}_0(s)\left(\{h_{(2)}, h_{(0)}\} + h_{(1)}^2\right) - \frac{1}{4}\mathrm{Tr}\int_{1/\Lambda^2}^{\infty} ds \int_0^s ds' \hat{K}_0(s-s') \\ & \times \left([\partial_\tau, h_{(1)}]\hat{K}_0(s')[\partial_\tau, h_{(1)}] + \{h_{(1)}, h_{(0)}\}\hat{K}_0(s')\{h_{(1)}, h_{(0)}\}\right).\end{aligned} \tag{7.45}$$

The introduction of the zeroth-order heat kernel $\hat{K}_0(s) = \exp\left(s(\partial_\tau^2 - h_{(0)}^2)\right)$ has turned out to be useful for this computation. For the imaginary part of the fermion determinant, $\mathcal{A}_I$, the cut-off $\Lambda$ is introduced by the substitution

$$\begin{aligned}&\left((\partial_\tau - h_{(0)} - h_{(1)})(\partial_\tau + h_{(0)} + h_{(1)})\right)^{-1} \\ &\qquad \longrightarrow -\int_{1/\Lambda^2}^{\infty} ds \, \exp\left(s(\partial_\tau - h_{(0)} - h_{(1)})(\partial_\tau + h_{(0)} + h_{(1)})\right)\end{aligned} \tag{7.46}$$

which is legitimate since the argument is negative definite, *i.e.* it converges for large momenta. The resulting imaginary part starts off at second order in the fluctuations

$$\mathcal{A}_I = \mathrm{Tr}\int_{1/\Lambda^2}^{\infty} ds \int_0^s ds' \hat{K}_0(s-s')\partial_\tau h_{(1)} \hat{K}_0(s') h_{(0)} h_{(1)} + \cdots . \tag{7.47}$$

To perform the temporal part of the functional trace in Euclidean space the meson fluctuations are Fourier transformed

$$\eta_a(\boldsymbol{r}, -i\tau) = \int_{-\infty}^{+\infty} \frac{d\omega}{2\pi} \tilde{\eta}_a(\boldsymbol{r}, i\omega) \mathrm{e}^{-i\omega\tau} \tag{7.48}$$

This transformation directly transfers to the Hamiltonians:

$$\begin{aligned} h_{(1)}(\boldsymbol{r}, -i\tau) &= \int_{-\infty}^{+\infty} \frac{d\omega}{2\pi} \tilde{h}_{(1)}(\boldsymbol{r}, i\omega) \mathrm{e}^{-i\omega\tau} \quad \text{and} \\ h_{(2)}(\boldsymbol{r}, -i\tau) &= \int_{-\infty}^{+\infty} \frac{d\omega}{2\pi} \int_{-\infty}^{+\infty} \frac{d\omega'}{2\pi} \tilde{h}_{(2)}(\boldsymbol{r}, i\omega, i\omega') \mathrm{e}^{-i(\omega+\omega')\tau} \end{aligned} \tag{7.49}$$

since the chiral transformation (7.42) is time independent. In order to project out the vacuum part of the fermion determinant, $\mathcal{A}_{\rm vac}$, one considers the limit of infinitely large Euclidean times. Noting that [70]

$$\langle\tau|\hat{K}_0(s)|\tau'\rangle = \frac{1}{\sqrt{4\pi s}}\exp(-sh_{(0)}^2)\exp\left(-\frac{(\tau-\tau')^2}{4s}\right) \tag{7.50}$$

the temporal part of the trace then amounts to carrying out Gaussian integrals involving the Fourier frequency $\omega$. The spatial part of the trace as well as the traces over Dirac and flavor indices are evaluated using the eigenstates of the static one-particle Hamiltonian $h_{(0)}$ (6.43). Finally, the frequency $\omega$ has to be continued back to Minkowski space in order to obtain physically relevant expressions. The second order (in meson fluctuations) contribution to the vacuum part of the fermion determinant finally results in [97]

$$\begin{aligned}\mathcal{A}^{(2)}_{\rm vac} &= \mathcal{A}^{(2)}_R + \mathcal{A}^{(2)}_I = \frac{N_C}{2}\int_{1/\Lambda^2}^\infty \frac{ds}{\sqrt{4\pi s}}\sum_\mu 2\epsilon_\mu {\rm e}^{-s\epsilon_\mu^2}\int_{-\infty}^{+\infty}\frac{d\omega}{2\pi}\langle\mu|\tilde{h}_{(2)}(\boldsymbol{r},\omega,-\omega)|\mu\rangle \\ &+\frac{N_C}{4}\int_{1/\Lambda^2}^\infty ds\sqrt{\frac{s}{4\pi}}\sum_{\mu\nu}\int_{-\infty}^{+\infty}\frac{d\omega}{2\pi}\langle\mu|\tilde{h}_{(1)}(\boldsymbol{r},\omega)|\nu\rangle\langle\nu|\tilde{h}_{(1)}(\boldsymbol{r},-\omega)|\mu\rangle \\ &\times\left\{\frac{{\rm e}^{-s\epsilon_\mu^2}+{\rm e}^{-s\epsilon_\nu^2}}{s} + [\omega^2-(\epsilon_\mu+\epsilon_\nu)^2]R_0(s;\omega,\epsilon_\mu,\epsilon_\nu) - 4\omega\epsilon_\nu R_1(s;\omega,\epsilon_\mu,\epsilon_\nu)\right\}.\end{aligned} \tag{7.51}$$

The information on the orderings of the operators in eqns. (7.45,7.47) is contained in the Feynman parameter integrals

$$R_i(s;\omega,\epsilon_\mu,\epsilon_\nu) = \int_0^1 x^i dx\,\exp\left(-s[(1-x)\epsilon_\mu^2 + x\epsilon_\nu^2 - x(1-x)\omega^2]\right) \tag{7.52}$$

which represent moments of the quark loop in the presence of the soliton.

Besides the polarized vacuum configuration also the explicit occupation of the valence quark level contributes to the action as long as the associated energy eigenvalue $\epsilon_{\rm val}$ is positive. Since no regularization is involved the computation is completely performed in Minkowski space. Treating the meson fluctuations as time-dependent perturbations the associated first order change $\delta\Psi_{\rm val}$ of the valence quark wave-function $\Psi_{\rm val}$ is obtained to be

$$\delta\Psi_{\rm val}(\boldsymbol{r},t) = \left(i\partial_t - h_{(0)}(\boldsymbol{r})\right)^{-1} h_{(1)}(\boldsymbol{r},t)\Psi_{\rm val}(\boldsymbol{r},t). \tag{7.53}$$

The corresponding contribution to the second order part of the action reads [97]

$$\begin{aligned}\mathcal{A}^{(2)}_{\rm val} &= -\eta_{\rm val}N_C\int_{-\infty}^{+\infty}\frac{d\omega}{2\pi}\Big(\langle{\rm val}|\tilde{h}_{(2)}(\boldsymbol{r},\omega,-\omega)|{\rm val}\rangle \\ &\qquad + \sum_{\mu\ne{\rm val}}\frac{\langle{\rm val}|\tilde{h}_{(1)}(\boldsymbol{r},\omega)|\mu\rangle\langle\mu|\tilde{h}_{(1)}(\boldsymbol{r},-\omega)|{\rm val}\rangle}{\epsilon_{\rm val}-\omega-\epsilon_\mu}\Big).\end{aligned} \tag{7.54}$$

Here $\eta_{\rm val} = 0,1$ again denotes the occupation number of the valence quark and anti-quark states.

The complete second order contribution to the action is then given by

$$\mathcal{A}^{(2)} = \mathcal{A}^{(2)}_{\rm vac} + \mathcal{A}^{(2)}_{\rm val} + \mathcal{A}^{(2)}_{\rm m}. \tag{7.55}$$

The mesonic contribution, $\mathcal{A}^{(2)}_{\rm m}$, is obtained by substituting the ansatz (7.39) into the expression for the mesonic part of the action (3.25)

$$\begin{aligned}\mathcal{A}^{(2)}_{\rm m} &= -\frac{1}{2}m_\pi^2 f_\pi^2 \int d^3r \int_{-\infty}^{+\infty}\frac{d\omega}{2\pi}\Big\{\cos\Theta\ \tilde{\boldsymbol{\eta}}(\omega)\cdot\tilde{\boldsymbol{\eta}}(-\omega) \\ &\qquad + \frac{1}{4}\Big(1+\frac{m_s}{m}\Big)\Big(\cos\Theta\ + \frac{m_s^0}{m^0}\Big)\sum_{\alpha=4}^{7}\tilde{\eta}_\alpha(\omega)\tilde{\eta}_\alpha(-\omega)\Big\} \qquad (7.56)\end{aligned}$$

wherein use has been made of the relation $G_1 = m^0 m/m_\pi^2 f_\pi^2$ (4.41). Obviously $\mathcal{A}^{(2)}$ contains terms of odd powers in $\omega$. These correspond to the imaginary part in Euclidean space and have the important property of removing the degeneracy between solutions with $\pm\omega$ [29]. In the Skyrme model these terms originate from the Wess–Zumino action [52] which is identical to the leading order term of the gradient expansion of the imaginary part [31], *cf.* section 4.1.

In general the second order contribution to the action can be written as a functional of the fluctuations $\eta_a$

$$\begin{aligned}\mathcal{A}^{(2)}[\eta_a] &= \frac{1}{2}\int_{-\infty}^{+\infty}\frac{d\omega}{2\pi}\Big\{\int d^3r\int d^3r'\ \Phi^{(2)}_{ab}(\omega;\boldsymbol{r},\boldsymbol{r}')\tilde{\eta}_a(\boldsymbol{r},-\omega)\tilde{\eta}_b(\boldsymbol{r}',\omega) \\ &\qquad + \int d^3r\ \Phi^{(1)}_{ab}(\boldsymbol{r})\tilde{\eta}_a(\boldsymbol{r},-\omega)\tilde{\eta}_b(\boldsymbol{r},\omega)\Big\}. \qquad (7.57)\end{aligned}$$

The local and bilocal kernels $\Phi^{(1),(2)}_{ab}$ have to be computed as mode sums involving the eigenstates and -functions of $h_{(0)}$ (6.43).

Using the Wigner–Eckart theorem it can be shown that due to the symmetry of the soliton under grand spin and parity transformations, the fluctuations decouple with respect to their grand spin and parity quantum numbers. *I.e.* the kernels $\Phi^{(1),(2)}_{ab}$ are diagonal in these quantum numbers. This property turns out to be helpful when investigating quantum corrections to the soliton mass [67] and hyperons in the bound state approach [99] to the NJL soliton.

## 7.4 Quantum corrections to the soliton mass

The action functional (7.57) of the meson fluctuations in the background field of the NJL chiral soliton has originally been derived to make possible a description of hyperons in the NJL model within the so–called bound state approach (*cf.* subsection 7.5.2). Very recently it has been shown in the two flavor reduction that the action (7.57) also allows one to estimate the quantum correction to the soliton mass [67]. Then the bilocal kernel $\Phi^{(2)}$ only depends on $\omega^2$, *i.e.* the contributions to the action corresponding to odd powers in the frequency vanish. This has the consequence that the solutions to the Bethe–Salpeter equation[f]

$$\int d^3r'\ \Phi^{(2)}_{ab}(\boldsymbol{r},\boldsymbol{r}',\omega)\tilde{\eta}_b(\boldsymbol{r}',\omega) + \Phi^{(1)}_{ab}(\boldsymbol{r})\tilde{\eta}_b(\boldsymbol{r},\omega) = 0 \qquad (7.58)$$

appear in pairs $\pm\omega_i$. Denoting the corresponding wave–functions which solve (7.58) by $\tilde{\eta}^{(i)}_a(\boldsymbol{r},\omega_i)$ the fluctuating field may be decomposed as

$$\boldsymbol{\eta}(\boldsymbol{r},t) = \sum_i \frac{1}{\sqrt{2\omega_i}}\left\{a_i\tilde{\boldsymbol{\eta}}^{(i)}(\boldsymbol{r},\omega_i)e^{i\omega_i t} + a_i^\dagger\tilde{\boldsymbol{\eta}}^{(i)}(\boldsymbol{r},\omega_i)e^{-i\omega_i t}\right\}. \qquad (7.59)$$

[f] A method for numerically solving this equation is provided in ref.[121].

After canonical quantization (*i.e.* assuming the commutation relations $\left[a_i, a_j^\dagger\right] = \delta_{ij}$) the Hamiltonian associated with the fluctuations is that of an harmonic oscillator

$$\mathcal{H} = \sum_i \omega_i \left(a_i^\dagger a_i + \frac{1}{2}\right). \tag{7.60}$$

This result (7.60) is quite non–trivial since the action (7.57) involves all orders of time derivatives rather than terminating at quadratic order as *e.g.* in the Skyrme model. This new feature leads to an involved energy functional (in coordinate space) as well as a complicated orthonormalization condition for the meson fluctuations. *E.g.* the normalization condition for a solution $\tilde{\boldsymbol{\eta}}(\boldsymbol{r}, \omega_i)$ to the Bethe–Salpeter equation reads

$$\int d^3r \int d^3r' \tilde{\eta}_a^{(i)}(\boldsymbol{r}, \omega_i) \left.\frac{\partial \Phi_2^{ab}(\boldsymbol{r}, \boldsymbol{r}', \omega)}{\partial \omega^2}\right|_{\omega=\omega_i} \tilde{\eta}_b^{(i)}(\boldsymbol{r}, \omega_i) = 1. \tag{7.61}$$

Unfortunately, the orthogonality condition for solutions with different frequencies cannot be presented in such a closed form.

It has also formally been shown that the vacuum contribution to the energy ($\sum_i \omega_i/2$) is obtained to be of the form (7.60) whenever the background field is static and the eigenvalues of the Bethe–Salpeter equation appear in pairs $\pm\omega_i$.

The form (7.60) implies the existence of an operator acting in the space spanned by the eigenstates of (7.58)

$$H^2 = H_0^2 + V \tag{7.62}$$

with eigenvalues $\pm\omega_i$. This operator can be expressed as the sum of the corresponding one in the absence of the soliton, $H_0^2$, and a "perturbation", $V$, which depends on the soliton. In ref.[25] it has been shown for the Skyrme model that a finite (renormalized) energy correction, $\triangle E$, is obtained from these operators via

$$\triangle E = \frac{1}{2} \mathrm{Tr} \left(H - H_0 - \frac{1}{2} H_0^{-1} V + \frac{1}{8} H_0^{-3} V^2\right) \tag{7.63}$$

which actually represents the generalization of the quantum corrections to the kink mass [122] to $3+1$ dimensions. Although the NJL model is quite different from the Skyrme model the trace may similarly be computed by expressing it in terms of the corresponding eigenvalues

$$\begin{aligned} \triangle E = \frac{1}{2} \sum_i \Bigg\{ \omega_i - \frac{1}{8} \sum_j \omega_j^{(0)} \left|\langle \tilde{\boldsymbol{\eta}}(\boldsymbol{r}, \omega_i) | \tilde{\boldsymbol{\eta}}^{(0)}(\boldsymbol{r}, \omega_j^{(0)}) \rangle\right|^2 & \\ \times \left[3 + 6 \left(\frac{\omega_i}{\omega_j^{(0)}}\right)^2 - \left(\frac{\omega_i}{\omega_j^{(0)}}\right)^4\right] \Bigg\}. & \end{aligned} \tag{7.64}$$

Here $\omega_j^{(0)}$ and $\tilde{\boldsymbol{\eta}}^{(0)}(\boldsymbol{r}, \omega_j^{(0)})$ denote the eigenfrequencies and –wave–functions to the Bethe–Salpeter equation (7.58) in the absence of the soliton. The overlap in eq (7.64) has occurred because the operators $H$ and $H_0$ act in distinct Hilbert spaces. A reasonable definition of these overlaps is gained by including the metric

$$\left.\frac{\partial \Phi_2^{ab}(\boldsymbol{r}, \boldsymbol{r}', \omega)}{\partial \omega^2}\right|_{\omega=\omega_i} = \mathcal{M}^{ab}(\boldsymbol{r}, \boldsymbol{r}', \omega_i) = \int d^3x \sum_{c=1}^{3} \left(\sqrt{\mathcal{M}}\right)^{ac} (\boldsymbol{r}, \boldsymbol{x}, \omega_i) \left(\sqrt{\mathcal{M}}\right)^{bc} (\boldsymbol{r}', \boldsymbol{x}, \omega_i) \tag{7.65}$$

into the wave–function

$$\phi^a(\boldsymbol{r},\omega_i) = \int d^3x \sum_{c=1}^{3} \left(\sqrt{\mathcal{M}}\right)^{ca} (\boldsymbol{x},\boldsymbol{r},\omega_i)\ \tilde{\eta}_c(\boldsymbol{x},\omega_i). \tag{7.66}$$

The modificatied wave–function $\boldsymbol{\phi}$ turns out to be independent of the parametrization (7.39) [123]. Finally the relevant overlap matrix element is given by

$$\langle \tilde{\boldsymbol{\eta}}(\boldsymbol{r},\omega_i)|\tilde{\boldsymbol{\eta}}^{(0)}(\boldsymbol{r},\omega_j^{(0)})\rangle := \int d^3r\ \boldsymbol{\phi}(\boldsymbol{r},\omega_i)\cdot\boldsymbol{\phi}^{(0)}(\boldsymbol{r},\omega_j^{(0)}) \tag{7.67}$$

where $\boldsymbol{\phi}^{(0)}$ denotes the analogue of $\boldsymbol{\phi}$ in the absence of the soliton.

From eq (7.64) it is intuitively clear that the zero modes ($\omega_i = 0$) provide the major contribution to $\triangle E$ because no counterpart exists in the absence of the soliton. The zero modes arise because the soliton breaks the rotational[g] and translational invariance. In that sense the zero modes may be regarded as Goldstone bosons. The associated wave–functions, $\tilde{\boldsymbol{\eta}}_{\rm z.m.}(\boldsymbol{r})$, correspond to a wave–function in the $P$–wave channel for the rotational zero mode while the translational zero mode contains $S$– and $D$–wave parts. According to these structures *ansätze* can be made for the wave functions in the zero mode channels. Then the Bethe–Salpeter equation reduces to (coupled) homogeneous integral equations for purely radial functions. As this procedure is quite technical we refer the reader to ref.[67] for details on this calculation as well as on the explicit construction of the modified wave function $\boldsymbol{\phi}$. As already noted the contribution of the zero modes to the energy correction

$$-\frac{3}{16}\sum_j \omega_j^{(0)} \left|\langle \tilde{\boldsymbol{\eta}}_{\rm z.m.}(\boldsymbol{r})|\tilde{\boldsymbol{\eta}}^{(0)}(\boldsymbol{r},\omega_j^{(0)})\rangle\right|^2 \tag{7.68}$$

is negative definite. Hence it automatically leads to the desired result of reducing the total energy.

Before presenting the numerical results one remark concerning the non–confining character of the NJL model has to be made. Obviously the transformation from Euclidean to Minkowski space is only well defined as long as the exponent in the Feynman parameter integrals (7.12) vanishes for large $s$ along the path connecting these two spaces. In the absence of the soliton this is only the case for $\omega < 2m$. Beyond this threshold the regularization functions develop imaginary parts which measure the decay of the meson fluctuations into quark–antiquark pairs. Consequently the quantum correction to the soliton mass which stems from the zero modes is estimated by the truncated sum

$$\triangle E_{\rm z.m.} = -\frac{3}{16}\sum_{\omega_j^{(0)}<2m} \omega_j^{(0)} \left|\langle \tilde{\boldsymbol{\eta}}_{\rm z.m.}(\boldsymbol{r})|\tilde{\boldsymbol{\eta}}^{(0)}(\boldsymbol{r},\omega_j^{(0)})\rangle\right|^2 . \tag{7.69}$$

In order to judge the quality of this truncation it is useful to define a sum of overlaps

$$\mathcal{S} = \sum_{\omega_j^{(0)}<2m} \left|\langle \tilde{\boldsymbol{\eta}}_{\rm z.m.}(\boldsymbol{r})|\tilde{\boldsymbol{\eta}}^{(0)}(\boldsymbol{r},\omega_j^{(0)})\rangle\right|^2 \tag{7.70}$$

which should approach unity if the model were insensible to the truncation.

[g] Due to the grand spin symmetry of the hedgehog *ansatz* coordinate space rotations and isospin transformations are equivalent.

Table 7.4: The quantum corrections to the soliton mass due to the rotational zero mode [67].

| | $m_\pi = 0$ | | | $m_\pi = 135\text{MeV}$ | | |
|---|---|---|---|---|---|---|
| $m$(MeV) | 400 | 500 | 600 | 400 | 500 | 600 |
| $\mathcal{S}$ | 0.88 | 0.94 | 0.95 | 0.83 | 0.89 | 0.91 |
| $\triangle E_{\rm z.m.}$(MeV) | -201 | -274 | -290 | -244 | -297 | -323 |

Table 7.5: The quantum corrections to the soliton mass due to the translational zero mode. The contributions stemming from the $S(l=0)$– and $D(l=2)$–waves are disentangled [67].

| | $m_\pi = 0$ | | | $m_\pi = 135\text{MeV}$ | | |
|---|---|---|---|---|---|---|
| $m$(MeV) | 400 | 500 | 600 | 400 | 500 | 600 |
| $\mathcal{S}$ | 0.32 | 0.41 | 0.52 | 0.28 | 0.37 | 0.46 |
| $\triangle E_{l=0}$(MeV) | -18 | -22 | -32 | -12 | -22 | -32 |
| $\triangle E_{l=2}$(MeV) | -127 | -140 | -207 | -82 | -128 | -187 |
| $\triangle E_{\rm z.m.}$(MeV) | -145 | -162 | -239 | -94 | -150 | -218 |

The numerical results obtained in ref.[67] for the contributions of the rotational and translational zero modes to the energy correction are displayed in tables 7.4 and 7.5, respectively. Also shown are the corresponding sums of overlaps, $\mathcal{S}$. The chiral limit ($m_\pi = 0$) as well as the physical case ($m_\pi = 135\text{MeV}$) are considered. Note that the number of states which lie below $\omega = 2m$ decreases as the pion mass increases. In case of the rotational zero mode $\mathcal{S}$ approaches unity close enough to consider $\triangle E \approx -250 \sim 300\text{MeV}$ as a reliable estimate of the quantum correction to the energy. The situation, however, is not as good for the translational zero mode. The corresponding wave–function is more strongly peaked at $r = 0$ than in case of the rotational zero mode. Therefore the Fourier transform acquires contributions from modes carrying higher frequencies which lie above $2m$. Hence $\mathcal{S}$ hardly reaches 0.5. The corresponding $\triangle E$ can then only be considered as a lower bound. As in the Skyrme model it turns out that the $D$–wave contributions strongly dominate the $S$–wave part.

In ref.[67] it has furthermore verified that the scattering modes provide a negligible contribution of a few MeV.

Taking the numbers displayed in tables 7.4 and 7.5 seriously (improving the model such as to overcome the problem of non–confinement would in any event also alter the predictions on the classical mass, moment of inertia, etc.) one obtains a mass formula for baryons

$$M = E_{\rm cl} + \triangle E + \frac{J(J+1)}{2\alpha^2} \qquad (7.71)$$

with $\triangle E$ estimated by the sum of the rotational and translational zero mode contributions. The last term in (7.71) arises from the semi–classical cranking procedure discussed in section 7.1. In general there are also quantum correction to this rotational term. However, these are of the order $\mathcal{O}(N_C^{-2})$ and hence they are omitted. The numerical results for the baryon masses are given in table 7.6. Reasonable agreement with the experimental data for the masses of the nucleon (939MeV) and the $\Delta$–resonance (1232MeV) are only obtained for constituent quark masses $m \approx 400\text{MeV}$. Unfortunately $\mathcal{S}$ then is as low as 0.4 for the translational zero mode.

As a conclusive statement of this section we would like to mention that the mass of the nucleon is significantly lower than the soliton mass due to quantum corrections. In general these corrections reduce the masses of baryons by a few hundred MeV as compared to the cranking result (7.20). The effect of the quantum corrections on observables other than the masses has not yet been studied in the NJL model.

Table 7.6: The predictions for the masses of the nucleon ($N$) and $\Delta$–resonance. The empirical data are 939MeV and 1232MeV, respectively. (Results taken from [67].)

| | $m_\pi = 0$ | | | $m_\pi = 135$MeV | | |
|---|---|---|---|---|---|---|
| $m$(MeV) | 400 | 500 | 600 | 400 | 500 | 600 |
| $E_{\rm cl}$(MeV) | 1212 | 1193 | 1166 | 1250 | 1221 | 1193 |
| $\triangle E$(MeV) | -302 | -436 | -525 | -338 | -448 | -538 |
| $\alpha^2$(1/GeV) | 6.26 | 4.73 | 3.87 | 5.80 | 4.17 | 3.43 |
| $M_N$(MeV) | 970 | 836 | 738 | 976 | 863 | 764 |
| $M_\Delta$(MeV) | 1210 | 1153 | 1126 | 1236 | 1223 | 1201 |

## 7.5 Hyperons

In this section we will describe the treatment of strange degrees of freedom within the NJL model of pseudoscalar fields. The main goal is to find a description of the hyperon spectrum in this model. As has already been pointed out, the projection of the soliton onto states with good spin and flavor quantum numbers can only be performed approximately. This is already the case for the (symmetric) two flavor model. For strange degrees of freedom the situation is even worse due to the presence of symmetry breaking. Calculations in Skyrme type models provide two different approaches which are frequently viewed as opposite limits of symmetry breaking. Both treatments represent approximations to the exact time dependent solution which is yet un–known. The first one represents a generalization of the zero mode quantization of SU(2) and requires the introduction of collective coordinates describing rotations in the whole SU(3) flavor space [26, 27]. We will therefore refer to this treatment as the collective approach. As these collective coordinates describe large amplitude fluctuations the restoring force is assumed to be small. Stated otherwise, the flavor symmetry breaking is considered to be small and an expansion in the parameters measuring symmetry breaking is performed. The collective treatment has undergone considerable improvement when Yabu and Ando [28] observed that the resulting collective Hamiltonian including symmetry breaking terms can be diagonalized exactly by numerical methods. Later on this treatment was seen to represent admixtures of states from higher dimensional SU(3) representations to the basic octet and decuplet states [111]. In the complementary treatment, which was initiated by Callan and Klebanov [29] and to which we will refer as the bound state approach, the starting point is to consider symmetry breaking large, although the treatment yields the correct results in the symmetric limit as well. Here only small amplitude fluctuations are allowed (*cf.* section 7.3). These fluctuations can be quantized canonically. The bound state approach heavily relies on the special feature that a solution to the Bethe–Salpeter equation (7.96) emerges in the zero–mode channel when the symmetry breaking is switched on. The eigenfrequency of this mode is different from zero but also significantly lower than the kaon mass, *i.e.* it represents a bound state. Additionally the bound state wave–function is well localized. *I.e.* the bound state is the "would–be" Goldstone boson of the strange flavor transformations in the soliton background. By construction the occupation number of this bound state is identical to the strangeness of the baryon under consideration. Finally the real zero modes corresponding to spin and isospin are treated within the collective approach yielding the hyperfine splitting and thus removing the degeneracy of baryons with identical strangeness as *e.g.* the $\Sigma$ and the $\Lambda$ [29].

Here we will discuss both approaches in the framework of the NJL model and critically compare the results.

### *7.5.1 Collective rotational approach*

In the collective approach symmetry breaking is considered to be small and consequently strange degrees of freedom are introduced as if they were zero modes. On top of this, effects due to symmetry breaking are treated by expanding the fermion determinant in terms of the difference of the constituent quark masses. It should be stressed that for the mesonic part of the action $\mathcal{A}_m$ (3.25) no expansion is performed.

According to the above discussion, collective coordinates for rotations are defined in the whole flavor space in order to approximate the time dependent solution. As the effects due to flavor symmetry breaking have to be taken into account we have to go beyond the *ansatz* for the two flavor case (7.1). We consider [62]

$$M(\boldsymbol{r},t) = R(t)\xi_0(\boldsymbol{r})R^\dagger(t)\langle\Sigma\rangle R(t)\xi_0(\boldsymbol{r})R^\dagger(t) \qquad R(t)\in \mathrm{SU}(3). \tag{7.72}$$

Here $\xi_0 = \exp\left(i\boldsymbol{\tau}\cdot\hat{\boldsymbol{r}}\Theta(r)/2\right)$ denotes the static soliton configuration. Obviously, only the pseudoscalar fields rotate in flavor space while the scalar fields are kept at their vacuum expectation values. For the ongoing exploration it is helpful to define eight angular velocities $\Omega_a$, $(a=1,..,8)$ which measure the time dependence of the flavor rotation[h]

$$\frac{i}{2}\sum_{a=1}^{8}\lambda_a\Omega_a = R^\dagger(t)\dot{R}(t). \tag{7.73}$$

These are the extensions of the previously introduced angular velocities (7.4) to the SU(3) flavor group. Again it is convenient to transform to the flavor rotating system in order to evaluate the fermion determinant: $q = Rq'$. This allows to eliminate the "outer" rotations in (7.72) at the expense of an induced rotational part

$$h_{rot} = \frac{1}{2}\sum_{a=1}^{8}\lambda_a\Omega_a. \tag{7.74}$$

In the rotating frame the Dirac operator acquires the form

$$i\beta D\!\!\!/' = i\partial_t - h_{(0)} - h_{rot} - h_{SB}. \tag{7.75}$$

Here $h_{(0)}$ represents the static one–particle Hamiltonian defined in eq (6.42). The introduction of the chiral transformation $\mathcal{T}$ (7.42) allows one to easily display the symmetry breaking part in the Dirac operator

$$\begin{aligned} h_{SB} &= \mathcal{T}\beta\left(R^\dagger\langle\Sigma\rangle R - \langle\Sigma\rangle\right)\mathcal{T}^\dagger \\ &= \frac{m-m_s}{\sqrt{3}}\mathcal{T}\beta\left(\sum_{i=1}^{3}D_{8i}\lambda_i + \sum_{\alpha=4}^{7}D_{8\alpha}\lambda_\alpha + (D_{88}-1)\lambda_8\right)\mathcal{T}^\dagger. \end{aligned} \tag{7.76}$$

[h] The dot indicates the derivative with respect to the time coordinate.

We have also indicated the SU(2) invariant pieces. Furthermore use has been made of the adjoint representation for the flavor rotations $D_{ab} = \frac{1}{2}\left(\lambda_a R \lambda_b R^\dagger\right)$.

The main task consists of expanding the fermion determinant $\mathcal{A}_F$ in terms of $h_{rot}$ and $h_{SB}$. In order to regularize $\mathcal{A}_F$ a continuation to Euclidean space is required. In this context it is important to take into account that $h_{rot}$ corresponds to the time component of an induced vector field. Hence $h_{rot}$ has to be considered as an anti–Hermitian quantity. The real part of $\mathcal{A}_F$ therefore contributes terms of even power in $\boldsymbol{\Omega}$ to the effective action. Since the expansion is constrained to the second order in $h_{rot} + h_{SB}$ we consider

$$\not{D}_E'^\dagger \not{D}_E' = -\partial_\tau^2 + h_{(0)}^2 + \{h_{(0)}, h_{SB}\} + h_{SB}^2 + [h_{(0)}, h_{rot}] - h_{rot}^2. \tag{7.77}$$

In general there could also be contributions from the commutator $[\partial_\tau, h_{SB}]$, however, these will contribute to the moments of inertia at second order in symmetry breaking and have to be discarded for consistency. Since the imaginary part of the fermion determinant, $\mathcal{A}_I$, contains only terms of odd powers in the angular velocity, it will receive contributions from the mixed terms of the form $h_{rot}h_{SB}$. Noting that the action is only expanded up to second order in the angular velocity, $\mathcal{A}_I$ may be obtained via

$$\begin{aligned}\mathcal{A}_I &= \frac{1}{2}\mathrm{Trlog}(\not{D}_E'^\dagger)^{-1}\not{D}_E' \\ &= \frac{1}{2}\mathrm{Tr}\left\{\left[\left(\partial_\tau - h_{(0)} - h_{SB}\right)\left(-\partial_\tau - h_{(0)} - h_{SB}\right)\right]^{-1}\{h_{rot}, h_{(0)} + h_{SB}\}\right\} + \ldots\end{aligned} \tag{7.78}$$

The time–dependence of $h_{SB}$ actually yields a non–vanishing commutator $[\partial_\tau, h_{SB}]$ which, however, will not contribute to $\mathcal{A}_I$ below $\mathcal{O}(\Omega_a^3)$ and may therefore be discarded for the further calculation. Although the imaginary part is finite the ongoing evaluation of $\mathcal{A}_I$ can be made consistent with the proper–time regularization of the real part. This is achieved by the replacement (7.46).

Now the functional trace can be evaluated. The temporal part is performed by introducing eigenstates[i] $\exp(i\omega_n\tau)$ of $\partial_\tau$ which satisfy anti–periodic boundary conditions in the Euclidean time interval $T$. The spatial part as well as the traces over Dirac and flavor indices are evaluated using eigenstates of the static one–particle Hamiltonian $h_{(0)}$ (6.43).

The leading term in the expansion of the fermion determinant is the vacuum contribution to the soliton mass and stems from the real part (*cf.* chapter 6). The imaginary part starts off with an expression related to the anti–commutator $\{h_{rot}, h_{(0)}\}$

$$\begin{aligned}&\frac{1}{2}\mathrm{Tr}\int_{1/\Lambda^2}^\infty ds\,\exp\left[s\left(\partial_\tau - h_{(0)}\right)\left(\partial_\tau + h_{(0)}\right)\right]\{h_{rot}, h_{(0)}\} \\ &\qquad = \frac{N_C}{2}T\sum_\mu \mathrm{sign}(\epsilon_\mu)\mathrm{erfc}\left(\left|\frac{\epsilon_\mu}{\Lambda}\right|\right)\langle\mu|h_{rot}|\mu\rangle.\end{aligned} \tag{7.79}$$

The sum over the eigenstates $|\mu\rangle$ of $h_{(0)}$ obviously projects out the grand spin zero part of $h_{rot}$ which is related to the eighth component of the angular velocity

$$\frac{N_C}{4\sqrt{3}}T\Omega_8\sum_\mu \mathrm{sign}(\epsilon_\mu)\mathrm{erfc}\left(\left|\frac{\epsilon_\mu}{\Lambda}\right|\right) = \frac{\sqrt{3}}{2}TB^{\mathrm{vac}}\Omega_8 \tag{7.80}$$

[i] $\omega_n = (2n+1)/T$ are the Matsubara frequencies, *cf.* chapter 6.

wherein $B^{\rm vac}$ denotes the contribution to the baryon number originating from the polarization of the Dirac sea. The remainder of the imaginary part is linear in both, the angular velocity and the constituent quark mass difference $m_s - m$ and we denote it by (see appendix B for the definition of $\Delta_{ab}$)

$$\frac{1}{2} T \Delta_{ab}^{\rm vac} \Omega_a D_{8b}. \tag{7.81}$$

The feature, which was already observed for the imaginary part, that the first order in the expansion of the real part in terms of $h_{(0)} + h_{SB}$ only allows for grand spin symmetric expressions also holds for the real part. Taking into account that $\mathcal{A}_R$ is even in the angular velocities there remains only one grand spin symmetric term involving $D_{88}$

$$\frac{1}{2} T \gamma^{\rm vac} (1 - D_{88}) \tag{7.82}$$

since the deviation of the $D$–matrix from unity is considered to be the dynamically relevant quantity (7.76). Furthermore the real part of the fermion determinant contains terms which are quadratic in either the angular velocity or the symmetry breaking

$$\frac{1}{2} T \Theta_{ab}^{\rm vac} \Omega_a \Omega_b, \qquad \frac{1}{2} T \Gamma_{ab}^{\rm vac} D_{8a} D_{8b}. \tag{7.83}$$

Again we refer to ref.[62] for the actual calculation of the quantities $\Theta_{ab}^{\rm vac}$ and $\Gamma_{ab}^{\rm vac}$ (see also section 7.1 for definitions and appendix D for the explicit expressions). Needless to mention that the resulting expression for the action has to be continued back to Minkowski space in order to extract the collective Lagrangian.

Up to this point we have considered the limit of large Euclidean times which has projected out the vacuum contribution to the action. Additionally the coefficients of the collective quantities $\Omega_a$ and $D_{8a}$ receive contributions from the explicit occupation of the valence quark level. These are obtained by consideration of the extented Dirac equation

$$\left( h_{(0)} + h_{rot} + h_{SB} \right) \Psi_{\rm val} = \epsilon_{\rm val} \Psi_{\rm val} \tag{7.84}$$

in stationary perturbation theory. Since the valence quark part of the action is not regularized there is no need to continue forth and back to Euclidean space. Thus the relevant calculations are completely performed in Minkowski space. The resulting expressions are displayed in appendix D. Finally there remains the contribution from the meson part of action $\mathcal{A}_m$. This can be evaluated straightforwardly by substituting the ansatz (7.72) into eq (3.25) yielding

$$\frac{1}{T} \mathcal{A}_m = -\frac{1}{2} \gamma^{\rm m} (1 - D_{88}) - \frac{1}{2} \Gamma_T^{\rm m} \sum_{i=1}^{3} D_{8i} D_{8i} - \frac{1}{2} \Gamma_S^{\rm m} \sum_{\alpha=4}^{7} D_{8\alpha} D_{8\alpha} - \frac{1}{2} \Gamma_8^{\rm m} (1 - D_{88} D_{88}) \tag{7.85}$$

with the coefficients containing the direct information on the current quark masses $m^0$ and $m_s^0$. Again we refer to appendix D for their explicit forms.

We have now collected almost all ingredients for the collective Lagrangian. We still need to discuss meson field components which vanish classically but are induced by the collective rotation into strange direction. Parametrizing the corresponding kaon fluctuation by (*cf.* the previous section 7.3)

$$\sum_{\alpha=4}^{7} \eta_\alpha(\boldsymbol{r}) \lambda_\alpha = \begin{pmatrix} 0 & K(\boldsymbol{r}) \\ K^\dagger(\boldsymbol{r}) & 0 \end{pmatrix} \tag{7.86}$$

the imaginary part of the Euclidean action provides couplings which are linear in both, the kaon field $K$ as well as the angular velocity $\Omega_\alpha$ $(\alpha = 4,..,7)$. Expanding the real part of the action in terms of $K$ only contains even powers of the kaon field. Hence a non–trivial solution for $K(\boldsymbol{r})$ may exist. A suitable ansatz which carries the appropriate flavor and parity quantum numbers reads [124]

$$K(\boldsymbol{r}) = \hat{\boldsymbol{r}} \cdot \boldsymbol{\tau} W(r) \begin{pmatrix} \Omega_4 - i\Omega_5 \\ \Omega_6 - i\Omega_7 \end{pmatrix}. \tag{7.87}$$

In general $W(r)$ is a complex radial function. The framework of the collective approach obviously requires to expand the action up to quadratic order in $\Omega_a$. This procedure adds

$$\frac{T}{2}\beta_I[W] \sum_{\alpha=4}^{7} \Omega_\alpha^2 \tag{7.88}$$

to the action. The equation of motion obtained from the variation $\delta\beta_I/\delta W(r) = 0$ for the radial function $W$ contains a source term stemming from the linear coupling to the collective rotation discussed aboved. This source term is completely given by the static (hedgehog) field. Hence the solution requires $W \neq 0$. It turns out that only the real part of $W$ is excited. In the NJL model up to now the excited fields have not been treated exactly but rather in the gradient expansion. Then the imaginary part of the action is approximated by the Wess–Zumino term. The real part of the action is approximated by the non–linear $\sigma$–model allowing, however, the kaon decay constant $f_K$ to be different from $f_\pi$. For $f_K$ the prediction of the NJL model is used (*cf.* table 4.1). Whenever the chiral angle shows up the self–consistent solution is substituted. The reader may consult ref.[62] for details on this calculation. The important result is that the induced part of the strange moment of inertia $\beta_I$ is not negligible. It should also be stressed that there are no analogous excited fields for the moment of inertia associated with rotations in iso–space since the imaginary part of the action vanishes identically in a two flavor model as long as only pion fields are included[j].

The derivation of the collective three flavor Lagrangian is now terminated and its final form may be discussed. Isospin and rotational invariance provide relations between certain components of the moment of inertia

$$\alpha^2 := \Theta_{11} = \Theta_{22} = \Theta_{33}, \qquad \beta^2 := \Theta_{44} = \Theta_{55} = \Theta_{66} = \Theta_{77} \qquad \text{and} \qquad \Theta_{88} = 0 \tag{7.89}$$

while all other components vanish. The last equation in (7.89) stems from the vanishing commutator $[h_{(0)}, \lambda_8]$. Analogous relations hold for $\Delta_{ab}$ and $\Gamma_{ab}$.

In the next step the collective Hamiltonian associated with the Lagrangian (D.13) has to be constructed. This is achieved by generalizing the treatment of section (7.1) to three flavors. This derivation is also described in appendix D; it results in the mass formula for baryon $B$ carrying spin $J$

$$E_B = E_{\text{tot}} + \frac{1}{2}(\frac{1}{\alpha^2} - \frac{1}{\beta^2})J(J+1) - \frac{3}{8\beta^2} + \frac{1}{2\beta^2}\epsilon_{SB}. \tag{7.90}$$

The moments of inertias $\alpha^2$ and $\beta^2$ may be found in appendix D as well. The quantity $\epsilon_{SB}$ denotes the eigenvalue of the SU(3) operator (D.18) which contains the information on the spin and strangeness of baryon $B$. As discussed in appendix D the eigenstates of the collective Hamiltonian leading to (7.90) are constrained to carry half–integer spin, *i.e.* they are fermions.

[j] An iso–singlet $\eta$–field, however, may yield a non–vanishing but small contribution to the non–strange moment of inertia.

Table 7.7: The mass differences of the low-lying $\frac{1}{2}^+$ and $\frac{3}{2}^+$ baryons with respect to the nucleon. We compare the predictions of the collective approach to the NJL model with the experimental data. The up-quark constituent mass $m$ is chosen such that the $\Delta$–nucleon mass difference is reproduced correctly, see text. The last column refers to the case that the symmetry breaker $\gamma$ is scaled by $(f_K^{\rm expt.}/f_K^{\rm pred.})^2$ All data (from ref.[99]) are in MeV.

| | $f_\pi$ fixed | $f_K$ fixed | Expt. | $f_K^{\rm corr.}$ |
|---|---|---|---|---|
| $\Lambda$ | 105 | 109 | 177 | 175 |
| $\Sigma$ | 148 | 151 | 254 | 248 |
| $\Xi$ | 236 | 243 | 379 | 396 |
| $\Delta$ | 293 | 293 | 293 | 291 |
| $\Sigma^*$ | 387 | 391 | 446 | 449 |
| $\Xi^*$ | 482 | 489 | 591 | 608 |
| $\Omega$ | 576 | 586 | 733 | 765 |

We are now enabled to discuss the numerical results on the hyperon mass spectrum. As already explained in section 4.3 the NJL model underestimates the ratio $f_K/f_\pi$ by about 15% for non–strange constituent quark masses of the order $m \approx 400$MeV. This shortcoming of the model is also expected to show up in the baryon sector. In order to demonstrate that the ratio $f_K/f_\pi$ rather than the absolute value for $f_K$ is the ingredient to which the results are sensitive we compare calculations for two different sets of parameters. First, $f_\pi = 93$MeV is kept at its empirical value and $m = 407$MeV is chosen such as to reproduce the experimental $\Delta$–nucleon mass difference (293MeV). For this set $f_K = 99.8$MeV is predicted, *i.e.* $f_K/f_\pi = 1.07$. The second set of parameters is fixed such as to correctly give $f_K = 114$MeV. Again $m = 433$MeV is determined by demanding the experimental $\Delta$–nucleon mass difference. Then $f_\pi = 104.9$MeV is increased considerably, however, the ratio $f_K/f_\pi = 1.09$ remains almost unaltered. The corresponding numerical results for the baryon mass difference are displayed in table 7.7. Obviously the change in the mass differences for the two sets of parameters is not larger than the change for the ratio $f_K/f_\pi$, *i.e.* 2%. As expected the SU(3) symmetry breaking in the baryon mass differences is considerable underestimated. It has been demonstrated that this effect is strongly correlated to problem of incorrectly predicting $f_K$. The leading order terms in a gradient expansion to the dominating symmetry breaking parameter $\gamma$ are given by (7.82,7.83)

$$
\begin{aligned}
\gamma^{\rm vac\ plus\ m}_{\rm grad.\ exp.} &= \frac{4}{3}\int d^3r\ \left\{(m_K^2 f_K^2 - m_\pi^2 f_\pi^2)(1-\cos\Theta) + \frac{f_K^2 - f_\pi^2}{2}\cos\Theta\left(\Theta'^2 + \frac{2\sin^2\Theta}{r^2}\right) + \ldots\right\} \\
&\approx \frac{4}{3} m_K^2 f_K^2 \int d^3r\ (1-\cos\Theta), \qquad (7.91)
\end{aligned}
$$

where the latter relation is verified numerically. From Skyrme model calculations it is well known [124] that this scaling of $\gamma$ with $f_K$ represents the most important dependence on the kaon decay constant. In order to estimate the effects of this scaling in the framework of the NJL soliton the corresponding quantities $(\gamma, \Gamma_{T,S,8})$ in the collective Lagrangian (D.13) are scaled by $(f_K^{expt.}/f_K^{pred.})^2 \approx (114/100)^2$ and the hyperon spectrum is re–evaluated [62]. Again $m$ is adjusted to reproduce $M_N - M_\Delta$. The results for the mass differences, which are also shown in the last column of table 7.7, agree excellently well with the experimental data. Of course, it should be stressed that this estimate has the purpose of relating the too low prediction for the SU(3) symmetry breaking in the baryon sector to the meson sector rather than being considered a restrictive prediction of the NJL model. We may conclude this section by stating that the NJL model in the collective approach provides the correct hyperon spectrum as far as the quantum numbers and the ordering of the hyperon states within a specified spin multiplet are concerned, however, the quantitative SU(3) breaking is underestimated due to the too small prediction for $f_K$.

It should be stressed that the collective approach together with the above described approximation is not sensitive to the absolute value of the strange current quark mass but rather to the ratio $m_s^0/m^0$ as can be seen from eqs (D.10, D.11 and D.12). This procedure avoids uncertainties related to the choice of a special regularization prescription because the ratio is known to be insensitive on the regularization prescription while the absolute values for the current quark masses are not[k].

There have been similar calculations by another group using the collective approach to the NJL model [125]. These authors choose not to perform the shift $S \rightarrow S - \hat{m}^0$ for the scalar fields (*cf.* section 3.2). Of course, when expanding the fermion determinant to all orders that treatment should be identical to the one presented here. However, this is not the case when approximations are involved. In ref.[125] furthermore flavor independent constituent quark masses ($m_s = m$) are assumed and the determinant is expanded in terms of $m_s^0$. It should be mentioned that an expansion of the action in $m_s^0$ in the sense of chiral perturbation theory is claimed not to converge for the "physical" value of $m_s^0$ [18]. In any event, it is obvious that this treatment is sensitive to the absolute value of $m_s^0$ and thus also to the regularization prescription. In order to accommodate the empirical value $m_s^0 \approx 150$MeV the authors of ref.[125] employ a double step proper-time prescription. The approximation ($m_s = m$) also implies the identity $f_K = f_\pi$ (4.44). It is therefore obvious that in that treatment the hyperon mass splittings cannot be described properly. In order to get reasonable agreement with the experimental spectrum $m_s^0$ had to be increased by about 20%. The relation between $m_s^0$ and the kaon mass $m_K$ is almost linear. The gradient expansion (7.91) reveals that this increase of $m_s^0$ just corresponds to substituting the physical kaon decay constant. Furthermore the double step regularization scheme of ref.[125] involves two more undetermined parameters. Not surprisingly this makes possible a further increase of the symmetry breaking parameter $\gamma$. It should be remarked that within the proper time regularization a consistent computation of the parameters of the collective Hamiltonian yields a larger $\gamma$ when the shift $S \rightarrow S - \hat{m}^0$ is actually performed[l]. This indicates that the results on the hyperon mass splittings are somewhat sensitive to the regularization prescription rather than the parametrization of the rotating fields.

### *7.5.2 Bound state approach*

The bound state approach represents an alternative treatment for the description of hyperons in soliton models. It has originally been applied to the Skyrme model [29] and later on experienced sizable extensions [126, 127, 128]. In the case of the NJL model the bound state approach provides, besides the description of hyperons, an apparent application of the formalism for treating mesonic fluctuations off the chiral soliton [97]. The general aspects of this formalism have been explained in section 7.3.

[k] For a compilation of relevant articles see ref.[18].

[l] Our numerical computations in the proper time scheme show that the $\mathcal{O}(m_s^1)$ contribution to $\gamma$ is about 10% larger when the shift is performed and different constituent quark masses are assumed.

As shown in ref.[99] and mentioned in section 7.3 the mesonic fluctuations decouple with respect to their grand spin and parity quantum numbers. Since kaons carry isospin 1/2 the grand spin of these fluctuations is half–integer. It is therefore sufficient to only consider kaon modes in eqs (7.40, 7.48) and disregard other pseudoscalar fluctuations *i.e.*

$$\tilde{\boldsymbol{\eta}}(\boldsymbol{r},\omega)=0 \quad \text{and} \quad \sum_{\alpha=4}^{7}\tilde{\eta}_\alpha(\boldsymbol{r},\omega)\lambda_\alpha=\begin{pmatrix} 0 & \tilde{K}(\boldsymbol{r},\omega) \\ \tilde{K}^\dagger(\boldsymbol{r},-\omega) & 0\end{pmatrix} \tag{7.92}$$

wherein $\tilde{K}(\boldsymbol{r},\omega)$ is a two–component isospinor. The bound state is expected to appear in exactly that channel which contains a zero mode in the symmetric ($m_s^0=m^0$) case. An infinitesimal vector transformation in strange direction may be associated with

$$K_0(\boldsymbol{r})=\hat{\boldsymbol{r}}\cdot\boldsymbol{\tau}U_0\begin{pmatrix}\sin\frac{\Theta(r)}{2}\\ 0\end{pmatrix}. \tag{7.93}$$

Here $U_0$ denotes an arbitrary $2\times 2$ space–time independent unitary matrix fixing the isospin orientation. This form of the "would–be" zero mode[m] suggests the following *ansatz* for the kaon bound state wave–function

$$\tilde{K}(\boldsymbol{r},\omega)=\hat{\boldsymbol{r}}\cdot\boldsymbol{\tau}\Omega(\omega,r) \qquad \text{with} \qquad \Omega(r,\omega)=\begin{pmatrix}a(r,\omega)\\ b(r,\omega)\end{pmatrix} \tag{7.94}$$

wherein $\Omega(r,\omega)$ is a two–component isospinor (not to be confused with the angular velocity $\boldsymbol{\Omega}$) which only depends on the radial coordinate $r$ and the frequency $\omega$. The angular dependence is completely given by $\hat{\boldsymbol{r}}\cdot\boldsymbol{\tau}$ and thus (7.94) represents a P–wave kaon. Due to isospin invariance the kernels (7.57) corresponding to $\Omega$ are unit matrices in iso–space

$$\begin{aligned}\mathcal{A}^{(2)}[\Omega] \;=\; & \int_{-\infty}^{+\infty}\frac{d\omega}{2\pi}\Big\{\int drr^2\int dr'r'^2\;\Phi^{(2)}(\omega;r,r')\Omega^\dagger(r,\omega)\Omega(r',\omega)\\ & +\int drr^2\;\Phi^{(1)}(r)\Omega^\dagger(r,\omega)\Omega(r,\omega)\Big\}.\end{aligned} \tag{7.95}$$

Explicit expressions for the kernels $\Phi^{(1,2)}$ are given in appendix E.

Varying (7.95) with respect to $\Omega$ yields a homogeneous linear integral equation

$$r^2\left\{\int dr'r'^2\Phi^{(2)}(\omega;r,r')\Omega(r',\omega)+\Phi^{(1)}(r)\Omega(r,\omega)\right\}=0 \tag{7.96}$$

which in fact is the Bethe–Salpeter equation for the kaon bound state in the soliton background. It is the analog of the bound state equation in the Callan–Klebanov approach [29] to the Skyrme model. The fact that the kernels $\Phi^{(1),(2)}$ are unit matrices in isospace leads to identical Bethe–Salpeter equations for the isospinor components $a(r,\omega)$ and $b(r,\omega)$ implying that both have the same radial dependence $\eta_\omega(r)$. Thus Fourier amplitudes $a_i(\omega)$ may be defined as

$$a(r,\omega)=\eta_\omega(r)a_1(\omega) \qquad \text{and} \qquad b(r,\omega)=\eta_\omega(r)a_2(\omega). \tag{7.97}$$

[m] Actually, the bound state is nothing but the "would–be" Goldstone boson for flavor rotations into strange direction.

In the process of quantization the Fourier amplitudes $a_i(\omega)$ acquire the status of creation- and annihilation operators.

In order to determine the normalization of $\eta_\omega(r)$ (which cannot be deduced from the Bethe–Salpeter equation (7.96)) it is helpful to compute the strangeness charge $S$ associated with the kaon bound state. At the microscopic level $S$ is defined as the expectation value

$$S = \int \mathcal{D}\bar{q}\mathcal{D}q \int d^3r \; q^\dagger(-\hat{S})q \, \exp(i \int d^4x \mathcal{L}) \tag{7.98}$$

wherein $\mathcal{L}$ is the NJL – Lagrangian of eq (3.1) and $\hat{S} = \text{diag}(0,0,1)$ is the projector onto strangeness. Note that in the standard convention a strange quark possesses strangeness $-1$. The expression (7.98) is expanded up to second order in the fluctuations. This calculation has been performed in ref.[99] and may be summarized as

$$S = \int \frac{d\omega}{2\pi} \int dr \int dr' \Phi_S(\omega; r, r') \eta^*_\omega(r) \eta_\omega(r') \left( a_1^\dagger(\omega) a_1(\omega) + a_2^\dagger(\omega) a_2(\omega) \right). \tag{7.99}$$

The symmetric bilocal kernel $\Phi_S(\omega; r, r')$ is given in appendix E. Since the strangeness charge has to be quantized to be integer–valued, eqn. (7.99) provides a suitable normalization condition for the radial function $\eta_\omega(r)$

$$\Big| \int dr \int dr' \Phi_S(\omega; r, r') \eta^*_\omega(r) \eta_\omega(r') \Big| = 1. \tag{7.100}$$

Obviously, the sign of the expectation value $S$ is determined by the dynamics. In nature only hyperons with negative strangeness are observed. Thus a major criterion which the NJL model has to satisfy is the existence of a bound state with negative strangeness while the Bethe–Salpeter equation (7.96) should not possess solutions with $|\omega| \le m_K$ and positive strangeness.

In figure 7.1 the radial dependence of the bound state wave–function normalized according to (7.100) is displayed for several values of the constituent quark mass $m$ while the corresponding bound state energy $\omega_0$ is listed in table 7.8.

Different occupations of the bound state remove the degeneracy of states with different strangeness. However, the system consisting of the static soliton and the kaon bound state has still to be projected onto states with good spin and isospin quantum numbers in order to also remove the degeneracy of baryons with identical strangeness, say the $\Sigma$ and the $\Lambda$. This projection is achieved by the usual cranking procedure for the exact zero modes, which correspond to isospin rotations. The associated collective coordinates are introduced analogously to the procedure explained in section 7.1, however, also the fluctuations have to be taken into account. This is accomplished by writing

$$M = R(t)\xi_0\xi_f\langle\Sigma\rangle\xi_f\xi_0R^\dagger(t) \qquad \text{with} \qquad R(t) \in SU(2). \tag{7.101}$$

Obviously, a global isospin rotation $U_g$ corresponds to the substitution $R(t) \to U_g R(t)$. This indicates that the total isospin is carried by $R(t)$. Since $\langle\Sigma\rangle$ commutes with the isospin generators the *ansatz* (7.101) is equivalent to

$$\xi_0 \to R(t)\xi_0R^\dagger(t) \qquad \text{and} \qquad \tilde{K} \to R(t)\tilde{K}. \tag{7.102}$$

This demonstrates that $\tilde{K}$ has lost its isospin. It should be remarked that although the *ansatz* (7.101) is quite different from the one assumed in the original Skyrme model calculation [29] the substitution (7.102) is identical. As for the two flavor version of the model angular velocities $\boldsymbol{\Omega}$ are introduced, see eq (7.4)

$$\frac{i}{2}\boldsymbol{\tau}\cdot\boldsymbol{\Omega} = R^\dagger(t)\dot{R}(t). \tag{7.103}$$

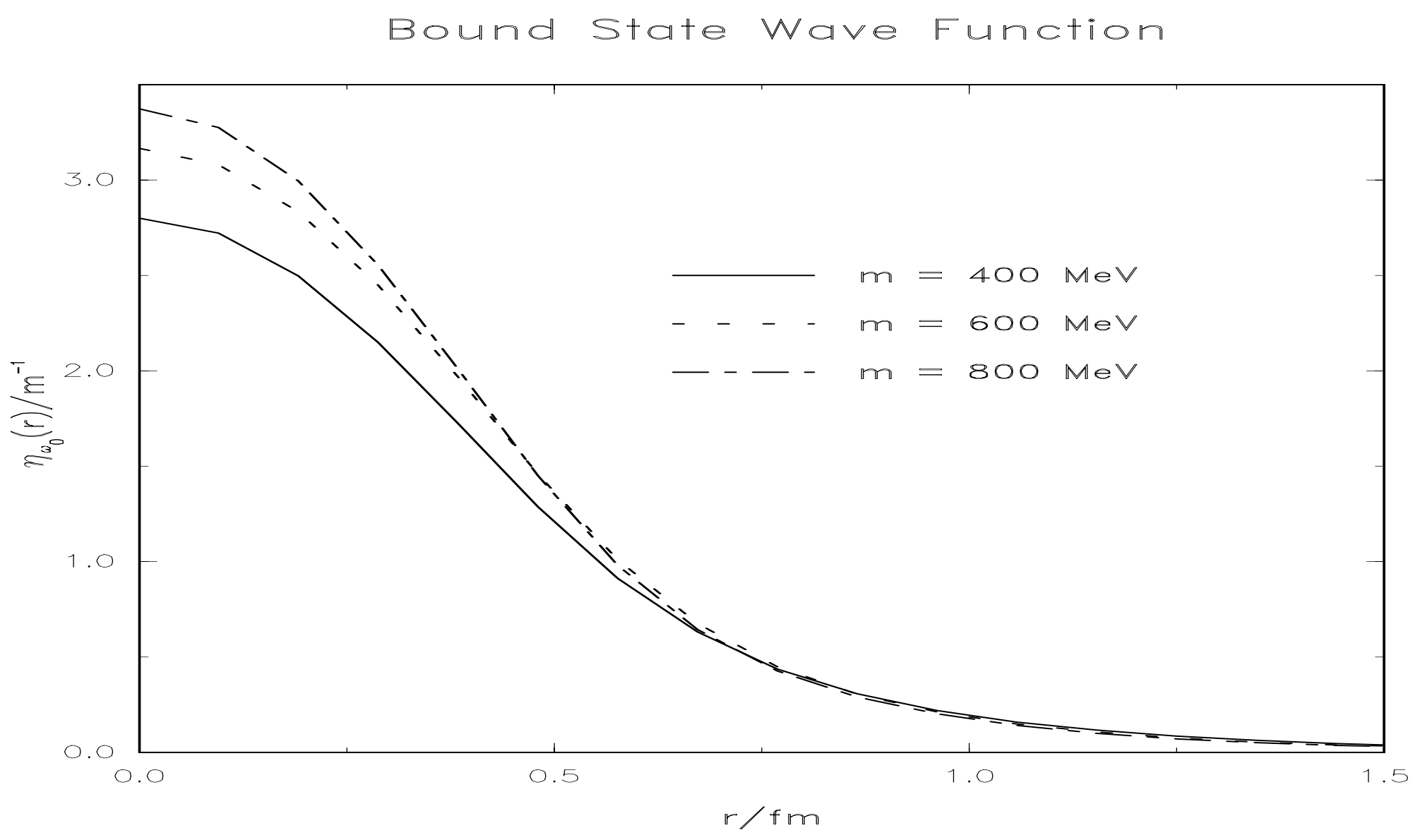

Figure 7.1: The radial dependence of the bound state wave–function $\eta_\omega(r)$ for various constituent quark masses $m$. $\eta_\omega(r)$ is normalized according to (7.100). (Taken from ref. [99].)

Although the collective rotations are not the only time–dependent field components when fluctuations are present, the identity

$$\left[\frac{i}{2}\tau_i, M\right] = -D_{ij}\frac{\partial\dot{M}}{\partial\Omega_j} \tag{7.104}$$

nevertheless holds. The rotation matrix $D_{ij}$ for two flavors is defined in section 7.1 before eq (7.19). Adopting the previous argumentation on Noether charges (*cf.* eqs (7.18, 7.19)) the total isospin is given by

$$I_i = -D_{ij}\frac{\partial L(R,\boldsymbol{\Omega})}{\partial\Omega_j} \tag{7.105}$$

wherein $L$ is the Lagrange function which is a functional of the chiral angle $\Theta$ as well as the kaon fluctuations $\tilde{K}$. It is furthermore instructive to take into account the hedgehog structure of the soliton and define the momentum conjugated to the angular velocity $\boldsymbol{\Omega}$

$$\boldsymbol{J}_\Theta = \frac{\partial L}{\partial\boldsymbol{\Omega}} \tag{7.106}$$

as the spin carried by the soliton. This leads to the identity $\boldsymbol{I}^2 = \boldsymbol{J}_\Theta^2$. Still $\boldsymbol{J}_\Theta$ needs to be related to the total spin. On the microscopic level the latter is defined as the expectation

value

$$\langle \boldsymbol{J} \rangle = \int D\bar{q}Dq \int d^3r \; q^\dagger \boldsymbol{J} q \; \exp\left(i\mathcal{A}_{\rm NJL}\right) \tag{7.107}$$

wherein $\boldsymbol{J}$ is the spin operator for a Dirac spinor and $\mathcal{A}_{\rm NJL} = \int d^4x \mathcal{L}_{\rm NJL}$ denotes the action associated with the NJL Lagrangian (3.1). Since the spin operator commutes with the iso-rotations $R(t)$ the transformation into the rotating frame $q = Rq'$ is straightforward

$$\langle \boldsymbol{J} \rangle = \int D\bar{q}'Dq' \int d^3r \; q'^\dagger \boldsymbol{J} q' \; \exp\left(i\mathcal{A}'_{\rm NJL}\right). \tag{7.108}$$

Here $\mathcal{A}'_{\rm NJL}$ represents the NJL action in the rotating frame which also contains the Coriolis term

$$\mathcal{A}'_{\rm NJL} = \int d^4x \left(\mathcal{L}_{\rm NJL} - \frac{1}{2} q'^\dagger \boldsymbol{\tau} \cdot \boldsymbol{\Omega} q'\right). \tag{7.109}$$

Substituting the definition of the grand spin $\boldsymbol{G}$ yields

$$\langle \boldsymbol{J} \rangle = \int D\bar{q}'Dq' \int d^3r \; q'^\dagger \left(\boldsymbol{G} - \frac{\boldsymbol{\tau}}{2}\right) q' \; \exp\left(i\mathcal{A}'_{\rm NJL}\right). \tag{7.110}$$

In this expression the soliton contribution to the spin $\boldsymbol{J}_\Theta$ may be identified by differentiating $\mathcal{A}'_{\rm NJL}$ with respect to the angular velocity $\boldsymbol{\Omega}$

$$\langle \boldsymbol{J} \rangle = \langle \boldsymbol{G} \rangle + \int D\bar{q}'Dq' \; \frac{1}{T}\frac{\partial \mathcal{A}'_{\rm NJL}}{\partial \boldsymbol{\Omega}} \; \exp\left(iA'_{\rm NJL}\right) = \langle \boldsymbol{G} \rangle + \boldsymbol{J}_\Theta. \tag{7.111}$$

Stated otherwise: the spin carried by the kaons

$$\boldsymbol{J}_K = \langle \boldsymbol{J} \rangle - \boldsymbol{J}_\Theta = \langle \boldsymbol{G} \rangle \tag{7.112}$$

is identical to the grand spin expectation value. This is generally expected based on the $1/N_C$ expansion for baryons, see *e.g.* ref.[129]. Since the eigenstates of the static Hamiltonian are also eigenstates of the grand spin operator, *i.e.* $G_3|\mu\rangle = M_\mu|\mu\rangle$ the result for the kaonic spin is easily obtainable. Repeating the calculation leading to $\mathcal{A}_F^{(2)}$, however, including the grand spin projection quantum number $M_\mu$ when taking matrix elements provides after a straightforward calculation [99] the third component $J_{K3}$, which in turn is proportional to $a_1^\dagger(\omega)a_1(\omega) - a_2^\dagger(\omega)a_2(\omega) = \sum_{ij} a_i^\dagger(\tau_3)_{ij}a_j$. Exploiting rotational invariance one finds

$$\boldsymbol{J}_K = -\frac{1}{2}\int \frac{d\omega}{2\pi} d(\omega) \left(\sum_{i,j=1}^{2} a_i^\dagger(\omega)\boldsymbol{\tau}_{ij}a_j(\omega)\right). \tag{7.113}$$

The spectral function $d(\omega)$ is given in appendix E. Here it is interesting to discuss a conceptual difference in comparison with Skyrme type models. In these models the bound state approach involves classical static fields as well as kaon fluctuations. The former have vanishing grand spin while the latter carry grand spin 1/2. This yields the identity $d(\omega) = 1$ [30]. In the NJL model the situation is different because the functional trace involves quark spinors with arbitrary grand spin. These spinors get polarized by the soliton field as well as the kaon bound state causing $d(\omega)$ to deviate from unity.

In the context of the collective quantization the main task now consists of determining the coupling of the kaon fluctuations to the angular velocities $\boldsymbol{\Omega}$. For this purpose the fermion determinant is first expanded in $\boldsymbol{\Omega}$

$$\mathcal{A}_{\mathcal{F}} = \mathcal{A}_{\mathcal{F}}(\boldsymbol{\Omega}=0) + \sum_{a=1}^{3} \Omega_a \frac{\partial \mathcal{A}_{\mathcal{F}}}{\partial \Omega_a}\Big|_{\boldsymbol{\Omega}=0} + \mathcal{O}(\boldsymbol{\Omega}^2) = \mathcal{A}_F^{(0)} + \mathcal{A}_F^{(1)} + \mathcal{O}(\boldsymbol{\Omega}^2). \tag{7.114}$$

where the contributions of order $\boldsymbol{\Omega}^2$ yield the non–strange moment of inertia $\alpha^2$ (7.15). Here the interest is on the linear term which, in contrast to the two flavor model, does not vanish but rather provides the coupling between the fluctuations and the collective rotations. In the second step of the calculation, $\mathcal{A}_F^{(1)}$ is therefore expanded up to quadratic order in the kaon fields. To this end the $\boldsymbol{\Omega}$–dependent terms of the Lagrange function may be extracted

$$L_{\boldsymbol{\Omega}} = \frac{1}{2}\alpha^2\boldsymbol{\Omega}^2 - \frac{1}{2}\int \frac{d\omega}{2\pi} c(\omega)\boldsymbol{\Omega}\cdot\left(\sum_{i,j=1}^{2} a_i^\dagger(\omega)\boldsymbol{\tau}_{ij}a_j(\omega)\right) + \cdots. \tag{7.115}$$

Again we refer to ref.[99] for details on this calculation while the explicit expression for the spectral function $c(\omega)$ is presented in appendix E.

As already mentioned hyperons with different strangeness are constructed by various occupations of the kaon bound state *i.e.* the solution to the Bethe–Salpeter equation (7.96) with $|\omega_0| < m_K$. Thus

$$c := c(\omega_0) \qquad \text{and} \qquad d := d(\omega_0) \tag{7.116}$$

are projected out from the spectral integrals (7.115,7.113). With this restriction to the bound state the collective Hamiltonian obtained from $L_{\boldsymbol{\Omega}}$ reads

$$H_{\boldsymbol{\Omega}} = \frac{1}{2\alpha^2}\left(\boldsymbol{J}_\Theta + \frac{c}{2}\sum_{i,j=1}^{2} a_i^\dagger \boldsymbol{\tau}_{ij} a_j\right)^2 = \frac{1}{2\alpha^2}\left(\boldsymbol{J}_\Theta + \chi \boldsymbol{J}_K\right)^2 \tag{7.117}$$

where the parameter $\chi = -c/d$ has been introduced. The arguments of the annihilation (creation) operators have been omitted. According to the above discussion $\boldsymbol{J}_\Theta$ and $\boldsymbol{J}_K$ may be related to spin $J$ and isospin $I$ quantum numbers of the considered baryon. This yields

$$H_{\boldsymbol{\Omega}} = \frac{1}{2\alpha^2}\left(\chi J(J+1) + (1-\chi)I(I+1) + (\frac{d\chi}{2})^2 S(S-2)\right) \tag{7.118}$$

because it can be shown that $\left(\sum_{i,j=1}^{2} a_i^\dagger \boldsymbol{\tau}_{ij} a_j\right)^2 = S(S-2)$ [30]. This term is already of fourth order in the bound state wave–function and has to be dropped for consistency. Each substitution of a non-strange valence quark by a strange one changes the valence quark contribution to the energy from $\epsilon_{\rm val}$ to $\epsilon_{\rm val} - \omega_0$. The mass formula for physical baryons thus becomes

$$M_B = E_{\rm cl} + S\omega_0 + \frac{1}{2\alpha^2}\left(\chi J(J+1) + (1-\chi)I(I+1)\right) \tag{7.119}$$

wherein $E_{\rm cl}$ is the classical energy (6.46). This expression is similar to the Skyrme model result [30], however, here the ratio $\chi = -c/d$ had to introduced with $d \neq 1$. The term linear in $\omega_0$ should also come out by performing a calculation similar to that leading to eq (7.60) when the terms of odd powers in $\omega$ are properly accounted for.

Table 7.8: Parameters for describing the hyperon spectrum as functions of the constituent mass $m$. Also listed are the empirical values which are obtained by the consideration of certain mass differences. (Data taken from ref.[99].)

| $m$(MeV) | 350 | 400 | 450 | 500 | empir. |
|---|---|---|---|---|---|
| $\omega_0$(MeV) | -207.1 | -182.6 | -163.6 | -148.8 | -189.5 |
| $c$ | -0.20 | -0.36 | -0.46 | -0.53 | — |
| $d$ | 0.90 | 0.89 | 0.89 | 0.89 | — |
| $\chi = -c/d$ | 0.22 | 0.40 | 0.52 | 0.60 | 0.62 |
| $\alpha^2(1/(\mathrm{GeV}))$ | 8.30 | 5.80 | 4.78 | 4.17 | 5.12 |

Before going into the details of the numerical results[n] it is important to note that bound states have only been obtained in the region $-m_K < \omega_0 < 0$ and that these carry negative strangeness according to eq (7.99). No bound state has been observed in the interval $0 < \omega < m_K$.

As in the collective treatment only the mass differences with respect to the nucleon mass are considered. In this case eq (7.119) contains three quantities $(\alpha^2, \omega_0, \chi)$ which determine seven mass differences. Thus the baryon mass formula (7.119) may be inverted relating $\alpha^2, \omega_0$ and $\chi$ to hyperon mass differences. These relations may be employed to obtain empirical data for $\alpha^2, \omega_0$ and $\chi$. We list these together with the calculated values for various constituent quark masses $m$ in table 7.8. The empirical data are reasonably well accommodated in the region $400\mathrm{MeV} \le m \le 450\mathrm{MeV}$ although $\chi$ is somewhat too small.

Again, the only free parameter $m$ is chosen to get a best fit to the experimental mass differences. The $\Delta$–nucleon mass difference is the same as in the two flavor model, $3/2\alpha^2$. In order to reproduce this mass splitting kaon fluctuations need not be considered. This simplification has been used in ref.[99] to fix $m = 430\mathrm{MeV}$ such that $\alpha^2 = 5.12/\mathrm{GeV}$. Then the six other mass differences are predicted. The results are listed in table 7.9. There also the case is presented when $f_K$ is fixed to its experimental value rather than $f_\pi$. This yields a different value for the cut-off $\Lambda$ and thus $m$ needs to be readjusted in order to get $\alpha^2 = 5.12/\mathrm{GeV}$. This does not significantly alter the predictions on the mass differences since the ratio $f_K/f_\pi$ is almost identical in both cases. Comparing with table 7.7 it is found that the bound state approach improves on the mass differences, especially for the spin 1/2 hyperons.

The bound state approach may be extended to the investigation of kaonic channels others than the P–wave. A prominent case is represented by the S–wave since it in principle allows one to explore the odd parity $\Lambda$ hyperon which is observed at 1405MeV experimentally. In the Skyrme model this state has been studied intensively, see *e.g.* refs.[29, 127]. The corresponding NJL model calculations are reported in ref.[130]. The kaon S–wave is described by the ansatz

$$\tilde{K}_S(\boldsymbol{r}, \omega) = \begin{pmatrix} a^S(r,\omega) \\ b^S(r,\omega) \end{pmatrix}. \tag{7.120}$$

Although $\tilde{K}_S$ possesses the same grand spin ($G = 1/2$) as the P–wave (7.94) these two channels nevertheless decouple due to the opposite parity. As for the P–wave, a Bethe–Salpeter equation for the radial function $\eta^S_\omega(r)$ in the Fourier decomposition

$$a^S(r,\omega) = a^S_1(\omega)\eta^S_\omega(r) \qquad \text{and} \qquad b^S(r,\omega) = a^S_2(\omega)\eta^S_\omega(r) \tag{7.121}$$

[n]The numerical method is explained in ref.[121].

Table 7.9: The mass differences of the low-lying $\frac{1}{2}^+$ and $\frac{3}{2}^+$ baryons with respect to the nucleon in the bound state approach to the NJL model. The up-quark constituent mass $m$ is chosen such that the $\Delta$–nucleon mass difference is reproduced. All data (from ref.[99]) are in MeV.

| | $f_\pi$ fixed | $f_K$ fixed | Expt. |
|---|---|---|---|
| $\Lambda$ | 132 | 137 | 177 |
| $\Sigma$ | 234 | 247 | 254 |
| $\Xi$ | 341 | 357 | 379 |
| $\Delta$ | 293 | 293 | 293 |
| $\Sigma^*$ | 374 | 375 | 446 |
| $\Xi^*$ | 481 | 485 | 591 |
| $\Omega$ | 613 | 622 | 733 |

can be derived. The solution of this equation determines the bound state energy, $\omega_S$. The quantization of spin and isospin proceeds as for the P–wave yielding $c_S$ and $d_S$, which denote the coupling of the S–wave bound state to the collective rotations and the contribution of the S–wave bound state to the spin, respectively. Following the quantization of the P–wave bound state it is then straightforward to derive the mass formula for the odd parity $\Lambda$

$$M(\text{odd parity } \Lambda) = E_{\text{cl}} - \omega_S + \frac{3\chi_S}{8\alpha^2} \qquad (7.122)$$

wherein the ratio $\chi_S = -c_S/d_S$ does not dependent on the normalization of $\eta_\omega^S(r)$.

Since the $\Lambda(1405)$ is only about 40MeV below the kaon–nucleon threshold the bound state eigenvalue for the S–wave is expected to lie at $\omega_S \approx -450$MeV. Fluctuations possessing energies as large as this value may raise problems because the NJL model is not a confining theory. In case the valence quark is only slightly bound, *i.e.* $\epsilon_{\text{val}} - m$ is small, kaon energies of several hundred MeV may scatter the occupied valence quark state into the strange continuum. This situation corresponds to having an undesired nucleon–strange quark threshold, $E_{Ns}$, below the kaon–nucleon threshold. In that case the second term on the *RHS* of eq (7.54) develops a pole, the position of which determines $E_{Ns}$. Since for kaon fluctuations the perturbation $\tilde{h}_{(1)}$ carries unit strangeness, $\epsilon_\mu$ in (7.54) is the eigenvalue of a free Dirac–Hamiltonian in the strange sector. Therefore $\epsilon_\mu \approx m_s$ is the smallest eigenvalue which determines $|E_{Ns}| \approx m_s - \epsilon_{\text{val}}$. It turns out that $|E_{Ns}|$ is a monotonously rising function of the non–strange constituent quark mass $m$ and that for $m \geq 450$MeV $|E_{Ns}|$ lies above the kaon–nucleon threshold [130]. The predictions obtained in this region on the odd parity $\Lambda$ hyperon are displayed in table 7.10.

The resulting mass difference between the odd parity $\Lambda$ hyperon and the nucleon is insensitive to the constituent mass $m$. As shown above, the $\Delta$–nucleon mass difference is reproduced for $m \approx 430$MeV. Then, unfortunately, the threshold $E_{Ns}$ lies slightly below the physical kaon–nucleon threshold. Nevertheless a reasonable description for both the relative position of this hyperon as well as the $\Delta$ can be obtained for $m \approx 450$MeV.

Table 7.10: Parameters for the baryon mass formula (7.122) and the prediction for the mass of the odd parity $\Lambda$ hyperon relative to the nucleon $M_{\Lambda_S} - M_N$ as functions of the constituent mass $m$. Also the resulting values for the $\Delta$ nucleon mass splitting are presented. (Data taken from ref.[130].)

| $m$ (MeV) | $\alpha^2$ (1/GeV) | $\omega_S$ (MeV) | $\chi$ | $\Lambda_S - N$ (MeV) | $\Delta - N$ (MeV) |
|---|---|---|---|---|---|
| 450 | 4.78 | -461 | 0.46 | 418 | 314 |
| 500 | 4.19 | -461 | 0.54 | 419 | 358 |
| 550 | 3.74 | -459 | 0.57 | 416 | 401 |
| 600 | 3.42 | -456 | 0.60 | 411 | 439 |

# 8 Summary

We have presented the description of baryons as chiral solitons within the NJL model, a microscopic model for the quark flavor dynamics. Within QCD we have stressed the role of chiral symmetry, its spontaneous breaking and its consequences in hadron physics as a basis for modeling the strong interaction. Furthermore, we have reviewed some important features of QCD in the limit of infinitely many color degrees of freedom. This provides the basic motivation for the picture that baryons emerge as soliton solutions of an effective meson theory. In the framework of QCD a realization of the soliton picture of baryons is not feasible. Therefore we have approximated the quark flavor dynamics by the NJL model which, like QCD, is invariant under global chiral transformations and breaks this symmetry dynamically, which is reflected by a non–vanishing quark condensate in the ground state.

Using functional integral techniques the NJL model can be converted as an effective meson theory. This effective meson theory is highly non–local. At low energies one can, fortunately, resort to an gradient expansion, which describes the low energy (light flavor) meson dynamics reasonably well. In leading order the gradient expansion of the regular parity part of the effective meson theory yields the gauged linear $\sigma$–model. The associated massive gauge bosons are interpreted as vector and axial–vector mesons. In the limit of infinitely large (axial) vector meson masses this approximation to the effective theory reduces to the Skyrme model. The gradient expansion of the irregular parity part of the effective meson theory, which involves only odd numbers of spatial components of Lorentz–vectors, provides the Wess–Zumino action in leading order. The latter result furthermore allows one to identify the baryon current with the topological current of the Skyrme model. It should be emphasized that this gradient expansion refers to the vacuum (or sea) part of the action only. Hence the Skyrmion picture implies that the baryon number is carried by the polarized Dirac sea of the quarks. Stated otherwise, a soliton of baryon number $B$ requires $N_C \cdot B$ valence quarks to be bound in the Dirac sea by the solitonic meson fields. This is an inherent assumption of all purely mesonic soliton models of baryons (Witten's conjecture). Restraining from the gradient expansion the NJL model allows one to test this conjectures. For this investigation we have considered the self–consistent soliton solutions of various field configurations in the NJL model. Although in order to describe a unit baryon number configuration explicit valence quarks have to be added to the chiral soliton of the pseudoscalar fields only, the baryon number is indeed carried by the Dirac sea once also the axial–vector meson fields are included. This result strongly supports the Skyrmion picture of the baryon. It furthermore provides a microscopic explanation for the fact that explicit valence quarks and (axial) vector meson degrees of freedom represent alternative approaches to describe baryons as solitons.

In order to simplify the investigation of static baryon properties we have resorted to the model containing only the pseudoscalar field while explicit valence quarks are present. In order to generate states with good spin and isospin quantum numbers we have adopted the semi–classical quantization scheme. In this scheme the time dependent coordinates, which parametrize the (iso) rotational zero modes, are quantized canonically. Since this treatment is analogous to the quantization of a rigid top it yields (iso) rotational bands of baryons. The energy splitting of the low–lying band members is of order $1/N_C$ while the soliton mass is of order $N_C$. Like in Skyrme type models the absolute values for the baryon masses are generally predicted too large if $O(N_C^0)$ quantum corrections corrections are neglected. We have estimated these corrections and found that they may considerably reduce the baryon masses towards their experimental values.

When extending this quantization prescription to the case of three flavors and including SU(3) symmetry breaking explicitly a reasonable description of the hyperons is obtained. Furthermore, an excellent description of hyperons within this approach is somewhat restricted because the NJL model predicts to small a ratio between the kaon and pion decay constants; a problem inherited from the meson sector. A alternative description of hyperons within soliton models relies on the existence of a strongly bound kaon meson in the background of the chiral soliton. As a matter of fact this meson becomes a zero mode if the SU(3) symmetry breaking is ignored. For the NJL model this treatment apparently provides a better description of the spectrum of the low–lying hyperons, especially for the $J^\pi = \frac{1}{2}^+$ states. When applying this approach to excited states one is confronted with the existence of unphysical quark–(anti)quark thresholds. These reflect the non–confining character of the NJL model. Of course, any attempt to extend or refine the NJL model in such a way that this problem is avoided would be interesting.

Nevertheless the NJL soliton model appears to be a well–established approach to study the low–energy properties of the nucleon. Especially it provides an apparent opportunity to compare the contributions of valence and sea quarks to various baryon observables.

## Acknowledgements

On of us (HR) would like to thank D. Ebert for the cooperation on the mesonic properties. We are grateful to U. Zückert and J. Schlienz for the cooperation on parts of the material presented in this report.

## Appendix A

In this appendix we demonstrate the calculation of the Jacobi determinant (4.59) using Fujikawa's functional integral method [49]. In order to make the computation of the change of the integral measure, reflected by the relation $J(\Theta) \neq 1$, more transparent we will use a few simplifications. First, we will only consider infinitesimal chiral transformations, *i.e.* small $\Theta$. Since the axial anomaly occurs only for the abelian subgroup $U_A(1)$ we assume that $\Theta$ is proportional to the unit matrix in flavor space, $\Theta \propto 1\!\!I_F$. This assumption will make the computation more feasible as $\Theta$ commutes with other flavor matrices. The whole calculation may be done retaining these commutators with the result that the traceless parts of $\Theta$ do not contribute. Second, we will work with a Dirac operator containing besides the usual kinetic and mass term only a vector field

$$i\not{D} = i(\not{\partial} + \not{V}) - m^0 =: i\not{d} - m^0. \tag{A.1}$$

In Euclidean space we are working with hermitian Dirac matrices and antihermitian vector fields the operator $i\not{d}$ is hermitian with respect to the ordinary scalar product. Therefore the eigenfunctions of the operator $i\not{d}$ can be chosen to form a complete and orthonormal set

$$\begin{aligned} i\not{d}\varphi_n(x) &= \lambda_n \varphi_n(x) \\ \int d^4x \varphi_n^\dagger(x)\varphi_m(x) &= \delta_{nm} \\ \sum_n \varphi_n(x)\varphi_n^\dagger(y) &= \delta(x-y). \end{aligned} \tag{A.2}$$

We expand the dynamical quark field, *i.e.* the functional integration variable, in terms of the functions $\varphi_n$

$$\begin{aligned} q(x) &= \sum_n a_n \varphi_n(x) \\ \bar{q}(x) &= \sum_n a_n^* \bar{\varphi}_n(x) \end{aligned} \tag{A.3}$$

The expansion coefficients $a_n$ and $a_n^*$ are independent anticommuting Grassmann variables. The integration measure can now be written as

$$\mathcal{D}\bar{q}\mathcal{D}q = \prod_m da_m da_m^* \tag{A.4}$$

The chirally transformed quark field (4.58) is given by

$$\begin{aligned} \chi(x) &= \sum_m \tilde{a}_m \varphi_m(x) = \sum_m a_m e^{i\Theta(x)\gamma_5}\varphi_m(x) \\ \bar{\chi}(x) &= \sum_m \bar{\varphi}_m(x)\tilde{a}_m^* = \sum_m \bar{\varphi}_m(x) a_m^* e^{i\Theta(x)\gamma_5}. \end{aligned} \tag{A.5}$$

where the expansion coefficients for the transformed quark field are related to the original ones by

$$\begin{aligned} \tilde{a}_m &= \sum_n C_{mn} a_n \\ \tilde{a}_m^* &= \sum_n a_n^* C_{nm} \\ C_{mn} &= \langle m | e^{i\Theta(x)\gamma_5} | n \rangle = \int d^4x \varphi_m^\dagger(x) e^{i\Theta(x)\gamma_5}\varphi_n(x) \\ &= \delta_{mn} + i\int d^4x \Theta(x)\varphi_m^\dagger(x)\gamma_5\varphi_n(x) + \mathcal{O}(\Theta^2). \end{aligned} \tag{A.6}$$

Here we have used the orthonormality of the eigenfunctions $\varphi_n$. As the sets $a_n, a_n^*$ and $\tilde{a}_m, \tilde{a}_m^*$ are Grassmann variables we obtain

$$J(\Theta) = (\mathrm{Det}C)^{-2}. \tag{A.7}$$

For infinitesimally small $\Theta$ the matrix $C$ is given by

$$C_{mn} = \delta_{mn} + \langle m|i\gamma_5\Theta(x)|n\rangle + \mathcal{O}(\Theta^2). \tag{A.8}$$

Defining the matrix

$$\epsilon_{mn} = \langle m|i\gamma_5\Theta(x)|n\rangle \tag{A.9}$$

and shorthandly writing $C = 1 + \epsilon$ the functional determinant of this matrix is evaluated to be in first order in $\Theta$

$$\begin{aligned} \mathrm{Det}C &= \mathrm{Det}(1+\epsilon) = \exp\Big(\mathrm{Tr}(\epsilon) + \mathcal{O}(\epsilon^2)\Big) \\ &= \exp\left(i\int d^4x\Theta(x)\sum_n \varphi_n^\dagger\gamma_5\varphi_n(x)\right) + \ldots\ . \end{aligned} \tag{A.10}$$

As it stands the operator appearing in the exponent is not well–defined. In order to attach a suitable definition we have to regularize (A.10) and to remove the regularization at the end of the calculation

$$\begin{aligned} B(x) &:= \lim_{\Lambda\to\infty}\sum_n \varphi_n^\dagger\gamma_5 e^{-(\lambda_n/\Lambda)^2}\varphi_n(x) \\ &= \lim_{\Lambda\to\infty}\mathrm{tr}\gamma_5\langle x|e^{-\not{d}^\dagger\not{d}/\Lambda^2}|x\rangle. \end{aligned} \tag{A.11}$$

To apply the heat kernel expansion we introduce the "proper time" $\tau = 1/\Lambda^2$ as an expansion parameter and use $A_0 = -\partial_\mu\partial^\mu + m^2$ as unperturbed operator. Expanding now the matrix element

$$\langle x|e^{-\tau\not{d}^\dagger\not{d}}|y\rangle = \langle x|e^{-\tau A_0}|y\rangle\sum_{k=0}^{\infty}h_k(x,y)\tau^k \tag{A.12}$$

we obtain

$$B(x) = \frac{1}{(4\pi)^2}\lim_{\tau\to 0}\sum_{k=0}^{\infty}\tau^{k-2}\mathrm{tr}(\gamma_5 h_k(x,x)). \tag{A.13}$$

Note that only the heat coefficients $h_0$, $h_1$ and $h_2$ in the sum (A.13) contribute to $B(x)$ because the coefficients of the higher order terms are suppressed by a factor $\tau^{k-2}$ and therefore vanish in the limit $\tau \to 0$ ($\Lambda \to \infty$). As

$$\begin{aligned} \mathrm{tr}\gamma_5 h_0(x,x) &= \mathrm{tr}\gamma_5 = 0 \\ \mathrm{tr}\gamma_5 h_1(x,x) &= -\frac{1}{2}\mathrm{tr}(\gamma_5\gamma_\mu\gamma_\nu)F_{\mu\nu} = 0 \end{aligned} \tag{A.14}$$

only one term contributes

$$\mathrm{tr}\gamma_5 h_2(x,x) = N_c\mathrm{tr}_F\tilde{F}^{\mu\nu}F_{\mu\nu}. \tag{A.15}$$

Here $F_{\mu\nu}$ is the field strength tensor corresponding to the vector field $V_\mu$ and $\tilde{F}_{\mu\nu}$ is the dual tensor. Note that a flavor trace has to be taken in (A.15). Eqs. (A.13) and (A.15) can now be combined to the result

$$B(x) = \frac{N_c}{16\pi^2}\mathrm{tr}_F \tilde{F}^{\mu\nu} F_{\mu\nu} \tag{A.16}$$

finally yielding the expression (4.61) for the Jacobian $J(\Theta)$:

$$J(\Theta) = (\mathrm{Det}C)^{-2} = \exp\left(-i\frac{N_c}{8\pi^2}\int d^4x\Theta(x)\mathrm{tr}_F(\tilde{F}F)\right). \tag{A.17}$$

Using the invariance of the classical action under (4.58) one can easily derive the anomalous Ward identity. Under infinitesimal chiral rotations the quark bilinear in the action transforms as

$$\begin{aligned}\bar{q}(i\not{d} - m^0)q &= \bar{\chi}e^{-i\Theta(x)\gamma_5}(i\not{d} - m^0)e^{-i\Theta(x)\gamma_5}\chi \\ &= \bar{\chi}(i\not{d} - m^0)\chi + \partial_\mu\Theta(x)\bar{\chi}\gamma^\mu\gamma_5\chi + 2im^0\Theta(x)\bar{\chi}\gamma_5\tilde{q} + \mathcal{O}(\Theta^2)\end{aligned} \tag{A.18}$$

which leads to the following identities for the effective quark action

$$\begin{aligned}\mathcal{A}_F &= \mathrm{Tr}\log(i\not{d} - m^0) = \log\int \mathcal{D}q\mathcal{D}\bar{q}\exp\left(i\int d^4x\bar{q}(i\not{d} - m^0)q\right) \\ &= \log J(\Theta) \\ &\quad + \log\int \mathcal{D}\chi\mathcal{D}\bar{\chi}\exp\left(i\int d^4x\bar{\chi}(i\not{d} - m^0)\chi + \partial_\mu\Theta(x)\bar{\chi}\gamma^\mu\gamma_5\chi + 2im^0\Theta(x)\bar{\chi}\gamma_5\chi\right).\end{aligned} \tag{A.19}$$

As the first equation is independent of $\Theta$ the last one has to be, too. Therefore,

$$0 = \frac{\delta\mathcal{A}_F}{\delta\Theta}\Big|_{\Theta=0} = -i\frac{N_c}{8\pi^2}\langle\mathrm{tr}_F(\tilde{F}F)\rangle - \partial_\mu\langle\bar{\chi}\gamma^\mu\gamma_5\chi\rangle + 2im^0\langle\bar{\chi}\gamma_5\chi\rangle. \tag{A.20}$$

Defining the axial singlet current and density

$$\begin{aligned}j_5^\mu &= \langle\bar{\chi}\gamma^\mu\gamma_5\chi\rangle \\ j_5 &= \langle\bar{\chi}\gamma_5\chi\rangle\end{aligned} \tag{A.21}$$

we finally arrive at the anomalous Ward identity (4.69)

$$\partial_\mu j_5^\mu = 2im^0 j_5 - i\frac{N_c}{8\pi^2}\mathrm{tr}_F\langle(\tilde{F}F)\rangle. \tag{A.22}$$

# Appendix B

In this brief appendix we present the explicit representation of the eigenstates $|\nu\rangle$ of the Dirac Hamiltonian $h$ (6.42,6.63) in coordinate space. As indicated in chapter 6 we first construct eigenstates to the grand spin operator

$$\boldsymbol{G} = \boldsymbol{l} + \frac{\boldsymbol{\sigma}}{2} + \frac{\boldsymbol{\tau}}{2}, \tag{B.1}$$

which is the vector sum of orbital angular momentum $\boldsymbol{l}$, spin $\boldsymbol{\sigma}/2$ and isospin $\boldsymbol{\tau}/2$. $\boldsymbol{G}$ acts on the quark spinors. These are obtained by first coupling states with orbital angular momentum $l$ with $s = 1/2$ states to form states carrying total angular momentum $j$ and projection $j_3$

$$|ljj_3\rangle = \sum_{m,\sigma_3} C^{jj_3}_{lm\frac{1}{2}\sigma_3} |lm\rangle |\frac{1}{2}\sigma_3\rangle. \tag{B.2}$$

Subsequently these states are coupled with isospin $t = 1/2$ states yielding the grand spin eigenstates

$$|ljGM\rangle = \sum_{j_3,\tau_3} C^{GM}_{jj_3\frac{1}{2}\tau_3} |ljj_3\rangle |\frac{1}{2}\tau_3\rangle. \tag{B.3}$$

This coupling scheme obviously leads to the selection rules (6.45). Since the upper and lower components of Dirac spinors transform oppositely under the parity transformation the general form of Dirac spinors, which are eigenstates of $\boldsymbol{G}$ and parity $\Pi = \pm 1$, are given by

$$\Psi^{(G,+)}_\nu = \begin{pmatrix} ig^{(G,+;1)}_\nu(r)|GG+\frac{1}{2}GM\rangle \\ f^{(G,+;1)}_\nu(r)|G+1G+\frac{1}{2}GM\rangle \end{pmatrix} + \begin{pmatrix} ig^{(G,+;2)}_\nu(r)|GG-\frac{1}{2}GM\rangle \\ -f^{(G,+;2)}_\nu(r)|G-1G-\frac{1}{2}GM\rangle \end{pmatrix} \tag{B.4}$$

$$\Psi^{(G,-)}_\nu = \begin{pmatrix} ig^{(G,-;1)}_\nu(r)|G+1G+\frac{1}{2}GM\rangle \\ -f^{(G,-;1)}_\nu(r)|GG+\frac{1}{2}GM\rangle \end{pmatrix} + \begin{pmatrix} ig^{(G,-;2)}_\nu(r)|G-1G-\frac{1}{2}GM\rangle \\ f^{(G,-;2)}_\nu(r)|GG-\frac{1}{2}GM\rangle \end{pmatrix}. \tag{B.5}$$

The second superscript labels the intrinsic parity $\Pi_{\rm intr}$ and enters the parity eigenvalue via $\Pi = (-1)^G \times \Pi_{\rm intr}$. $\Pi_{\rm intr}$ represents a good quantum number since both grand spin and parity operators commute with $h$. Furthermore, the grand spin invariance is very helpful when choosing *ansätze* for meson fields other than the chiral field, see section 6.2. The diagonalization (6.43,6.59) yields the radial functions $g^{(G,+;1)}_\nu(r)$, *etc.* as linear combinations of spherical Bessel functions (the solutions to the free problem). *E.g.*

$$g^{(G,+;1)}_\nu(r) = \sum_k V_{\nu k}[\Theta] N_k \sqrt{1 + m/E_{kG}}\; j_G(k_{kG} r),$$
$$f^{(G,+;1)}_\nu(r) = \sum_k V_{\nu k}[\Theta] N_k {\rm sgn}(E_{kG}) \sqrt{1 - m/E_{kG}}\; j_{G+1}(k_{kG} r) \tag{B.6}$$

where the eigenvectors[a] $V_{\nu k}[\Theta]$ are obtained by diagonalizing $h$. In case that the Hamiltonian is not Hermitian (6.63) these eigenvectors are complex. $N_k$ are normalization constants. $E_{kG} = \pm\sqrt{m^2 + k^2_{kG}}$ denote the energy eigenvalues in the absence of the soliton. The momentum eigenvalues $k_{kG}$ are subject to the boundary condition $j_G(k_{kG}D) = 0$ [81]. There are also other boundary conditions discussed in the literature [62, 98] and the pertinent boundary condition depends on the problem under consideration.

[a]Grand spin and parity indices are omitted.

# Appendix C

In this appendix the equations of motion for the meson fields as taken from ref.[73, 75] are listed. These equations result from the stationary condition $\delta E/\delta\varphi$ for the Minkowski energy functional (6.39). The functional derivatives of the one–particle energy eigenvalues $\epsilon_\nu$ are extracted from the Euclidean Dirac Hamiltonian (6.63).

It is appropriate to define densities[a]

$$
\begin{aligned}
\rho(\boldsymbol{x},\boldsymbol{y}) &= N_c\left\{\rho^{\rm val}(\boldsymbol{x},\boldsymbol{y})+\rho^{\rm vac}(\boldsymbol{x},\boldsymbol{y})\right\} \quad \text{and} \quad b(\boldsymbol{x},\boldsymbol{y}) = N_c\left\{b^{\rm val}(\boldsymbol{x},\boldsymbol{y})+b^{\rm vac}(\boldsymbol{x},\boldsymbol{y})\right\}\\
\rho^{\rm val}(\boldsymbol{x},\boldsymbol{y}) &= \eta_{\rm val}\left\{\Re\psi_{\rm val}(\boldsymbol{x})\bar\psi_{\rm val}(\boldsymbol{y})+\Im\psi_{\rm val}(\boldsymbol{x})\bar\psi_{\rm val}(\boldsymbol{y})\right\}\\
\rho^{\rm vac}(\boldsymbol{x},\boldsymbol{y}) &= \sum_\nu\left\{f_R(\epsilon_\nu/\Lambda)\Re\psi_\nu(\boldsymbol{x})\bar\psi_\nu(\boldsymbol{y})+f_I(\epsilon_\nu/\Lambda)\Im\psi_\nu(\boldsymbol{x})\bar\psi_\nu(\boldsymbol{y})\right\}\\
b^{\rm val}(\boldsymbol{x},\boldsymbol{y}) &= \eta_{\rm val}\left\{\Re\psi_{\rm val}(\boldsymbol{x})\psi^\dagger_{\rm val}(\boldsymbol{y})-\Im\psi_{\rm val}(\boldsymbol{x})\psi^\dagger_{\rm val}(\boldsymbol{y})\right\}\\
b^{\rm vac}(\boldsymbol{x},\boldsymbol{y}) &= \sum_\nu\left\{f_I(\epsilon_\nu/\Lambda)\Re\psi_\nu(\boldsymbol{x})\psi^\dagger_\nu(\boldsymbol{y})-f_R(\epsilon_\nu/\Lambda)\Im\psi_\nu(\boldsymbol{x})\psi^\dagger_\nu(\boldsymbol{y})\right\}.
\end{aligned}
\tag{C.1}
$$

Real ($\Re$) and imaginary ($\Im$) parts refer to the expansion coefficients $V_{\nu k}[\varphi]$ of the free basis (*cf.* eq (B.6)). The regulator functions

$$
f_R(\epsilon_\nu/\Lambda) = \begin{cases} -\frac{1}{2}{\rm sgn}(\epsilon_\nu^R){\rm erfc}\left(\left|\frac{\epsilon_\nu}{\Lambda}\right|\right) & \mathcal{A}_I \quad \text{not regularized,}\\[2ex] -\frac{1}{2}{\rm sgn}(\epsilon_\nu^R){\rm erfc}\left(\left|\frac{\epsilon_\nu}{\Lambda}\right|\right)+\frac{1}{\sqrt{\pi}}(\epsilon_\nu^I/\Lambda){\rm exp}(-(\epsilon_\nu^R/\Lambda)^2), & \mathcal{A}_I \quad \text{regularized}\end{cases} \tag{C.2}
$$

$$
f_I(\epsilon_\nu/\Lambda) = -\frac{1}{2}{\rm sgn}(\epsilon_\nu^R)\begin{cases} 1, & \mathcal{A}_I \quad \text{not regularized}\\[2ex] {\rm erfc}\left(\left|\frac{\epsilon_\nu}{\Lambda}\right|\right) & \mathcal{A}_I \quad \text{regularized}\end{cases} \tag{C.3}
$$

reflect the derivative of the energy functional with respect to the single quark energy eigenvalues. The specific form of the equations of motion $\delta E/\delta\varphi$ finally becomes

$$\Phi(r) = \frac{m^0}{m}{\rm cos}\Theta(r) - \frac{m^0}{m_\pi^2 f_\pi^2}{\rm tr}\int\frac{d\Omega}{4\pi}\left({\rm cos}\Theta(r)+i\gamma_5\boldsymbol{\tau}\cdot\hat{\boldsymbol{r}}{\rm sin}\Theta(r)\right)\rho(\boldsymbol{x},\boldsymbol{x}) \tag{C.4}$$

$$\sin\Theta(r) = \frac{m}{m_\pi^2 f_\pi^2}{\rm tr}\int\frac{d\Omega}{4\pi}\left({\rm sin}\Theta(r)-i\gamma_5\boldsymbol{\tau}\cdot\hat{\boldsymbol{r}}{\rm cos}\Theta(r)\right)\rho(\boldsymbol{x},\boldsymbol{x}) \tag{C.5}$$

$$\omega(r) = \frac{g_V^2}{4m_V^2}{\rm tr}\int\frac{d\Omega}{4\pi}\, b(\boldsymbol{x},\boldsymbol{x}) \tag{C.6}$$

$$G(r) = -\frac{g_V^2}{4m_V^2}{\rm tr}\int\frac{d\Omega}{4\pi}\left((\gamma\times\hat{\boldsymbol{r}})\cdot\boldsymbol{\tau}\right)\rho(\boldsymbol{x},\boldsymbol{x}) \tag{C.7}$$

$$F(r) = -\frac{g_V^2}{4m_V^2}{\rm tr}\int\frac{d\Omega}{4\pi}\,\beta\left(3(\boldsymbol{\sigma}\cdot\hat{\boldsymbol{r}})(\boldsymbol{\tau}\cdot\hat{\boldsymbol{r}})-(\boldsymbol{\sigma}\cdot\boldsymbol{\tau})\right)\rho(\boldsymbol{x},\boldsymbol{x}) \tag{C.8}$$

$$H(r) = \frac{g_V^2}{4m_V^2}{\rm tr}\int\frac{d\Omega}{4\pi}\,\beta\left((\boldsymbol{\sigma}\cdot\hat{\boldsymbol{r}})(\boldsymbol{\tau}\cdot\hat{\boldsymbol{r}})-(\boldsymbol{\sigma}\cdot\boldsymbol{\tau})\right)\rho(\boldsymbol{x},\boldsymbol{x}). \tag{C.9}$$

Here the trace is taken with respect to Dirac and flavor indices only. Note that ${\rm tr}\int d\Omega\; b(\boldsymbol{r},\boldsymbol{r})$ represents the baryon charge density. In eq (C.6) isospin invariance has been assumed, *i.e.* $m_\omega = m_\rho = m_V$.

[a] $\psi_\nu(\boldsymbol{x})$ represents the coordinate space representation of $|\nu\rangle$.

## Appendix D

In this appendix we present the explicit expressions for the prefactors of the collective operators which are relevant for the description of the hyperons as described in section 7.5.1.

The explicit expression for $\Delta_{ab}$ may readily be found in ref.[62] where the corresponding calculation is described

$$\Delta_{ab}^{\mathrm{vac}} = \frac{-N_C}{2\sqrt{3}}\Delta M \sum_{\mu\nu} f_\Gamma(\epsilon_\mu,\epsilon_\nu;\Lambda)\langle\mu|\lambda_a|\nu\rangle\langle\nu|\mathcal{T}\beta\lambda_b\mathcal{T}^\dagger|\mu\rangle \tag{D.1}$$

with the cut-off function

$$f_\Gamma(\epsilon_\mu,\epsilon_\nu;\Lambda) = \frac{\mathrm{sgn}(\epsilon_\mu)\mathrm{erfc}\left(\left|\frac{\epsilon_\mu}{\Lambda}\right|\right) - \mathrm{sgn}(\epsilon_\nu)\mathrm{erfc}\left(\left|\frac{\epsilon_\nu}{\Lambda}\right|\right)}{\epsilon_\mu-\epsilon_\nu}. \tag{D.2}$$

The contribution of the Dirac sea to the coefficient $\gamma$ reads

$$\gamma^{\mathrm{vac}} = -\frac{2N_C\Delta M}{\sqrt{3}}\sum_\mu \mathrm{sgn}(\epsilon_\mu)\mathrm{erfc}\left(\left|\frac{\epsilon_\mu}{\Lambda}\right|\right)\langle\mu|\mathcal{T}\beta\lambda_8\mathcal{T}^\dagger|\mu\rangle. \tag{D.3}$$

The second order terms in the expansion of the vacuum contribution to the fermion determinant may be expressed as double mode sums over the eigenstates of the static Dirac Hamiltonian (6.42)

$$\Theta_{ab}^{\mathrm{vac}} = \frac{N_C}{4}\sum_{\mu\nu} f_\Theta(\epsilon_\mu,\epsilon_\nu;\Lambda)\langle\mu|\lambda_a|\nu\rangle\langle\nu|\lambda_b|\mu\rangle \tag{D.4}$$

where the cut-off function $f_\Theta(\epsilon_\mu,\epsilon_\nu;\Lambda)$ is defined in (7.14). The symmetry breaking terms are found to be

$$\Gamma_{ab}^{\mathrm{vac}} = \frac{N_C}{3}(m-m_s)^2\sum_{\mu\nu} f_\Gamma(\epsilon_\mu,\epsilon_\nu;\Lambda)\langle\mu|\mathcal{T}\beta\lambda_a\mathcal{T}^\dagger|\nu\rangle\langle\nu|\mathcal{T}\beta\lambda_b\mathcal{T}^\dagger|\mu\rangle \tag{D.5}$$

with the cut-off function $f_\Gamma(\epsilon_\mu,\epsilon_\nu;\Lambda)$ being defined in (D.2).

The analogous expressions associated with the explicit occupation of the valence quark level are found to be

$$\gamma^{\mathrm{val}} = -\frac{2}{3}N_C(m-m_s)\eta_{\mathrm{val}}\langle\mathrm{val}|\mathcal{T}\beta\mathcal{T}^\dagger|\mathrm{val}\rangle, \tag{D.6}$$

$$\Theta_{ab}^{\mathrm{val}} = \frac{N_C}{2}\eta_{\mathrm{val}}\sum_{\mu\neq\mathrm{val}}\frac{\langle\mathrm{val}|\lambda_a|\mu\rangle\langle\mu|\lambda_b|\mathrm{val}\rangle}{\epsilon_\mu-\epsilon_{\mathrm{val}}}, \tag{D.7}$$

$$\Delta_{ab}^{\mathrm{val}} = \frac{N_C}{\sqrt{3}}(m-m_s)\eta_{\mathrm{val}}\sum_{\mu\neq\mathrm{val}}\frac{\langle\mathrm{val}|\lambda_a|\mu\rangle\langle\mu|\mathcal{T}\beta\lambda_b\mathcal{T}^\dagger|\mathrm{val}\rangle}{\epsilon_\mu-\epsilon_{\mathrm{val}}}, \tag{D.8}$$

$$\Gamma_{ab}^{\mathrm{val}} = -\frac{2}{3}N_C(m-m_s)^2\eta_{\mathrm{val}}\sum_{\mu\neq\mathrm{val}}\frac{\langle\mathrm{val}|\mathcal{T}\beta\lambda_a\mathcal{T}^\dagger|\mu\rangle\langle\mu|\mathcal{T}\beta\lambda_b\mathcal{T}^\dagger|\mathrm{val}\rangle}{\epsilon_\mu-\epsilon_{\mathrm{val}}}, \tag{D.9}$$

wherein $|\mathrm{val}\rangle$ denotes the valence quark state $\Psi_{\mathrm{val}}$. Finally the coefficients appearing in the mesonic part of the collective action (7.85) are given by

$$\gamma^{\mathrm{m}} = \frac{8\pi}{9}m_\pi^2 f_\pi^2\left[\left(\frac{m_s^0}{m^0}-1\right)\left(2+\frac{m_s}{m}\right)+\left(2+\frac{m_s^0}{m^0}\right)\left(\frac{m_s}{m}-1\right)\right]\int drr^2(1-\cos\Theta), \tag{D.10}$$

$$\Gamma_T^{\rm m} = -\Gamma_8^{\rm m} = \frac{8\pi}{9} m_\pi^2 f_\pi^2 \left(\frac{m_s^0}{m^0} - 1\right) \left(\frac{m_s}{m} - 1\right) \int dr r^2 \left(1 - \cos\Theta\right), \tag{D.11}$$

$$\Gamma_S^{\rm m} = \frac{8\pi}{3} m_\pi^2 f_\pi^2 \left(\frac{m_s^0}{m^0} - 1\right) \left(\frac{m_s}{m} - 1\right) \int dr r^2 \left(1 - \cos\left(\frac{\Theta}{2}\right)\right). \tag{D.12}$$

These various contributions finally add up to the collective Lagrangian which may be cast into the form

$$\begin{aligned} L = & -E_{\rm tot} + \frac{1}{2}\alpha^2 \sum_{i=1}^{3} \Omega_i^2 + \frac{1}{2}\beta^2 \sum_{\alpha=4}^{7} \Omega_\alpha^2 - \frac{\sqrt{3}}{2} B \Omega_8 - \frac{1}{2}\gamma(1 - D_{88}) \\ & -\frac{1}{2}\Gamma_T \sum_{i=1}^{3} D_{8i} D_{8i} - \frac{1}{2}\Gamma_S \sum_{\alpha=4}^{7} D_{8\alpha} D_{8\alpha} - \frac{1}{2}\Gamma_8 (1 - D_{88} D_{88}) \\ & +\Delta_T \sum_{i=1}^{3} \Omega_i D_{8i} + \Delta_S \sum_{\alpha=4}^{7} \Omega_\alpha D_{8\alpha}. \end{aligned} \tag{D.13}$$

The coefficients are sums of the quantities listed above

$$\begin{aligned} \alpha^2 &= \Theta_{33}^{\rm val} + \Theta_{33}^{\rm vac}, \quad \beta^2 = \Theta_{44}^{\rm val} + \Theta_{44}^{\rm vac} + \beta_I, \quad B = \eta_{\rm val} + B^{\rm vac} \\ \gamma &= \gamma^{\rm val} + \gamma^{\rm vac} + 2\Gamma_{88}^{\rm val} + 2\Gamma_{88}^{\rm vac} + \gamma^{\rm m}, \quad \Gamma_T = \Gamma_{33}^{\rm vac} + \Gamma_{33}^{\rm vac} + \Gamma_T^{\rm m}, \\ \Gamma_S &= \Gamma_{44}^{\rm vac} + \Gamma_{44}^{\rm vac} + \Gamma_S^{\rm m}, \quad \Gamma_8 = -\Gamma_{88}^{\rm vac} - \Gamma_{88}^{\rm vac} + \Gamma_8^{\rm m}, \\ \Delta_T &= \Delta_T^{\rm vac} + \Delta_T^{\rm vac}, \quad \Delta_S = \Delta_S^{\rm vac} + \Delta_S^{\rm vac}. \end{aligned} \tag{D.14}$$

Whenever necessary, the contribution of the trivial ($\xi_0 = 1$) field configuration has to be subtracted since the fermion determinant is normalized accordingly.

In order to obtain the collective Hamiltonian from (D.13) it is important to note that the time-derivative of the collective rotations $\dot{D}_{lb} = f_{bcd}\Omega_c D_{ld}$ only affects the right indices of the adjoint representation (see also eq (7.17)). Hence the momenta associated with $\partial L/\partial\Omega_a$ are the right generators $R_a$ of SU(3). These obey the algebra $[R_a, R_b] = -f_{abc}R_c$. Noting that $[R_a, R] = -R\lambda_a$ which also implies that $[R_a, D_{lb}] = -if_{abc}D_{lc}$ one easily verifies that

$$[\frac{i}{2}R_a, \xi(\boldsymbol{x}, t)] = -\frac{\partial\dot{\xi}(\boldsymbol{x}, t)}{\partial\Omega_a}. \tag{D.15}$$

Thus the quantization prescription reads

$$R_a = -\frac{\partial L}{\partial \Omega_a} = \begin{cases} -(\alpha^2\Omega_a + \Delta_T D_{8a}) = -J_a, & \text{a=1,2,3} \\ -(\beta^2\Omega_a + \Delta_S D_{8a}), & \text{a=4,..,7} \\ \frac{\sqrt{3}}{2}B, & \text{a=8} \end{cases} \tag{D.16}$$

wherein $J_i$ ($i = 1, 2, 3$) denote the spin operators. This relation is the flavor SU(3) extension of (7.18).

The Hamiltonian operator:

$$H = -\sum_{a=1}^{8} R_a \Omega_a - L \tag{D.17}$$

may be diagonalized exactly by generalizing [131] the Yabu-Ando [28] approach to more complicated symmetry breaking terms. In the original Yabu-Ando approach only the $\gamma(1 - D_{88})$ type symmetry breaking term was considered. It should be noted that the constraint $R_8 - \sqrt{3}B/2 = 0$ commutes with the Hamiltonian and therefore is first class. Hence the baryon wave–functions live on the hypersphere $R_8 = \sqrt{3}/2$ for $B = 1$. On this hypersphere only states with half–integer spin exist. Thus the Hamiltonian has the appreciated feature that its eigenstates are fermions [52, 26, 27]. The corresponding energy eigenvalue for baryon $B$ is found to be given by the formula (7.90). The quantity $\epsilon_{SB}$ appearing in that equation represents the eigenvalue of the SU(3) operator

$$\begin{aligned} C_2 + \beta^2\gamma(1 - D_{88}) + \beta^2(\Delta_T/\alpha^2)\textstyle\sum_{i=1}^3 D_{8i}(2R_i + \Delta_T D_{8i}) + \Delta_S \sum_{\alpha=4}^7 D_{8\alpha}(2R_\alpha + \Delta_S D_{8\alpha}) \\ +\beta^2\Gamma_8(1 - D_{88}D_{88}) + \beta^2\Gamma_T \textstyle\sum_{i=1}^3 D_{8i}D_{8i} + \beta^2\Gamma_S \sum_{\alpha=4}^7 D_{8\alpha}D_{8\alpha}, \end{aligned} \tag{D.18}$$

where $C_2 = \sum_{a=1}^8 R_a^2$ denotes the quadratic Casimir operator of SU(3). Left SU(3) generators are constructed via $L_a = \sum_{b=1}^8 D_{ab}R_b$ with $I_a = L_a$, $(a = 1, 2, 3)$ and $Y = 2L_8/\sqrt{3}$ being the isospin and hypercharge operators, respectively. A pertinent parametrization of the collective rotations in terms of eight "Euler angles" is given by [28]

$$R = R_I(\alpha, \beta, \gamma)e^{-i\nu\lambda_4}R_J(\alpha', \beta', \gamma')e^{-i(\rho/\sqrt{3})\lambda_8}. \tag{D.19}$$

Here $R_I$ represents pure isospin transformations and, due to the hedgehog structure of the soliton, $R_J$ corresponds to spatial rotations. For the parametrization (D.19) the generators $R_a$ are expressed as differential operators in the "Euler angles" $\alpha, \beta, ..., \rho$. Employing the explicitly forms of the $R_a$, as given in appendix A of ref.[131], the operator (D.18) can be formulated as a second order differential operator for the "Euler angle" $\nu$. This suggests the *ansatz* for the eigenfunctions [28]

$$\Psi_{YII_3,JJ_3}(R) = (-1)^{J-J_3} \sum_{M_L M_R} D^{(I)*}_{I_3 M_L}(\alpha, \beta, \gamma) f_{M_L M_R}(\nu) e^{iY_R\rho} D^{(J)*}_{M_R - J_3}(\alpha', \beta', \gamma'), \tag{D.20}$$

where the $D's$ denote the SU(2) Wigner functions. The sum over the intrinsic projection numbers $M_{L,R}$ is subject to the condition $M_L - M_R = Y/2 - \sqrt{3}$ for $B = 1$. In order to finally obtain the eigenvalue $\epsilon_{SB}$ (and thus the baryon spectrum) a system of coupled second–order differential equations for the isoscalar functions $f_{M_L M_R}(\nu)$ has to be integrated. The $f_{M_L M_R}(\nu)$ decouple for the various isospin channels and the integration is accomplished by standard (numerical) techniques.

# Appendix E

Here we briefly display the kernels $\Phi^{(1,2)}$ entering the Bethe–Salpeter equation (7.96) which represents the starting point for the bound state description of the hyperons. Also presented are the kernel $\Phi_S$ for the strangeness charge (7.99) as well as quantities relevant for the semi–classical quantization of the kaon bound state. These expressions are computed in ref.[97] and the results are readily taken over.

The local kernel $\Phi^{(1)}(r)$ acquires contributions from the meson part of the action as well as those terms involving $h_{(2)}$

$$\begin{aligned}\Phi^{(1)}(r) =& -\frac{\pi}{2}m_\pi^2 f_\pi^2\left(1+\frac{m_s}{m}\right)\left(\cos\Theta+\frac{m_s^0}{m^0}\right)\\ &-\frac{N_C}{4}\eta_{\rm val}(m+m_s)\int\frac{d\Omega}{4\pi}\psi_{\rm val}^\dagger(\boldsymbol{r})u_0(\boldsymbol{r})\beta u_0(\boldsymbol{r})\psi_{\rm val}(\boldsymbol{r}) \\ &-\frac{N_C}{4}(m+m_s)\int_{1/\Lambda^2}^\infty\frac{ds}{\sqrt{4\pi s}}\int\frac{d\Omega}{4\pi}\Big\{\sum_{\mu=ns}\epsilon_\mu e^{-s\epsilon_\mu^2}\psi_\mu^\dagger(\boldsymbol{r})u_0(\boldsymbol{r})\beta u_0(\boldsymbol{r})\psi_\mu(\boldsymbol{r})\\ &\qquad\qquad+\sum_{\rho=s}\epsilon_\rho e^{-s\epsilon_\rho^2}\psi_\rho^\dagger(\boldsymbol{r})\beta\psi_\rho(\boldsymbol{r})\Big\}.\end{aligned} \tag{E.1}$$

The integral $\int(d\Omega/4\pi)$ indicates that the average with regard to the internal degrees of freedom has been taken. For convenience the unitary, self-adjoint transformation matrix $u_0$, which represents a modified form of the chiral rotation (7.42) has been introduced

$$u_0(\boldsymbol{r}) = i\beta\gamma_5\hat{\boldsymbol{r}}\cdot\boldsymbol{\tau}\mathcal{T} = \beta\left(\sin\frac{\Theta}{2}-i\gamma_5\hat{\boldsymbol{r}}\cdot\boldsymbol{\tau}\cos\frac{\Theta}{2}\right). \tag{E.2}$$

Obviously $u_0(\boldsymbol{r})$ acts as a unit matrix on the strange spinors. The bilocal kernel $\Phi^{(2)}(\omega;r,r')$ originates from the terms quadratic in $h_{(1)}$ and is symmetric in $r$ and $r'$

$$\begin{aligned}\Phi^{(2)}(\omega;r,r') =& -\frac{N_C}{4}(m+m_s)^2\int\frac{d\Omega}{4\pi}\int\frac{d\Omega'}{4\pi}\Big\{\eta_{\rm val}\sum_{\rho=s}\frac{\psi_{\rm val}^\dagger(\boldsymbol{r})u_0(\boldsymbol{r})\psi_\rho(\boldsymbol{r})\psi_\rho^\dagger(\boldsymbol{r}')u_0(\boldsymbol{r}')\psi_{\rm val}(\boldsymbol{r}')}{\epsilon_{\rm val}-\omega-\epsilon_\rho}\\ &-\sum_{\substack{\mu=ns\\ \rho=s}}\psi_\mu^\dagger(\boldsymbol{r})u_0(\boldsymbol{r})\psi_\rho(\boldsymbol{r})\psi_\rho^\dagger(\boldsymbol{r}')u_0(\boldsymbol{r}')\psi_\mu(\boldsymbol{r}')\mathcal{R}_{\mu,\rho}(\omega)\Big\}\end{aligned} \tag{E.3}$$

The regulator function appearing eq (E.3) has been obtained as

$$\begin{aligned}\mathcal{R}_{\mu,\rho}(\omega) =& \int_{1/\Lambda^2}^\infty ds\sqrt{\frac{s}{\pi}}\Big\{\frac{{\rm e}^{-s\epsilon_\mu^2}+{\rm e}^{-s\epsilon_\rho^2}}{s}+[\omega^2-(\epsilon_\mu+\epsilon_\rho)^2]R_0(s;\omega,\epsilon_\mu,\epsilon_\rho)\\ &-2\omega\epsilon_\rho R_1(s;\omega,\epsilon_\mu,\epsilon_\rho)+2\omega\epsilon_\mu R_1(s;\omega,\epsilon_\rho,\epsilon_\mu)\Big\}.\end{aligned} \tag{E.4}$$

The Feynman parameter integrals are defined in eq (7.52)

$$R_i(s;\omega,\epsilon_\mu,\epsilon_\nu) = \int_0^1 x^i dx\ \exp\left(-s[(1-x)\epsilon_\mu^2+x\epsilon_\nu^2-x(1-x)\omega^2]\right). \tag{E.5}$$

Similarly the bilocal kernel $\Phi_S(\omega;r,r')$ for the strangeness charge (7.99) is given by

$$\Phi_S(\omega;r,r') = -\eta_{\rm val}\int\frac{d\Omega}{4\pi}\int\frac{d\Omega'}{4\pi}\sum_{\rho=s}\frac{\psi_{\rm val}^\dagger(\boldsymbol{r})u_0(\boldsymbol{r})\psi_\rho(\boldsymbol{r})\psi_\rho(\boldsymbol{r}')u_0(\boldsymbol{r}')\psi_{\rm val}(\boldsymbol{r}')}{(\epsilon_{\rm val}-\omega-\epsilon_\rho)^2} \tag{E.6}$$

$$+\int\frac{d\Omega}{4\pi}\int\frac{d\Omega'}{4\pi}\int_{1/\Lambda^2}^{\infty}\frac{ds}{\sqrt{4\pi s}}\sum_{\rho=s}e^{-s\epsilon_\rho^2}\Big\{(1-2s\epsilon_\rho^2)\psi_\rho(\boldsymbol{r})^\dagger\beta\psi_\rho(\boldsymbol{r})-s\epsilon_\rho\psi_\rho(\boldsymbol{r})^\dagger\psi_\rho(\boldsymbol{r})\Big\}\delta(\boldsymbol{r}-\boldsymbol{r}')$$

$$-\int\frac{d\Omega}{4\pi}\int\frac{d\Omega'}{4\pi}\int_{1/\Lambda^2}^{\infty}ds\sqrt{\frac{s}{4\pi}}\sum_{\substack{\mu=ns\\ \rho=s}}\psi_\mu(\boldsymbol{r})^\dagger u_0(\boldsymbol{r})\psi_\rho(\boldsymbol{r})\psi_\rho(\boldsymbol{r}')^\dagger u_0(\boldsymbol{r}')\psi_\mu(\boldsymbol{r}')(\epsilon_\mu+\epsilon_\rho-\omega)$$

$$\times\Big\{R_0(s;\omega,\epsilon_\mu,\epsilon_\rho)-s(\omega+\epsilon_\mu+\epsilon_\rho)\Big((\epsilon_\rho-\omega)R_1(s;\omega,\epsilon_\mu,\epsilon_\rho)+\omega R_2(s;\omega,\epsilon_\mu,\epsilon_\rho)\Big)\Big\}.$$

To conclude this appendix we also list the spectral functions $c(\omega)$ and $d(\omega)$ which enter the relation between the spin and kaon spin expectation values (*cf.* eqs (7.113) and (7.115)). It is appropriate to list the contributions stemming from the explicit occupation of the valence quark level and the polarized Dirac sea separately. In what follows $\tilde{h}_{(i)}$ are understood as the perturbative parts of the Dirac Hamiltonian (7.43) with the properly normalized solution $\eta_\omega(r)$ to the Bethe–Salpeter equation (7.96,7.100) substituted

$$\begin{aligned}
c(\omega) &= c_{\rm val}(\omega)+c_{\rm vac}(\omega)\\
c_{\rm val}(\omega) &= \eta_{\rm val}\sum_{\mu=ns}\Big[\frac{\langle{\rm val}|\tilde{h}_{(2)}(\omega,-\omega)|\mu\rangle\langle\mu|\tau_3|{\rm val}\rangle}{\epsilon_{\rm val}-\epsilon_\mu}\ +\ \text{h. c.}\Big]\\
&\quad+\eta_{\rm val}\sum_{\substack{\mu=ns\\ \rho=s}}\Big[\frac{\langle{\rm val}|\tilde{h}_{(1)}(\omega)|\rho\rangle\langle\rho|\tilde{h}_{(1)}(-\omega)|\mu\rangle\langle\mu|\tau_3|{\rm val}\rangle}{(\epsilon_{\rm val}-\epsilon_\mu)(\epsilon_{\rm val}-\omega-\epsilon_\rho)}\ +\ \text{h. c.}\Big]\\
c_{\rm vac}(\omega) &= -\int_{1/\Lambda^2}^{\infty}\frac{ds}{\sqrt{4\pi s}}\Bigg\{\sum_{\mu,\nu=n.s.}\langle\mu|\tau_3|\nu\rangle\langle\nu|\tilde{h}_{(2)}(\omega,-\omega)|\mu\rangle\frac{\epsilon_\mu e^{-s\epsilon_\mu^2}-\epsilon_\nu e^{-s\epsilon_\nu^2}}{\epsilon_\mu-\epsilon_\nu}\\
&\quad+\sum_{\substack{\mu,\nu=n.s.\\ \rho=s.}}\langle\mu|\tau_3|\nu\rangle\langle\nu|\tilde{h}_{(1)}(\omega)|\rho\rangle\langle\rho|\tilde{h}_{(1)}(-\omega)|\mu\rangle\Big[\epsilon_\nu\frac{e^{-s\epsilon_\mu^2}-e^{-s\epsilon_\nu^2}}{\epsilon_\mu^2-\epsilon_\nu^2}+s\mathcal{R}_{\mu,\nu,\rho}(\omega)\Big]\Bigg\}.
\end{aligned}\tag{E.7}$$

The relevant regulator function is given by a Feynman parameter integral

$$\begin{aligned}
\mathcal{R}_{\mu,\nu,\rho}(\omega) &= \int_0^1 dx\Bigg\{(\omega+\epsilon_\nu+\epsilon_\rho)\exp\Big(-s[(1-x)\epsilon_\nu^2+x\epsilon_\rho^2-x(1-x)\omega^2]\Big)\\
&+\Big(2(1-x)\omega^3+[\epsilon_\mu+\epsilon_\nu]\omega^2-[2(1-x)(\epsilon_\nu+\epsilon_\rho)(\epsilon_\mu+\epsilon_\rho)-(\epsilon_\mu-\epsilon_\nu)^2]\omega\\
&+(\epsilon_\mu+\epsilon_\nu)(\epsilon_\mu+\epsilon_\rho)(\epsilon_\nu+\epsilon_\rho)\Big)\frac{e^{-sx\epsilon_\mu^2}-e^{-sx\epsilon_\nu^2}}{\epsilon_\mu^2-\epsilon_\nu^2}\exp\Big(-s[(1-x)\epsilon_\rho^2-x(1-x)\omega^2]\Big)\Bigg\}.
\end{aligned}\tag{E.8}$$

Finally $d(\omega)$ is found to be

$$\begin{aligned}
d_{\rm val}(\omega) &= -2\eta_{\rm val}\sum_{\rho=s}M_\rho\frac{\langle{\rm val}|\tilde{h}_{(1)}(\mathbf{r},\omega)|\rho\rangle\langle\rho|\tilde{h}_{(1)}(\mathbf{r},-\omega)|{\rm val}\rangle}{(\epsilon_{\rm val}-\omega-\epsilon_\rho)^2}\\
d_{\rm vac}(\omega) &= 2\int_{1/\Lambda^2}^{\infty}\frac{ds}{\sqrt{4\pi s}}\sum_{\mu=ns}M_\mu e^{-s\epsilon_\mu^2}\Big\{(1-2s\epsilon_\mu^2)\ \langle\mu|\tilde{h}_{(2)}(\mathbf{r},\omega,-\omega)|\mu\rangle\\
&\qquad\qquad -s\epsilon_\mu\ \langle\mu|\tilde{h}_{(1)}(\mathbf{r},\omega)\tilde{h}_{(1)}(\mathbf{r},-\omega)|\mu\rangle\Big\}\\
&\quad-2\int_{1/\Lambda^2}^{\infty}ds\sqrt{\frac{s}{4\pi}}\sum_{\mu\nu}M_\nu\langle\mu|\tilde{h}_{(1)}(\mathbf{r},\omega)|\nu\rangle\langle\nu|\tilde{h}_{(1)}(\mathbf{r},-\omega)|\mu\rangle(\epsilon_\mu+\epsilon_\nu-\omega)\\
&\times\Big\{R_0(s;\omega,\epsilon_\mu,\epsilon_\nu)-s(\omega+\epsilon_\mu+\epsilon_\nu)\Big((\epsilon_\nu-\omega)R_1(s;\omega,\epsilon_\mu,\epsilon_\nu)+\omega R_2(s;\omega,\epsilon_\mu,\epsilon_\nu)\Big)\Big\}.
\end{aligned}\tag{E.9}$$

# List of Figures

# List of Tables